# Collaboration in Coworking Spaces:
# Impact on Firm Innovativeness and Business Models

JP International Management of Technology and Innovation

Prof. Dr. Tim Kessler

Submitted to:

Prof. Dr. Tim Kessler

Universität Bayreuth

Summer Term 2017

Submitted by:

Steven Moore

Matriculation Number: 1247286

Hermannstr. 111, 12051 Berlin

Phone: 0152-29440210

E-mail: S5stmoor@stmail.uni-bayreuth.de

Berlin, 07 July 2017

## ABSTRACT

**Purpose:** The purpose of this paper is to contribute to the debate as to whether collaboration in coworking spaces contributes to firm innovativeness and impacts the business models of organizations in a positive manner.

**Methodology:** This paper includes primary data from 75 organizations in 17 coworking spaces and uses quantitative research methods. The methodology includes multiple statistical methods, such as principal component analysis, correlation analysis as well as linear and binary regression analysis.

**Results**: The results show a positive interrelation between collaboration and innovation, indicating that coworkers are able to improve their innovative capabilities by making use of strategic partnerships in coworking spaces. Further, this study shows that business models are significantly affected by the level of collaboration in coworking spaces, which suggests that coworking is a promoting force for business model development or business model innovation.

**Contributions:** The paper contributes to management literature and represents the first empirical investigations which focuses on the effects of collaboration on a firm-level in coworking spaces.

**Practical implications:** The results indicate that organizations in coworking spaces should embrace a collaborative mindset and should actively seek out collaborative alliances and partnerships, as doing such is shown to increase their innovativeness and/or develop their business model.

**Future Research:** Future research should focus on the antecedents of collaboration or could investigate the effects of collaboration in coworking spaces on a community level.

**Paper type:** Master Thesis

## ACKNOWLEDGEMENTS

I would like to express my great appreciation to Professor Dr. Kessler for his valuable and constructive suggestions during the planning and development of this paper. His time and guidance has been immeasurably valuable and his persistent critique has been a strong motivator during this process. I would also like to extend my gratitude to all managers of the coworking spaces that distributed my survey. In addition, I would like to thank all organizations which have dedicated their valuable time to participate in my study.

# TABLE OF CONTENT

## LIST OF FIGURES

## LIST OF TABLES

# LIST OF ABBREVIATION

| | |
|---|---|
| OECD | Organization for Economic Co-operation and Development |

# CHAPTER 1: INTRODUCTION

Office design is continuously changing. Yahoo notoriously revoked mobile privileges and Google´s office space is more reminiscient of a gigantic creative playground than an actual office (DellaPelle, 2016). Facebook recently put several thousand of its employees into a gigantic mile-long-room set up with multiple snack bars and cafeterias, hoping engineers, sales-people and marketing specialists would mingle and it will lure them into increased interactions. Large and successful corporations are consistently designing or redesigning office space in accordance to their company's needs. Nowadays, for many businesses to be successful, it requires new creative ideas and novel solutions which evolve out of a heterogenous group of knowledge workers. When Telenor, a Norwegian telecommunication company, in 2003 designed its company´s Oslo headquarter with incorporated "hot desking" where spaces could effortless be reconfigured for different tasks and teams, it was way ahead of its time. Later Telenor indicated that the elected design has significantly improved internal communication and has helped them to shift from a state-run monopoly to a competitive multinational corporation. The impact of office design on firms´ performance, however, is also underlined by quantitative data. Studies have examined the effects of office design of a pharmaceutical company which intended to increase its sales volume. In this study 50 executives who were responsible for almost $ 1 billion in annual sale were deployed with sociometric badges to measure the level of interaction with other coworkers. The results showed that when salespeople increased their interactions by 10 percent, sales accordingly grew by 10 percent. Based on this knowledge, the company invested hundreds of thousands of dollars into an office redesign. The firm´s strategy was to rip off multiple coffee stations - at the time the company had one coffee machine for every six employees - and to build fewer bigger stations, only one for every 120 employees. The increased interactions across different departments eventually caused sales to rise by 20 percent which corresponded to an overall amount of $200 million, rapidly justifying the invested capital in the redesign (Waber, Magnolfi, & Lindsay, 2014). However, there have also been less successful examples, and understanding how design, technology and physical work spaces interplay is an incredibly complex challenge. Workers themselves have tackled this challenge in earnest. Technologists, programmers and creative individuals desired to work outside a confining regular office, but also aimed to avoid the professional isolation of home-offices (Kojo & Nenonen, 2014; Parrino, 2015). They demanded work spaces that adapt to their needs rather than vice versa and have established what is now known as coworking spaces. (Waber et al., 2014)

Since then, the new office concept has been globally spreading with an impressive annual growth rate and by the end of 2017 the global number of coworking spaces will rise up to nearly 13,800 spaces (Foertsch, 2017; Gandini, 2015).

Coworking spaces are shared, flexible and collaborative office spaces used by a diverse group of professionals that rent office space on a weekly, monthly or permanent basis. It was only recently, however, that the concept of coworking also attracted scholars´ attention. First studies aimed to provide a general conceptualization of coworking, by offering definitions, classifications and configurations of the new work concept as well as illustrating the differences between coworking spaces and regular work environments (Spinuzzi, 2012; Tadashi, 2013). Soon studies spilled over to other research areas, such as economics and urban development (Avdikos & Kalogeresis, 2016; Capdevila, 2013; Fuzi, 2015; Gandini, 2015; Moriset, 2013), sociology and psychology (Parrino, 2015; Schopfel, Roche, & Hubert, 2015; Soerjoatmodjo, Bagasworo, Joshua, Kalesaran, & van den Broek, K. F., 2015) and management literature (Bouncken & Reuschl, 2016; Surman, 2013). According to prior studies, coworking positively impacts urban development (Moriset, 2013) and is a driving force for entrepreneurship in regions with sparse entrepreneurial movements (Fuzi, 2015). Alternative studies show that coworking facilitates knowledges exchanges (Parrino, 2015), organizational learning (Bouncken & Reuschl, 2016) and motivates individuals by making their work more meaningful (Spreitzer, Garrett, & Bacevice, 2015). However, scholars have also adopted a more critical view on coworking, arguing that the new concept increasingly pulls workers into freelance occupations with high job uncertainty and insecurity (Avdikos & Kalogeresis, 2016).

Scholars and practitioners have attributed coworking to several social, entrepreneurial and business-related services that exceed the offer of mere office space. Coworking spaces are designed to grant members access to external resources which they otherwise would not have access to (Parrino, 2015; Soerjoatmodjo et al., 2015). Due to frequent and ongoing interactions with a multitude of members, coworking spaces are expected to increase the probability of fruitful encounters that may lead to unexpected chance discoveries. This scenario has been referred to as serendipity, and importantly, scholars have argued that the production of serendipitous occurrences is a core principle of coworking spaces (Moriset, 2013). Other business-related benefits of coworking supposedly include the promotion of business model development and innovation (Bouncken & Reuschl, 2016). The study of Capdevila (2015) argues that coworking spaces unite heterogenous knowledge bases, promote mutual problem solving processes and in this way contribute to innovation on multiple levels. While coworking spaces actively use these argumentations for marketing purposes and to attract customers, quantitative prove in academic research is still lacking. Therefore, this study aims to partially fill this void in academic literature by providing one of the first systematic empirical investigations. Specifically, the study is focused on the effects of collaboration in coworking spaces on a firm-level and aims to conclusively answer the four research questions:

*H 1:* Collaboration in coworking spaces contributes to the access of external resources.

*H 2:* Collaboration in coworking spaces contributes to serendipitous occurrences.

*H 3:* Collaboration in coworking spaces contributes to firm innovativeness.

*H 4:* Collaboration in coworking spaces impacts organizations´ business models.

To answer these questions, the paper encompasses seven chapters. While the first part of the second chapter aims to generally unfold the concept of coworking by providing definitions, background information and an extensive literature review, the second part provides information on the collaborative behavior in coworking spaces. The focus of the third chapter is on the interrelation between collaboration and innovation. After conceptualizing the terms innovation and innovativeness, it consequently illustrates in what ways collaboration and networking have proven to impact firm innovativeness. The fourth chapter initially outlines the functions, components and overarching dimensions of business models and introduces business model innovation, an alternative form of innovation. In the second part of the fourth chapter, I will argue in what ways collaboration obtains the potential to impact business models. The fifth chapter presents the first empirical investigations on the effects of collaboration on innovation and business models in coworking spaces. It includes the collection of primary data from 75 organizations in 17 coworking spaces. Within the first section, I will argue how collaboration in coworking spaces may impact firm innovativeness and organizations´ business models, indicating that it may be a promoting force for business model development or business model innovation. After that the methodology will be outlined and the results will be explored in detail. The sixth chapter represents the discussion of this paper. It includes the contributions this study has made to management literature, considers practical implications, acknowledges the limitations of this empirical investigation and provides inquiries for future research. Finally, the seventh chapter concludes the paper by summarizing the core results and providing an outlook on the potential of coworking.

# CHAPTER 2: COLLABORATION IN COWORKING SPACES

## 2.1 Coworking Spaces

### *2.1.1 Definition*

Recently, a new form of working environment has emerged. Coworking is a globally expanding phenomenon which is commonly found in urban areas and that is increasingly attracting the attention of both practitioners and scholars alike. Spinuzzi (2012) was one of the first scholars who attempted to conceptualize coworking. According to his study, coworking can be described by the pure co-presence of professionals working alongside other unaffiliated professionals for a fee. More recently, however, scholars have been adding further characteristics to the definition of coworking, indicating that coworking spaces provides more to its members than mere office space. TABLE 1 provides an overview of present definitions in literature.

**TABLE 1: Definition of Coworking Spaces**

| Author | Definition |
|---|---|
| Bouncken and Reuschl (2016: 6) | "(…) coworking-spaces provide their individual or institutional users a flexible and highly autonomous use of both office and social space that eases the direct personal interaction among the coworking-users for social, learning, cultural and business related interests." |
| Capdevila (2013: 3) | "Coworking-spaces are defined as localized spaces where independent professionals work sharing resources and are open to share their knowledge with the rest of the community." |
| Fuzi (2015: 468) | "Co-working spaces are shared, proactive and community-oriented workspaces rented by a diverse group of professionals from different sectors. They emphasize intangible factors and social aspects including entrepreneurial networking, mentoring (from fellow members, hosts and networks) through flexible, informal settings, which enhance possession, access and use of different forms of capital (social, human and financial)." |
| Hallwright and Brady (2016: 41) | "Coworking occurs when personnel from two or more organisations share the same office or work space with the intention of sharing resources, information, and building an understanding of each other's overarching goals." |
| Parrino (2015: 265) | "We identified a working definition of coworking space based on three traits (…)<br>1. the co-localisation of various coworkers within the same work environment;<br>2. the presence of workers heterogeneous by occupation and/or sector in which they operate and/or organizational status and affiliation (freelancers in the strict sense, microbusiness, employees or self-employed workers);<br>3. the presence (or not) of activities and tools designed to stimulate the emergence of relationships and collaboration among coworkers." |
| Spinuzzi (2012: 399) | Coworking spaces are "open-plan office environments in which they work alongside other unaffiliated professionals for a fee." |

Scholars, for instance, have described the physical space of coworking spaces in greater detail. Coworking spaces are described as shared, proactive and community-oriented work spaces (Fuzi, 2015). A community provides members with emotional and professional support and helps to remedy professional isolation by enabling members to build or sustain relationships within those spaces (Capdevila, 2013; Fuzi, 2015; Hallwright & Brady, 2016). Alternatively, Bouncken and Reuschl (2016) highlight the flexibility of coworking spaces and emphasize that coworking grants its members with a high level of autonomy. Coworkers usually can access and use the office infrastructure and outside amenities in an autonomous and flexible manner which enables members to independently self-regulate their work.

New definitions have also added descriptive details about the members using coworking spaces. Members of coworking spaces can be individual or institutional users (Bouncken & Reuschl, 2016) and commonly are a diverse group of professionals heterogenous by occupation, sector, organizational status and affiliation (Capdevila, 2013; Parrino, 2015). Coworking spaces are used by self-employed workers, entrepreneurs and micro-businesses, but also by employees of established firms and consultants (Gandini, 2015; Parrino, 2015). While freelancers often consider coworking a solution to professional isolation, larger corporations may seek it out for investment opportunities, or might take advantage of current entrepreneurial trends (Fuzi, 2015).

Additionally, scholars increasingly included certain social, entrepreneurial and business-related aspects to the definitions of coworking spaces (Kojo & Nenonen, 2014). Coworking spaces emphasize social and intangible aspects that are designed to encourage personal interactions, networking, mentoring and collaboration (Bouncken & Reuschl, 2016; Fuzi, 2015; Parrino, 2015). Because of these interactive activities amongst members, coworking spaces provide access to social or human capital. Coworker not only share physical equipment, such as printers, office materials, technological devices, but also knowledge, expertise, information or skills with each other (Schopfel et al., 2015).

While reviewing existing definitions, it stands out that coworking spaces are a rather complex phenomenon which is difficult to uniformly conceptualize. This study adds to management literature and defines coworking spaces as shared, flexible and collaborative office spaces used by a diverse group of professionals that offer its members multiple social, entrepreneurial and business-related opportunities or services.

### *2.1.2 Drivers of the Evolution of Coworking Spaces*

Discussions as to the origins of coworking often refer to software programmer Brad Neuberg, who established the first coworking space in San Francisco in 2005, while aiming to combine the structure and commodity of a regular office job with the freedom and independence of a freelance occupation (Kojo & Nenonen, 2014; Parrino, 2015). Coworking spaces are also often referred to as “third places”, because they describe a place alternative to classical office and home-office spaces, where members freely convene and socialize (Moriset, 2013). Since its origins, coworking spaces showed a significant global diffusion and an impressive annual growth rate (Gandini, 2015). The coworking movement is reported to have increased from around 75 spaces in 2007 to roughly 8,700 in 2015 (see FIGURE 1). According to the online editorial Deskmag there will be approximately 13,800 spaces in operation worldwide by the end 2017 (Foertsch, 2017).

**FIGURE 1: Global Numbers of Coworking Spaces between 2007-2017**

**(Source: Foertsch, 2017)**

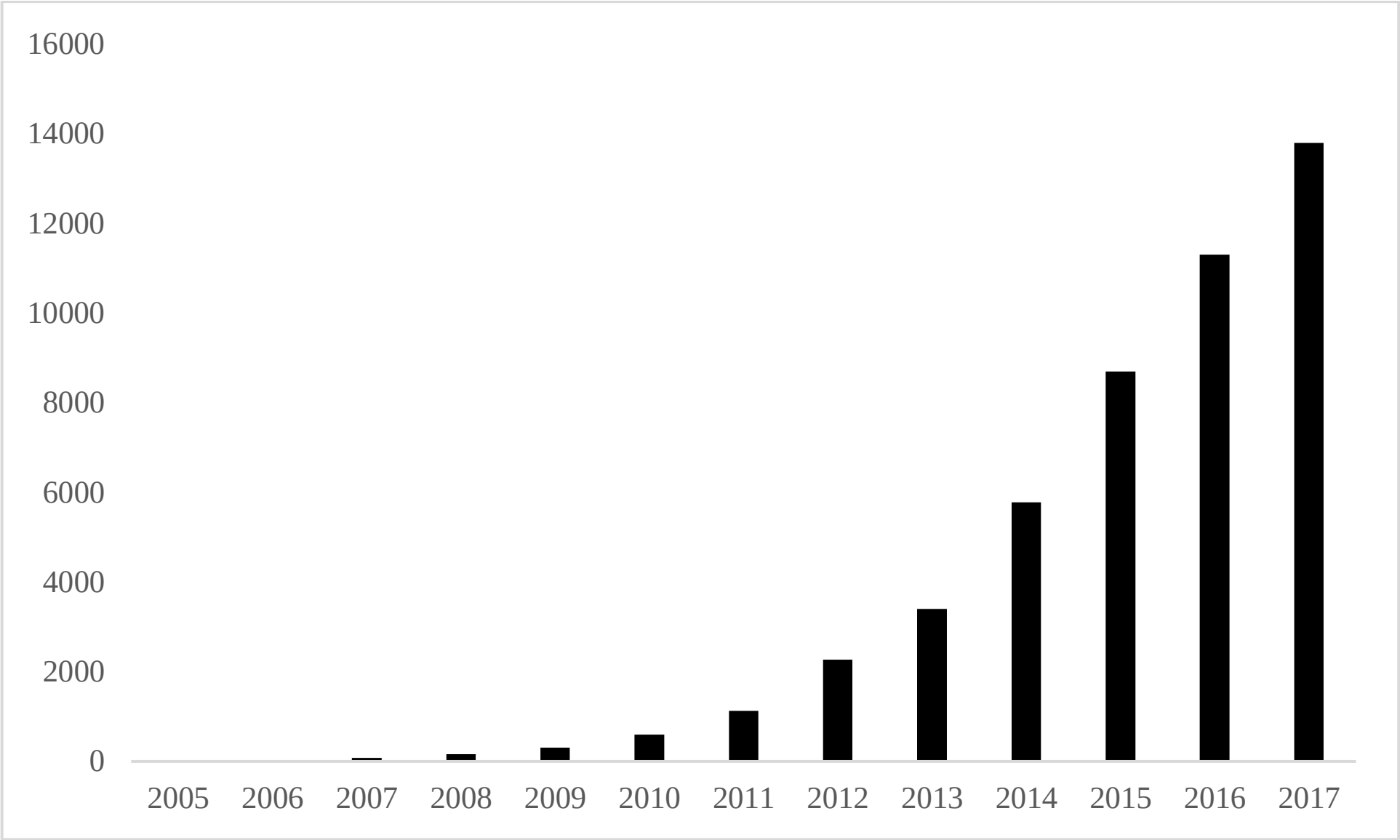

It is worth mentioning, however, that the concept behind coworking is not novel. Shared office spaces in other forms have been present for decades. Emerged in the 1980´s, telecentres or telecommuting centres offer public office space equipped with computers and technology to their members (Kojo & Nenonen, 2014). Moreover, Regus, a multinational real estate company founded in 1989, provides workspace and offices for rent, too. While most coworking spaces rent and re-rent their spaces, Regus owns their properties. Though the concept behind coworking is not novel, the rise and success of coworking is strongly supported by contemporary shifts in economy and labor market (Moriset, 2013). The demand for creative knowledge processing and creation skills amongst workers – often referred to as knowledge economy - has increased within the twenty-first century. Advances in technology enable white-collar workers to do their jobs from virtually anywhere (Kojo & Nenonen, 2014; Spreitzer et al., 2015). This in combination with the global economic crises of 2007-2008, which transformed the practices into a highly individualized labor market, strongly supported the rise of coworking (Gandini, 2015). Further, people nowadays increasingly tend to apply a collaborative consumption of goods and services, often referred to as "sharing economy". The sharing economy describes a system where individuals share underutilized resources in peer-to-peer networks (Cohen & Kietzmann, 2014). This trend has been present in various forms, including crowd funding, apartment- or car-sharing (Bouncken & Reuschl, 2016). Overall, these shifts have let coworking become a globally diffused phenomenon with an unknown potential.

### *2.1.3 Literature Review*

Coworking is currently not only attracting strong attention from practitioners, but also increasingly gaining importance in academic research. TABLE 2 provides an overview of the literature review this study has conducted. First studies on coworking aimed to conceptualize the new office model in general. Spinuzzi (2012) conducted a qualitative case study based upon a twenty-month study of nine coworking sites in the area around Austin (Texas). His study provides several definitions for coworking, classifies coworking into different configurations and reveals various intentions of coworkers. The study of Tadashi (2013) represents an additional attempt to generally explain the concept of coworking. Tadashi, however, is focused on elaborating on the differences between coworking spaces and other work environments, such as corporate offices or incubation facilities.

Research studies soon spilled over to other areas, such as economic and urban development (Avdikos & Kalogeresis, 2016; Capdevila, 2013; Fuzi, 2015; Gandini, 2015; Moriset, 2013), sociology (Parrino, 2015; Schopfel et al., 2015; Soerjoatmodjo et al., 2015) and management literature (Bouncken & Reuschl, 2016; Surman, 2013).

Scholars found out that the rising number of coworking spaces has positive impacts on urban development. Moriset (2013) explains the global spread of coworking spaces in major business cities and argues that coworking promotes the development of creative cities. Fuzi (2015) makes a similar argument and empirically explores whether coworking promotes entrepreneurship in regions with sparse entrepreneurial movements by providing support and infrastructure to organizations. Capdevila (2015) argues that coworking spaces increase local innovation by symbiotically bringing together creative individuals and innovative firms. However, studies have also shown a more critical view on the rise of coworking (Avdikos & Kalogeresis, 2016; Gandini, 2015). Although Avdikos and Kalogeresis (2016) acknowledge that coworking embeds workers into resourceful networks, the study argues that coworking increasingly pulls workers into freelance occupations. Many individuals entering this milieu of low pay and high job insecurity might lead to high unemployment rates for cities.

Further research has investigated the impact of coworking spaces on individuals from a sociological or psychological perspective. A crucial study in this field is a project of the University of Michigan, conducted by Spreitzer et al. (2015). Although the research is still ongoing, Spreitzer et al. have already released some insights, showing that coworking motivates individuals by making their work more meaningful and productive. Other studies in this context focus on the exchange of knowledge in coworking spaces. Two studies, for instance observe how the role of proximity stimulates knowledge exchanges amongst coworkers (Parrino, 2015; Soerjoatmodjo et al., 2015).

Lastly, studies can be found within management literature. Surman´s paper (2013) argues that coworking enables innovators and entrepreneurs to succeed by providing a strong and supportive community. Bouncken and Reuschl (2016) provided a model which assumes that coworking increases organizational learning, eventually improving entrepreneurial performance. The paper argues that coworking spaces offer the technological and social structure that is required for freelancers, entrepreneurs and micro-businesses to innovate.

While reviewing current literature on coworking two aspects stand out. First, especially because coworking represents a very recent phenomenon, there are still large voids in academic research. Further, while a large proportion of studies are either literature reviews (Bouncken & Reuschl, 2016; Gandini, 2015; Kojo & Nenonen, 2014; Moriset, 2013; Surman, 2013; Tadashi, 2013) or are grounded on qualitative research methods and case studies (Capdevila, 2015; Parrino, 2015; Soerjoatmodjo et al., 2015; Spinuzzi, 2012), only very few studies include the collection primary data through quantitative research methods (Avdikos & Kalogeresis, 2016; Fuzi, 2015). This study contributes to management literature and further attempts to partially close the lack of quantitative research studies. The purpose of this study is to systematically and empirically analyze the collaboration behavior in coworking spaces and to investigate if collaboration in coworking spaces impacts the innovativeness and business models of organizations within such spaces. Therefore, this study represents an attempt to close a void in academic literature that previous studies within management literature have specifically called out for (Bouncken & Reuschl, 2016).

**TABLE 2: Literature Review on Coworking**

| Area of Research | Author | Method | Contributions |
|---|---|---|---|
| Coworking Literature | Kojo and Nenonen (2014) | Literature Review | Based on an analysis of academic literature, the study investigates the evolution of coworking spaces. It explains the main drivers for the development, but also provides perspectives and implications for designers or facility managers. |
| | Spinuzzi (2012) | Qualitative Research Methods | The study conceptualizes the concept of coworking by providing multiple definitions from various standpoints and by outlining the intentions of coworkers. Based on the findings, the study provides configurations and frameworks which ought to explain coworkers´ motivation and behavior in coworking spaces. |
| | Tadashi (2013) | Literature Review | This study provides definitions on coworking and investigates the behavior of coworkers. It specifically reveals that freelancers, entrepreneurs, or members of corporation use coworking spaces for different reasons and show a varying behavior within coworking spaces. |
| Literature on Labor and Urban Development | Avdikos and Kalogeresis (2016) | Qualitative and quantitative Research Methods | The study analyzes how coworking affects the socio-economic profile and working conditions of the Greek designer market. The results show that designers in coworking spaces work long hours under poor salary conditions. Although coworking embeds designers into business networks, within those spaces they work under specific precarious working conditions. |
| | Capdevila (2015) | Qualitative Research Methods | The study is based on twenty-one interviews with managers of coworking. It argues that coworking spaces promote local innovation by acting as intermediaries between innovative firms and creative individuals. |
| | Fuzi (2015) | Case Studies with qualitative and quantitative Research Methods | Based on two case studies in Wales, the paper provides an empirical exploration of whether coworking spaces can promote entrepreneurship in regions with sparse entrepreneurial activities. It explains different models of coworking spaces and evaluates their efficiency of promoting entrepreneurship. |
| | Gandini (2015) | Literature Review | The study provides a current review on the existing literature of coworking. Further, the study argues whether coworking may be a threat to cities by increasingly pulling workers into low payed freelance occupation. |
| | Moriset (2013) | Literature Review | A thorough overview of the historical development and background of coworking is provided. In addition, the study illustrates the global spread of coworking in major business cities and argues whether coworking spaces may promote the creation of creative cities. |
| Management Literature | Bouncken and Reuschl (2016) | Literature Review | The study proposes a model for further research. The model assumes that coworking spaces promote individual self-efficacy and create a trustworthy community for their members. In this way coworking increases organizational learning, eventually improving entrepreneurial performance. |
| | Surman (2013) | Literature Review | This paper claims that coworking enables social innovation by providing a strong and supportive community which builds social capital and enables innovators and entrepreneurs to succeed. The observations and arguments are based a model. |
| Literature on Sociology and Psychology | Parrino (2015) | Case Studies with qualitative Research Methods | The study assesses if the local proximity of coworkers in coworking spaces stimulate knowledge exchanges among them. Organizational and social proximity were found to be crucial for stimulating collaboration and knowledge exchanges. |
| | Schopfel et al. (2015) | Case Studies | The proposal of this study is that academic libraries should include features of coworking spaces to become diverse and innovative learning centres and to meet its social responsibility. |
| | Soerjoatmodjo et al. (2015) | Qualitative Research Methods | The results of this study show that coworking spaces are work settings that accumulate diverse knowledge resources and support knowledge sharing amongst members. |

## 2.2 The Collaborative Behavior in Coworking Spaces

### *2.2.1 Collaboration*

The term collaboration has gained attention throughout a variety of research disciplines, such as environmental science (Plummer & Fennell, 2007; Selin & Chavez, 1995), communication (Keyton, Ford, & Smith, 2008), sociology (Appley & Winder, 1977; Graham & Barter, 1999) or management (Kenis & Knoke, 2002; Sundaramurthy & Lewis, 2003). Collaborative processes result in various individual or collective outcomes, which encompass ideas, feedback, shared understandings, satisfaction, personal growth or tangible products, increased performance or innovation. Yet, the overall goal of a collaborative alliance can also be the collaboration process itself. (Bedwell, Wildman, DiazGranados, Salazar, Kramer, & Salas, 2012)

Within management research the term collaboration takes many forms, including interorganizational coalitions, strategic alliances, joint ventures or research and development centers (Alexiev, Volberda, & van den Bosch, 2016; Bedwell et al., 2012). Considering the wide array of research attention and to avoid bias in literature, it is crucial to provide a unified and comprehensive definition for the term collaboration. This study refers to the definition of Bedwell et al. (2012), as his study represents a thorough multidisciplinary literature review on the term collaboration. Accordingly, this study "(…) define[s] collaboration as an evolving process whereby two or more social entities actively and reciprocally engage in joint activities aimed at achieving at least one shared goal" (Bedwell et al., 2012: 130). Four features should particularly be highlighted in this definition. First, in accordance with the majority of literature, the definition conceptualizes collaboration as an **evolving and dynamic process** rather than a prescribed state or static structure (Gray, 1989; Keyton et al., 2008). Second, collaboration is perceived as a process that involves through interaction among **at least two social entities**, whereby entities may refer to individuals, groups, organizations, or even counties and societies (Longoria, 2005). Third, collaboration cannot exclusively be one-sided, but instead has to occur **actively and reciprocally**. Although this does not imply that the engagement or participation of each entity must be equal, one party controlling or dictating another party cannot be perceived as collaboration either. Hence, it is necessary that all involved parties are active and contribute sufficiently towards their joint goals. Finally, **the existence of at least one shared goal** is a key element of collaboration (Wood & Gray, 1991), as without such mutual goal there would be no reason for entities to collaborate. (Bedwell et al., 2012)

### *2.2.2 Collaboration in Coworking Spaces*

A central feature of coworking is the community. The coworking community is described as a dynamic and proactive network that unites a multitude of individuals and institutions (Fuzi, 2015). Coworkers contribute to the community, but also reciprocally receive from the community. The community also represents the overall reason why coworkers engage in coworking. For some coworkers, a community may help to overcome isolation and to feel a sense of belonging or appreciation, for others it may be a driver to extend their professional network. Although Freelancers, entrepreneurs and micro-businesses compose a large proportion of members, recently larger corporations have also started to engage in coworking (Gandini, 2015; Parrino, 2015). Firms, such as KPMG or Ernst and Young for instance, are entering coworking spaces actively seeking for like-minded collaborative partners, new customers or access to startups and entrepreneurs that may inspire or generate new ideas. In return, these firms offer their expertise and consultancy services. Hence, due to the community coworking provides the ideal prerequisites for collaboration while fulfilling all required features of this study´s collaboration definition.

Moreover, there already is sufficient evidence in literature that coworking spaces stimulate collaboration and interaction amongst members. Spinuzzi (2012) gives insights into the desired benefits of coworkers and reveals that coworkers specifically join coworking spaces with the hopes to interact with external entities and to obtain partnerships, feedback and referrals. Further, the study of Avidikos & Kalogeresis (2016) analyzes how the collaborative environment of coworking spaces effect freelancers and entrepreneurs within the Greek designer market. Based on a sufficient survey, the results show that freelance designers that work in coworking spaces are more strongly embedded into networks. Other studies in this field focus on the exchange of knowledge in coworking spaces, revealing that coworkers frequently interact, collaborate and share information with each other (Parrino, 2015; Soerjoatmodjo et al., 2015).

However, the coworking community allows various forms of collaboration, ranging from frequent informal interactions to working together on projects and addressing shared problems in strict hierarchical structures (Bedwell et al., 2012; Ryu, 2014). Hence, scholars have started to explore different configurations of collaboration in respect to coworking. Parrino (2015) studied the knowledge flows of coworkers in two coworking spaces and found two different scenarios. While in one space interaction among coworkers has been scarce or episodic and the space has been characterized by an absence of collaborative relationships, the second space revealed a completely different picture. Within this space, all respondents reported frequent knowledge exchanges with other members and collaborative relationships helped coworkers professionally to create new job opportunities and to expand their networks. The study is based on two qualitative case studies, which allow to understand the mechanism of knowledge spillovers, but the findings, as the study acknowledges itself, are not formally generalizable. Spinuzzi´s study (2012) also indicates that different forms of collaborative behavior occur in coworking spaces. Based upon a twenty-month research, his study differentiates between the ‘good-neighbor’ and ‘good-partner configuration’. ‘Good-partners’ are specialists that link up into momentary collaborations to attack shared problems. Alternatively, ‘good neighbors’ prefer working in parallel and their aim is it to build and sustain neighborly relationships through frequent informal interactions. Overall, these studies represent the first steps into exploring the collaborative behavior of coworkers, a research inquiry which so far has not been fully resolved.

# CHAPTER 3: COLLABORATION AND THE IMPACT ON FIRM INNOVATIVENESS

## 3.1. Innovation

### *3.1.1 Definition*

The first attempt to conceptualize the term innovation was made by Schumpeter in his book 'Theory of economic development' (Schumpeter, 2012). The subject of innovation is considerably complex and depending on the context the term has adopted various meanings on a variety of scales. Some of the definitions proposed in literature are provided in TABLE 3. Defining innovation precisely is important, though, as it delineates what attributes will be included when an organization attempts to measure innovation or firm innovativeness. (Dewangan & Godse, 2014; Edison, bin Ali, & Torkar, 2013)

**TABLE 3: Definition of Innovation**

| Author | Definition |
|---|---|
| Acs and Audretsch (1988: 678) | "Innovation is a process that begins with an invention, proceeds with the development of the inventions, and results in the introduction of a new product, process or service to the market-place." |
| Bjerke and Johansson (2015: 221) | "Innovation is a process where different types of knowledge and competencies are combined." |
| Damanpour (1992: 376) | "Innovation is defined as the adoption of an idea or behaviour whether a system, policy, program, device, process, product or service that is new to the adopting organisation." |
| Dewangan and Godse (2014: 536) | "Innovation = Invention + Exploitation<br>In this definition, invention implies conceiving and developing the idea into a workable application, whereas exploitation entails the process of commercialization and reaping the benefits." |
| Dibrell, Davis, and Craig (2008: 205) | "Innovations vary in complexity and can range from minor changes to existing products, processes, or services to breakthrough products, and processes or services that introduce first-time features or exceptional performance." |
| Fruhling and Keng (2007: 136) | "Innovation is an idea, practice or object that is perceived as new to an individual or another unit of adoption." |
| Geiger and Cashen (2002: 68) | "Innovation refers to the creation of new product within the firm." |
| Palmberg (2004: 1) | "Innovation is defined as a technologically new or significantly enhanced product compared to the firm´s previous product which has been commercialized on the market." |
| Pittaway, Robertson, Munir, Denyer, and Neely (2004: 144) | "Innovation is the successful exploitation of ideas, into new products, processes, services or business practices, and is a critical process for achieving the two complementary business goals of performance and growth, which in turn will help to close the productivity gap." |

Some preexisting definitions of innovation are cushioned specifically to products (Geiger & Cashen, 2002; Palmberg, 2004), whereas other attempts are comparatively broad allowing a wide range of potential innovations (Acs & Audretsch, 1988; Damanpour, 1992; Dibrell et al., 2008; Pittaway et al., 2004). I believe that a general definition of innovation should not be tailored to one object and include multiple forms of innovation as firms should not limit their innovation activities only on product development. Regarding a general definition on innovation, this study refers to the Oslo Manual guidelines developed by the Organization for Economic Co-operation and Development (OECD). These guidelines have been established in collaboration with 30 countries worldwide and have specifically been elaborated for collecting and interpreting innovation data.

> *"An innovation is the implementation of a new or significantly improved product (good or service), or process, [or] a new marketing method (...) in business practices, workplace organization or external relationship" (OECD, 2005: 46)*

The OECD definition necessitates two requirements for a feature to be an innovation. The minimum requirement for an innovation is that it is at least novel or significantly improved to the firm. Therefore, slight improvements of existing products, processes, marketing methods, or organizational methods cannot be considered innovative outcomes. Alternatively, however, it is not necessary that the innovation is new to the market or industry of operation. Moreover, an innovation must be implemented which prevents pure ideas or concepts from being considered as innovations. A product for instance, is implemented once it has been introduced to the market. New or significantly improved processes or marketing methods must be in actual use to be implemented.

While several previous studies, however, found that an innovation also must be internally invented, I do not consider the invention process as a necessary feature to account for an innovation (Acs & Audretsch, 1988; (Dewangan & Godse, 2014). Innovation processes have significantly changed and have recently become more dynamic. Historically the majority of innovation activities occurred internally and firms aimed to protect their technologies and intellectual properties in order to remain competitive. Nowadays, however, we are entering a world of open innovation in which companies permanently make use of external knowledge for their innovation activities. Firms commonly diffuse innovations from external organizations and may consider licensing a technology or product instead of inventing and developing it themselves. (Chesbrough, 2006; Chesbrough & Brunswicker, 2013; Kline, 2003)

### *3.1.2 Classifications*

There are several ways how innovative outcomes can be classified and categorizations vary according to the research studies. Three categorizations, however, are consistently present in literature and should be briefly explained.

Firstly, the degree of novelty refers to how novel an innovation is and commonly the three levels are differentiated: **new to the firm, new to the market, new to the world.** The minimum requirement for an innovation is that it must be new to the firm. Thus, regardless whether a product, process or marketing method has already been implemented by other organizations, it must contain a degree of novelty to the firm to be considered an innovation (Berger & Revilla Diez, 2006; Trauffler & Tschirky, 2007). When an organization is the first to implement an innovation to the market of business operation, it can be considered new to the market. In this case the degree of novelty is evaluated on a macroeconomic level from outside the company (Acs & Audretsch, 1988; OECD, 2005). Innovations new to the world refer to a qualitatively greater degree of novelty than new to the market and imply that the innovation is first introduced to all markets, national and international. (Berger & Revilla Diez, 2006; Edison et al., 2013; OECD, 2005)

Secondly, based on the degree of innovation, innovative outcomes can be classified into **incremental and radical**. While incremental innovations describe slight improvements or minor changes within a given frame of solutions, radical refers to a change of frame or doing something radically different from previous conditions. The major difference between the two forms is whether the particular innovation is perceived as a continuous modification of previously accepted practices or whether it represents a unique and discontinuous approach (Chandy & Tellis, 1998; Norman & Verganti, 2014). Notably, the degree of innovation can refer to a firm or market perspective. A new product, process or marketing innovation can be radically different from pre-existing ones of the firm as well as the market.

Thirdly, based on the impact of an innovation, a classification into **disruptive and non-disruptive** can be made. Though a firm perspective is not excluded, this scale commonly is perceived from a market perspective. Disruptive innovations have significant impacts on the market and on the economic activities of firms on that particular market (OECD, 2005). Disruptive innovations may transform the pre-existing structure of markets or even create totally new markets. However, whether or not an innovation is disruptive might not be apparent until long after the innovation has been implemented. (Assink, 2006; Herrmann, Tomczak, & Befurt, 2006; Norman & Verganti, 2014; OECD, 2005)

In literature, the outlined distinctions are not always clearly stated and the terms ‘radical’ and ‘disruptive’ are often used interchangeably (Edison et al., 2013; Norman & Verganti, 2014). A few examples, however, should illustrate that there is a difference between radical and disruptive. When Apple released iTunes, the product has not only been novel to the market, but also has been radically different from pervious practices. Apple has introduced a new way to download or manage music and eventually Apple´s new software has significantly impacted multiple pre-existing markets. Thus, Apple´s iTunes revealed to be both, a radical and disruptive innovation. Edison´s development of the light bulb has been similarly disruptive, resulting in a major revolution in home and business. Yet, Edison´s new product has not been significantly different from previous products on the market. Instead, Edison made minor improvements to pre-existing bulbs by extending the bulb´s life time and providing all the necessary infrastructure, including distribution systems, the indoor wiring or the sockets to hold the light bulbs. Hence, although Edison´s efforts had a disruptive impact on the market, regarding the degree of innovation, the product was not radical. (Norman & Verganti, 2014)

#### *3.1.3 Types of Innovation and Firm Innovativeness*

While this study chose an overall definition of innovation which is comparatively broad and allows multiple forms of innovation, in a second step, various forms of innovative outcomes can be differentiated. There are three types of innovation that should be laid out in detail, as they are relevant for this study.

- **Product Innovation**: This type of innovation refers to the introduction of a good or service that is new or significantly improved with regards to its characteristics or intended uses. It includes architecture structure, technology, components and materials, incorporated software, user friendliness or other functional characteristics. (Amara, Landry, & Doloreux, 2009; Assink, 2006; Geiger & Cashen, 2002; OECD, 2005; Schumpeter, 2012; Singh & Singh, 2009)
- **Process Innovation**: This type of innovation refers to the implementation of a new or significantly improved production or delivery method that changes the way how products are created or distributed. This includes significant changes in techniques, equipment and/or software. 8(Acs & Audretsch, 1988; OECD, 2005)
- **Marketing Innovation**: This type of innovation refers to implementation of a new marketing method involving significant changes in product design or packaging, product placement, promotion or pricing. It includes exploiting new market opportunities or repositions of innovations. (Amara et al., 2009; OECD, 2005; Singh & Singh, 2009)

The distinction between the presented forms of innovations sometimes may be difficult. For instance, there might be borderline cases between product and marketing innovations. The main distinguishing factor is whether the product´s functions or user characteristics have changed. For example, if clothes are equipped with a waterproof technology, it is a product innovation. Alternatively, if only the shape of clothes is changed to previous versions solely to address new customer groups, this would correspond to a marketing innovation. While it appears to be easy to distinguish products and processes, the differentiation between services and processes may be less distinctive in certain cases. If an innovation involves new characteristics of a service that are offered directly to customers, it accounts to a product innovation. If an innovation, however, involves new methods, equipment or skills used to support the establishment of a service, it is a process innovation. Lastly, in some cases it may also be difficult to differentiate process and marketing innovations. Both forms may involve moving goods or information, but their purposes and intentions differ. Marketing innovations aim to increase sales volumes or market shares, whereas process innovations aim to reduce unit costs or improve product quality. If an organization implements new logistics to address the same customer but at lower costs, it is a process innovation. If this organization, however, implements new sales channels to address new customer segments, this would be a marketing innovation. (OECD, 2005)

In practice, the outlined innovation forms may be strongly interrelated, one resulting the another. A firm that introduces a new product may require new sales channels or might be forced to change internal processes, too. Thus, it can easily occur that a firm is a product, process and marketing innovator simultaneously. Whether or not a firm is innovative can be assessed in several ways. Historically the term innovativeness has been associated with the extend of innovation activities and the research and development costs have commonly been used as an indicator. The basic definition, however, refers to whether a firm has released or implemented innovations within a specific time frame. For this study, an innovative firm is one that has implemented either a new or significantly improved product, process or marketing method during the period under review. (Edison et al., 2013; OECD, 2005)

### 3.2 The Effects of Networking and Collaboration on Innovation

The locus of innovation increasingly shifts from an individual organization to the overall network the firm is embedded in. Several breakthrough innovations result from the contribution of numerous actors closely working together in a network (Powel, Koput, & Smith-Doerr, 1996). Hence, as knowledge is more and more distributed across organizations, firms recognize an increasing demand to formally and informally network and collaborate with other firms to improve their innovative capabilities (Bougrain & Haudeville, 2002). Across numerous sectors, networking behavior is identified to sufficiently drive firms´ innovative output and competitiveness (Ahuja, 2000; Powel et al., 1996). Industries, in which a relevant network has had a significant impact on innovation, encompass service industries (Elg & Johansson, 1997; Knights, Murray, & Willmont, 1993), manufacturing industry (Biemans, 1991; Grotz & Braun, 1997; Hyun, 1994; Izushi, 1997; Shaw, 1993) and high-tech industry (Frenken, 2000; Romijn & Albaladejo, 2002; Streb, 2003). Some of the benefits a relevant network provides to firms are risk sharing (Grandori, 1997), obtaining access to new markets and technologies (Grandori & Soda, 1995), speeding product development (Almeida & Kogut, 1999), pooling complimentary resources and skills (Eisenhardt & Schoonhoven, 1996), or obtaining access to external knowledge (Cooke, 1996). Researchers Gemünden, Ritter, and Heydebreck (1996), in a study found that firms were nearly 20 percent more likely to have product improvements, and 10 percent more likely to release novel products when firms began utilizing and applying specific forms of networking. Notably, there is also evidence in literature that those firms which do not engage in formal or informal networking activities limit their knowledge base long-term and reduce their ability to find resourceful collaborative alliances (Shaw, 1993). Following the results of previous research, a resourceful and diverse network can significantly contribute to firm innovativeness and may be a vehicle to find collaborative alliances for innovation activities. Prior research provided sufficient evidence that inter-organizational collaboration within innovation activities also impacts firms´ innovative performance positively (Faems, van Looy, & Debackere, 2005; Fitjar & Rodriguez-Pose, 2014; Hardwick, Anderson, & Cruickshank, 2013; Stuart, 2000). The innovative capabilities of firms that strategically partner with suppliers, customers (Hippel, 1995; Shaw, 1996), universities and research institutions (Gerwin, Kumar, & Pal, 1992; Santoro, 2000), industry competitors (Dodgson, 1993; Hamel, 1991) or lead user (Hippel, Thomke, & Sonnack, 1999; Quinn, 1985) is significantly higher and more effective than of firms that do not take advantage of collaboration (Faems et al., 2005). There are several reasons for that. For instance, strategic alliances may help to spread research and development costs among different organizations, resulting in a considerable reduction of risk associated with the particular innovation projects (Hagedoorn, 2002; Veugelers, 1998). Further, inter-organizational alliances may be a vehicle to acquire resources that are particularly required to turn innovation projects into commercial success (Hagedoorn, 1993; Powel et al., 1996; Teece, 1986). Though a large portion of the studies predominantly focus on larger corporations (Alexiev et al., 2016; Fitjar & Rodriguez-Pose, 2014), more recent studies have also demonstrated that a relevant network with collaborative alliances are beneficial for entrepreneurial ventures, too (Bjerke & Johansson, 2015).

Overall, regardless of firm size, networking and collaboration can be valuable and may have positive impacts on the innovative performance of firms (Larson, 1991; Liao & Welsch, 2003; Pittaway et al., 2004).

# CHAPTER 4: COLLABORATION AND THE IMPACT ON BUSINESS MODELS

## 4.1 Business Models and Business Model Innovation

### *4.1.1 Business Models*

*Definitions and Functions*

Business models have increasingly become popular in both management literature and in corporate practice (Spieth & Schneider, 2016; Zott, Amit, & Massa, 2011). Scholars and practitioners, however, have been using the term without having achieved mutual agreement on the definition of the term business model (George & Bock, 2011; Spieth, Ricart, & Schneckenberg, 2014). TABLE 4 provides an overview of some definitions found in literature and, despite various perceptions and definitions, scholars have mostly equipped business models with three core functions. Business models describe an organization´s way of doing business. They also represent an instrument to facilitate or take advantage of opportunities in the market and lastly, business models are necessary to commercialize new ideas and technologies. (Spieth & Schneider, 2016)

**TABLE 4: Definition of Business Model**

| Author | Definition |
|---|---|
| Chesbrough (2006) | “The business model is a useful framework to link ideas and technologies to economic outcomes.” |
| Chesbrough and Rosenbloom (2002) | A business model is “the heuristic logic that connects technical potential with the realization of economic value”. |
| Linder and Cantrell (2000) | “The business model is the organization’s core logic for creating value.” |
| Osterwalder and Pigneur (2013) | “A business model describes the rationale of how an organization creates, delivers and captures value.” |
| Skarzynski and Gibson (2008) | “The business model is a conceptual framework for identifying how a company creates, delivers, and extracts value. It typically includes a whole set of integrated components, all of which can be looked on as opportunities for innovation and competitive advantage”. |
| Teece (2010) | “A business model articulates the logic, the data and other evidence that support a value proposition for the customer, and a viable structure of revenues and costs for the enterprise delivering that value.” |

The first function of business models is to describe a firm´s business logic. Business models are acknowledged to explain a company´s organizational and financial business architecture (Chesbrough & Rosenbloom, 2002; Teece, 2010). The narrative description of a firm´s business model contains two major advantages. First, it helps to reduce the complexity of a firm´s business logic while decomposing it into manageable units. Further, the narrations potentially allow to deduct or classify successful business models in the market that may serve as role models for others (Casadesus-Masanell & Ricart, 2010; George & Bock, 2011). While the first function of business models emphasizes a rather static perspective, the second and third functions are more transformative and dynamic. The second function of business models is to facilitate or exploit opportunities of innovation. Business models are the primarily vehicle for creating and capturing opportunities in the market that may lead to monetary success. In this sense, business models are a linkage between an organization and the overall market the organization operates in (Schindehutte, Morris, & Kocak, 2008). The third function of business models is to commercialize products and technologies (Chesbrough, 2006). Thereby, business models translate an innovation into commercial or monetary success. The economic value of an innovation, however, remains latent until it is commercialized by a business model. This implies that the same product or technology commercialized in two ways will yield different economic or monetary returns. Thus, firms and managers are continually challenged to find or develop appropriate business models for their innovations (Chesbrough, 2010; Teece, 2010). It must be recognized, though, that each of these perceptions serve distinct yet complimentary functions of business models and the varying perceptions underline the complexity of the term business models (Spieth & Schneider, 2016).

*Components and Dimensions*

Scholars have not only defined business models differently, but also assigned them a varying number of components and dimensions. A literature review by Morris, Schindehutte, and Allen (2005) showed that the number of business model components range between four and eight. Nowadays, one of the most popular tools to describe an organization´s business model is Osterwalder and Pigneur´s (2013) Business Model Canvas (Bertels, Koen, & Elsum, 2015; Chesbrough, 2010). The canvas, which originally derived from Osterwalder´s dissertation (2004), illustrates a business model with nine interrelated components in a simple one-sheet format, illustrated in FIGURE 2 (García-Gutiérrez & Martínez-Borreguero, 2016). The canvas proved to be a useful tool not only to define an existing business model but also for reconfigurations or the development of completely novel business model designs. Therefore, Osterwalder and Pigneur (2013) introduced several techniques and suggestions how to efficiently apply the canvas (García-Gutiérrez & Martínez-Borreguero, 2016). Due to its widespread use and popularity, this study also relies on the nine components of the Business Model Canvas.

**FIGURE 2: The Business Model Canvas**

**(Source: Innovatively, 2017)**

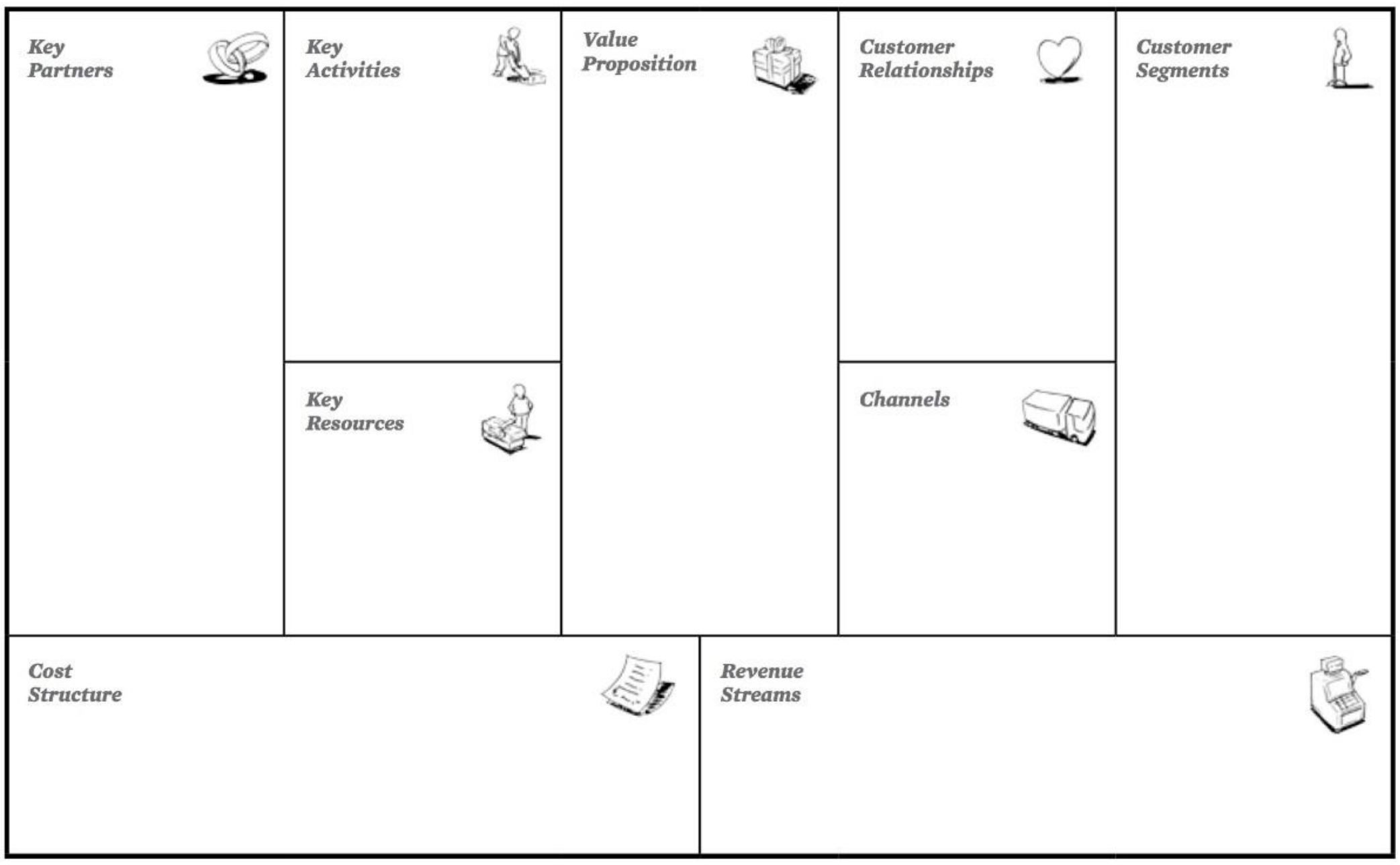

Business models are complex and strongly interrelated constructs. Therefore, authors have not only started to categorize business models into elements, but also integrated those elements to overarching units. An extensive literature review by Spieth and Schneider (2016) showed that, despite different labelling (Clauss, 2016), the majority of business model understandings comprise the three main integrative dimensions: **(1) value offering, (2) value creation architecture and (3) revenue model** (Chesbrough & Rosenbloom, 2002; Teece, 2010; Yunus, Moingeon, & Lehmann-Ortega, 2010).  TABLE 5 summarizes this three-dimensional business model structure. The first dimension, an organization´s value offering, consists of a firm´s value proposition, its customer segments, its customer relationships and its channels. The value offering contains the portfolio of products and services that are offered to customer segments. A firm´s value proposition is commonly delivered through distribution channels with the goal to build or sustain customer relationships. The value offering dimension can be considered as a ‘front stage’ with direct contact to customer. Alternatively, the value creation processes occur ‘backstage’ (Acs & Audretsch, 1988) and are captured by an organization´s key resources, key activities and key partners. It describes how an organization realizes or creates its value offering along the value chain. Finally, the third dimension, an organization´s revenue model, describes how a firm´s value offerings is converted into profits. The revenue model is particularly crucial to ensure an organization´s sustainability. It takes a firm´s revenue streams and cost structure into account. Although there are several different classifications in literature, this study follows the most commonly used three-dimensional classification as it is crucial to find common ground and make studies on business models comparable. (Clauss, 2016; Spieth & Schneider, 2016)

**TABLE 5: Business Model Dimensions**

| Dimension | Elements | Description |
|---|---|---|
| Value Offering | Value Proposition | "The Value Proposition Building Block describes the bundle of products and services that create value for a specific Customer Segment." (Osterwalder & Pigneur, 2013: 22) |
| | Customer Segments | "The Customer Segments Building Block defines the different groups and people or organizations an enterprise aims to reach and serve." (Osterwalder & Pigneur, 2013: 20) |
| | Customer Relationships | "The Customer Relationships Building Block describes the types of relationships a company establishes with specific Customer Segments." (Osterwalder & Pigneur, 2013: 28) |
| | Channels | "The Channel Building Block describes how a company communicates with and reaches its Customer Segments to deliver a Value Proposition." (Osterwalder & Pigneur, 2013: 26) |
| Value Creation Architecture | Key Activities | "The Key Activities Building Block describes the most important things a company must do to make its business model work." (Osterwalder & Pigneur, 2013: 36) |
| | Key Resources | "The Key Resources Building Block describes the most important assets required to make a business model work." (Osterwalder & Pigneur, 2013: 34) |
| | Key Partners | "The Key Partnership Building Block describes the network of suppliers and partners that make the business model work." (Osterwalder & Pigneur, 2013: 38) |
| Revenue Model | Revenue Streams | "The Revenue Streams Building Block represents the cash a company generates from each Customer Segment." (Osterwalder & Pigneur, 2013: 30) |
| | Cost Structure | "The Cost Structure describes all costs incurred to operate a business model." (Osterwalder & Pigneur, 2013: 40) |

### *4.1.2 Business Model Innovation*

*Definition and Types*

The previous chapter conceptualized business models, outlined their functions as well as definitions. However, research on business model recently shifted from a rather static to a more dynamic perspective. In times where product lifespans increasingly decrease and research and development costs permanently rise, organizations cannot avoid taking alternative forms of innovation into account. In fact, literature points out that organizations can glean at least as much value from innovating their business model as they can from innovative products or services (Amit & Zott, 2012; Chesbrough, 2010; García-Gutiérrez & Martínez-Borreguero, 2016; Johnson, Christensen, & Kagermann, 2008). Business model innovation is a complimentary form of innovation which considers the business model instead of products or processes as subject of innovation (Clauss, 2016; Zott et al., 2011). Importantly, it is an essential and unavoidable phenomenon for firms, as it addresses questions, how they can capture the full potential of their internal resources, capabilities or innovations. (Spieth & Schneider, 2016; Teece, 2010)

Few attempts to define business model innovation have been made in literature. Among these, business model innovation has been defined as “the discovery of a fundamentally different business model in an existing business” (Markides, 2006: 20), “the search of new logics…and ways to create or capture value” (Casadesus-Masanell & Zhu, 2013: 464), or as the situation when “a firm adopts a novel approach to commercializing its underlying assets” (Gambardella & McGahan, 2010: 263).

Business model innovation represents a complex form of innovation as it can have multiple triggers and sources. In literature, multiple forms of business model innovation have been differentiated. Depending on the degree of change, there are two different types of business model innovation: **business model design and business model reconfiguration.** The first corresponds to building a novel business model or newly formed organization from scratch. In most cases, this refers to entrepreneurship, but it can also apply to larger corporations in forms of spin-offs, joint ventures or internal innovation projects. The second form, business model reconfiguration, corresponds to the modification of an existing business model (García-Gutiérrez & Martínez-Borreguero, 2016; Stampfl, 2016). Due to technological and environmental changes, reconfiguring a firm´s business model has become an increasingly important task for managers. Firms are forced to adopt an effective attitude toward business model experimentation and must learn to deal with high level of uncertainty to permanently renew growth and profits (Chesbrough, 2010). Moreover, business models have also been classified into **radical and incremental.** While incremental business model innovation are adjustments or improvements of an existing business model, radical refers to a significant change of a business model (Laudien & Daxböck, 2015). Lastly, business model innovation has also been characterized by the degree of novelty. Similar to product or service innovation, business models can also be **new to the firm, new to the market or new to the world**. While new to the world business models are a rather rare, business models that are new to the market have become of high interest for practitioners as they are able to create a sufficient competitive advantage to firms. Business models which are new to the firm are the most common form, yet in most cases not sufficient to create a sustainable competitive advantage. (Berends, Smits, Reymen, & Podoynitsyna, 2016; Stampfl, 2016)

*Measurement*

The last section showed that business model innovation can take on different forms and the phenomenon is rather complex and diverse. In accordance, scholars have applied different scales and methods to measure business model innovation. While studies have focused on the design of new business models (Amit & Zott, 2012; García-Gutiérrez & Martínez-Borreguero, 2016; Zott & Amit, 2007), others have dedicated attention to finding useful frameworks for business model reconfigurations (Clauss, 2016; Spieth & Schneider, 2016). Zott and Amit´s research (2007) investigates how business model design affects the success of entrepreneurial firms. Specifically, their work analyzed the interrelation between the efficiency and novelty of business models and compared it to the overall performance of entrepreneurial firms.

Other recent studies of scholars, such as Clauss (2016) and Spieth and Schneider (2016) focus on business model reconfiguration. Their purpose is to find a validated measurement scale for business model innovation on a new to a firm level. Although both studies classify a business model into three dimensions, the two studies used different guidelines of how much change it requires to occur in those dimensions to count as a business model innovation. For Clauss (2016) it requires changes in all three overarching dimensions of a business model. For Spieth and Schneider (2016) it requires only one of the business model dimensions (value offering, value architecture, revenue model) to change. In accordance, Spieth and Schneider (2016) differentiate three basic types of business model innovation: (1) value offering innovation, (2) value architecture innovation and (3) revenue model innovation. Each form can occur independently, even though changes in one dimension are likely to affect other dimensions of a business model. This study follows the definition of Spieth and Schneider (2016) and determines that it needs one of the three overarching dimensions (value offering, value creation architecture, revenue model dimension) to change to account for an business model innovation.

### 4.2 The Effects of Collaboration on Business Models

Recent research demonstrates that business model innovation is a promising approach for firms to improve their sustainability and competitiveness. In general, however, the concept of business models and business model innovation are comparatively new, and there remains multiple unresolved gaps in academic research. Scholars have recently started to investigate the antecedents of business model innovation and highlighted that it may be triggered by multiple aspects including both internal and external factors (Bucherer, Eisert, & Gassmann, 2012; Demil & Lecocq, 2010).

In general several researchers agree on the importance of collaboration and networking for business model innovation (Frankenberger, Weiblen, Csik, & Gassmann, 2013; Giesen, Riddleberger, Christner, & Bell, 2010). Previous studies have outlined the potentials collaboration activities offer firms, indicating that it is able to impact the components of an organizations´ business model. For instance, literature points out that collaboration offers the potential of risk sharing and may generate economies of scale and scope (Katz & Shapiro, 1985; Shapiro & Varian, 1999). Although this research has been conducted outside the scope of business models - as it was only until later the term has been introduced – the results yet indicate that firms can reduce their costs or change their cost structure by engaging in collaboration. Moreover, collaboration enables firms to access information, external resources and technologies, new capabilities or even new markets (Gulati, Nohria, & Zaheer, 2000). Other benefits include enhanced transaction efficiency (Kogut, 2000) and shortened time to market products or services (Gulati et al., 2000). The effects of collaboration clearly show the potential it obtains to impact various business model components. Organizations may obtain new partnerships or resources through collaboration or collaborative alliances may enable access to new customer segments or even enable the invention of new products both resulting in novel revenue streams. The study of Laudien and Daxböck (2015) has been one of the first that explicitly included collaboration and network partners in the context of business model innovation. Based on multiple case studies, their study provided evidence that collaboration with partners and customers is a crucial element for business model innovation. As an interesting side effect, their findings show that business model innovation often is an unintended process rather than a conscious decision.

Overall, though not many studies have investigated the concrete effects of collaboration on the business model components or overarching dimensions, there is still enough proof in academic literature that collaboration obtains the potential to significantly impact business models.

# CHAPTER 5: EMPIRICAL STUDY

## 5.1 Framework and Hypothesis

### *5.1.1 Collaboration in Coworking Spaces*

The empirical study has three main goals. Firstly, the study aims to shed light into the collaborative behavior of coworkers by differentiating configurations of collaboration in coworking spaces. Second, it thrives to analyze how collaboration in coworking spaces contributes to firm innovativeness. Therefore, it takes the access to external resources and serendipitous occurrences into account, as they have been identified to be crucial regarding innovation activities. Lastly, the study aims to provide insights in whether coworking may obtain the potential to impact business models by offering increased opportunities for collaboration.

Previous literature provided sufficient evidence that coworking spaces are collaborative work environments that simulate frequent interaction and knowledge exchanges amongst members (Parrino, 2015). Coworking spaces embed members into a heterogeneous network and individuals or institutions engage in coworking because they are specifically seeking for a community or hoping to obtain partnerships, feedback or referrals from other members (Spinuzzi, 2012). Studies have shown that coworking spaces are shared and pro-active work spaces in which members frequently exchange knowledge and information with each other (Parrino, 2015; Soerjoatmodjo et al., 2015). Therefore, the coworking community creates a work environment that is characterized by high level of trust and thus provides ideal prerequisites for interaction, networking or collaboration. However, the coworking environment allows multiple forms of collaboration and scholars have started to develop configurations of collaboration (Parrino, 2015; Spinuzzi, 2012). Based on Spinuzzi´s differentiation between 'good neighbor' and 'good partner', this study also differentiates two major types of collaboration in coworking spaces, namely 'social interaction' and 'strategic collaboration'. The former generally is more informal and not goal-oriented. It refers to the level coworkers interact, communicate and build/sustain relationships with other members in the community. The latter refers to how commonly coworkers strategically make use of coworking spaces by entering into partnerships, integrate other members into their business, or co-develop with other members in the space. Hence, this form of collaboration is more business-oriented and commonly is initiated to achieve specific objectives. Particularly, the first part of this study is interested in whether these two distinct forms of collaboration exist in coworking spaces and whether they occur mutually or independently from each other.

#### *5.1.2 Collaboration in Coworking Spaces and the Impact on Firm Innovativeness*

Recent literature claims that coworking may be a driver for innovation (Bouncken & Reuschl, 2016). Innovative outcomes arise out of a pool of resources and require the combination of different sources of knowledge and competencies (Bjerke & Johansson, 2015). Coworking spaces are shared, pro-active and community oriented work environments that enclose its members into diverse and heterogeneous networks. Ongoing frequent interactions and knowledges exchanges amongst workers can provide a source of inspiration or lead to fruitful encounters and unexpected discoveries (Moriset, 2013). Capdevila´s study (2015) demonstrates that coworking spaces can contribute to the dynamics of innovation. According to his study, collective innovation activities require platforms that stimulate the involved actors to efficiently communicate and collaborate. Coworking spaces are such platforms and can contribute to innovation on different levels. On an individual level, members can support each other, collaborate and can advance in their profession. On a firm level, the explorative and exploitative practices in coworking spaces represent an outside source for inspiration and may allow the diffusion of resources or innovations amongst members. On a community level, coworking spaces can be perceived as innovation networks with heterogeneous knowledges bases that can compete with the innovation activities of larger corporations. Lastly, on a local level coworking spaces are platforms that bring together firms of different size. Capdevila´s main argument is that coworking spaces unite creative individuals ("the underground") with larger firms ("the upperground") that obtain the resources for innovation projects which will eventually contribute to the innovative performance of a city or district.

In this context, the local and social proximity of individuals and organizations in coworking spaces have gained academic attention. Parrino (2015) analyzed the theoretical frameworks of proximity and knowledge exchanges in coworking spaces. His study proved that the co-presence of coworkers in shared environments is a stimulator for knowledge spillovers amongst members. Therefore, the exchange of tacit knowledge includes social and cultural components, requiring direct contact and intimate trust between participants. Hence, coworking spaces are not only considered geographical spaces, but also social spaces where coworkers build or sustain relationships and permanently exchange knowledge with the rest of the community (Capdevila, 2013). Due to the local and social proximity of members, coworking spaces provide their members an ideal vehicle to access external intangible resources, such as knowledge or skills, in order to expand their capabilities and competencies. Notably, previous research provided evidence that resources play a key role for innovation processes. Studies have shown that the access to resources may turn innovation projects into success and may affect the innovativeness of firms positively (Bjerke & Johansson, 2015; Hagedoorn, 1993). This is the case for larger corporations (Faems et al., 2005; Powel et al., 1996), but especially for smaller businesses since small firms increasingly face limitations regarding internal resources. (Bjerke & Johansson, 2015).

Importantly, organizations may use collaborative alliances to get access to resources which cannot be produced internally. Hence, it is likely that collaboration in coworking spaces may provide an explorative or exploitative practice that allows the spillover or diffusion of resources. Relating to this, the study proposes the first hypothesis:

*H 1:* Collaboration in coworking spaces contributes to the access of external resources.

Another potential value of coworking discussed in literature is “serendipity”, which refers to the opportunity to make unexpected discoveries entirely by chance (Moriset, 2013). The notion of serendipity and the importance of unexpected occurrences for innovation has been noted in management literature for many years (Lester & Piore, 2006). Coworking spaces create a community characterized by a strong culture of openness, accessibility and sharing. Within literature, coworking spaces are reckoned to be “serendipity accelerators”, as they represent dynamic networks that bring together individuals and institutions of different backgrounds (Moriset, 2013; Spreitzer et al., 2015). Frequent formal and informal interactions with a multitude of different actors in these networks, could facilitate unexpected encounters, through which coworkers can make entirely new discoveries or gain entirely new insights. In accordance to this, the study proposes the following hypothesis:

*H 2:* Collaboration in coworking spaces contributes to serendipitous occurrences.

Lastly, prior research investigated various forms of collaboration and provided evidence that interfirm alliances impact the innovative performance of firms positively (Baum, Calabrese, and Silverman, 2000; Rogers, 2004; Shan, Walker, and Kogut, 1994; Stuart, 2000; Faems, van Looy, Debackere, 2005; Fitjar & Rodriguez-Pose, 2014). Firms may collaborate with suppliers, customers (Hippel, 1995; Shaw, 1996), universities and research institutions (Gerwin et al., 1992; Santoro, 2000), industry competitors (Dodgson, 1993; Hamel, 1991) or lead user (Hippel et al., 1999; Quinn, 1985). Importantly, all forms of collaborations have proved to significantly improve the innovativeness of firms (Faems et al., 2005; Larson, 1991; Pittaway et al., 2004). Literature on coworking revealed that, due to the local and social proximity, coworking spaces offer members increased opportunities for networking and collaboration. Studies have also shown that coworkers particularly join the coworking movement because they are seeking for collaborative alliances and partners for joint problem solving or innovation processes (Parrino, 2015; Spinuzzi, 2012). Thus, in a similar manner to collaborative manufacturing or collaborative research and development activities, which have proven to contribute to firm innovativeness (Bjerke & Johansson, 2015; Fitjar & Rodriguez-Pose, 2014), collaboration in coworking spaces may also impact the innovative capabilities of organizations positively (Capdevila, 2015). Hence, the study proposes the following third hypothesis:

*H 3:* Collaboration in coworking spaces contributes to firm innovativeness.

**FIGURE 3: Impact on Firm Innovativeness: Framework and Hypotheses**

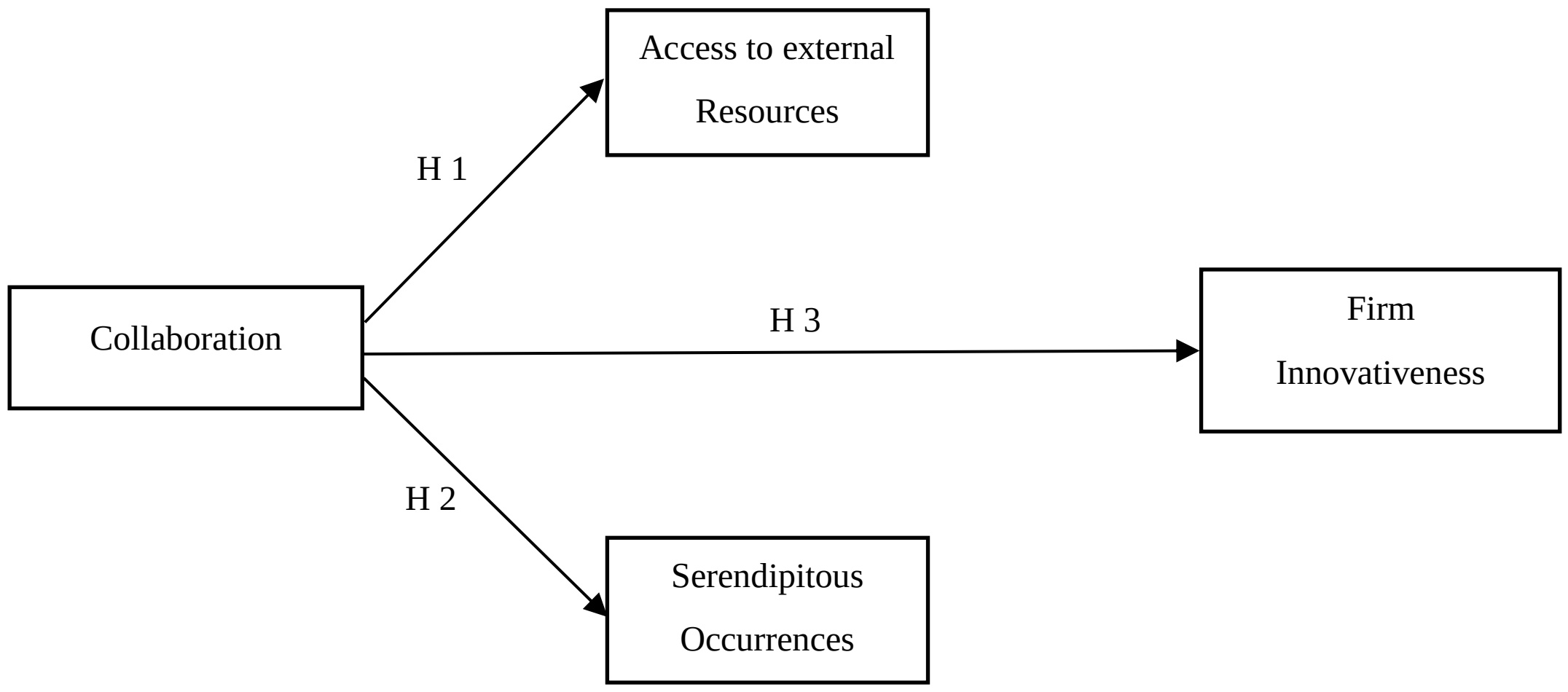

### *5.1.3 Collaboration in Coworking Spaces and the Impact on Business Models*

When engaging in coworking, organizations are put into a new collaborative work environment while being offered multiple entrepreneurial and business-related opportunities and services. Due to this, scholars have increasingly questioned whether coworking may contribute to business model development or even may be stimulator for business model innovation (Bouncken & Reuschl, 2016). Yet, until today, there is little known how coworking may impact business models. Furthermore, transferring this field of research into the scope of coworking brings along several difficulties. Organizations in coworking spaces are completely diverse and can range from self-employed individuals over entrepreneurial firms to larger corporations. Hence, while some organizations may be concerned with designing a new business model, others may already have established a well working model and be looking for ways to reconfigure it. Other organizations may not even consider changing their business model at all, as they operate in industries where it simply is not necessary.

Regardless of the problematics, however, previous work points out that collaboration can impact an organization´s business model positively (Amit & Zott, 2001). Collaboration can provide access to essential information, markets and technologies (Baum, Calabrese, & Silverman, 2000; Gulati et al., 2000), offer the potential of risk sharing, generate economies of scale and scope (Katz & Shapiro, 1985; Shapiro & Varian, 1999) or facilitate organizational learning (Anand & Khanna, 2000; Dyer & Nobeoka, 2000). Other benefits include enhanced transaction efficiency (Kogut, 2000) and shortened time to market products or services (Gulati et al., 2000). According to Laudien and Daxböck (2015) collaboration with network partners and customer is a key element for successful business model innovation. Hence, scholars have argued that coworking spaces also offer the necessary collaborative and social structure to foster business model development or facilitate business model innovation (Bouncken & Reuschl, 2016). In accordance with the above information, I propose the following hypothesis:

*H 4:* Collaboration in coworking spaces impacts organizations´ business models.

**FIGURE 4: Impact on Business Model: Framework and Hypotheses**

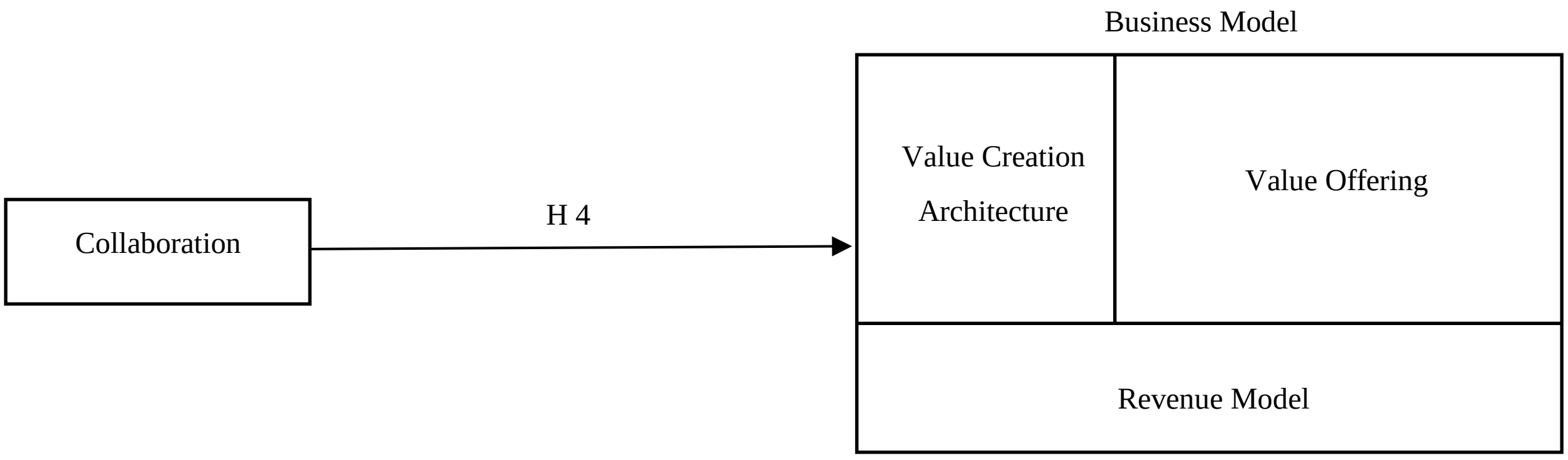

## 5.2 Methodology

### *5.2.1 Survey Structure and Analytics*

To test the hypotheses of the study, primary data was collected with a survey. The survey was addressed to organizations of all sizes in coworking spaces. It contained skip logic so that, depending on whether participants worked individually or within a team, they were addressed accordingly. The overall survey consisted of four parts (see APPENDIX A: Survey for further details).

After a short letter which explained the overall purpose and structure of the survey, the first part contained introductory questions, such as what industry participants worked in, what affiliation participants they obtained, what coworking space they worked in and when they or their team have joined the particular coworking space. Other questions addressed how many different coworking spaces respondents have already worked in, whether the coworking space was their primary workspace and how many hours they or their team regularly work in the coworking space.

The second part of the survey consisted of five questions, three analyzing the collaborative behavior of participants and two which were concerned with the benefits of coworking. Within this section participants indicated whether they have socially interacted or have strategically collaborated with other members or whether they or their team have entered into strategic partnerships during their coworking engagement. Furthermore, participants indicated whether they believed that coworking helped their organization to get access to external resources or whether coworking stimulated serendipitous occurrences.

The third part the survey contained questions on innovation. Participants were asked if their organization has released any innovation while engaging in coworking. The question differentiated three types of innovation, product, process and marketing innovation, and consistently provided participants with examples for each innovation. Also, if participants indicated to have released an innovation of any kind, they were also asked to evaluate the novelty of the innovation.

The fourth and last part of the survey was concerned with the impact of coworking on the business model of organizations. Specifically, participants were asked if coworking has improved or changed any of the nine building stones of their organizations´ business model. Participants indicated whether coworking helped their organization to find new collaborative partnerships, whether it helped to significantly improve internal processes by making them more innovative and/or efficient, or whether their organization has gained new technical, intellectual or financial resources through coworking. Further, participants were asked if coworking helped to improve their existing value proposition or if it helped to find entirely new value propositions, if it improved or changed distribution channels, if coworking helped them to expand or find new customer segments, or if participants believed that their organization increased customer retention or strengthened customer relationship through coworking. Lastly, participants of the survey indicated whether their organization has significantly improved the existing revenue model or whether coworking helped to develop new revenue streams as well as whether they found ways to save costs through coworking.

Regarding the analytics, the study used multiple statistical methods, including principal component analysis, bivariate correlation as well as linear and binary regression analysis conducted with IBM SPSS 23. During the entire process multiple sources of literature have been used as an overall guidance for the analysis (Backhaus, Erichson, Plinke, & Weiber, 2016; Chatterjee & Hadi, 2012; Fabrigar & Wegener, 2012; Montgomery, Peck, & Vining, 2013). The main goal of this study was to find causal relationships by using different forms of regression analysis. Therefore, the next sections will outline the relevant dependent, independent and control variables for the analysis in more details.

### *5.2.2 Dependent Variables*

*Access to external Resources (AccRes)*

In total, the study contains four dependent variables. The first variable is “Access to external Resources (AccRes)” which is operationalized with three items on a seven-point Likert scale, ranging from “1 - strongly agree” to “7 - strongly disagree”. The scale also included “8 - I do not know”. Respondents were asked if they have gained access to external resources which they otherwise would not have access to (AccRes_1), if they gained access to external knowledge (AccRes_2) and if they gained access to external skills (AccRes_3). To access the internal consistency of the items a reliability test was conducted. The Cronbach´s Alpha showed a value of .895, indicating strong internal consistency of the measures as it exceeds .700 (Chatterjee & Hadi, 2012; Montgomery et al., 2013). One item (AccRes_4) had to be deleted, as it significantly decreased the validity of the scale. Respondents´ answers of the three items were scored and aggregated.

*Serendipitous Occurrences (SerOcc)*

SerOcc is measured with four items on a seven-point Likert scale ranging from “1 - strongly agree” to “7 - strongly disagree”. Again, the scale also included “8 - I do not know”. Participants were asked whether they gained entirely new ideas and inspirations (SerOcc_1), if they gained entirely new impressions and thoughts (SerOcc_2), if they have made entirely new discoveries (SerOcc_3) or if they have discovered entirely new opportunities (SerOcc_4). The Cronbach´s Alpha was .926 demonstrating high reliability between the four items (Chatterjee & Hadi, 2012; Montgomery et al., 2013). The four items were again scored and aggregated.

*Firm Innovativeness (FirInn)*

The measurement of firm innovativeness is based on the previous study of Bjerke and Johansson (2015) and Faems et al. (2005). Respondents were asked whether their organization has launched any new innovations during the engagement in coworking. The study included product and service, marketing and process innovations. To ensure a common understanding, participants were given a general definition of innovation and examples for all three innovation types extracted from the OECD (2005) guidelines, which have particularly been developed for studies on innovation. Accordingly to the guidelines, a product or service innovation "(…) includes significant improvements in technical specifications, components and materials, incorporated software, user friendliness or other functional characteristics" (OECD, 2005: 48). "A process innovation is the implementation of a new or significantly improved production or delivery method. This includes significant changes in techniques, equipment and/or software" (OECD, 2005: 49). Lastly, "a marketing innovation is the implementation of a new marketing method involving significant changes in product design or packaging, product placement, product promotion or pricing" (OECD, 2005: 49). Based on their indications, organizations were classified into "innovative" and "non-innovative" using a dummy variable.

*Impact on Business Model (BusMod)*

To operationalize the impact of coworking on the business models, the study followed constructs of previous studies, which provided validated measurement scales for business models (Clauss, 2016; Spieth & Schneider, 2016). Respondents were asked if coworking affected each of the nine building stones of their business model. Therefore, each item was measured on a seven-point Likert scale from "1 - strongly agree to "7 - strongly disagree" with a "8 - I do not know" option. By aggregating the components "Key Partners", "Key Resources" and "Key Activities" the impact of coworking on the "Value Creation Architecture (ValCreArc)" was conducted. A Cronbach´s Alpha of .885 showed a strong internal reliability. In a similar manner, the four components "Value Proposition", "Customer Segments", "Customer Relationships" and "Channels" were aggregated to illustrate the impact of coworking on the "Value Offering (ValOff)" dimension. Again, with .882 the Cronbach´s Alpha was significantly above the recommended .700 (Chatterjee & Hadi, 2012; Montgomery et al., 2013). Further, the aggregation of the items "Revenue Streams" and "Cost Structure" to scores assessed the impact on the “Revenue Model (RevMod)” dimension. The variable showed a Cronbach´s Alpha of .528 which was lower than .700. Finally, the three dimensions “Value Creation Architecture”, “Value Offering” and “Revenue Model” were aggregated to assess how significant the concept of coworking affected respondents´ business models (BusMod). High scores indicated that coworking had a strong impact on the business model and vice versa. All respondents which indicated a “0 - I do not know” were eliminated. A Cronbach´s Alpha of .914 showed that the nine building stones measured the impact on business models in a reliable manner (Chatterjee & Hadi, 2012; Montgomery et al., 2013).

### *5.2.3 Independent Variables*

*Social Interaction (SocInt)*

The empirical study includes three independent variables which demonstrate different forms of collaboration in coworking spaces. The first variable “Social Interaction (SocInt)” was measured with four items on a seven-point Likert scale. Participants were asked if they or their team regularly interact with other members (SocInt_1), if they regularly communicate with other members (SocInt_2), if they regularly share opinions with other embers (SocInt_3) or if they regularly build or sustain relationships with other members (SocInt_4) in the space. Depending on the level of agreement, the four items were aggregated and a Cronbach´s Alpha of .930 demonstrated high internal reliability of the scale.

*Strategic Collaboration (StrCol)*

“Strategic Collaboration (StrCol)” is operationalized with four items on a seven-point Likert scale from “1 - strongly agree” to “7 - strongly disagree” including “8 - I do not know”. For this variable, participants were asked whether they have entered partnerships with other members (StrCol_1), whether they have integrated other members into their business (StrCol_2), whether they have co-developed projects with other members (StrCol_3) or whether they have collaborated with other members to achieve specific goals (StrCol_4). The four items were aggregated and the Cronbach´s Alpha showed an excellent reliability for the scale ($\alpha = .944$).

*Strategic Partnerships (StrPar)*

The third independent variable is “Strategic Partnerships (StrPar)”. The variable is based on a previous study on collaboration and innovation conducted outside the scope of coworking (Faems et al., 2005). Respondents specified whether they have entered into specific partnerships with (1) competitors of the same industry, (2) people or firms of different industries, (3) suppliers, (4) customers, (5) research institutions, or (6) consultants. A binary or dummy variable was conducted. Organizations which have not entered into partnerships were coded with “0”, whereas organizations that have were coded with “1”.

### 5.2.4 Control Variables

*Firm Size (FirSiz)*

Two variables are included to control for possible confounding effects. Since Schumpeter´s theory of economic development the relationship between firm size and the innovative performance has been discussed. In line with other studies on innovation (Faems et al., 2005; Fitjar & Rodriguez-Pose, 2014) the number of employees is used as an indicator. For this study, respondents were asked whether they work individually as a freelancer or solopreneur, whether they work for a small organization with two to nine employees or for an established company with 10 or more workers A similar distinction was made within prior studies on innovation (Bjerke & Johansson, 2015).

*Duration (Dur)*

The second control variable is concerned with the length of respondents coworking engagement. Therefore, participants were asked when their organization have joined the respective coworking space and the overall duration was recorded in months. To fulfil the requirements of a normal distribution the variable was logarithmized.

### 5.2.5 Data Collection and Sample

The study collected primary data by distributing a survey in several coworking spaces. To secure that all coworking spaces allowed opportunities for interaction and collaboration among members, the spaces were visited in person. In total, 53 coworking spaces in three different countries were visited. Seven spaces were located in Vancouver (Canada), six spaces in Portland Oregon (United States of America), 28 in Berlin (Germany) and 12 in Munich (Germany). At every coworking space I was given a tour and talked to a manager of the space. Out of the 53 spaces, 17 spaces agreed on distributing the survey to their members. The coworking spaces that participated were the HiVE, L´Atelier, The Amp, Creative Coworkers in Vancouver (Canada), Hatch Innovation and NXT Industries in Portland Oregon (United States), Fritz46, ESDIP, co.up, Pulsraum, b+office, Rainmaking Loft, betahaus and Space Shack in Berlin (Germany) as well as Coworking Holzschuh, Impact Hub and Mates in Munich (Germany). In all spaces, the survey was administered by e-mail to one person of each organization accompanied by a letter explaining the purpose of the study and encouraging participation. Participation was also stimulated by a reminder note a few weeks after the initial distribution. In total, the survey was sent out to 686 organizations. Between December 20, 2016 and May 29, 2017 survey responses were obtained from 132 organizations. After excluding incomplete surveys, usable responses covered 75 organizations which corresponds to an overall response rate of 10.9 percent. Regarding firm size and industry background, the sample was very diverse. 29.3 percent worked as freelancer or solopreneur, 45.3 percent represented a small organization between two and nine workers and 25.3 percent worked for an established company with more than ten workers. Classified by their industry, the organizations represented information/tech-industry (30.7%), arts/entertainment (13.3%), business services (13.3%), financial and insurance services (6.7%), education and healthcare (4.0%), wholesale/retail (2.7%), construction (2.7%), real estate (1.3%) and other sectors (25.3%) (see FIGURE 5).

**FIGURE 5: Overview of the Sample Composition**

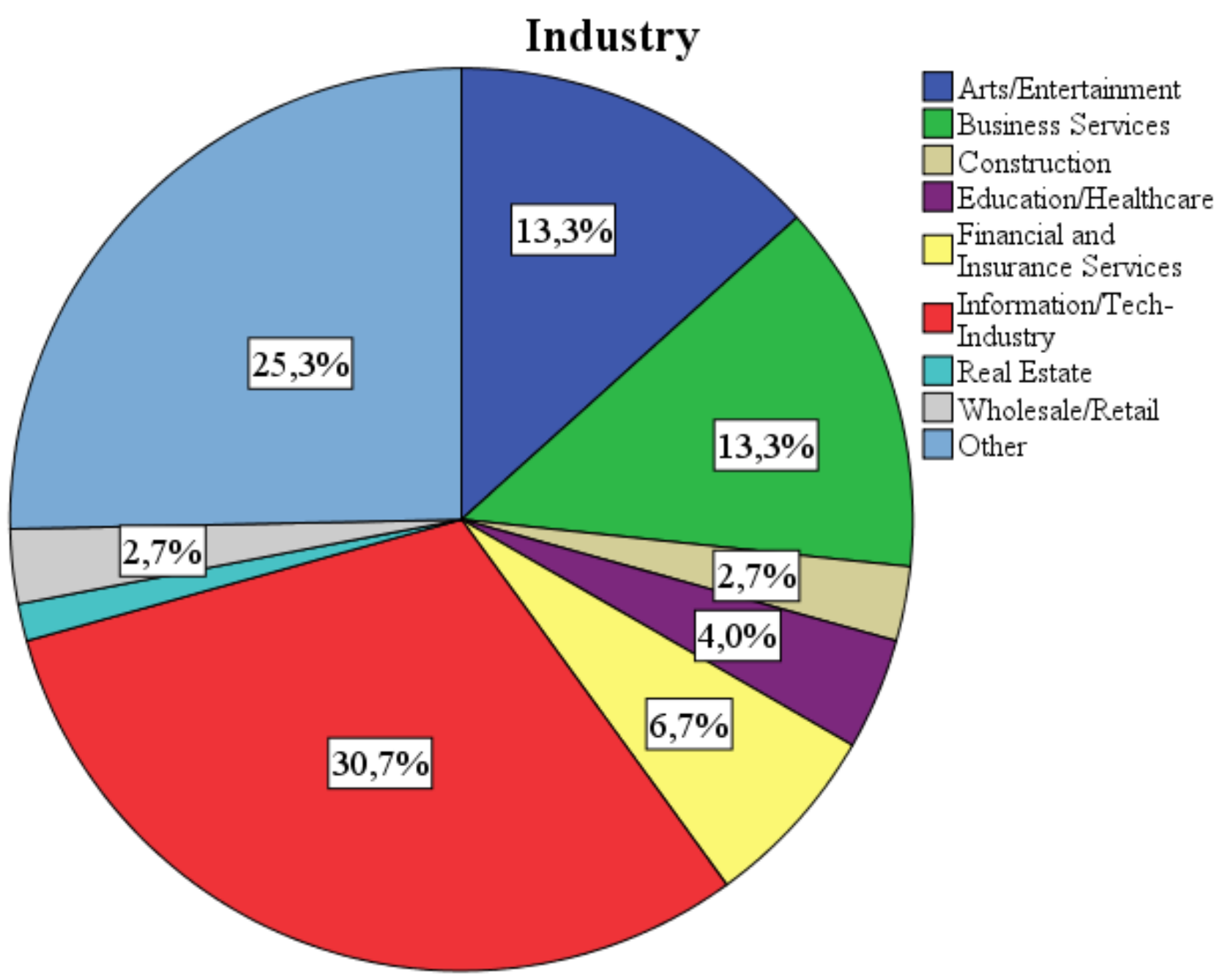

## 5.3 Results

### *5.3.1 Collaboration in Coworking Spaces*

In an initial step, the study aims to shed light into the collaboration behavior of coworkers. The study differentiated between a social and a strategic form of collaboration and while coworkers tend to be quite socially interactive, they are more reluctant regarding strategic collaborations and partnerships. For instance, when participants were asked whether they have entered into strategic partnerships with competitors, firms of same industry, suppliers, customers, research institutions or consultants, in total, 54.7 percent indicated to have never engaged in such alliances. Instead, 45.3 percent of the respondents partnered at least with of one those entities. An even better comparison was provided by the variables social interaction (SocInt) and strategic collaboration (StrCol), as both were measured with four items on a seven-point Likert scale from “1 - strongly agree” to “7 - strongly disagree” including a “8 - I do not know” option. For the former, participants were asked whether they regularly communicate, interact, share opinions and build or sustain relationships with other members in the space. Alternatively, the latter variable included questions whether coworkers have entered partnerships, integrated others into their business, co-developed projects, or collaborated with other members to achieve specific goals. Following the descriptive statistics, the level of agreement substantially differed regarding those two variables. While, on average, 42.6 percent of all participating coworkers agreed or even strongly agreed to the statements concerning SocInt, almost half of the respondents (48.3%) disagreed or even strongly disagreed to the four items representing StrCol.

To provide more thorough insights, the study conducted an exploratory factor analysis or principal component analysis. This type of statistical method is used to uncover the underlying structure of variables with the overarching goal to identify the relationships between them (Fabrigar & Wegener, 2012). The goal is to find out whether the variables SocInt and StrCol formed components, indicating that they stand for distinct variables. The principal component analysis used a varimax rotation method and extracted variables based on eigenvalues greater than one. TABLE 6 provides the results of the exploratory factor analysis. The Kaiser-Meyer-Olkin Measure of Sample Adequacy (KMO) shows a value of .844, whereas the Bartlett´s Test of Sphericity shows a significant Chi-square of 564.979 (p-value < .000), both indicating a significant factor model. The rotated component matrix nicely separated the variables StrCol_1 to StrCol_4 from the variables SocInt_1 to SocInt_4 into two components. The matrix shows high factor loadings for all items ranging from .932 to .859 for StrCol and from .914 to .844 for the variable SocInt. The loadings on the non-corresponding factors, however, are comparatively low, none exceeding .354. This clearly indicates that SocInt and StrCol represent two measures with a comparatively low relationship across the variables. Overall, the factor model explained 84.7% of the variance cumulatively.

**TABLE 6: Principal Component Analysis (SocInt & StrCol)**

| | Component | | Model Summary | | |
|---|---|---|---|---|---|
| | 1 | 2 | Kaiser-Meyer-Olkin Measure of Sampling Adequacy | Bartlett´s Test of Sphericity | |
| | | | | Chi-square | Sig. |
| StrCol_1 | .932 | .234 | | | |
| StrCol_2 | .890 | .174 | | | |
| StrCol_3 | .884 | .239 | | | |
| StrCol_4 | .859 | .354 | | | |
| SocInt_2 | .200 | .914 | | | |
| SocInt_3 | .303 | .880 | | | |
| SocInt_1 | .170 | .868 | | | |
| SocInt_4 | .307 | .844 | | | |
| | | | .844 | 564.979 | .000 |

### *5.3.2 Collaboration in Coworking Spaces and the Impact on Firm Innovativeness*

*Collaboration and the Access to external Resources*

After exploring the collaboration behavior of coworkers, the study´s four hypotheses shall be tested. The first two hypothesis of this study are concerned with whether collaboration in coworking spaces contributes to the access of external resources (H1) or serendipitous occurrences (H2), as they may be a key factor for the success of innovation activities. The study conducted a principal component analysis with a varimax rotation to test the reliability and validity of the variables AccRes and SerOcc. The factor model, which is presented in TABLE 7, extracted values based on eigenvalues greater than 0.7. The Kaiser-Meyer-Olkin Measure (KMO) shows a value of .855 and the Bartlett´s Test of Sphericity revealed an approximate Chi-square of 472.503, which is highly significant ($p < .000$). Reviewing the components, both variables SerOcc and AccRes obtain high factor loadings and comparatively low loadings on the non-correspondent factors. The factor loadings for SerOcc lied between .932 and .859, and for AccRes between .914 and .868. The model explained a cumulative variance of 82.92% which is more than acceptable value according to literature (Fabrigar & Wegener, 2012).

**TABLE 7: Principal Component Analysis (AccRes & SerOcc)**

| | Component | | Model Summary | | |
|---|---|---|---|---|---|
| | 1 | 2 | Kaiser-Meyer-Olkin Measure of Sampling Adequacy | Bartlett´s Test of Sphericity | |
| | | | | Chi-square | Sig. |
| SerOcc_2 | .932 | .234 | | | |
| SerOcc_4 | .890 | .174 | | | |
| SerOcc_1 | .884 | .239 | | | |
| SerOcc_3 | .859 | .354 | | | |
| AccRes_1 | .200 | .914 | | | |
| AccRes_2 | .303 | .880 | | | |
| AccRes_3 | .170 | .868 | | | |
| | | | .855 | 472.503 | .000 |

To validate H1, the study uses a hierarchical linear regression analysis with AccRes as a dependent variable and the three variables representing collaboration as independent variables (SocInt, StrCol, StrPar). The analysis also included the two control variables firm size (FirSiz) and duration of coworking engagement (Dur). TABLE 8 presents the bivariate Pearson correlations for all construct scales. The correlation between AccRes and the three indicators of collaboration all show very significant two-tailed values, ranging from .399 to .569. In contrast, none of the control variables show a significant correlation with the dependent variable AccRes.

**TABLE 8: Correlation Analysis (AccRes)**

| | AccRes | SocInt | StrCol | StrPar | FirSiz | Dur |
|---|---|---|---|---|---|---|
| AccRes | 1 | | | | | |
| SocInt | .538** | 1 | | | | |
| StrCol | .569** | .523** | 1 | | | |
| StrPar | .399** | .318** | .757** | 1 | | |
| FirSiz | .165 | .211 | .158 | .129 | 1 | |
| Dur | .204 | .287* | .093 | .197 | .044 | 1 |

**. Correlation is significant at the 0.01 level (2-tailed).

*. Correlation is significant at the 0.05 level (2-tailed).

N=75

To analyze whether collaboration directly affects AccRes multiple linear regression models are conducted. TABLE 9 represents the results of the regression analysis. Model 1a includes the control variables to capture their effect on the dependent variable. The overall model is not significant, indicating that the control variables do not affect AccRes in a significant manner.

Further, none of the coefficients are significant at a .05 level. Model 1b then adds the three independent variables SocInt, StrCol and StrPar to the analysis, as all forms of collaboration have shown a significant correlation with AccRes. The model is significant ($p < .000$) and explains 36.8 percent of the variance of the dependent variable. Two of the three independent coefficients are particularly significant for AccRes. While SocInt ($p = .015$) positively impacts the dependent variable with a beta value of .243, the effect of StrCol is even stronger on AccRes ($\beta = .292$; $p = .009$). The last regression model (model 1c) excluded the non-significant independent variable StaPar to show the direct contribution of SocInt and StrCol on AccRes. Model 1c is significant and explains 37.6 percent of the variance of the dependent variable (Ad $R^2 = .367$; $p < 000$) and further the overall model remains stable in respect to the two control variables. The coefficients SocInt ($\beta = .249$; $p < .010$) and StrCol ($\beta = .266$; $p < .000$) both positively contribute to AccRes.

To confirm the first hypothesis, however, the prerequisites for regression analysis must be evaluated. For each model, the study reviewed the regression analyses for signs of non-linearity, additivity and heteroscedasticity Further, violations of the statistical interdependence and the normality of the error distribution were conducted. The scatterplots of the dependent and independent variables do not indicate signs of non-linearity and only slight signs of heteroscedasticity (see APPENDIX C: Collaboration and the Access to external Resources). Further, the Durban-Watson Tests for all three models indicated that the residuals are independent, as the values are within the acceptable range of 1.5 and 2.5. Model 1a showed a Durban-Watson value of 1.937, model 1b of 1.675and for model 1c the value was 1.690. By reviewing the histograms and normal quantile plots, the study analyzed whether the errors of the residuals violated the normal distribution assumption. The graphs showed that the errors were nicely normally distributed. Lastly, the collinearity diagnostics showed VIF values continually below the recommended value of ten. The study could also not find any indication to reject the multicollinearity assumption. (Chatterjee & Hadi, 2012; Montgomery et al., 2013)

Consequently, H1 of this study can be confirmed. Collaboration in coworking spaces can contribute to the access of external resources.

**TABLE 9: Hierarchical Linear Regression Analysis (AccRes)**

| | | Coefficients | | | Model Summary | | | |
|---|---|---|---|---|---|---|---|---|
| Model | | **B** | **Sig.** | **VIF** | **R** | **Ad R²** | **F** | **Sig.** |
| 1a | **Control** | | | | | | | |
| | FirSiz | .578 | .176 | 1.002 | | | | |
| | Dur | 2.003 | .089 | 1.002 | | | | |
| | | | | | .256 | .040 | 2.533 | .086 |
| 1b | **Independent** | | | | | | | |
| | SocInt | .243 | .015 | 1.601 | | | | |
| | StrCol | .292 | .009 | 3.125 | | | | |
| | StrPar | -.420 | .753 | 2.550 | | | | |
| | **Control** | | | | | | | |
| | FirSiz | .134 | .704 | 1.052 | | | | |
| | Dur | .885 | .387 | 1.167 | | | | |
| | | | | | .641 | .368 | 9.629 | .000 |
| 1c | **Independent** | | | | | | | |
| | SocInt | .249 | .010 | 1.529 | | | | |
| | StrCol | .266 | .000 | 1.388 | | | | |
| | **Control** | | | | | | | |
| | FirSiz | .130 | .711 | 1.050 | | | | |
| | Dur | .806 | .413 | 1.095 | | | | |
| | | | | | .640 | .376 | 12.167 | .000 |

Dependent Variable: AccRes

*Collaboration and Serendipitous Occurrences*

The second hypothesis is concerned with whether collaboration in coworking spaces contributes to serendipitous occurrences amongst members, as they may initiate innovation activities. Therefore, TABLE 10 provides the correlation between SerOcc with all independent and control variables. The analysis shows that none of the control variables have a significant correlation with the dependent variable SerOcc. In addition, all forms of collaboration significantly correlated with SerOcc at a .01 level. The coefficients range between .438 and .611 with StrCol obtaining the highest correlation to SerOcc.

**TABLE 10: Correlation Analysis (SerOcc)**

| | **SerOcc** | **SocInt** | **StrCol** | **StrPar** | **FirSiz** | **Dur** |
|---|---|---|---|---|---|---|
| SerOcc | 1 | | | | | |
| SocInt | .438** | 1 | | | | |
| StrCol | .611** | .523** | 1 | | | |
| StrPar | .483** | .318** | .757** | 1 | | |
| FirSiz | .139 | .211 | .158 | .129 | 1 | |
| Dur | .117 | .287* | .093 | .197 | .044 | 1 |

**. Correlation is significant at the 0.01 level (2-tailed).

*. Correlation is significant at the 0.05 level (2-tailed).

N=75

In a similar manner to the previous section, the hierarchical linear regression analysis included three models to asses if collaboration contributes to serendipitous occurrences within coworking spaces (see TABLE 11). Therefore, model 2a included both control variables, FirSiz and Dur to investigate whether the dependent variable SerOcc is impacted by the controls. The outcome shows that neither the overall model ($p = .316$) nor any control coefficients are significant.  The second model (model 2b) included all three variables of collaboration as independent variables. The regression model is significant and explains 35.2 percent of the variance of SerOcc (Ad $R^2 = .352$; $p < .000$). Reviewing each predictive coefficient, one variable significantly affects SerOcc. StrCol, which has also shown the strongest correlation amongst the variables, predominately contributes to the dependent variable SerOcc ($\beta = 388$; $p < .007$). Hence, the variable StrCol together with all control variables formed the third model. Model 2c shows that for every increase of the independent variable StrCol, the dependent variable SerOcc increases by .501. The overall model is highly significant ($p < .000$) and explains 35.2 percent of the variance in SerOcc (Ad $R^2 = .352$).

Again, all prerequisites for linear regression analysis were observed and checked. The scatterplots of SerOcc suggest variance homogenic residuals and show no concerns to reject the linearity assumption. The histograms indicated that the errors show strong tendencies of a normal distribution (see APPENDIX C: Collaboration and Serendipitous Occurrences). The three Durban-Watson Tests of the models are within the acceptable range of 1.5 and 2 - the Durban-Watson value of model 2a is 2.425, of model 2b is 2.240 and for model 2c the value is 2.314. Multicollinearity can be assumed, as the collinearity diagnostics show variance inflation factors (VIF) between 1.002 and 3.144, none of them exceeding the value of ten. (Chatterjee & Hadi, 2012; Montgomery et al., 2013)

Consequently, H2 can be confirmed, too. One form of collaboration, namely strategic collaboration, can contribute to serendipitous occurrences within coworking spaces.

**TABLE 11: Hierarchical Linear Regression Analysis (SerOcc)**

| | | **Coefficients** | | | **Model Summary** | | | |
|---|---|---|---|---|---|---|---|---|
| Model | | **B** | **Sig.** | **VIF** | **R** | **Ad $R^2$** | **F** | **Sig.** |
| 2a | **Control** | | | | | | | |
| | FirSiz | .622 | .253 | 1.002 | | | | |
| | Dur | 1.420 | .342 | 1.002 | | | | |
| | | | | | .178 | .005 | 1.171 | .316 |
| 2b | **Independent** | | | | | | | |
| | SocInt | .172 | .169 | 1.601 | | | | |
| | StrCol | .388 | .007 | 3.125 | | | | |
| | StrPar | .852 | .616 | 2.550 | | | | |
| | **Control** | | | | | | | |
| | FirSiz | .096 | .831 | 1.052 | | | | |
| | Dur | .139 | .915 | 1.167 | | | | |
| | | | | | .629 | .352 | 9.031 | .000 |
| 2c | **Independent** | | | | | | | |
| | StrCol | .501 | .000 | 1.033 | | | | |
| | **Control** | | | | | | | |
| | FirSiz | .193 | .664 | 1.027 | | | | |
| | Dur | .760 | .530 | 1.010 | | | | |
| | | | | | .615 | .352 | 14.400 | .000 |

Dependent Variable: SerOcc

*Collaboration and Firm Innovativeness*

The third hypothesis of this study is to find out whether collaboration in coworking spaces can contribute to the innovativeness of firms. Respondents have been asked if their organization has launched any product, process, or marketing innovation during their engagement in coworking. Accordingly, respondents then were categorized into innovative and non-innovative firms. The descriptive statistics reveal that out of the 75 organizations, 39 have launched either a product, process or marketing innovation, whereas 36 indicated to not have launched any innovation. Hence, 52 percent of the organizations were innovative, leaving 48 percent non-innovative organizations. All descriptive statistics are outlined in APPENDIX B: Descriptive Statistics.

TABLE 12 presents the binary correlations between FirInn and all independent and control variables. It shows that FirInn correlates with two independent variables, namely StrCol and StrPar. All other correlation coefficients are not significant. The variables AccRes and SerOcc, which ought to represent potential values for innovation activities, neither show significant correlations with FirInn. The correlation coefficients of StrCol and StrPar, however, are considerably strong and significant at a .01 level. The table also shows that the two variables obtain a strong interrelation with each other (r = .757), as both variables represent strategic forms of collaboration.

**TABLE 12: Correlation Analysis (FirInn)**

| | FirInn | SocInt | StrCol | StrPar | AccRes | SerOcc | FirSiz | Dur |
|---|---|---|---|---|---|---|---|---|
| FirInn | 1 | | | | | | | |
| SocInt | .125 | 1 | | | | | | |
| StrCol | .401** | .523** | 1 | | | | | |
| StrPar | .500** | .318** | .757** | 1 | | | | |
| AccRes | .141 | .538** | .569** | .399** | 1 | | | |
| SerOcc | .215 | .438** | .611** | .483** | .723** | 1 | | |
| FirSiz | .002 | .211 | .158 | .129 | .165 | .139 | 1 | |
| Dur | -.030 | .287* | .093 | .197 | .204 | .117 | .044 | 1 |

**. Correlation is significant at the 0.01 level (2-tailed).

*. Correlation is significant at the 0.05 level (2-tailed).

N=75

To test H3 a binary logistic regression analysis was run with FirInn as a dependent and StrPar as an independent variable. TABLE 13 presents the results of the binary logistic regression model. According to the model summary, the model is highly significant ($p < .000$) with a Chi-square value of 21.775. The $R^2$ values of Cox & Snell and Nagelkerke show that the regression model explains between 25.2 and 33.6 percent of the variance within the dependent variable. This is a considerable amount, as only one independent variable (StrPar) is included into the model. The descriptive classification together with the Hosmer and Lemeshow Test also underline the significance of the regression model. The descriptive classification show that the model classified 80.6 percent of the non-innovative and 69.2 percent of the innovative firms correctly. This adds up to overall percentage of 74.4 percent. Further, the Hosmer and Lemeshow Test is non-significant with a Chi-square of 5,771, also indicating a good fit of the overall regression model. Reviewing each coefficient in the model, both control variables (FirSiz and Dur) do not impact the innovativeness of firms (p = .575; p = .205). However, the independent variable StrPar obtains a positive beta value of 2.487 and is highly significant, showing a p-value smaller .000. The Exp(B) value, often referred to as odds ratio, supports that StrPar significantly impacts FirInn. It shows that firms which have collaborated with at least one partner are approximately 12-times as likely to be innovative. (Backhaus et al., 2016)

The high significance of the regression model together with the strong beta values of the variable StrPar, clearly indicates that the null hypothesis should be rejected. As a result, the study confirms H3, indicating that collaboration in coworking spaces can impact the innovativeness of firms positively.

**TABLE 13: Binary Logistic Regression Analysis (FirInn)**

| | | **Coefficients** | | | | | **Model Summary** | | | |
|---|---|---|---|---|---|---|---|---|---|---|
| | | | | | **95% of Exp(B)** | | | | | |
| Model | | **B** | **Sig.** | **Exp(B)** | **Lower** | **Upper** | **Chi-square** | **Sig** | **Cox & Snell $R^2$** | **Nagel-kerke $R^2$** |
| 3 | **Independent** | | | | | | | | | |
| | StrPar | 2.487 | .000 | 12.020 | 3.719 | 38.847 | | | | |
| | **Control** | | | | | | | | | |
| | FirSiz | -.126 | .575 | .882 | .568 | 1.368 | | | | |
| | Dur | -.822 | .205 | .439 | .123 | 1.567 | | | | |
| | | | | | | | 21.775 | .000 | .252 | .336 |

Dependent Variable: FirInn

### *5.3.3 Collaboration in Coworking Spaces and the Impact on Business Models*

*Collaboration and the Impact on the Value Creation Architecture Dimension (ValCreArc)*

The third part of this study is dedicated to the fourth hypothesis of the empirical framework. It is concerned with whether forms of collaboration in coworking spaces impact organizations´ business models. Therefore, the study divides a firm´s business model into three dimensions: The value creation architecture (ValCreAr), the value offering (ValOff) and revenue model (RevMod). Participants were asked how significant they felt coworking impacted each dimension of their business model. Specifically, the study is interested in whether respondents, that show high levels of collaboration, also indicate that their business mode has changed more significantly due to coworking. To test the hypothesis linear regression analysis were conducted.

First, the study investigates how respondents felt coworking changed the value creation architecture dimension of their business, which includes key partnerships, key resources and key activities. TABLE 14 presents the correlations between the impact on the value creation architecture dimension with all independent and control variables. According to the analysis, all three variables regarding collaboration significantly correlate with ValCreArc. The variables StrCol (r = .748) and StrPart (r = .618) show the strongest correlation with the dependent variable. Moreover, none of the control variables appear to have a significant impact on ValCreArc, as both are not significant at a .05 level.

**TABLE 14: Correlation Analysis (ValCreArc)**

| | ValCreArc | SocInt | StrCol | StrPar | FirSiz | Dur |
|---|---|---|---|---|---|---|
| ValCreArc | 1 | | | | | |
| SocInt | .413** | 1 | | | | |
| StrCol | .748** | .523** | 1 | | | |
| StrPar | .618** | .318** | .757** | 1 | | |
| FirSiz | .117 | .211 | .158 | .129 | 1 | |
| Dur | .074 | .287* | .093 | .197 | .044 | 1 |

**. Correlation is significant at the 0.01 level (2-tailed).

*. Correlation is significant at the 0.05 level (2-tailed).

N=75

The study uses hierarchical linear regression analysis to assess if any independent or control variables impact the value creation architecture dimension positively. The analysis included three models, each with ValCreArc as the dependent variable (see TABLE 15). Model 4a captures the effect of the two control variables on the dependent variable. Neither the overall model (p = .512) nor any of the coefficients are significant. Therefore, the control variables appear to not particularly affect the value creation architecture. Model 4b includes all forms of collaboration as independent variables to the regression model. This model is significant with a p-value smaller than .000 and an adjusted R-square value of .529. Reviewing the coefficients, however, the model appears to be dominated by one variable. With a beta value of .409 and a p-value smaller than .000, StrCol is the only variable which significantly contributes to the dependent variable. Finally, within model 4c the direct contribution of StrCol to ValCreArc is presented. The model is significant with a p value below .000, it is robust against all control variables and explains 54.0 percent of the variance within the dependent variable (Ad $R^2$ = .540). The variable StraCol shows a strong and very significant beta value of .409 ($p < .000$).

**TABLE 15: Hierarchical Linear Regression Analysis (ValCreArc)**

| | | Coefficients | | | Model Summary | | | |
|---|---|---|---|---|---|---|---|---|
| Model | | **B** | **Sig.** | **VIF** | **R** | **Ad R²** | **F** | **Sig.** |
| 4a | **Control** | | | | | | | |
| | FirSiz | .421 | .333 | 1.002 | | | | |
| | Dur | .701 | .557 | 1.002 | | | | |
| | | | | | .136 | -.009 | .676 | .512 |
| 4b | **Independent** | | | | | | | |
| | SocInt | .047 | .579 | 1.601 | | | | |
| | StrCol | .409 | .000 | 3.125 | | | | |
| | StrPar | 1.267 | .269 | 2.550 | | | | |
| | **Control** | | | | | | | |
| | FirSiz | -.033 | .912 | 1.052 | | | | |
| | Dur | -.271 | .756 | 1.167 | | | | |
| | | | | | .753 | .536 | 18.109 | .000 |
| 4c | **Independent** | | | | | | | |
| | StrCol | .497 | .000 | 1.033 | | | | |
| | **Control** | | | | | | | |
| | FirSiz | -.005 | .987 | 1.027 | | | | |
| | Dur | .046 | .954 | 1.010 | | | | |
| | | | | | .748 | .540 | 30.005 | .000 |

Dependent Variable: ValCreArc

*Collaboration and the Impact on the Value Offering Dimension (ValOff)*

Second, the study is interested in the impact of collaboration on the value offering dimension. Organizations´ value offering is described by its value proposition, its customer relationships, its customer segments and its channels. TABLE 16 shows the correlation analysis and reveals that all three independent variables are significant with p-values smaller than .001. Again, the two strategic forms of collaboration, StrCol (r = .635) and StrPar (r = .567), show the highest correlations amongst the variables. In addition, the control variable Dur correlates significantly with ValOff at a .05 level (r = .251).

**TABLE 16: Correlation Analysis (ValOff)**

| | ValOff | SocInt | StrCol | StrPar | FirSiz | Dur |
|---|---|---|---|---|---|---|
| ValOff | 1 | | | | | |
| SocInt | .302** | 1 | | | | |
| StrCol | .635** | .509** | 1 | | | |
| StrPar | .567** | .312** | .763** | 1 | | |
| FirSiz | -.051 | .183 | .128 | .101 | 1 | |
| Dur | .251* | .274* | .076 | .188 | .020 | 1 |

**. Correlation is significant at the 0.01 level (2-tailed).

*. Correlation is significant at the 0.05 level (2-tailed).

N=72

The linear regression analysis includes four models (see TABLE 17). Model 4d and 4e present the impact of the control variables on the dependent variable, whereas model 4f and 4g are supposed to demonstrate the impact of the independent variables on the value offering dimension. Model 4f shows that out of all independent variables, StrCol is the only variable with a significant p-value at a .005 level. SocInt (p = .590) and StrPar (p = .393) do not appear affect the dependent variable, as their beta values are not significant. Model 4g illustrates the direct impact of StrCol on ValOff. The model is highly significant with a p-value smaller than .000 and an adjusted R-square value of .440. StrCol significantly impacts the variable ValOff with a beta-value of .512. However, one of the control variable also impacts the dependent variable within model 4g. With a beta-value of 2.502, the control coefficient Dur also shows a direct and significant contribution to ValOff (p = .024). Hence, model 4e elaborated the impact of Dur on ValOff. Within model 4e, the variable Dur reveals a significant beta value of 3.057 (p = .034). However, despite its significance, the impact of Dur on ValOff is rather low. Model 4e shows that the control variable Dur only explains 4.9 percent of the variance within ValOff. Consequently, approximately 39.1 percent of the variance can be referred to the variable StrCol.

Further, a linear regression analysis with StrCol as a dependent and Dur as the independent variable was conducted to assure that Dur did not impact StrCol in the first place. The overall model was not significant (p = .427) and the beta value of the coefficient Dur was comparatively low (β = .093) and more importantly, not significant either (p = .427). As a result, though the control variable Dur significantly impacts the ValOff, most variance of the variable can be explained through the variable StrCol.

**TABLE 17: Hierarchical Linear Regression Analysis (ValOff)**

| | | Coefficients | | | Model Summary | | | |
|---|---|---|---|---|---|---|---|---|
| Model | | **B** | **Sig.** | **VIF** | **R** | **Ad $R^2$** | **F** | **Sig.** |
| 4d | **Control** | | | | | | | |
| | FirSiz | -.251 | .634 | 1.000 | | | | |
| | Dur | 3.070 | .034 | 1.000 | | | | |
| | | | | | .257 | .039 | 2.436 | .095 |
| 4e | **Control** | | | | | | | |
| | Dur | 3.057 | .034 | 1.000 | | | | |
| | | | | | .251 | .049 | 4.695 | .034 |
| 4f | **Independent** | | | | | | | |
| | SocInt | -.061 | .590 | 1.555 | | | | |
| | StrCol | .460 | .001 | 3.146 | | | | |
| | StrPar | 1.352 | .393 | 2.604 | | | | |
| | **Control** | | | | | | | |
| | FirSiz | -.582 | .160 | 1.038 | | | | |
| | Dur | 2.478 | .039 | 1.163 | | | | |
| | | | | | .688 | .434 | 11.871 | .000 |
| 4g | **Independent** | | | | | | | |
| | StrCol | .512 | .000 | 1.022 | | | | |
| | **Control** | | | | | | | |
| | FirSiz | -.613 | .133 | 1.017 | | | | |
| | Dur | 2.502 | .024 | 1.006 | | | | |
| | | | | | .681 | .440 | 19.568 | .000 |

Dependent Variable: ValOff

*Collaboration and the Impact on the Revenue Model Dimension (RevMod)*

The “Revenue Model” dimension consists of an organization´s revenue streams together with its cost structure. In a similar manner to previous sections, the impact of collaboration on this dimension is elaborated. In an initial step, TABLE 18 presents the correlations of all independent and control variables with the revenue model dimension. While the variables SocInt and FirSiz do not significantly correlate with the dependent variable, the correlation coefficient of StrCol (r = .492), StrPar (r = .443) and Dur (r = .260) are significant at least at a .05 level.

**TABLE 18: Correlation Analysis (RevMod)**

| | **RevMod** | **SocInt** | **StrCol** | **StrPar** | **FirSiz** | **Dur** |
|---|---|---|---|---|---|---|
| RevMod | 1 | | | | | |
| SocInt | .165 | 1 | | | | |
| StrCol | .492** | .509** | 1 | | | |
| StrPar | .443** | .312** | .761** | 1 | | |
| FirSiz | .002 | .185 | .131 | .112 | 1 | |
| Dur | .260* | .275* | .078 | .195 | .027 | 1 |

**. Correlation is significant at the 0.01 level (2-tailed).

*. Correlation is significant at the 0.05 level (2-tailed).

N=73

The hierarchical linear regression analysis, presented in TABLE 19, consists of four models. The first two model, model 4g and 4h show the impact of the control variables on RevMod, revealing that Dur significantly contributes to the dependent variable. Within model 4i the variable Dur has a beta value of 1.560 and a p-value of .026. The model displays that 5.5 percent of the variance in RevMod can be explained by Dur. Despite its significance (p-value = .26), the impact of Dur on the revenue model dimension is comparatively low. Model 4k includes the variable StrCol to the investigation, as it is the only significant variable (see model 4j). StrCol reveals to be a strong driver for the revenue model dimension with a highly significant beta-value of .191. In total, model 4k explains 26.6 percent of the variance within the dependent variable ($AdR^2$ = .266). Comparing this to model 4k, which has shown the direct contribution of Dur on RevMod, roughly 21 percent of the variance within RevMod can be explained solely by StrCol.

In summary, StrCol has turned out to be the only form of collaboration which significantly impacted each dimension of organization´s business model. This variable has the greatest influence on the value creation architecture, followed by the value offering and revenue model dimension.

**TABLE 19: Hierarchical Linear Regression Analysis (RevMod)**

| | | Coefficients | | | Model Summary | | | |
|---|---|---|---|---|---|---|---|---|
| Model | | **B** | **Sig.** | **VIF** | **R** | **Ad $R^2$** | **F** | **Sig.** |
| 4h | **Control** | | | | | | | |
| | FirSiz | -.012 | .961 | 1.001 | | | | |
| | Dur | 1.561 | .027 | 1.001 | | | | |
| | | | | | .261 | .041 | 2.549 | .085 |
| 4i | **Control** | | | | | | | |
| | Dur | 1.560 | .026 | 1.000 | 1,875 | | | |
| | | | | | .260 | .055 | 5.168 | .026 |
| 4j | **Independent** | | | | | | | |
| | SocInt | -.092 | .143 | 1.555 | | | | |
| | StrCol | .212 | .004 | 3.132 | | | | |
| | StrPar | .251 | .772 | 2.600 | | | | |
| | **Control** | | | | | | | |
| | FirSiz | -.104 | .649 | 1.040 | | | | |
| | Dur | 1.568 | .019 | 1.167 | | | | |
| | | | | | .568 | .272 | 6.376 | .000 |
| 4k | **Independent** | | | | | | | |
| | StrCol | .191 | .000 | 1.023 | | | | |
| | **Control** | | | | | | | |
| | FirSiz | -.150 | .507 | 1.018 | | | | |
| | Dur | 1.345 | .030 | 1.006 | | | | |
| | | | | | .544 | .266 | 9.687 | .000 |

Dependent Variable: RevMod

*Collaboration and the Impact on the Business Model (BusMod)*

Finally, this section includes the entire business model to the investigation. Previous chapters pointed out that an organization´s business model is a complex and interdependent system that is described by nine building stones. TABLE 20 illustrates the interdependency of the nine components of a business model, by showing that most variables have a strong correlation with each other.

**TABLE 20: Correlation Business Model Components**

| | **1** | **2** | **3** | **4** | **5** | **6** | **7** | **8** | **9** |
|---|---|---|---|---|---|---|---|---|---|
| 1 Key Activities | 1 | | | | | | | | |
| 2 Key Resources | .660** | 1 | | | | | | | |
| 3 Key Partnerships | .629** | .710** | 1 | | | | | | |
| 4 Value Proposition | .627** | .605** | .614** | 1 | | | | | |
| 5 Customer Segments | .587** | .597** | .647** | .657** | 1 | | | | |
| 6 Customer Relationships | .513** | .478** | .483** | .628** | .639** | 1 | | | |
| 7 Channels | .513** | .723** | .687** | .659** | .765** | .566** | 1 | | |
| 8 Revenue Streams | .460** | .554** | .588** | .688** | .709** | .655** | .630** | 1 | |
| 9 Cost Structure | .269* | .339** | .277* | .407** | .156 | .455** | .284* | .364** | 1 |

**. Correlation is significant at the 0.01 level (2-tailed).

*. Correlation is significant at the 0.05 level (2-tailed).

N=75

The previous sections analyzed the three overarching dimensions of organizations´ business models and concluded that one variable has a very significant impact on all three dimensions. StrCol consistently showed significant and strong beta values within all previous regression models. Thus, it stands to reason that StrCol also has a strong contribution to how respondents judged the change of their business model due to coworking. Therefore, TABLE 21 presents the correlations amongst all variables, showing that StrCol and BusMod are strongly interrelated with a highly significant correlation coefficient of .708. Moreover, the variables SocInt and StrPar both are significant with r-values of .340 and .617. The analysis also shows that none of the control variables significantly correlate with BusMod.

**TABLE 21: Correlation Analysis (BusMod)**

| | BusMod | SocInt | StrCol | StrPar | FirSiz | Dur |
|---|---|---|---|---|---|---|
| BusMod | 1 | | | | | |
| SocInt | .340** | 1 | | | | |
| StrCol | .708** | .509** | 1 | | | |
| StrPar | .617** | .312** | .763** | 1 | | |
| FirSiz | .017 | .183 | .128 | .101 | 1 | |
| Dur | .203 | .274* | .076 | .188 | .020 | 1 |

**. Correlation is significant at the 0.01 level (2-tailed).

*. Correlation is significant at the 0.05 level (2-tailed).

N=72

The results of the hierarchical linear regression analysis presented in TABLE 22 show three regression models. Model 4l shows the effects of the control variables on BusMod and suggests that none of the controls significantly impact the dependent variable. As assumed, StrCol turned out to be a strong promoter for how recipients felt their business model has improved or changed due to coworking. Model 4m shows that out of all independent variables, StrCol is the only significant coefficient with a beta-value of 1.076 and a p-value below .000. Model 4n shows the direct contribution of StrCol to the dependent variable. StrCol shows a highly significant beta value of 1.206, which is robust against all controls as none of them are significant. Moreover, the overall regression model explains 50.8 percent of the variance within BusMod. Considering the fact, that only one variable is included in this analysis, this can be perceived as a strong impact. The F-value of 25.447 and the p-value of .000 indicate that model 4m is highly significant.

However, to confirm that collaboration can impact the business models of organizations in coworking spaces, the prerequisites for linear regression model must be fulfilled. Therefore, for each regression model the key assumptions for regression models were assessed. Most crucially, the scatterplots showed no signs to reject the linearity or additivity assumptions. The statistical interdependence of errors can be assumed, because the Durban-Watson values for all model lied within the acceptable range between 1.5 and 2.5. The values ranged between 1.714 and 2.322 for all regression models. Further, the collinearity diagnostics showed that none of the variance inflation factors (VIF) exceeded the value of ten (the factors ranged between 1.000 and 3.177). Lastly, the study made sure that none of the models severely violated the assumptions of normally distributed errors. The histograms and normal probability plots showed that the errors of the four variables ValCreArc, ValOff, RevMod and BusMod were not perfectly normally distributed. However, because some references on regression analysis do not list normally distributed errors among the key assumptions and because the assumption has not been violated severely, the study saw no need to transform the variables (Chatterjee & Hadi, 2012; Montgomery et al., 2013). The APPENDIX provides all supporting graphics of the regression analysis (see APPENDIC C: Collaboration and the Impact on Business Models).

To reject the null hypothesis, I defined that collaboration must at least impact one overarching dimension (value creation architecture, value offering, revenue model) of the business model. The previous section showed that StrCol was able to affect all three dimensions together with the overall business model. As a result, H4 can also be confirmed and the study concludes that collaboration can impact organizations´ business models in coworking spaces.

**TABLE 22: Hierarchical Linear Regression Analysis (BusMod)**

| | | **Coefficients** | | | **Model Summary** | | | |
|---|---|---|---|---|---|---|---|---|
| Model | | **B** | **Sig.** | **VIF** | **R** | **Ad $R^2$** | **F** | **Sig.** |
| 4l | **Control** | | | | | | | |
| | FirSiz | .120 | .916 | 1.000 | | | | |
| | Dur | 5.247 | .090 | 1.000 | | | | |
| | | | | | .203 | .014 | 1.486 | .233 |
| 4m | **Independent** | | | | | | | |
| | SocInt | -.102 | .649 | 1.555 | | | | |
| | StrCol | 1.076 | .000 | 3.146 | | | | |
| | StrPar | 3.041 | .335 | 2.604 | | | | |
| | **Control** | | | | | | | |
| | FirSiz | -.683 | .406 | 1.038 | | | | |
| | Dur | 3.756 | .113 | 1.163 | | | | |
| | | | | | .734 | .504 | 15.412 | .000 |
| 4n | **Independent** | | | | | | | |
| | StrCol | 1.206 | .000 | 1.022 | | | | |
| | **Control** | | | | | | | |
| | FirSiz | -.734 | .364 | 1.017 | | | | |
| | Dur | 3.910 | .075 | 1.006 | | | | |
| | | | | | .727 | .508 | 25.447 | .000 |

Dependent Variable: BusMod

# CHAPTER 6: DISCUSSION

## 6.1 Contributions

This study sought to understand whether organizations can enhance their innovativeness and improve their business models by taking advantage of collaboration in coworking spaces. This idea has become an often noted topic in academic literature (Bouncken & Reuschl, 2016), yet few researchers so far have attempted to investigate the benefits of coworking on firm innovativeness and business models. Particularly scarce are studies using quantitative research methods within this field. This study is one of the first systematic empirical investigation which is concerned with the collaborative behavior of coworkers and helps to fill this oft-noted and yet unresolved gap in management research. The results offer various important theoretical and empirical extensions to prior literature.

The study explored how intensively and in what ways collaboration takes place in coworking spaces. Though few scholars have already stepped into this void in academic research (Avdikos & Kalogeresis, 2016; Parrino, 2015, 2015; Soerjoatmodjo et al., 2015; Spinuzzi, 2012), several of these studies are based on a few case studies that mostly used qualitative methods for their investigations. This study provided new and solid measurement scales for collaboration in coworking spaces which are likely to be used for further research. Further, this study showed that coworkers tend to apply different forms collaborative behavior - one socially and one strategically characterized – and these forms appear to be not as connected as initially expected. Overall, while coworkers tend to socially interact very strongly with each other, the majority of coworkers indicated that they did not engage in strategic collaborative alliances or partnerships. Underlined by a strongly significant exploratory factor analysis, the study has shown that these two variables represent two distinct components with low loadings on the non-corresponding factors. Therefore, the study builds upon the findings of previous studies (Parrino, 2015; Spinuzzi, 2012). Spinuzzi´s study (2012) previously proposed two configurations of coworking spaces in academic literature, which he referred to as the “good-neighbor” and the good-partner” configuration. While “good neighbors” build and sustain neighborly relationships through frequent informal interactions, “good-partners” link up into momentary collaborative alliances to attack shared problems. This study provided quantitative prove for these forms of collaboration, but has also showed that coworkers increasingly make use of the more social form of collaboration.

The study also provided insights into how collaboration may help coworkers to grant access to external resources and to gain serendipitous occurrences. The spillover of tacit knowledge in coworking spaces has gained previous attention in literature (Parrino, 2015; Soerjoatmodjo et al., 2015) and, as management literature has pointed out, collaboration can be a vehicle to acquire distinct tangible and intangible resources that are required for innovation activities (Hagedoorn, 1993; Powel et al., 1996; Teece, 1986). The study empirically shows that collaboration can catalyze superior access to resources and can contribute to serendipitous occurrences. Therefore, this study extended the findings of Parrino´s study (2015). Unlike Parrino´s study, which had been based on qualitative research methods, this study has shown a positive interrelation between social interaction or strategic collaboration and the access to resources by using quantitative methods. However, it must be acknowledged that this study could not find any connection to firm innovativeness, leaving the concrete interrelation between resources and serendipity to innovation unresolved. Further, the study acknowledges that the scales used to measure the access to resources and serendipity require further improvements as the factor analysis extracted eigenvalues greater than 0.7.

Perhaps the most important contribution of the current study has been to conclusively show a positive interrelation between collaboration and firm innovativeness. These findings indicate that ventures in coworking spaces can enhance their innovative capabilities by making use of strategic partnerships. The methodology and measurement sales used for this investigation were based on previous studies (Bjerke & Johansson, 2015; Faems et al., 2005), which have shown a positive relationship between collaboration and firm innovativeness outside the scope of coworking. Hence, this study has successfully transferred this distinct interrelation from general management literature to coworking literature. The results showed that the probability for organizations to be innovative were approximately 12-times higher if they have entered into at least one strategic partnership. The results have been highly significant, and especially because the measurement scales have been proven in other areas of research, the contribution of this study´s results are considerable. Noteworthy, the links between collaboration and firm innovativeness were not significantly affected by either firm size or an organization´s duration of coworking engagement. Thus, the findings may yield another important implication, whereby firms of different characteristics or size may reach the same innovative capabilities when obtaining support from collaborative partnerships in coworking spaces. This argument is in line with the results of previous studies outside the scope of coworking as well (Bjerke & Johansson, 2015).

Lastly, the study provides one of the very first empirical examination of coworking and the impact on business models. Coworking spaces provide several services to its members and individuals or organizations entering these spaces put themselves into a new collaborative and heterogeneous working environment. The study revealed that the more organizations strategically collaborated the greater the impact of coworking on their business model had been judged. This was the case for all three dimensions of a business model, the value creation architecture, the value offering and the revenue model, as well as for the overall business model. As an interesting fact, the descriptive results showed that a high number of coworkers disagreed that coworking has had a significant impact on their business model (see APPENDIX B: Impact on Business Model (BusMod)). Thus, it appears that coworking only provides the required services that may impact business models, but coworkers themselves are in charge to take advantage of the benefits by actively reaching out for strategic collaborative opportunities. When used accordingly, however, the results of this study show that the coworking concept can significantly impact business models, indicating collaboration in coworking spaces represents a driver for business model development or a trigger for business model innovation.

### 6.2 Practical Implications

The results of this study imply multiple practical implications on different levels. On an individual level, coworkers should be aware about the potentials coworking offers, from both a social and a business perspective. Coworkers should develop a collaborative mindset and are strongly encouraged to frequently communicate with other members, to constantly build new relationships as well as to actively seek out for collaborative alliances.

This argument can also be translated to an organizational level. In accordance to individuals, ventures in coworking spaces must be open for (spontaneous) strategic alliances and need to embrace or develop a collaborative culture, too. Organizations are encouraged to build resourceful networks and trustworthy connections, both within and outside of the space, which will allow them to mutually solve problems and eventually increase their innovativeness. However, this also implies that organizations remedy the burdens for successful collaboration. Organizations and their executives cannot solely protect their resources, but need to freely share the firm´s knowledge and intellectual properties with others. An excessively exploitative behavior is very likely to prevent successful collaborative alliances in coworking spaces. Instead, organizations need be willing to learn from others and maybe even align their strategic goals with others. Moreover, the results indicated that organization must transfer social interaction into results. If organizations expect to increase their innovativeness or develop their business model, ongoing frequent interaction will not be enough. To achieve such goals, it requires strategic partnerships, or mutually elaborated and executed projects with other coworkers.

On a community level, managers of coworking spaces must proceed indoctrinating a sense of community to establish a working environment characterized by trust and goodwill. Many spaces offer their members several opportunities to connect with others, such as mutual breakfasts, outside events or organized workshops. However, from my experience, managers` intentions mostly lie on providing a comfortable and interactive workspace to their members. Rarely they are focused on nurturing individuals and ventures from a business perspective, which, as the results of this study show, remains great potentials for innovation.  A crucial feature for the acceleration of collaboration is providing the relevant infrastructure for it. This includes internal communication tools or networking platforms which enable members to connect and exchange knowledge.  Out of the 53 coworking spaces that have been visited for this study, many did not provide the described infrastructure. Coworking spaces may also consider including incubator or accelerator programs into their spaces, as such programs are likely to enhance strategic collaboration. Out of all visited spaces, only one provided such service. Moreover, coworking spaces could invent and implement new products that help to enhance collaboration. For instance, platforms, where members can enter their skills and preferences could be developed and in this way, members could effortlessly match up into strategic partnerships or could directly hire members for temporary or non-temporary work. Finally, I believe that managers should continually observe and control the composition of members in coworking spaces. Firms tend to increasingly attract people of same or similar skills and backgrounds, yet it requires a certain heterogeneity and diversity for collaboration and innovation to occur.

### 6.3 Limitations

The study represents an attempt to disentangle the under-explored influence of collaboration on firm innovativeness and business models in coworking spaces. Particularly, it focuses on whether two forms of collaboration, a social and a strategic form, may contribute to the access of external resources, the occurrence of serendipitous situations or whether firms may derive superior benefits for innovations from collaboration in coworking spaces. Further, this study investigated into whether or not coworking obtains an improving or changing influence on the business models and if collaboration represents a promoting force for this particular change. The main aim of the study has been to step into a large void in academic research and to provide quantitative primary data, which so far has been lacking. The objective of this study was pursued by providing new measurement scales for collaboration in coworking spaces and the outcomes were supported by multiple regression analyses. The results revealed that strategic forms of collaboration represented a strong driver for innovation and changes within organizations´ business models. However, the generalizability of these findings implied some trade-offs which must be acknowledged.

First, as expected, the sample was characterized by a high diverse collection of firms. The ability to innovate, however, naturally varies amongst different types of firms and across different industries. Thus, to account for firm differences that could impact the relationship of interests, firm size was included as a control variable. However, though initially planned, industry could not be used as a control variable, as a high proportion of coworkers indicated to work in other industries (25.3%). In many studies on innovation the industry of firms is included as a control variable and hence could have biased the outcome of this study.

Second, the overall sample size included 75 organization, which is below the overall recommendation for regression analysis (Backhaus et al., 2016). High efforts were made to receive more recipients, though. All coworking spaces were visited in person, for instance, because this has proven to be a more effective method to convince managers to distribute the survey than contacting them by phone. Yet, out of 53 spaces only 17 were willing to distribute the survey to its members, mostly due to an extraordinary high number of requests.

Third, the data on collaboration and innovation was obtained from one single person for each firm. While some scholars question the validity of studies which rely on single informants (Lant, Milliken, & Batra, 1992), the approach has been commonly applied in innovation studies (Jennings & Young, 1990; Starbuck & Mezias, 1996). Usually the survey then is selectively addressed to CEO´s or top administrators, where there is little dispute that they can provide reliable information about the organizational occurrences. In this case, however, the survey was randomly forwarded to one person of each firm. Due to the fact, that the largest proportion of respondents (74.6%) recorded to be either one-person businesses or small firms with 2-9 employees, it can be assumed that every employee could provide trustworthy statements on behalf of their organization. However, it still remains a factor which could have biased the results of the study.

Fourth, to ensure comparability, guidelines on innovation studies recommend specifying the observation period for questions on innovation. The OECD (2005) for instance, recommends that the length of observation should uniformly lie between one and three years. Coworking spaces, however, are a comparatively new phenomenon which in addition are characterized by a high fluctuation rate amongst members. It appears to be extraordinary difficult to adhere to these guidelines within the context of coworking. To avoid strong biasing effects of time, however, the duration was included as a control variable.

Lastly, within innovation studies it is commonly problematic to classify what accounts to an innovation and what should instead be considered as an incremental improvement. Especially for questionnaires this is problematic, as no overall guidance during the process is possible. To minimize this bias, respondents were provided with both an overall definition of innovation and definitions for each type of innovation. Yet, participants may have had different opinions what accounts to an innovation and what not, which could have impacted the study´s results negatively.

Despite its limitations, it is crucial to acknowledge, however, that this study was the very first systematic empirical investigation that provided insights into how collaboration in coworking spaces contributes to firm innovativeness and impacts business models.

### 6.4 Future Research

The results of this study suggest several new directions of academic inquiry. First, this study highlighted the importance of differentiating forms of collaboration within coworking spaces. The study distinguished between a social and a strategic form of collaboration and the results clearly demonstrated that these forms have had different effects on innovativeness and business models. Future research may assure the existence of these two categories of collaboration with other statistical methods, such as F-Tests of difference in group means. Though the study of Freel and Jong (2009) is not customized to the context of coworking, it may provide a valuable orientation for such investigation. Further research may also find new measurement scales for coworkers´ collaborative behavior or may specifically focus on the tie formation processes within coworking spaces, similarly to investigations which had been conducted outside the context of coworking (Elfring & Hulsink, 2007).

The results of this study showed that collaboration in coworking spaces foster the access to external resources and serendipitous occurrences, which in literature are often discussed potential values for innovation activities. This study, however, could not find a significant contribution of external resources or serendipity to firm innovativeness in coworking spaces, leaving the concrete interrelation between these variables unexplored.

Nevertheless, the study represents an initial attempt to clarify whether collaboration in coworking spaces contributes to firm innovativeness and a positive business model development. While the present study has discovered a number of robust regression models, the scope of this line of inquiry could be extended in various directions. Future research may include other variables which contribute to innovativeness or impact business model development in coworking spaces. Studies may also build upon the findings of this study, by establishing structural equation models which explain more variance in firm innovativeness - the study only explained 25.2 to 33.6 percent of the variance in firm innovativeness. Research may also specify types of innovations in coworking spaces and may dedicate attention to what kind of innovation require collaborative alliances. Freel and Jong (2009) for instance, showed that firms intensively seek out for collaborative partnerships if the particular innovation is new to the market and requires new competencies which cannot be produced internally. Other studies may investigate the antecedents of collaboration, such as the environment or firm characteristics (Alexiev et al., 2016) or may investigate other effects of inter-organizational collaboration discussed in literature, such as organizational learning (Hardy, Phillips, & Lawrence, 2003).

Lastly, this study represented an initial attempt to discover the impact of coworking on business models. The results revealed the more organizations strategically collaborate in coworking spaces the higher they have judged the impact of coworking on their business model. However, the results still leave large voids in academic research that require further academic inquiries. Studies in this context may focus on business model design and find out whether coworking positively affects the novelty or efficiency of business models design. Similar investigations were made by Zott and Amit (2007) who showed that the novelty of business models substantially contributes to an increased entrepreneurial performance. Studies could also analyze the business model development of entrepreneurial organizations in coworking spaces and compare it to similar organizations outside of coworking spaces. This may provide insights into whether the services of coworking spaces beneficially affect the development of business models in comparison to non-members.

# CHAPTER 7: CONCLUSION

The overall goal of this study has been to examine whether and in what ways collaboration in coworking spaces can contribute to firm innovativeness and positively affect business models. The concept of coworking represents a newly emerged form of work space that is rapidly spreading and increasingly gaining attention by practitioners and scholars. Although the concept of a shared working environment is not new – telecommuting centers for instance have been present for decades – it appears that the impressive growth rate and global diffusion of coworking spaces are occurring due to contemporary shifts in economy and labor market (Moriset, 2013). The increasing demand for knowledge workers within the twenty-first century and current advances in technology, together with the rise of the so called sharing economy, where individuals generally tend to share underutilized resources in peer-to-peer networks, all strongly supported the rise of coworking (Bouncken & Reuschl, 2016; Cohen & Kietzmann, 2014; Gandini, 2015; Kojo & Nenonen, 2014). Coworking spaces can be defined as shared, flexible and collaborative office spaces used by a diverse group of professionals that offer its members multiple social, entrepreneurial and business-related opportunities or services. Coworking embeds its members into a resourceful community, which may help individuals to overcome professional isolation, but also offers members increased possibilities to extend their network and find collaborative alliances for mutual problem solving and innovation activities (Capdevila, 2015; Fuzi, 2015). While a great deal is known about the effects of collaboration in general, comparatively little research has been applied to the scope of coworking. Therefore, this study particularly sought to provide quantitative prove for the effects of collaboration in coworking spaces and aimed to answer four research questions.

First the study has been concerned with whether collaboration in coworking spaces grants members access to external resources (H1) as management literature highlights that resources play a key role for innovation processes and especially small firms increasingly face limitations regarding internal resources (Bjerke & Johansson, 2015). Second, the study investigated whether collaboration may be a vehicle for members to find or experience serendipitous occurrences (H2). Third, the study aimed to answer if collaboration significantly affects firm innovativeness (H3). Finally, if collaboration positively impacts the business models of organizations within coworking spaces (H4), as business model innovation has become a relevant alternative form of innovation in academic research. To answer these four questions the study internationally distributed a survey and within five months collected primary data from 75 organizations in multiple coworking spaces.

The results offered various important contributions and theoretical extensions to prior literature. The study has empirically demonstrated that collaboration in coworking spaces can catalyze superior access to external resources and contribute to serendipity. However, the most important contribution of this study has been to conclusively show that collaborative alliances in coworking spaces positively impact firm innovativeness. Supported by a strongly significant regression analysis, the study has revealed that the probability for organizations to be innovative has been approximately 12-times higher if they had entered into at least one strategic partnership during their engagement in coworking. Hence, the findings indicate that organizations can enhance their innovative capabilities by making use of collaborative alliances in coworking spaces and are consequently encouraged to actively seek out for such partnerships. Notably, the interrelation between strategic partnerships and firm innovativeness has not significantly been influenced by firm size, which may yield another interesting implication. Smaller organizations may be able to reach similar levels of innovativeness in coworking spaces if they engage in strategic partnerships and obtain the support that is required for their innovation activities. Finally, the study examined whether frequent social interactions or strategic collaborations with other members may improve or change the business model of organizations in coworking spaces. The results have shown that while social interactions with other coworkers revealed to not significantly contribute to improvements or changes of business models, strategic collaboration has been significant for all overarching dimensions of the business model. The more intensively organizations strategically collaborated, the higher they have judged the impact of coworking on their business model, suggesting that collaboration in coworking spaces is a promoting force for business model development or even business model innovation. Interestingly, the results have also revealed that while overall coworkers interact and communicate very frequently with each other, strategic collaborative alliances occur much more rarely. Hence, it appears that coworkers so far are not fully reaping the full potential of what coworking spaces offer from an entrepreneurial or business perspective.

Overall, this study has been one the very first systematic empirical investigation that has provided insights into how collaboration in coworking spaces contributes to firm innovativeness and impacts business models. However, the concept of coworking brings along several difficulties for innovation studies. Members of coworking spaces are extremely heterogenous regarding firm size and commonly represent multiple industries. The ability to innovate, however, naturally varies amongst different types of firms and across different industries (Fitjar & Rodriguez-Pose, 2014). Further, while a large proportion of coworkers represented entrepreneurial organizations that are concerned with the design of a novel business model, the sample also included larger corporations with an existing model. Hence, the diversity of organizations in coworking spaces significantly complicates matters for innovation studies.

Nevertheless, the results of this study imply that organizations in coworking spaces should embrace a collaborative culture and are encouraged to build a resourceful network by actively seeking out for strategic alliances. Strategic collaborative alliances and partnerships with organizations of the same industry, competitors, suppliers, customers, research institutions, or consultants will eventually allow coworkers to increase their innovativeness and improve or reconfigure their business models, which may likely yield financial returns or improve competitiveness.

The results of this study also suggest several new directions of academic inquiry. Future research may include other variables which contribute to or are affected by innovativeness or business model development, such as entrepreneurial performance, and establish conclusive structural equation models. Studies could also specify types of innovations in coworking spaces and may dedicate attention to what kind of innovative outcomes specifically require collaborative alliances. This may provide interesting insights, as studies outside the scope of coworking have shown that innovation which are new to the market increasingly require new competencies and collaborative partnerships (Freel & Jong, 2009). Future studies could also examine the role of coworking in respect to entrepreneurship by narrowly focusing on business model design and business model development. Helpful studies for such investigation were conducted by Zott and Amit (2007), who investigated the efficiency and novelty of business model design and showed that novelty substantially contributes to an increased entrepreneurial performance. Finally, future research could dedicate attention to coworking on a community level. Studies could investigate the level of interactions and collaborations of entire coworking communities and compare the number and quality of innovation from spaces to either other spaces or even larger corporations. This may provide insights into whether coworking spaces can be perceived as innovation networks, which previous scholars have argued (Capdevila, 2015).

Overall, this study can be considered as a first step into elaborating the benefits of coworking spaces, leaving the concrete value of coworking for innovation unknown. However, knowing how substantial the impacts of office (re)designs of larger firms, such as those from Google, Facebook or Telenor are, it appears firms have not fully explored or exploited the entire potential of coworking.

**REFERENCES**

Acs, Z., & Audretsch, D. 1988. Innovation in large and small firms: an empirical analysis. ***American Economic Review***, 4: 678.

Ahuja, G. 2000. The duality of collaboration: inducements and opportunities in the formation of interfirm linkages. ***Strategic Management Journal***, 21: 317–343.

Alexiev, A., Volberda, H., & van den Bosch, F. 2016. Interorganizational collaboration and firm innovativeness: Unpacking the role of the organizational environment. ***Journal of Business Research***, 69(2): 974–984.

Almeida, P., & Kogut, B. 1999. Localization and knowledge and the mobility of engineers in regional networks. ***Management Science***, 45: 905–917.

Amara, N., Landry, R., & Doloreux, D. 2009. Patterns of innovation in knowledge intensive business services. ***Service Industries Journal***, 29(4): 407–430.

Amit, R., & Zott, C. 2001. Value Creation in E-Business. ***Strategic Management Journal***, 22(6/7): 493–520.

Amit, R., & Zott, C. 2012. Creating value through business model innovation. ***MIT Sloan Management Review***, 53(3): 40–50.

Anand, B., & Khanna, T. 2000. Do firms learn to create value? The case of alliances. ***Strategic Management Journal***, 21(3): 295–315.

Appley, D., & Winder, A. 1977. An evolving definition of collaboration and some implications for the world of work. ***The Journal of Applied Behavioral Science***, 13(3): 279–291.

Assink, M. 2006. Inhibitors of disruptive innovation capability: a conceptual model. ***European Journal of Innovation Management***, 9(2): 215–233.

Avdikos, V., & Kalogeresis, A. 2016. Socio-economic profile and working conditions of freelancers in co-working spaces and work collectives: Evidence from the design sector in Greece. ***Area***, 49(1): 35–42.

Backhaus, K., Erichson, B., Plinke, W., & Weiber, R. 2016. ***Multivariate Analysemethoden***: ***Eine anwendungsorientierte Einführung*** (14th ed.). Berlin, Heidelberg: Springer Gabler.

Baum, J., Calabrese, T., & Silverman, B. S. 2000. Don’t Go It Alone: Alliance Network Composition and Startups’ Performance in Canadian Biotechnology. ***Strategic Management Journal***, 21(3): 267-294.

Bedwell, W., Wildman, J., DiazGranados, D., Salazar, M., Kramer, W., & Salas, E. 2012. Collaboration at work: An integrative multilevel conceptualization. ***Human Resource Management Review***, 22(2): 128–145.

Berends, H., Smits, A., Reymen, I., & Podoynitsyna, K. 2016. Learning while (re)configuring: Business model innovation processes in established firms. ***Strategic Organization***, 14(3): 181–219.

Berger, M., & Revilla Diez, J. 2006. Technological capabilities and innovation in Southeast Asia: results from innovation surveys in Singapore, Penang and Bangkok. ***Technology and Society***, 11(1): 109–148.

Bertels, H., Koen, P., & Elsum, I. 2015. Business models outside the core: Lessons learned from success and failure. ***Research Technology Management***, 58(2): 20–29.

Biemans, W. 1991. User and third-party involvement in developing medical equipment innovations. ***Technovation***, 11: 163–182.

Bjerke, L., & Johansson, S. 2015. Patterns of innovation and collaboration in small and large firms. ***The Annals of Regional Science***, 55(1): 221–247.

Bougrain, F., & Haudeville, B. 2002. Innovation, collaboration and SMEs internal research capacities. ***Research Policy***, 31: 735–747.

Bouncken, R., & Reuschl, A. 2016. Coworking-spaces: How a phenomenon of the sharing economy builds a novel trend for the workplace and for entrepreneurship. ***Review of Managerial Science***: 1–18.

Bucherer, E., Eisert, U., & Gassmann, O. 2012. Towards systematic business model innovation: lessons from product innovation management. ***Creativity and Innovation Management***, 21(2): 183–198.

Capdevila, I. 2013. Knowledge dynamics in localized communities: coworking spaces as microclusters. ***SSRN***: 1–18.

Capdevila, I. 2015. Co-working Spaces and the localised Dynamics of Innovation in Barcelona. ***International Journal of Innovation Management***, 19(3): 1–28.

Casadesus-Masanell, R., & Ricart, J. 2010. From strategy to business models and to tactics. ***Long Range Plan***, 43(2/3): 195–215.

Casadesus-Masanell, R., & Zhu, F. 2013. Business model innovation and competitive imitation: the case of sponsor-based business models. ***Strategic Management Journal***, 34(4): 464–482.

Chandy, R., & Tellis, G. 1998. Organizing for radical product innovation: the overlooked role of willingness to cannibalize. ***Journal of Marketing Research***, 35(4): 474–487.

Chatterjee, S., & Hadi, A. 2012. ***Regression analysis by example*** (5th ed.). Hoboken, NJ: Wiley.

Chesbrough, H. 2006. ***Open business models***: ***How to thrive in the new innovation landscape***. Boston, Mass.: Harvard Business School Press.

Chesbrough, H. 2010. Business Model Innovation: Opportunities and Barriers. ***Long Range Planning***, 43(2-3): 354–363.

Chesbrough, H., & Brunswicker, S. 2013. ***Managing Open Innovation in large firms.*** Frauenhofer Institut.

Chesbrough, H., & Rosenbloom, S. 2002. The role of the business model in capturing value from innovation: evidence from Xerox Corporation's technology spinoff companies. ***Industrial and Corporate Change***, 11(3): 529–555.

Clauss, T. 2016. Measuring business model innovation: conceptualization, scale development, and proof of performance. ***R&D Management***, 00(00): 1–19.

Cohen, B., & Kietzmann, J. 2014. Ride On! Mobility Business Models for the Sharing Economy. ***Organization & Environment***, 27(3): 279–296.

Cooke, P. 1996. The new wave of regional innovation networks: analysis, characteristics and strategy. ***Small Business Economics***, 8: 159–171.

Damanpour, F. 1992. Organizational size and innovation. ***Organization Studies***, 13(3): 375–402.

DellaPelle, D. 2016. ***Physical extensions of corporate culture.*** Cornell University, ILR School. Ithaca, New York.

Demil, B., & Lecocq, X. 2010. Business model evolution: in search of dynamic consistency. ***Long Range Planning***, 43(2-3): 227–246.

Dewangan, V., & Godse, M. 2014. Towards a holistic enterprise innovation performance measurement system. ***Technovation***, 34(9): 536–545.

Dibrell, C., Davis, P., & Craig, J. 2008. Fueling innovation through information technology in SMEs. ***Journal of Small Business Management***, 46(2): 203–218.

Dodgson, M. 1993. ***Technological collaboration in industry***: ***Strategy, policy and internationalization in innovation***. London: Routledge.

Dyer, J. H., & Nobeoka, K. 2000. Creating and managing a high-performance knowledge-sharing network: the Toyota case. ***Strategic Management Journal***, 21(3): 345–367.

Edison, H., bin Ali, N., & Torkar, R. 2013. Towards innovation measurement in the software industry. ***Journal of Systems and Software***, 86(5): 1390–1407.

Eisenhardt, K., & Schoonhoven, C. 1996. Resource-based view of strategic alliance formation: strategic and social effects in entrepreneurial firms. ***Organization Science***, 7: 136–150.

Elfring, T., & Hulsink, W. 2007. Networking by Entrepreneurs: Patterns of Tie Formation in Emerging Organizations. ***Organization Studies***, 28(12): 1849–1872.

Elg, U., & Johansson, U. 1997. Decision making in inter-firm networks as a political process. ***Organization Studies***, 18: 361–384.

Fabrigar, L. R., & Wegener, D. T. 2012. ***Exploratory factor analysis***. Oxford, New York: Oxford Univ. Press.

Faems, D., van Looy, B., & Debackere, K. 2005. Interorganizational Collaboration and Innovation: Toward a Portfolio Approach. ***The Journal of Product Innovation and Management***, 22: 238–250.

Fitjar, R., & Rodriguez-Pose, A. 2014. The geographical dimension of innovation collaboration: Networking and innovation in Norway. ***Urban Studies***, 51(12): 2572–2595.

Foertsch, C. More than one million people will work in coworking spaces in 2017; http://www.deskmag.com/en/the-complete-2017-coworking-forecast-more-than-one-million-people-work-from-14000-coworking-spaces-s, 17 Jul 2017.

Frankenberger, K., Weiblen, T., Csik, M., & Gassmann, O. 2013. The 4I–framework of business model innovation: a structured view on process phases and challenges. ***International Journal of Product Development***, 18(3): 249–273.

Freel, M., & Jong, J. de 2009. Market novelty, competence-seeking and innovation networking. ***Technovation***, 29(12): 873–884.

Frenken, K. 2000. A complexity approach to innovation networks: The case of the aircraft industry (1909-1997). ***Research Policy***, 29: 257–272.

Fruhling, A., & Keng, S. 2007. Assessing organizational innovation capability and its effect on e-commerce initiatives. ***Journal of Computer Information Systems***, 48(1): 133–145.

Fuzi, A. 2015. Co-working spaces for promoting entrepreneurship in sparse regions: The case of South Wales. ***Regional Studies***, 2(1): 462–469.

Gambardella, A., & McGahan, A. 2010. Business-model innovation: general purpose technologies and their implications for industry structure. ***Long Range Plan***, 43(2/3): 262–271.

Gandini, A. 2015. The rise of co-working spaces: A literature review. ***ephemera: theory & politics in organization***, 15(1): 193–205.

García-Gutiérrez, I., & Martínez-Borreguero, J. 2016. The Innovation Pivot Framework: Fostering Business Model Innovation in Startups. ***Research-Technology Management***, 59(5): 48–56.

Geiger, S. W., & Cashen, L. 2002. A multidimensional examination of slack and its impact on innovation. ***Journal of Managerial Issues***, 14(1): 68.

Gemünden, H. G., Ritter, T., & Heydebreck, P. 1996. Network configuration and innovation success: an empirical analysis in German high-tech industries. ***International Journal of Research in Marketing***, 13: 449–462.

George, G., & Bock, A. 2011. The business model in practice and its implications for entrepreneurship research. ***Entrepreneurship Theory and Practice***, 35(1): 83–111.

Gerwin, D., Kumar, V., & Pal, S. 1992. Transfer of Advanced Manufacturing Technology from Canadian Universities to Industry. ***Technology Transfer***, 12(2): 57–67.

Giesen, E., Riddleberger, E., Christner, R., & Bell, R. 2010. When and how to innovate your business model. ***Strategy & Leadership***, 38(4): 17–26.

Graham, J. R., & Barter, K. 1999. Collaboration: A social work practice method. , 6–13. ***Families in Society***: 6–13.

Grandori, A. 1997. An organizational assessment of interfirm coordination modes. ***Organization Studies***, 18: 897–925.

Grandori, A., & Soda, G. 1995. Inter-firm networks: antecedents, mechanisms and forms. ***Organization Studies***, 16: 183–214.

Gray, B. 1989. ***Collaborating: Finding common ground for multiparty problems***. San Francisco, CA: Jossey-Bass.

Grotz, R., & Braun, B. 1997. Territorial or trans-territorial networking: spatial aspects of technology-oriented co-operation within the German mechanical engineering industry. ***Regional Studies***, 31: 545–557.

Gulati, R., Nohria, N., & Zaheer, A. 2000. Strategic Networks. ***Strategic Management Journal***, 21(3): 203–215.

Hagedoorn, J. 1993. Understanding the Rationale of Strategic Technology Partnering: Interorganizational Modes of Cooperation and Industry Differences. ***Strategic Management Journal***, 14(5): 371–385.

Hagedoorn, J. 2002. Inter-firm R&D Partnerships: An Overview of Major Trends and Patterns since 1960. ***Research Policy***, 31(4): 477–492.

Hallwright, J., & Brady, K. 2016. 2015 Resilience Roundtable: co-working as a way of enhancing 2015 Resilience Roundtable: co-working as a way of enhancing collaboration in post-disaster environments. ***Australian Journal of Emergency Management***, 31(1): 41–45.

Hamel, G. 1991. Competition for Competence and inter-partner Learning within international Strategic Alliances. ***Strategic Management Journal***, 12(4): 83–103.

Hardwick, J., Anderson, A., & Cruickshank, D. 2013. Trust formation processes in innovative collaborations. ***European Journal of Innovation Management***, 16(1): 4–21.

Hardy, C., Phillips, N., & Lawrence, T. 2003. Resources, Knowledge and Influence: The Organizational Effects of Interorganizational Collaboration. ***Journal of Management Studies***, 40(2): 321–347.

Herrmann, A., Tomczak, T., & Befurt, R. 2006. Determinants of radical product innovations. ***European Journal of Innovation Management***, 9(1): 20–43.

Hippel, E. v. 1995. ***The sources of innovation***. New York, NY: Oxford Univ. Press.

Hippel, E. von, Thomke, S., & Sonnack, M. 1999. Creating Breakthroughs at 3M. ***Harvard Business Review***, 4: 47–57.

Hyun, J. H. 1994. Buyer supplier relations in the European automobile component industry. ***Long Range Planning***, 27(2): 66–75.

Innovatively. Business Model Canvas; https://wordpress-innovately.rhcloud.com/business-model-canvas/, 17 Jul 2017.

Izushi, H. 1997. Conflict between two industrial networks: technological adaptation and inter-firm relationships in the ceramics industry in Seto, Japan. ***Regional Studies***, 31: 117–129.

Jennings, D. F., & Young, D. M. 1990. An empirical comparison between objective and subjective measures of the product innovation domain of corporate entrepreneurship. ***Entrepreneurial Theory & Practice***, 15: 53–66.

Johnson, M. W., Christensen, C. M., & Kagermann, H. 2008. Reinventing your business model. ***Harvard Business Review***, 86(12): 50–59.

Katz, M. L., & Shapiro, C. 1985. Network externalities, competition, and compatibility. ***American Economic Review***, 75: 424–440.

Kenis, P., & Knoke, D. 2002. How organizational field networks shape interorganizational tie-formation rates. ***Academy of Management Review***, 27: 275–293.

Keyton, J., Ford, D. J., & Smith, F. I. 2008. A meso-level communicative model of collaboration. ***Communication Theory***, 18: 376–406.

Kline, D. 2003. Sharing the Corporate Crown Jewels. ***MIT Sloan Management Review***, 44(3): 88–93.

Knights, D., Murray, F., & Willmont, H. 1993. Networking as knowledge work. A study of strategic interorganizational development in the financial services industry. ***Journal of Management Studies***, 30: 975–996.

Kogut, B. 2000. The network as knowledge: generative rules and the emergence of structure. ***Strategic Management Journal***, 21(2): 405–425.

Kojo, I., & Nenonen, S. 2014. Evolution of co-working places: Drivers and possibilities. ***Intelligent Buildings International***, 6: 1–13.

Lant, T. K., Milliken, F. J., & Batra, B. 1992. The role of managerial learning and interpretation in strategic persistence and reorientation: An empirical exploration. ***Strategic Management Journal***, 13: 585–608.

Larson, A. 1991. Partner networks: leveraging external ties to improve entrepreneurial performance. ***Journal of Business Venturing***, 6: 173–188.

Laudien, S., & Daxböck, B. 2015. ***Antecedents and Outcomes of Collaborative Business Model Innovation.*** XXVI ISPIM Conference – Shaping the Frontiers of Innovation. Budapest, Hungary.

Lester, R. K., & Piore, M. J. 2006. ***Innovation***: ***The missing dimension***. Cambridge, Mass, London: Harvard University Press.

Liao, J., & Welsch, H. 2003. Social capital and entrepreneurial growth aspiration: a comparison of technology- and non-technology-based nascent entrepreneurs. ***Journal of High Technology Management Research***, 14: 149–170.

Linder, J., & Cantrell, S. 2000. ***Changing business models: surveying the landscape.*** Accenture Institute for Strategic Change.

Longoria, R. A. 2005. Is inter-organizational collaboration always a good thing? ***Journal of Sociology and Social Welfare***, 32(3): 123–138.

Markides, C. 2006. Disruptive innovation: in the need for better theory. ***Journal of Product Innovation Management***, 23(1): 19–25.

Montgomery, D. C., Peck, E. A., & Vining, G. G. 2013. ***Introduction to Linear Regression Analysis*** (5th ed.). Hoboken, New Jersey: Wiley.

Moriset, B. 2013. ***Building new places of the creative economy***: ***The rise of coworking spaces.*** University of Lyon. Lyon.

Morris, M., Schindehutte, M., & Allen, J. 2005. The entrepreneur's business model: toward a unified perspective. ***Journal of Business Research***, 58(6): 726–735.

Norman, D., & Verganti, R. 2014. Incremental and Radical Innovation: Design Research vs. Technology and Meaning Change. ***Massachusetts Institute of Technology Design Issues***, 30(1): 78–96.

OECD 2005. ***Oslo Manual***: ***Guidelines for collecting and interpreting innovation data***. Paris: OECD Publishing.

Osterwalder, A. 2004. ***The business model ontology: A proposition in a design science approach.*** PhD dissertation. Université de Lausane.

Osterwalder, A., & Pigneur, Y. 2013. ***Business model generation***: ***A handbook for visionaries, game changers, and challengers***. New York: Wiley&Sons.

Palmberg, C. 2004. The sources of innovations—looking beyond technological opportunities. ***Economics of Innovation & New Technology***, 13(2): 1.

Parrino, L. 2015. Coworking: Assessing the role of proximity in knowledge exchange. ***Knowledge Management Research & Practice***, 13(3): 261–271.

Pittaway, L., Robertson, M., Munir, K., Denyer, D., & Neely, A. 2004. Networking and innovation: a systematic review of the evidence. ***International Journal of Management Reviews***, 5/6(3&4): 137–168.

Plummer, R., & Fennell, D. 2007. Exploring co-management theory: Prospects for sociobiology and reciprocal altruism. ***Journal of Environmental Management***, 85: 944–955.

Powel, W., Koput, K., & Smith-Doerr, L. 1996. Interorganizational Collaboration and the Locus of Innovation: Networks of Learning in Biotechnology. ***Administrative Science Quarterly***, 41(1): 116–145.

Quinn, J. B. 1985. Managing Innovation: Controlled Chaos. ***Harvard Business Review***, 63(3): 73–84.

Romijn, H., & Albaladejo, M. 2002. Determinants of innovation capability in small electronics and software firms in southeast England. ***Research Policy***, 31: 1053–1067.

Ryu, S. 2014. Networking Partner Selection and Its Impact on the Perceived Success of Collaboration. ***Public Performance & Management Review***, 37(4): 632–657.

Santoro, M. D. 2000. Success Breeds Success: The Linkage between Relationship Intensity and Tangible Outcomes in Industry–University Collaborative Ventures. ***Journal of High Technology Management Research***, 11(2): 255–273.

Schindehutte, M., Morris, M. H., & Kocak, A. 2008. Understanding market-driving behavior: the role of entrepreneurship, 46(1): 4–26.

Schopfel, J., Roche, J., & Hubert, G. 2015. Co-working and innovation: New concepts for academic libraries and learning centres. ***New Library World***, 116(1/2): 67–78.

Schumpeter, J. A. 2012. ***The theory of economic development***: ***An inquiry into profits, capital, credit, interest, and the business cycle*** (16th ed.). New Brunswick, NJ: Transaction Publ.

Selin, S., & Chavez, D. 1995. Developing a collaborative model for environmental planning and management. ***Environmental Management***, 19(2): 189–195.

Shapiro, C., & Varian, H. R. 1999. ***Information Rules: A Strategic Guide to the Network Economy***. Boston, MA: Harvard Business School Press.

Shaw, B. 1993. Formal and informal networks in the UK medical equipment industry. ***Technovation***, 13: 349–365.

Shaw, B. 1996. User/Supplier Links and Innovation. In M. Dodgson & R. Rothwell (Eds.), ***The handbook of industrial innovation***. Cheltenham: Elgar.

Singh, A., & Singh, V. 2009. Innovation in services: design and management. ***African Journal of Business Management***, 3(12): 871–878.

Skarzynski, P., & Gibson, R. 2008. ***Innovation to the core: a blueprint for transforming the way your company innovates***. Boston: Harvard Business School Press.

Soerjoatmodjo, G. W. L., Bagasworo, D. W., Joshua, G., Kalesaran, T., & van den Broek, K. F. 2015. ***Sharing Workspace, Sharing Knowledge: Knowledge Sharing Amongst Entrepreneurs in Jakarta Co-Working Spaces.*** Pembangunan Jaya University. Tangerang Selatan, Indonesia.

Spieth, P., Ricart, J. E., & Schneckenberg, D. 2014. Business model innovation—state of the art and future challenges for the field. ***R&D Management Journal***, 44(3): 237–247.

Spieth, P., & Schneider, S. 2016. Business model innovativeness: Designing a formative measure for business model innovation. ***Journal of Business Economics***, 86(6): 671–696.

Spinuzzi, C. 2012. Working Alone Together: Coworking as Emergent Collaborative Activity. ***Journal of Business and Technical Communication***, 26(4): 399–441.

Spreitzer, G., Garrett, L., & Bacevice, P. 2015. Should Your Company Embrace Coworking? ***MIT Sloan Management Review***, 57(1): 27–29.

Stampfl, G. 2016. ***The Process of Business Model Innovation***. Wiesbaden: Springer Fachmedien Wiesbaden.

Starbuck, W. H., & Mezias, J. M. 1996. Opening Pandora’s box: Studying the accuracy o f manager’s perceptions. ***Journal of Organizational Behavior***, 17: 99–177.

Streb, J. 2003. Shaping the national system of interindustry knowledge exchange: vertical integration, licensing and repeated knowledge transfer in the German plastics industry. ***Research Policy***, 32: 1125–1140.

Stuart, T. E. 2000. Interorganizational Alliances and the Performance of Firms: A Study of Growth and Innovation Rates in a High-Technology Industry. ***Strategic Management Journal***, 21(8): 791–811.

Sundaramurthy, C., & Lewis, M. 2003. Control and collaboration: Paradoxes of governance. ***Academy of Management Review***, 28: 397–415.

Surman, T. 2013. Building Social Entrepreneurship through the Power of Coworking. ***Innovation: Technology, Governance, Globalization***, 8(3/4): 189–195.

Tadashi 2013. ***What is Coworking? A Theoretical Study on the Concept of Coworking*** no. 265. Hokkaido University. Sapporo, Japan.

Teece, D. J. 1986. Profiting from Technological Innovation: Implications for Integration, Collaboration, Licensing, and Public Policy. ***Research Policy***, 15(6): 285–305.

Teece, D. J. 2010. Business models, business strategy and innovation. ***Long Range Plan***, 43(2/3): 172–194.

Trauffler, G., & Tschirky, H. 2007. ***Sustained innovation management***: ***Assimilating radical and incremental innovation management***. Basingstoke, England: Palgrave Macmillan in association with the European Institute for Technology and Innovation Management.

Veugelers, R. 1998. Collaboration in R&D: An Assessment of Theoretical and Empirical Findings. ***Economist***, 149(3): 419–443.

Waber, B., Magnolfi, J., & Lindsay, G. 2014. Workspaces That Move People: Today’s offices don’t encourage us to mingle-but that’s what creativity and productivity demand. ***Harvard Business Review***, 69(October): 69–77.

Wood, D. J., & Gray, B. 1991. Toward a comprehensive theory of collaboration. ***Journal of Applied Behavioral Science***, 27(2): 139–162.

Yunus, M., Moingeon, B., & Lehmann-Ortega, L. 2010. Building social business models: lessons from the Grameen experience. ***Long Range Plan***, 43(2/3): 308–325.

Zott, C., & Amit, R. 2007. Business Model Design and the Performance of Entrepreneurial Firms. ***Organization Science***, 18(2): 181–199.

Zott, C., Amit, R., & Massa, L. 2011. The business model: recent developments and future research. ***Journal of Management***, 37(4): 1019–1042.

# APPENDICES

## APPENDIX A: Survey

The survey was addressed to one person of each organization. Respondents represented their organization. After a short letter of introduction, which explained the purpose and overall structure, the survey included 16 questions. Q1-2 were uniform for everybody. For Q3-16, respondents followed two different paths. Therefore, Q2 asked respondents if they work individually as a freelancer or solopreneur, if they work for a small organization (2-9) workers, or if they work for an established organization (more than 10 workers). Path A addressed freelancers and solopreneurs, whereas path B addressed small organizations and established companies.  Importantly, however, the questions Q3-16 were completely identical, apart from the fact that individual organizations were addressed in first-person narrative ("I…") and small organizations and established companies talked on behalf their team ("Our team…").

Many questions were measured on a seven-point-Likert scale. For all questions, the following Likert scale was used:

| **(1) Strongly agree** | **(2) Agree** | **(3) Some-what agree** | **(4) Neither agree nor disagree** | **(5) Some-what disagree** | **(6) Disagree** | **(7) Strongly disagree** | **(8) I do not know** |
|---|---|---|---|---|---|---|---|
| • | • | • | • | • | • | • | • |

The following section shows the entire survey.

## Letter of Introduction

Dear participants,

this research project is conducted in collaboration with the University of Bayreuth in Germany. With the survey, we are trying to collect research data on coworking spaces. The survey consists of four parts:

Part I: Introduction

Part II: Networking and Collaboration Behavior

Part III: Innovation

Part IV: Business Model

The duration should not extend 10 minutes.

Your data will be anonymous and confidential.

Thank you for your help and contribution!

## Part I: Introduction

**Q1: What industry do you work in?**

**Q2: What do you work as/for?**

(If multiple apply please select the option you spend the majority of your time with while working at the coworking space)

- I work as a freelancer or solopreneur.
- I work for a small organization (2-9 workers).
- I work for an established company (more than 10 workers).

Path A: Freelancer or Solopreneur

**Q3: How many different coworking spaces have you worked in?**

**Q4: What coworking space do you work in?**

**Q5: When have you joined this coworking space?**

Year

Month

**Q6: Is the coworking space your primary workplace?**

- Yes
- No

**Q7: How many hours a week do you regularly work at the coworking space?**

## Part II: Networking and Collaboration Behavior

**Q8: The following questions address how frequently you interact and network with other members at the coworking space. How much do you agree with the following statements?**

| | (1) | (2) | (3) | (4) | (5) | (6) | (7) | (8) |
|---|---|---|---|---|---|---|---|---|
| I regularly communicate with other members. | • | • | • | • | • | • | • | • |
| I regularly interact with other members. | • | • | • | • | • | • | • | • |
| I regularly share opinions with other members. | • | • | • | • | • | • | • | • |
| I build or sustain relationships with members. | • | • | • | • | • | • | • | • |

**Q9: The following questions address how diverse the members are you interact with at the coworking space. How much do you agree with the following statements?**

| | (1) | (2) | (3) | (4) | (5) | (6) | (7) | (8) |
|---|---|---|---|---|---|---|---|---|
| I interact with members of various occupations. | • | • | • | • | • | • | • | • |
| I interact with people of various backgrounds. | • | • | • | • | • | • | • | • |
| I interact with members that have various skills. | • | • | • | • | • | • | • | • |
| I interact with members of different industries. | • | • | • | • | • | • | • | • |
| I mostly interact with members that have the same occupation. | • | • | • | • | • | • | • | • |

**Q10: The following questions address in what way you collaborate with other members. How much do you agree with the following statements?**

| | (1) | (2) | (3) | (4) | (5) | (6) | (7) | (8) |
|---|---|---|---|---|---|---|---|---|
| I enter partnerships with other members. | • | • | • | • | • | • | • | • |
| I integrate other members into my business. | • | • | • | • | • | • | • | • |
| I co-develop projects with other members. | • | • | • | • | • | • | • | • |
| I collaborate with other members to achieve specific goals. | • | • | • | • | • | • | • | • |

**Q11: Have you entered into strategic partnerships or collaborated with any of the following due to coworking:**

- Competitors of the same industry
- People or firms of different industries
- Suppliers
- Customers
- Research institutions
- Consultants
- None of them

**Q12: The following questions refer to the benefits of coworking. How much do you agree with the following statements?**

| | (1) | (2) | (3) | (4) | (5) | (6) | (7) | (8) |
|---|---|---|---|---|---|---|---|---|
| I gained access to external resources which I otherwise would not have had access to. | • | • | • | • | • | • | • | • |
| I gained access to external knowledge. | • | • | • | • | • | • | • | • |
| I gained access to external skills. | • | • | • | • | • | • | • | • |
| I gained access to external intellectual properties. | • | • | • | • | • | • | • | • |

**Q13: How much do you agree with the following statements?**

| | (1) | (2) | (3) | (4) | (5) | (6) | (7) | (8) |
|---|---|---|---|---|---|---|---|---|
| I gained entirely new ideas and inspirations. | • | • | • | • | • | • | • | • |
| I gained entirely new impressions and thoughts. | • | • | • | • | • | • | • | • |
| I made entirely new discoveries. | • | • | • | • | • | • | • | • |
| I discovered entirely new opportunities. | • | • | • | • | • | • | • | • |

## Part III: Innovation

**Q14: An Innovation is the implementation of a new or significantly improved product, service, process, or marketing method. The following questions address whether you have launched any innovations during your coworking engagement.**

- I have introduced new or significantly improved products or services. (A product or service innovation includes significant improvements in technical specifications, components and materials, incorporated software, user friendliness or other functional characteristics)
- I have introduced new or significantly improved processes. (A process innovation is the implementation of a new or significantly improved production or delivery method. This includes significant changes in techniques, equipment and/or software)
- I have introduced new or significantly improved marketing methods. (A marketing innovation is the implementation of a new marketing method involving significant changes in product design or packaging, product placement, promotion or pricing)
- None of them apply.

If None of them apply. Is Selected, Then Skip To Part IV: Business Model The followin...

**Q15: Innovations can be new to a firm or even new to a market. Please select the appropriate option.**

- The innovations were only new to the firm.
- The innovations were also new to the market.

## Part IV: Business Model

**Q16: The following questions address what impact coworking had on the following nine building stones of your business model. How much do you agree with the following statements?**

| | (1) | (2) | (3) | (4) | (5) | (6) | (7) | (8) |
|---|---|---|---|---|---|---|---|---|
| (1) Key Partners: Through coworking I found new collaborative partnerships that helped me to develop my business. | • | • | • | • | • | • | • | • |
| (2) Key Activities: Coworking helped me to significantly improve my internal processes by making them more innovative and/or efficient or coworking changed my internal processes completely. | • | • | • | • | • | • | • | • |
| (3) Key Resources: Through coworking I was able to include new technical, intellectual or financial resources into my business. | • | • | • | • | • | • | • | • |
| (4) Value Proposition: Through coworking I significantly improved my existing products/services or coworking helped me to find new value propositions to address unmet customer needs. | • | • | • | • | • | • | • | • |
| (5) Channels: Through coworking I increased the efficiency of my existing distribution and communication channels or coworking helped me to find new channels for my value propositions. | • | • | • | • | • | • | • | • |
| (6) Customer Segments: Through coworking I was able to expand my customer segments or coworking helped me to address new customer segments within the market. | • | • | • | • | • | • | • | • |
| (7) Customer Relationships: Through coworking I significantly increased customer retention or strengthened customer relationships. | • | • | • | • | • | • | • | • |
| (8) Revenue Streams: Through coworking I significantly improved my existing revenue model or coworking helped me to develop new revenue streams. | • | • | • | • | • | • | • | • |
| (9) Cost Structure: Through coworking I found ways to save costs (e.g. through economies of scale and scope) or coworking initiated a complete change within the structure of my cost mechanism. | • | • | • | • | • | • | • | • |

Path B: Small organization (2-9 workers) and established company (more than 10 workers)

**Q3: How many different coworking spaces have you and your team worked in?**

**Q4: What coworking space do you and your team work in?**

**Q5: When has your team first joined this coworking space?**

Year

Month

**Q6: Is the coworking space the primary workplace for your team?**

- Yes
- No

**Q7: How many hours a week does your team regularly work at the coworking space?**

## Part II: Networking and Collaboration Behavior

**Q8: The following questions address how frequently your team interacts and networks with other members or organizations. How much do you agree with the following statements?**

| | (1) | (2) | (3) | (4) | (5) | (6) | (7) | (8) |
|---|---|---|---|---|---|---|---|---|
| Our team regularly communicates with other members. | • | • | • | • | • | • | • | • |
| Our team regularly interacts with other members. | • | • | • | • | • | • | • | • |
| Our team regularly shares opinions with other members. | • | • | • | • | • | • | • | • |
| Our team builds or sustains relationships with other members. | • | • | • | • | • | • | • | • |

**Q9: The following questions address how diverse the members or organizations are your team interacts with. How much do you agree with the following statements?**

| | (1) | (2) | (3) | (4) | (5) | (6) | (7) | (8) |
|---|---|---|---|---|---|---|---|---|
| Our team interacts with members of various occupations. | • | • | • | • | • | • | • | • |
| Our team interacts with people of various backgrounds. | • | • | • | • | • | • | • | • |
| Our team interacts with members that have various skills. | • | • | • | • | • | • | • | • |
| Our team interacts with members of different industries. | • | • | • | • | • | • | • | • |
| Our team interacts with members of different industries. | • | • | • | • | • | • | • | • |

**Q10: The following questions address in what way your team collaborates with other members or organizations. How much do you agree with the following statements?**

| | (1) | (2) | (3) | (4) | (5) | (6) | (7) | (8) |
|---|---|---|---|---|---|---|---|---|
| Our team enters partnerships with other members. | • | • | • | • | • | • | • | • |
| Our team integrates other members into internal business practices. | • | • | • | • | • | • | • | • |
| Our team co-develops projects with other members. | • | • | • | • | • | • | • | • |
| Our team collaborates with other members to achieve specific goals. | • | • | • | • | • | • | • | • |

**Q11: Has your organization entered into strategic partnerships or collaborated with any of the following due to coworking:**

- Competitors of the same industry
- People or firms of different industries
- Suppliers
- Customers
- Research institutions
- Consultants
- None of them

**Q12: The following questions refer to the benefits of coworking. How much do you agree with the following statements?**

| | (1) | (2) | (3) | (4) | (5) | (6) | (7) | (8) |
|---|---|---|---|---|---|---|---|---|
| Our team gained access to external resources which we otherwise would not have had access to. | • | • | • | • | • | • | • | • |
| Our team gained access to external knowledge. | • | • | • | • | • | • | • | • |
| Our team gained access to external skills. | • | • | • | • | • | • | • | • |
| Our team gained access to external intellectual properties. | • | • | • | • | • | • | • | • |

**Q13: How much do you agree with the following statements?**

| | (1) | (2) | (3) | (4) | (5) | (6) | (7) | (8) |
|---|---|---|---|---|---|---|---|---|
| Our team gained entirely new ideas and inspirations. | • | • | • | • | • | • | • | • |
| Our team gained entirely new impressions and thoughts. | • | • | • | • | • | • | • | • |
| Our team made entirely new discoveries. | • | • | • | • | • | • | • | • |
| Our team discovered entirely new opportunities. | • | • | • | • | • | • | • | • |

## Part III: Innovation

**Q14: An Innovation is the implementation of a new or significantly improved product, service, process, or marketing method. The following questions address whether your organization has launched innovations during the coworking engagement.**

- Our organization has introduced new or significantly improved products or services. (A product or service innovation includes significant improvements in technical specifications, components and materials, incorporated software, user friendliness or other functional characteristics)
- Our organization has introduced new or significantly improved processes. (A process innovation is the implementation of a new or significantly improved production or delivery method. This includes significant changes in techniques, equipment and/or software)
- Our organization has introduced new or significantly improved marketing methods. (A marketing innovation is the implementation of a new marketing method involving significant changes in product design or packaging, product placement, promotion or pricing)
- None of them apply.

If None of them apply. Is Selected, Then Skip To Part IV: Business Model The followin...

**Q15: Innovations can be new to a firm or even new to a market. Please select the appropriate option.**

- The innovations were only new to the firm.
- The innovations were also new to the market.

## Part IV: Business Model

**Q16: The following questions address what impact coworking had on the following nine building stones of your organization's business model. How much do you agree with the following statements?**

| | (1) | (2) | (3) | (4) | (5) | (6) | (7) | (8) |
|---|---|---|---|---|---|---|---|---|
| (1) Key Partners: Through coworking our organization found new collaborative partnerships that helped us to develop our business. | • | • | • | • | • | • | • | • |
| (2) Key Activities: Coworking helped our organization to significantly improve our internal processes by making them more innovative and/or efficient or coworking changed our internal processes completely. | • | • | • | • | • | • | • | • |
| (3) Key Resources: Through coworking our organization was able to include new technical, intellectual or financial resources into our business. | • | • | • | • | • | • | • | • |
| (4) Value Proposition: Through coworking our organization significantly improved existing products/services or coworking helped us to find new value propositions to address unmet customer needs. | • | • | • | • | • | • | • | • |
| (5) Channels: Through coworking our organization increased the efficiency of our existing distribution and communication channels or coworking helped us to find new channels for our value propositions. | • | • | • | • | • | • | • | • |
| (6) Customer Segments: Through coworking our organization was able to expand our customer segments or coworking helped us to address new customer segments within the market. | • | • | • | • | • | • | • | • |
| (7) Customer Relationships: Through coworking our organization significantly increased customer retention or strengthened customer relationships. | • | • | • | • | • | • | • | • |
| (8) Revenue Streams: Through coworking our organization significantly improved our existing revenue model or coworking helped us to develop new revenue streams. | • | • | • | • | • | • | • | • |
| (9) Cost Structure: Through coworking our organization found ways to save costs (e.g. through economies of scale and scope) or coworking initiated a complete change within the structure of our cost mechanism. | • | • | • | • | • | • | • | • |

## APPENDIX B: Descriptive Statistics

The following sections outlines the descriptive statistics. All forms of organizations were treated uniformly. Thus, path A and B from the survey were re-merged.

### *Dependent Variables*

*Access to external Resources (AccRes)*

This variable refers to Q12.

| | **AccRes_1** | | **AccRes_2** | | **AccRes_3** | | **AccRes_4** | |
|---|---|---|---|---|---|---|---|---|
| | **N** | **%** | **N** | **%** | **N** | **%** | **N** | **%** |
| (1) Strongly agree | 6 | 8.0 | 4 | 5.3 | 4 | 5.3 | 0 | 0.0 |
| (2) Agree | 25 | 33.3 | 31 | 41.3 | 31 | 41.3 | 9 | 12.0 |
| (3) Somewhat agree | 18 | 24.0 | 18 | 24.0 | 16 | 21.3 | 15 | 20.0 |
| (4) Neither agree nor disagree | 9 | 12.0 | 5 | 6.7 | 7 | 9.3 | 19 | 25.3 |
| (5) Somewhat disagree | 1 | 1.3 | 3 | 4.0 | 3 | 4,0 | 4 | 5.3 |
| (6) Disagree | 14 | 18.7 | 12 | 16.0 | 13 | 17.3 | 20 | 26.7 |
| (7) Strongly Disagree | 2 | 2.7 | 2 | 2.7 | 1 | 1.3 | 7 | 9.3 |
| (8) I do not know | 0 | 0.0 | 0 | 0.0 | 0 | 0.0 | 1 | 1.3 |
| Total | 75 | 100.0 | 75 | 100.0 | 75 | 100.0 | 75 | 100.0 |
| Mean | 3.32 | | 3.31 | | 3.23 | | 4.48 | |
| Standard deviation | 1.694 | | 1.694 | | 1.624 | | 1.630 | |

*Serendipitous Occurrences (SerOcc)*

This variable refers to Q13.

| Serendipitous Occurrences | SerOcc_1 | | SerOcc_2 | | SerOcc_3 | | SerOcc_4 | |
|---|---|---|---|---|---|---|---|---|
| | N | % | N | % | N | % | N | % |
| (1) Strongly agree | 6 | 8.0 | 6 | 8.0 | 4 | 5.3 | 8 | 10.7 |
| (2) Agree | 20 | 26.7 | 25 | 33.3 | 17 | 22.7 | 17 | 22.7 |
| (3) Somewhat agree | 25 | 33.3 | 18 | 24.0 | 19 | 25.3 | 18 | 24.0 |
| (4) Neither agree nor disagree | 8 | 10.7 | 9 | 12.0 | 13 | 17.3 | 10 | 13.3 |
| (5) Somewhat disagree | 4 | 5.3 | 11 | 14.7 | 9 | 12.0 | 10 | 13.3 |
| (6) Disagree | 11 | 14.7 | 5 | 6.7 | 11 | 14.7 | 10 | 13.3 |
| (7) Strongly Disagree | 1 | 1.3 | 1 | 1.3 | 2 | 2.7 | 2 | 2.7 |
| (8) I do not know | 0 | 0.0 | 0 | 0.0 | 0 | 0.0 | 0 | 0.0 |
| Total | 75 | 100.0 | 75 | 100.0 | 75 | 100.0 | 75 | 100.0 |
| Mean | 3.28 | | 3.17 | | 3.63 | | 3.47 | |
| Standard deviation | 1.547 | | 1.474 | | 1.575 | | 1.663 | |

*Firm Innovativeness (FirInn)*

This variable refers to Q14.

| | Yes | | No | |
|---|---|---|---|---|
| | N | % | N | % |
| Product or Service Innovation | 27 | 36.0 | 48 | 64.0 |
| Process Innovation | 18 | 24.0 | 57 | 76.0 |
| Marketing Innovation | 14 | 18.7 | 61 | 81.3 |
| No Innovation | 36 | 48.0 | 39 | 52.0 |

Respondents were classified into ‘0 – Not innovative and ‘1 – Innovative’.

| Firm Innovativeness | N | % |
|---|---|---|
| Not innovative | 36 | 48.0 |
| Innovative | 39 | 52.0 |
| Total | 75 | 100.0 |

*Impact on Business Model (BusMod)*

This variable refers to Q16.

| Value Creation Architecture | Key Partners | | Key Activities | | Key Resources | |
|---|---|---|---|---|---|---|
| | N | % | N | % | N | % |
| (1) Strongly agree | 2 | 2.7 | 2 | 2.7 | 2 | 2.7 |
| (2) Agree | 17 | 22.7 | 10 | 13.3 | 13 | 17.3 |
| (3) Somewhat agree | 11 | 14.7 | 15 | 20.0 | 14 | 18.7 |
| (4) Neither agree nor disagree | 10 | 13.3 | 22 | 29.3 | 15 | 20.0 |
| (5) Somewhat disagree | 3 | 4.0 | 6 | 8.0 | 6 | 8.0 |
| (6) Disagree | 25 | 33.3 | 16 | 21.3 | 20 | 26.7 |
| (7) Strongly Disagree | 7 | 9.3 | 4 | 5.3 | 5 | 6.7 |
| (8) I do not know | 0 | 0.0 | 0 | 0.0 | 0 | 0.0 |
| Total | 75 | 100.0 | 75 | 100.0 | 75 | 100.0 |
| Mean | 4.31 | | 4.12 | | 4.20 | |
| Standard deviation | 1.860 | | 1.551 | | 1.693 | |

| Value Offering | Value Proposition | | Channels | | Customer Segments | | Customer Relationships | |
|---|---|---|---|---|---|---|---|---|
| | N | % | N | % | N | % | N | % |
| (1) Strongly agree | 2 | 2.7 | 1 | 1.3 | 10 | 13.3 | 2 | 2.7 |
| (2) Agree | 8 | 10.7 | 5 | 6.7 | 8 | 10.7 | 1 | 1.3 |
| (3) Somewhat agree | 14 | 18.7 | 17 | 22.7 | 16 | 21.3 | 16 | 21.3 |
| (4) Neither agree nor disagree | 11 | 14.7 | 17 | 22.7 | 9 | 12.0 | 19 | 25.3 |
| (5) Somewhat disagree | 10 | 13.3 | 5 | 6.7 | 26 | 34.7 | 4 | 5.3 |
| (6) Disagree | 22 | 29.3 | 23 | 30.7 | 5 | 6.7 | 27 | 36.0 |
| (7) Strongly Disagree | 6 | 8.0 | 6 | 8.0 | 1 | 1.3 | 4 | 5.3 |
| (8) I do not know | 2 | 2.7 | 1 | 1.3 | 0 | 0.0 | 2 | 2.7 |
| Total | 75 | 100.0 | 75 | 100.0 | 75 | 100.0 | 75 | 100.0 |
| Mean | 4.59 | | 4.57 | | 4.69 | | 4.72 | |
| Standard deviation | 1.733 | | 1.595 | | 1.585 | | 1.556 | |

| Revenue Model | Revenue Streams | | Cost Structure | |
|---|---|---|---|---|
| | **N** | **%** | **N** | **%** |
| (1) Strongly agree | 1 | 1.3 | 6 | 8.0 |
| (2) Agree | 1 | 1.3 | 14 | 18.7 |
| (3) Somewhat agree | 11 | 14.7 | 12 | 16.0 |
| (4) Neither agree nor disagree | 23 | 30.7 | 15 | 20.0 |
| (5) Somewhat disagree | 5 | 6.7 | 5 | 6.7 |
| (6) Disagree | 22 | 29.3 | 19 | 25.3 |
| (7) Strongly Disagree | 10 | 13.3 | 3 | 4.0 |
| (8) I do not know | 2 | 2.7 | 1 | 1.3 |
| Total | 75 | 100.0 | 75 | 100.0 |
| Mean | 4.95 | | 3.97 | |
| Standard deviation | 1.524 | | 1.823 | |

| **Business Model** | **(1) Strongly agree** | | **(2) Agree** | | **(3) Somewhat agree** | | **(4) Neither agree nor disagree** | | **(5) Somewhat disagree** | | **(6) Disagree** | | **(7) Strongly Disagree** | | **(8) I do not know** | |
|---|---|---|---|---|---|---|---|---|---|---|---|---|---|---|---|---|
| | **N** | **%** | **N** | **%** | **N** | **%** | **N** | **%** | **N** | **%** | **N** | **%** | **N** | **%** | **N** | **%** |
| Key Partners | 2 | 2.7 | 17 | 22.7 | 11 | 14.7 | 10 | 13.3 | 3 | 4.0 | 25 | 33.3 | 7 | 9.3 | 0 | 0.0 |
| Key Activities | 2 | 2.7 | 10 | 13.3 | 15 | 20.0 | 22 | 29.3 | 6 | 8.0 | 16 | 21.3 | 4 | 5.3 | 0 | 0.0 |
| Key Resources | 2 | 2.7 | 13 | 17.3 | 14 | 18.7 | 15 | 20.0 | 6 | 8.0 | 20 | 26.7 | 5 | 6.7 | 0 | 0.0 |
| Value Proposition | 2 | 2.7 | 8 | 10.7 | 14 | 18.7 | 11 | 14.7 | 10 | 13.3 | 22 | 29.3 | 6 | 8.0 | 2 | 2.7 |
| Channels | 1 | 1.3 | 5 | 6.7 | 17 | 22.7 | 17 | 22.7 | 5 | 6.7 | 23 | 30.7 | 6 | 8.0 | 1 | 1.3 |
| Customer Segments | 10 | 13.3 | 8 | 10.7 | 16 | 21.3 | 9 | 12.0 | 26 | 34.7 | 5 | 6.7 | 1 | 1.3 | 0 | 0.0 |
| Customer Relationships | 2 | 2.7 | 1 | 1.3 | 16 | 21.3 | 19 | 25.3 | 4 | 5.3 | 27 | 36.0 | 4 | 5.3 | 2 | 2.7 |
| Revenue Streams | 1 | 1.3 | 1 | 1.3 | 11 | 14.7 | 23 | 30.7 | 5 | 6.7 | 22 | 29.3 | 10 | 13.3 | 2 | 2.7 |
| Cost Structure | 6 | 8.0 | 14 | 18.7 | 12 | 16.0 | 15 | 20.0 | 5 | 6.7 | 19 | 25.3 | 3 | 4.0 | 1 | 1.3 |

## ***Independent Variables***

*Social Interaction (SocInt)*

This variable refers to Q8.

| **Social Interaction** | **SocInt_1** | | **SocInt_2** | | **SocInt_3** | | **SocInt_4** | |
|---|---|---|---|---|---|---|---|---|
| | **N** | **%** | **N** | **%** | **N** | **%** | **N** | **%** |
| (1) Strongly agree | 15 | 20.0 | 14 | 18.7 | 9 | 12.0 | 9 | 12.0 |
| (2) Agree | 24 | 32.0 | 22 | 29.3 | 14 | 18.7 | 20 | 26.7 |
| (3) Somewhat agree | 19 | 25.3 | 22 | 29.3 | 24 | 32.0 | 19 | 25.3 |
| (4) Neither agree nor disagree | 3 | 4.0 | 6 | 8.0 | 9 | 12.0 | 7 | 9.3 |
| (5) Somewhat disagree | 10 | 13.3 | 7 | 9.3 | 11 | 14.7 | 9 | 12.0 |
| (6) Disagree | 3 | 4.0 | 3 | 4.0 | 5 | 6.7 | 9 | 12.0 |
| (7) Strongly Disagree | 1 | 1.3 | 1 | 1.3 | 3 | 4.0 | 2 | 2.7 |
| (8) I do not know | 0 | 0.0 | 0 | 0.0 | 0 | 0.0 | 0 | 0.0 |
| Total | 75 | 100.0 | 75 | 100.0 | 75 | 100.0 | 75 | 100.0 |
| Mean | 2.76 | | 2.77 | | 3.35 | | 3.29 | |
| Standard deviation | 1.496 | | 1.429 | | 1.590 | | 1.667 | |

*Strategic Collaboration (StrCol)*

This variable refers to Q10.

| **Strategic Collaboration** | **SraCol_1** | | **StrCol_2** | | **StrCol_3** | | **StrCol_4** | |
|---|---|---|---|---|---|---|---|---|
| | **N** | **%** | **N** | **%** | **N** | **%** | **N** | **%** |
| (1) Strongly agree | 3 | 4.0 | 1 | 1.3 | 1 | 1.3 | 3 | 4.0 |
| (2) Agree | 9 | 12.0 | 8 | 10.7 | 11 | 14.7 | 17 | 22.7 |
| (3) Somewhat agree | 16 | 21.3 | 13 | 17.3 | 13 | 17.3 | 12 | 16.0 |
| (4) Neither agree nor disagree | 7 | 9.3 | 10 | 13.3 | 8 | 10.7 | 7 | 9.3 |
| (5) Somewhat disagree | 4 | 5.3 | 4 | 5.3 | 5 | 6.7 | 2 | 2.7 |
| (6) Disagree | 26 | 34.7 | 31 | 41.3 | 26 | 34.7 | 24 | 32.0 |
| (7) Strongly Disagree | 10 | 13.3 | 8 | 10.7 | 10 | 13.3 | 10 | 13.3 |
| (8) I do not know | 0 | 0.0 | 0 | 0.0 | 1 | 1.3 | 0 | 0.0 |
| Total | 75 | 100.0 | 75 | 100.0 | 75 | 100.0 | 75 | 100.0 |
| Mean | 4.57 | | 4.77 | | 4.71 | | 4.33 | |
| Standard deviation | 1.847 | | 1.681 | | 1.814 | | 1.968 | |

*Strategic Partnerships (StrPar)*

This variable refers to Q11.

| Strategic Partnerships | Yes | | No | |
|---|---|---|---|---|
| | N | % | N | % |
| Competitors of the same industry | 11 | 14,7 | 64 | 85,3 |
| People or firms of different industries | 17 | 22,7 | 58 | 77,3 |
| Suppliers | 7 | 9,3 | 68 | 90,7 |
| Customers | 7 | 9,3 | 68 | 90,7 |
| Research institutions | 7 | 9,3 | 68 | 90,7 |
| Consultants | 14 | 18,7 | 61 | 81,3 |
| None of them | 41 | 54,7 | 34 | 45,3 |

Respondents were classified into ‘0 – No partnerships’ and ‘1 – partnerships’.

| Strategic Partnerships | N | % |
|---|---|---|
| No Partnerships | 41 | 54,7 |
| Partnerships | 34 | 45,3 |
| Total | 75 | 100,0 |

### ***Control Variables***

*Firm Size (FirSiz)*

This variable refers to Q2.

| Firm Size | N | % |
|---|---|---|
| I work for an established company (more than 10 workers). | 19 | 25,3 |
| I work as a freelancer or solopreneur. | 22 | 29,3 |
| I work for a small organization (2-9 workers). | 34 | 45,3 |
| Total | 75 | 100,0 |

*Duration (Dur)*

This variable refers to Q5.

| Year | N | % |
|---|---|---|
| 2017 | 15 | 20.0 |
| 2016 | 30 | 40.0 |
| 2015 | 13 | 17.3 |
| 2014 | 10 | 13.3 |
| 2013 | 1 | 1.3 |
| 2012 | 3 | 4.0 |
| 2011 | 3 | 4.0 |
| Total | 75 | 100.0 |

| Month | N | % |
|---|---|---|
| 1 | 5 | 6.7 |
| 2 | 4 | 5.3 |
| 3 | 5 | 6.7 |
| 4 | 4 | 5.3 |
| 5 | 13 | 17.3 |
| 6 | 5 | 6.7 |
| 7 | 5 | 6.7 |
| 9 | 16 | 21.3 |
| 10 | 6 | 8.0 |
| 11 | 8 | 10.7 |
| 12 | 4 | 5.3 |
| Total | 75 | 100.0 |

Based on the year and month, the duration of coworking engagement was calculated.

| **Duration** | **N** | **%** |
|---|---|---|
| 1 month | 2 | 2.7 |
| 2 months | 6 | 8.0 |
| 3 months | 7 | 9.3 |
| 4 months | 8 | 10.7 |
| 5 months | 6 | 8.0 |
| 6 months | 3 | 4.0 |
| 7 months | 3 | 4.0 |
| 8 months | 4 | 5.3 |
| 9 months | 2 | 2.7 |
| 10 months | 1 | 1.3 |
| 11 months | 2 | 2.7 |
| 13 months | 1 | 1.3 |
| 15 months | 1 | 1.3 |
| 16 months | 3 | 4.0 |
| 17 months | 2 | 2.7 |
| 19 months | 1 | 1.3 |
| 20 months | 2 | 2.7 |
| 21 months | 1 | 1.3 |
| 22 months | 1 | 1.3 |
| 23 months | 1 | 1.3 |
| 26 months | 1 | 1.3 |
| 28 months | 3 | 4.0 |
| 31 months | 1 | 1.3 |
| 32 months | 3 | 4.0 |
| 34 months | 1 | 1.3 |
| 35 months | 1 | 1.3 |
| 36 months | 1 | 1.3 |
| more than 36 months | 7 | 9.3 |
| Total | 75 | 100.0 |

***Other Variables***

The following sections outlines all questions, which have not been used as independent, dependent or control variables. These includes Q1, Q3, Q4, Q6, Q7, Q9 and Q15.

**Q1:**

| **Industry** | **N** | **%** |
|---|---|---|
| Arts/Entertainment | 10 | 13.3 |
| Business Services | 10 | 13.3 |
| Construction | 2 | 2.7 |
| Education/Healthcare | 3 | 4.0 |
| Financial and Insurance Services | 5 | 6.7 |
| Information/Tech-Industry | 23 | 30.7 |
| Real Estate | 1 | 1.3 |
| Wholesale/Retail | 2 | 2.7 |
| Other | 19 | 25.3 |
| Total | 75 | 100.0 |

**Q3:**

| **Number of Coworking Spaces** | **N** | **%** |
|---|---|---|
| 1 | 40 | 53.3 |
| 2 | 18 | 24.0 |
| 3 | 8 | 10.7 |
| 4 | 1 | 1.3 |
| More than 4 | 8 | 10.7 |
| Total | 75 | 100.0 |

**Q4:**

Due to an explicit request of one coworking space, these details cannot be shared.

**Q6:**

| **Primary Workspace** | **N** | **%** |
|---|---|---|
| Yes | 63 | 84.0 |
| No | 12 | 16.0 |
| Total | 75 | 100.0 |

**Q7:**

| **Hours** | **N** | **%** |
|---|---|---|
| 2 | 1 | 1.3 |
| 4 | 3 | 4.0 |
| 5 | 2 | 2.7 |
| 6 | 1 | 1.3 |
| 8 | 2 | 2.7 |
| 10 | 2 | 2.7 |
| 15 | 1 | 1.3 |
| 16 | 2 | 2.7 |
| 20 | 6 | 8.0 |
| 25 | 2 | 2.7 |
| 27 | 1 | 1.3 |
| 30 | 11 | 14.7 |
| 32 | 2 | 2.7 |
| 35 | 3 | 4.0 |
| 38 | 1 | 1.3 |
| 39 | 1 | 1.3 |
| 40 | 26 | 34.7 |
| 42 | 1 | 1.3 |
| 45 | 2 | 2.7 |
| 50 | 2 | 2.7 |
| 60 | 2 | 2.7 |
| Above 65 | 1 | 1.3 |
| Total | 75 | 100.0 |
| Mean | 32.25 | |
| Standard deviation | 14.181 | |

**Q9:**

| | Item_1 | | Item_2 | | Item_3 | | Item_4 | | Item_5 | |
|---|---|---|---|---|---|---|---|---|---|---|
| | N | % | N | % | N | % | N | % | N | % |
| (1) Strongly agree | 18 | 24.0 | 19 | 25.3 | 25 | 33.3 | 19 | 25.3 | 1 | 1.3 |
| (2) Agree | 35 | 46.7 | 34 | 45.3 | 33 | 44.0 | 32 | 42.7 | 6 | 8.0 |
| (3) Somewhat agree | 14 | 18.7 | 14 | 18.7 | 9 | 12.0 | 13 | 17.3 | 10 | 13.3 |
| (4) Neither agree nor disagree | 2 | 2.7 | 2 | 2.7 | 3 | 4.0 | 5 | 6.7 | 2 | 2.7 |
| (5) Somewhat disagree | 1 | 1.3 | 2 | 2.7 | 2 | 2.7 | 2 | 2.7 | 11 | 14.7 |
| (6) Disagree | 4 | 5.3 | 3 | 4.0 | 2 | 2.7 | 2 | 2.7 | 32 | 42.7 |
| (7) Strongly Disagree | 1 | 1.3 | 1 | 1.3 | 1 | 1.3 | 2 | 2.7 | 13 | 17.3 |
| (8) I do not know | 0 | 0.0 | 0 | 0.0 | 0 | 0.0 | 0 | 0.0 | 0 | 0.0 |
| Total | 75 | 100.0 | 75 | 100.0 | 75 | 100.0 | 75 | 100.0 | 75 | 100.0 |
| Mean | 2.32 | | 2.29 | | 2.12 | | 2.37 | | 5.19 | |
| Standard deviation | 1.327 | | 1.303 | | 1.273 | | 1.383 | | 1.617 | |

**Q15:**

| Novelty of Innovation | N | % |
|---|---|---|
| New to the firm | 18 | 24.0 |
| New to the market | 21 | 28.0 |
| Overall | 39 | 52.0 |
| Missing | 36 | 48.0 |
| Total | 75 | 100.0 |

# APPENDIX C: Supporting Graphics of Regression Analysis

## *Collaboration and the Access to external Resources*

## Model 1a:

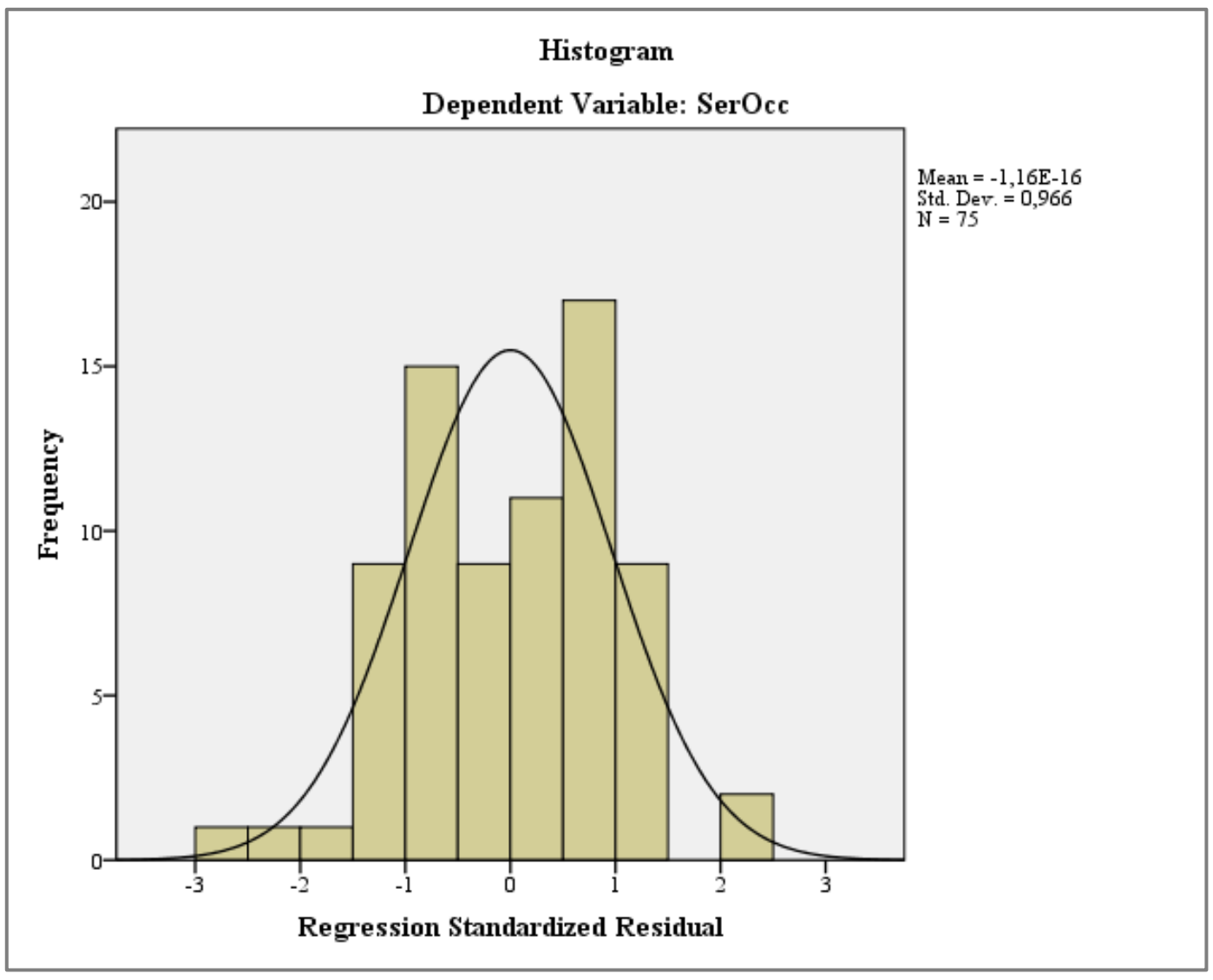

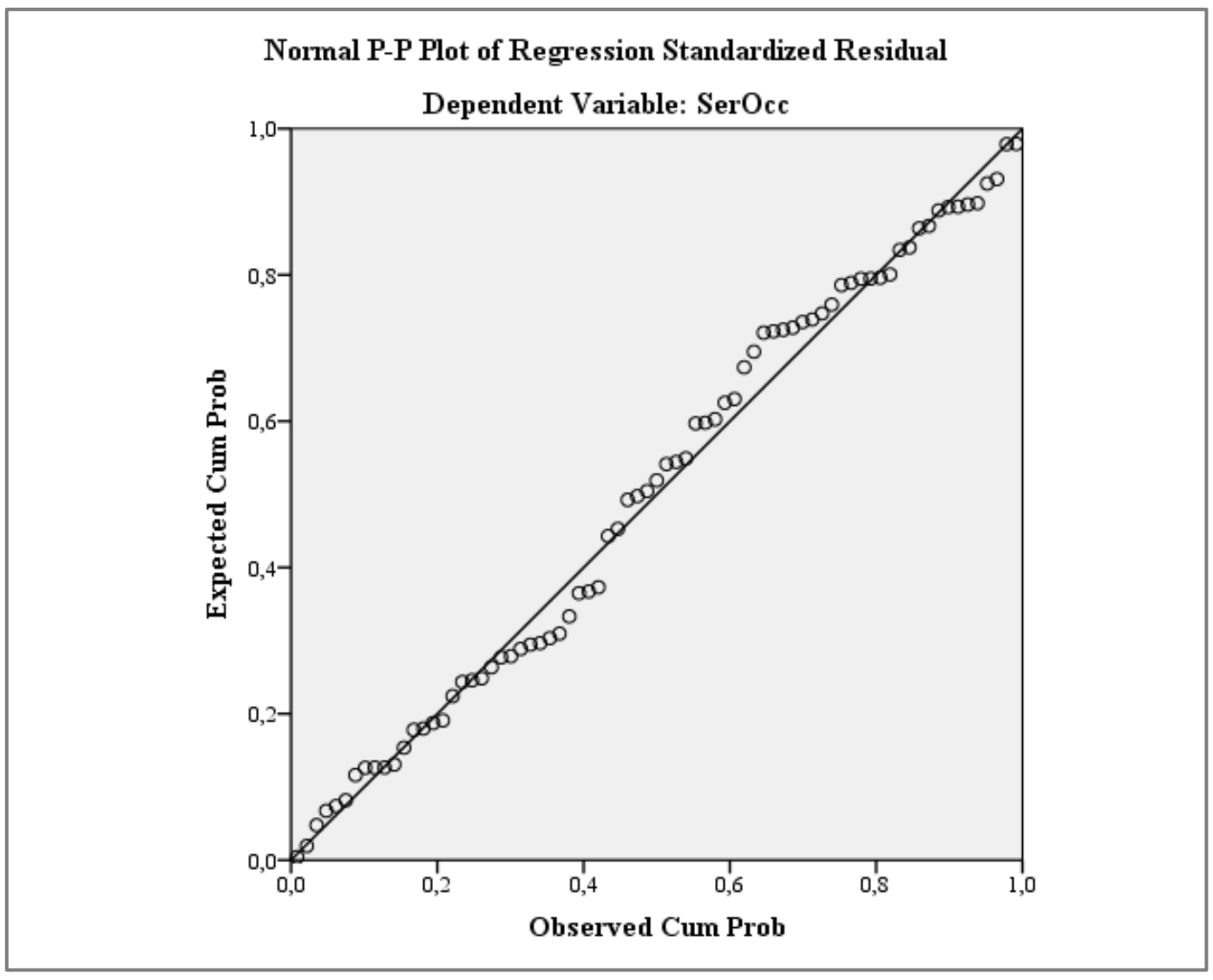

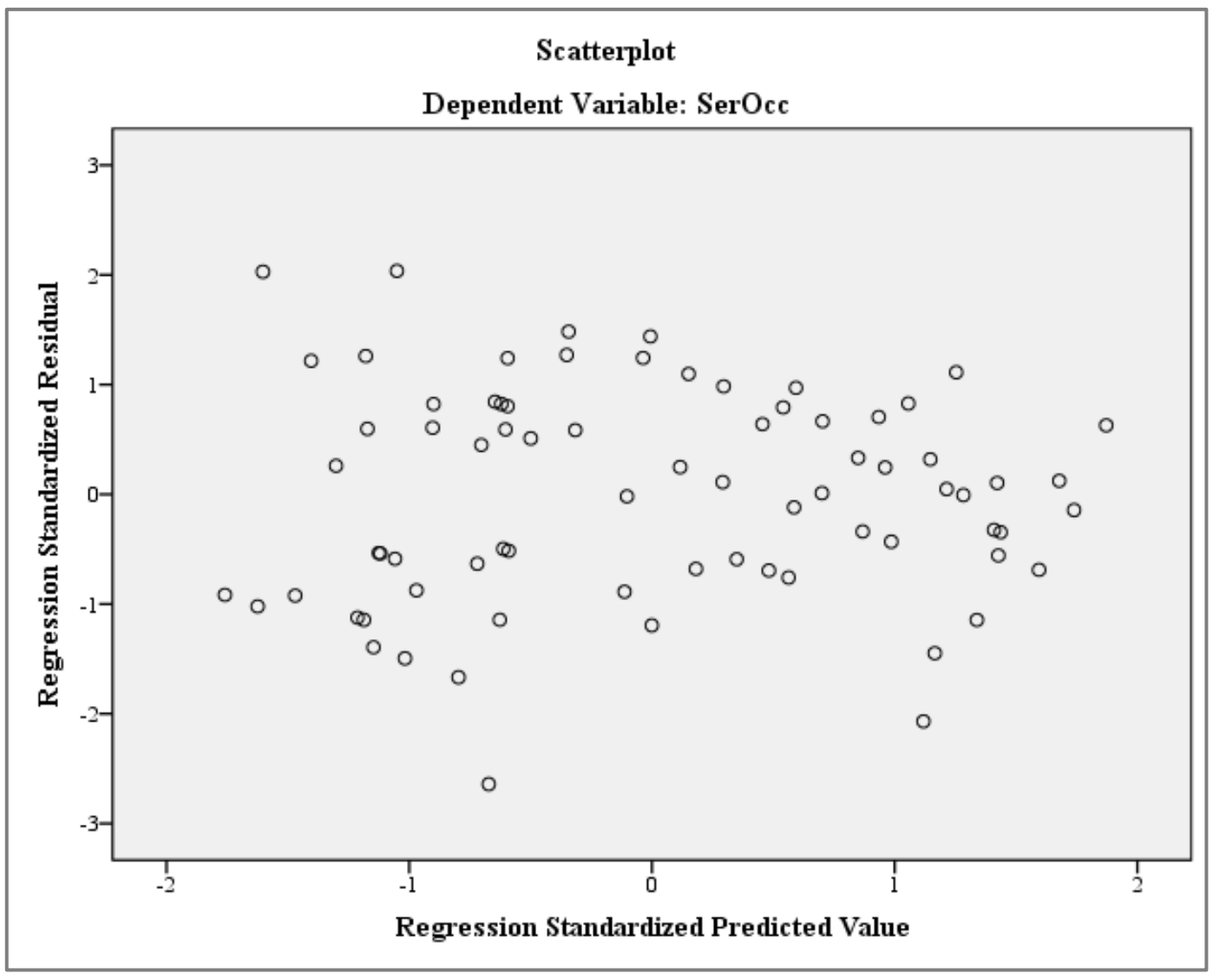

**Model 1b:**

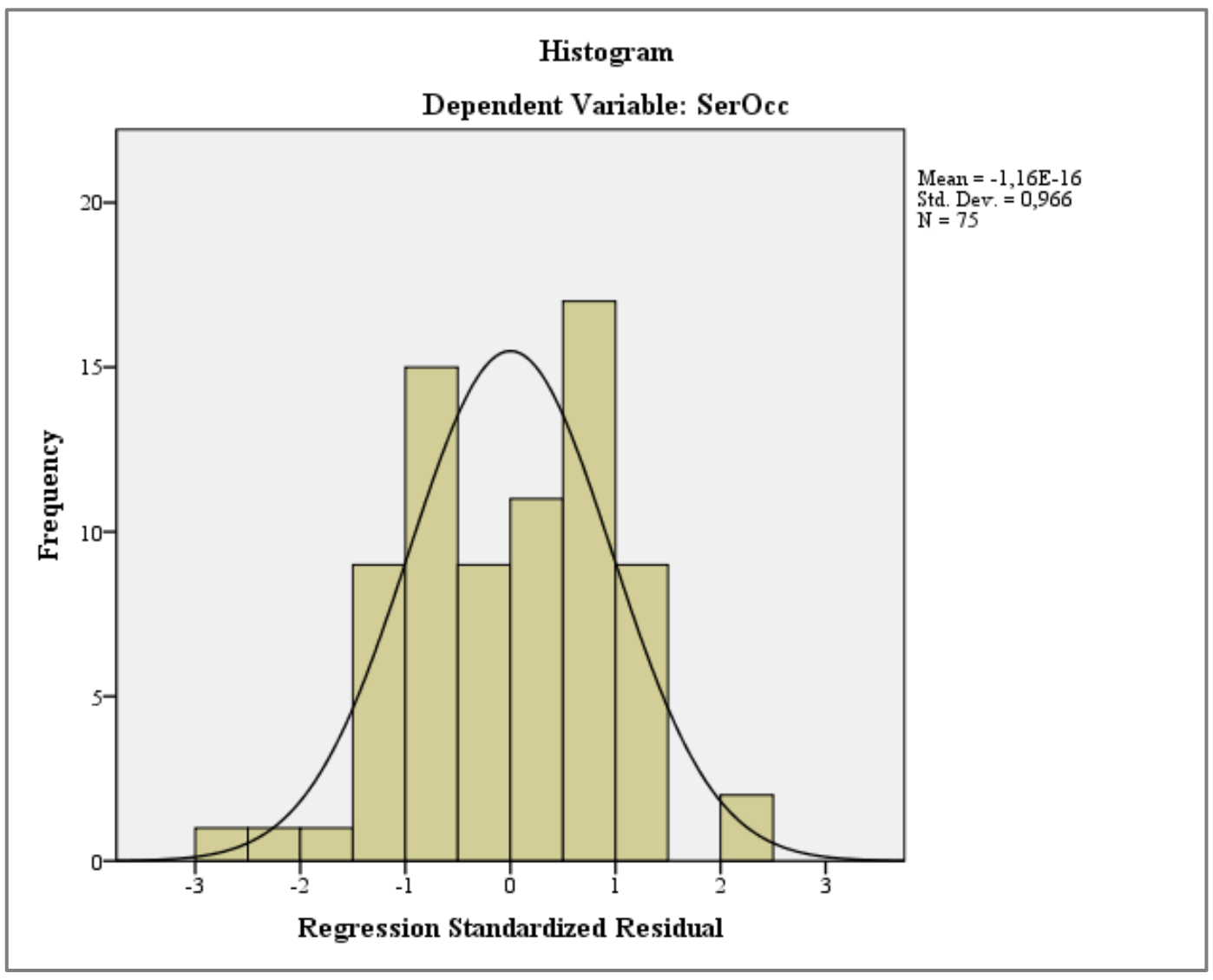
Histogram
Dependent Variable: SerOcc
Mean = -1,16E-16
Std. Dev. = 0,966
N = 75
Frequency
0
5
10
15
20
-3
-2
-1
0
1
2
3
Regression Standardized Residual

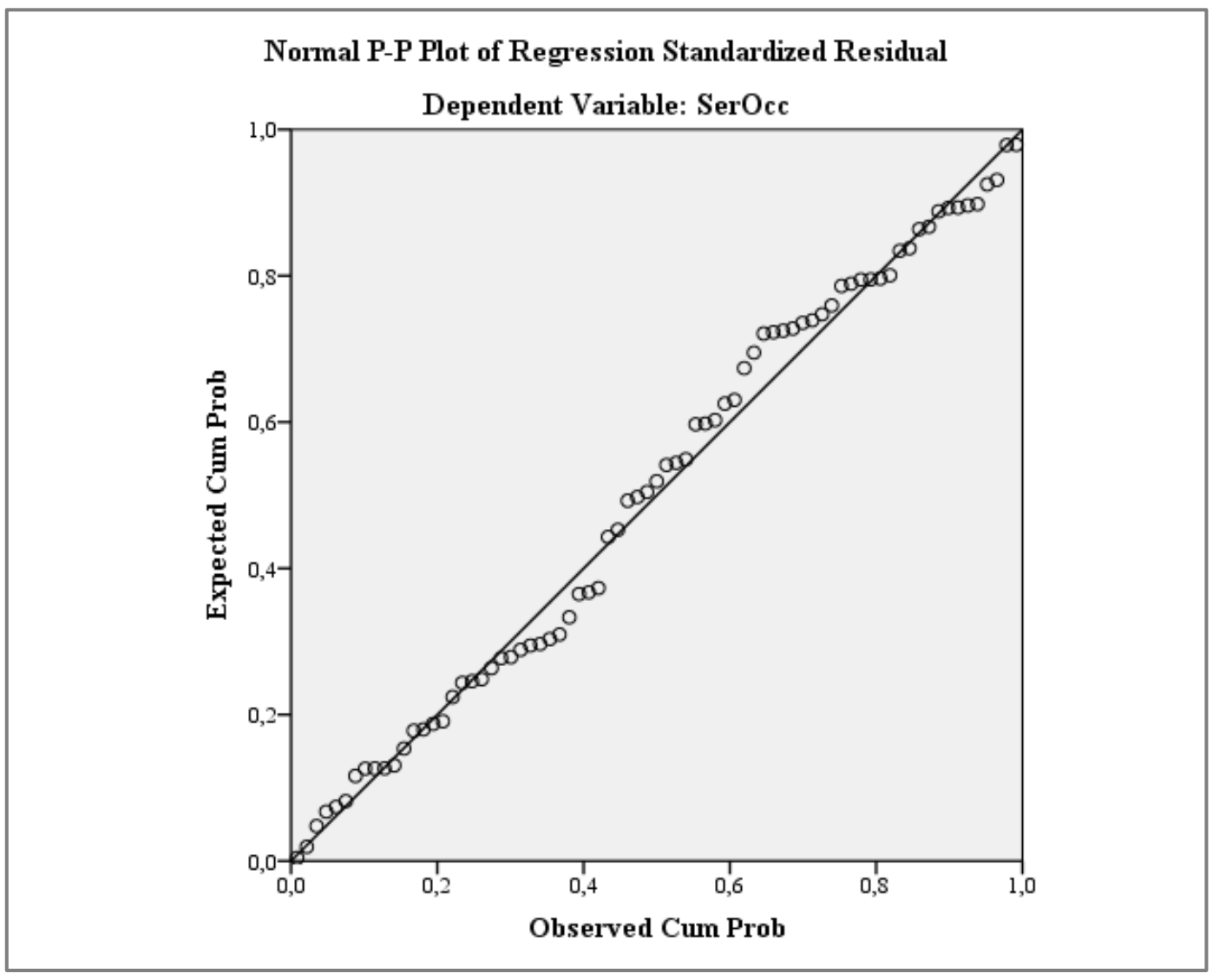
Normal P-P Plot of Regression Standardized Residual
Dependent Variable: SerOcc
Expected Cum Prob
0,0
0,2
0,4
0,6
0,8
1,0
Observed Cum Prob

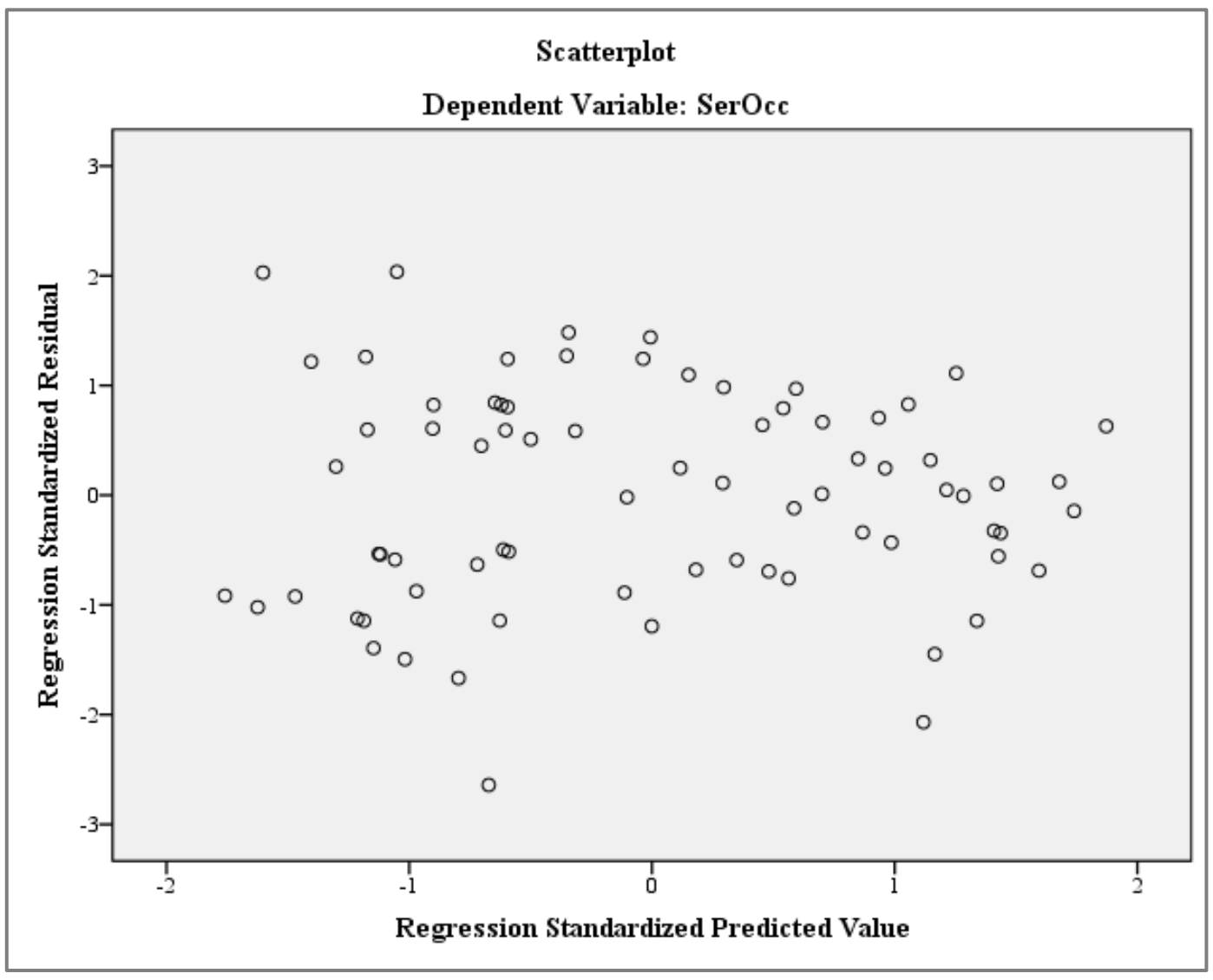
Scatterplot
Dependent Variable: SerOcc
Regression Standardized Residual
-3
-2
-1
0
1
2
3
Regression Standardized Predicted Value

**Model 1c:**

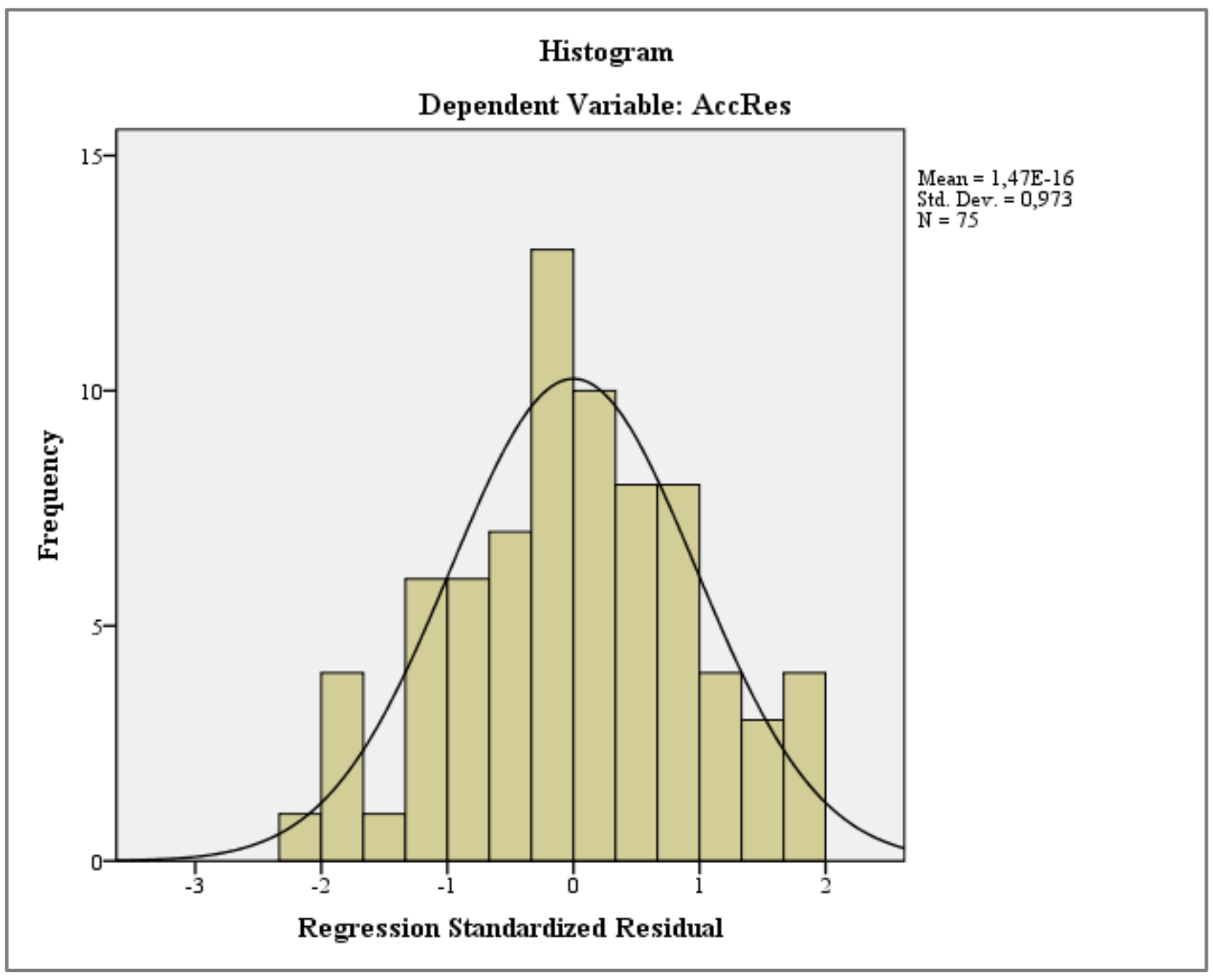
Histogram
Dependent Variable: AccRes
Mean = 1,47E-16
Std. Dev. = 0,973
N = 75
Frequency
Regression Standardized Residual

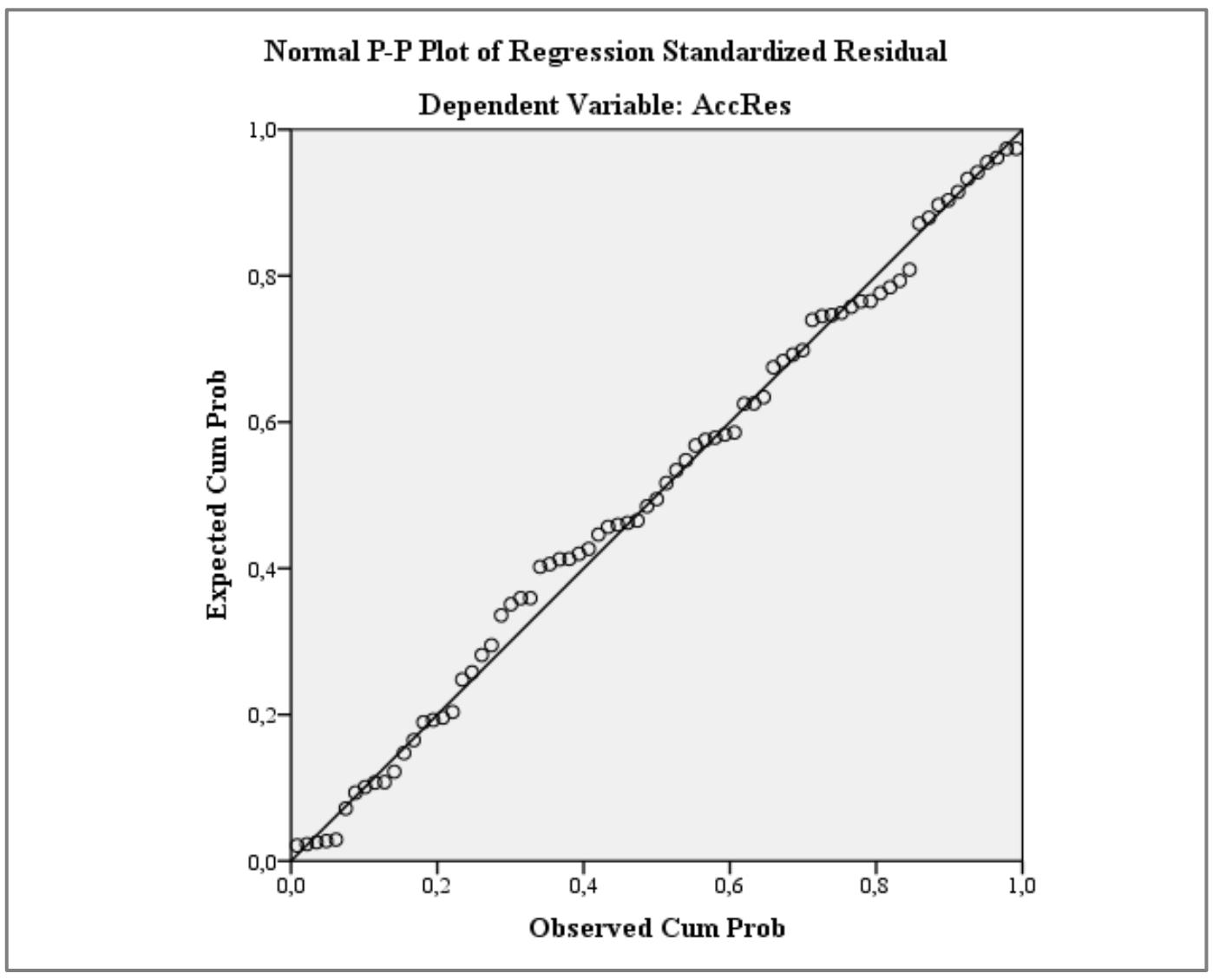
Normal P-P Plot of Regression Standardized Residual
Dependent Variable: AccRes
Expected Cum Prob
Observed Cum Prob

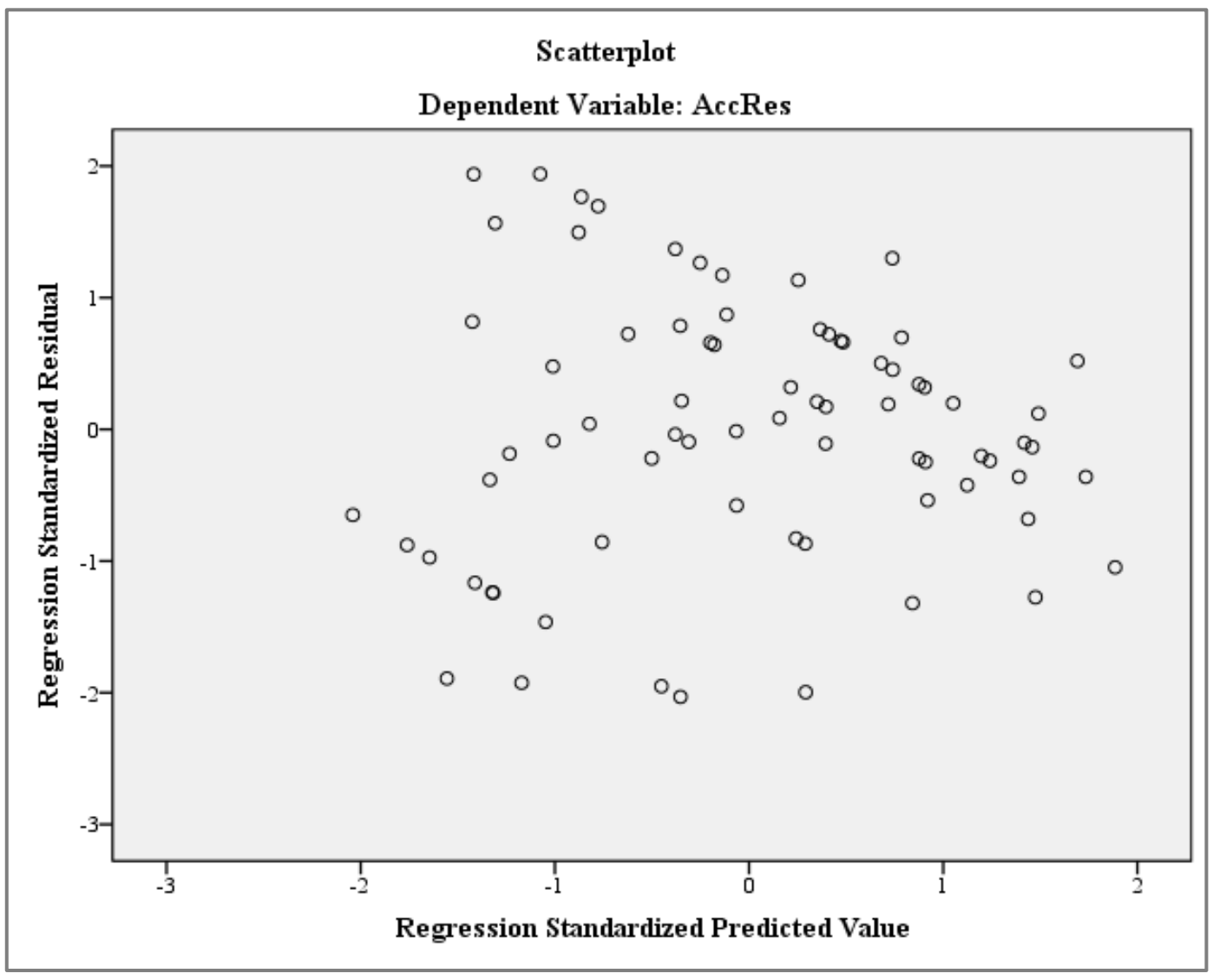
Scatterplot
Dependent Variable: AccRes
Regression Standardized Residual
Regression Standardized Predicted Value

***Collaboration and Serendipitous Occurrences***

**Model 2 a:**

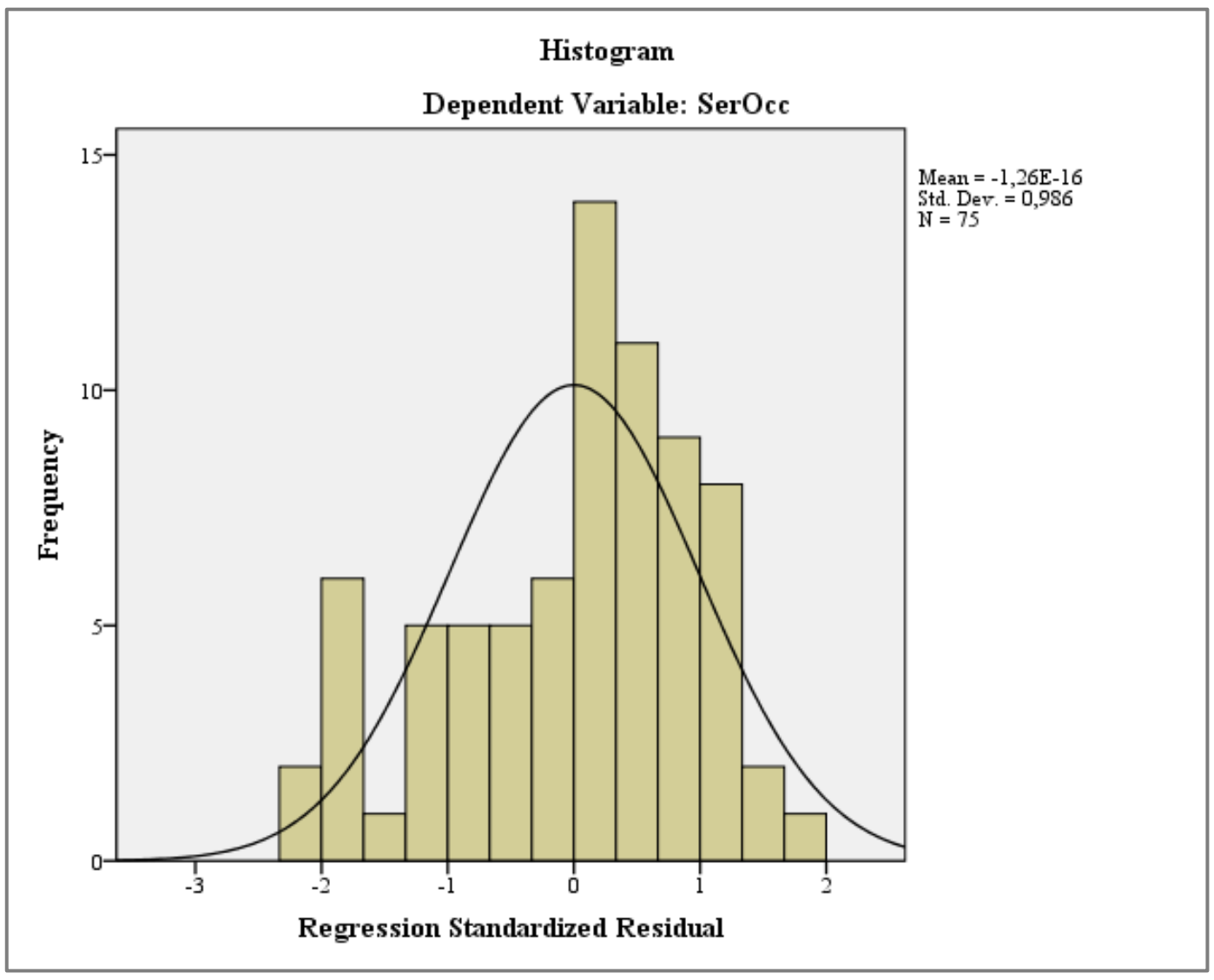

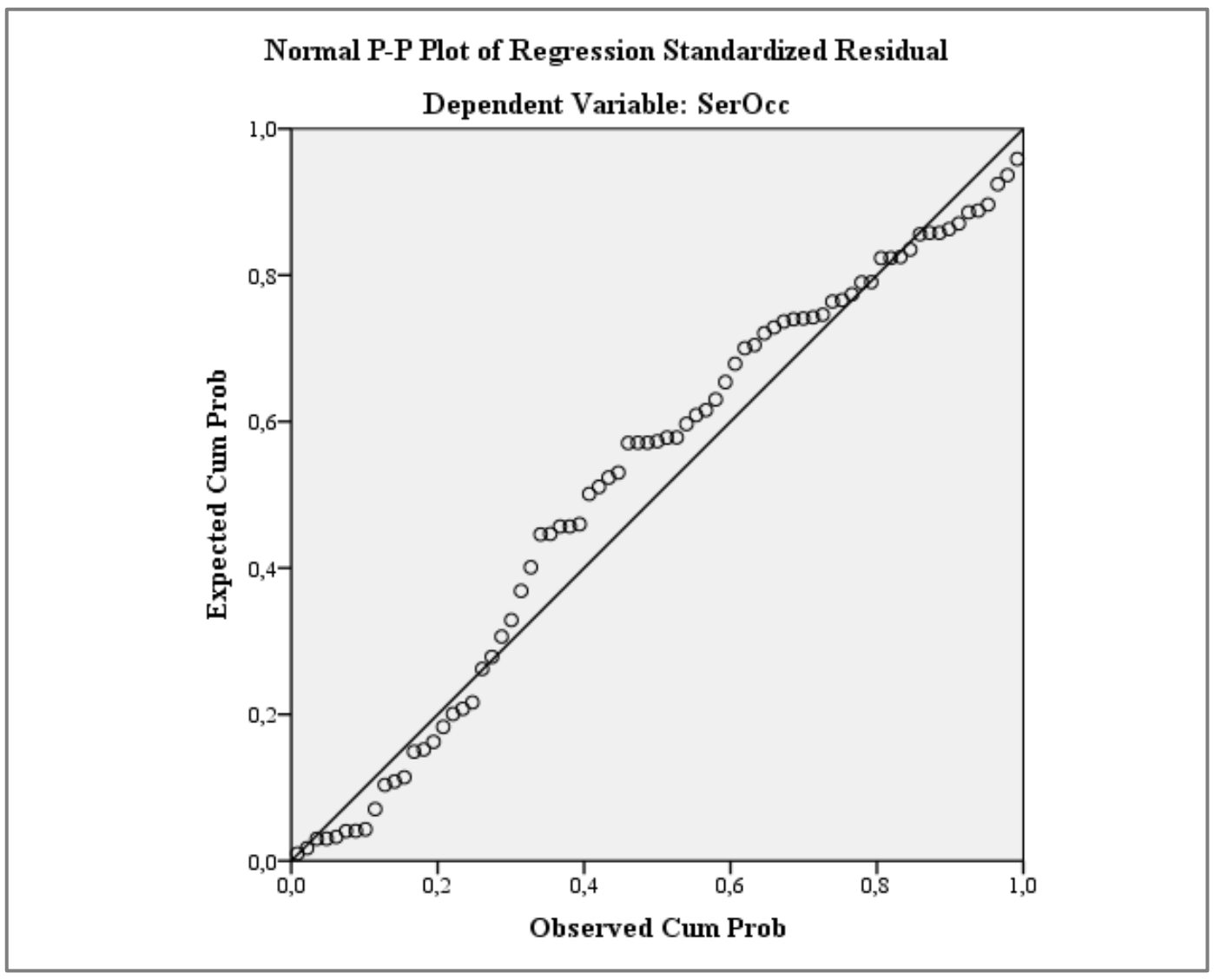

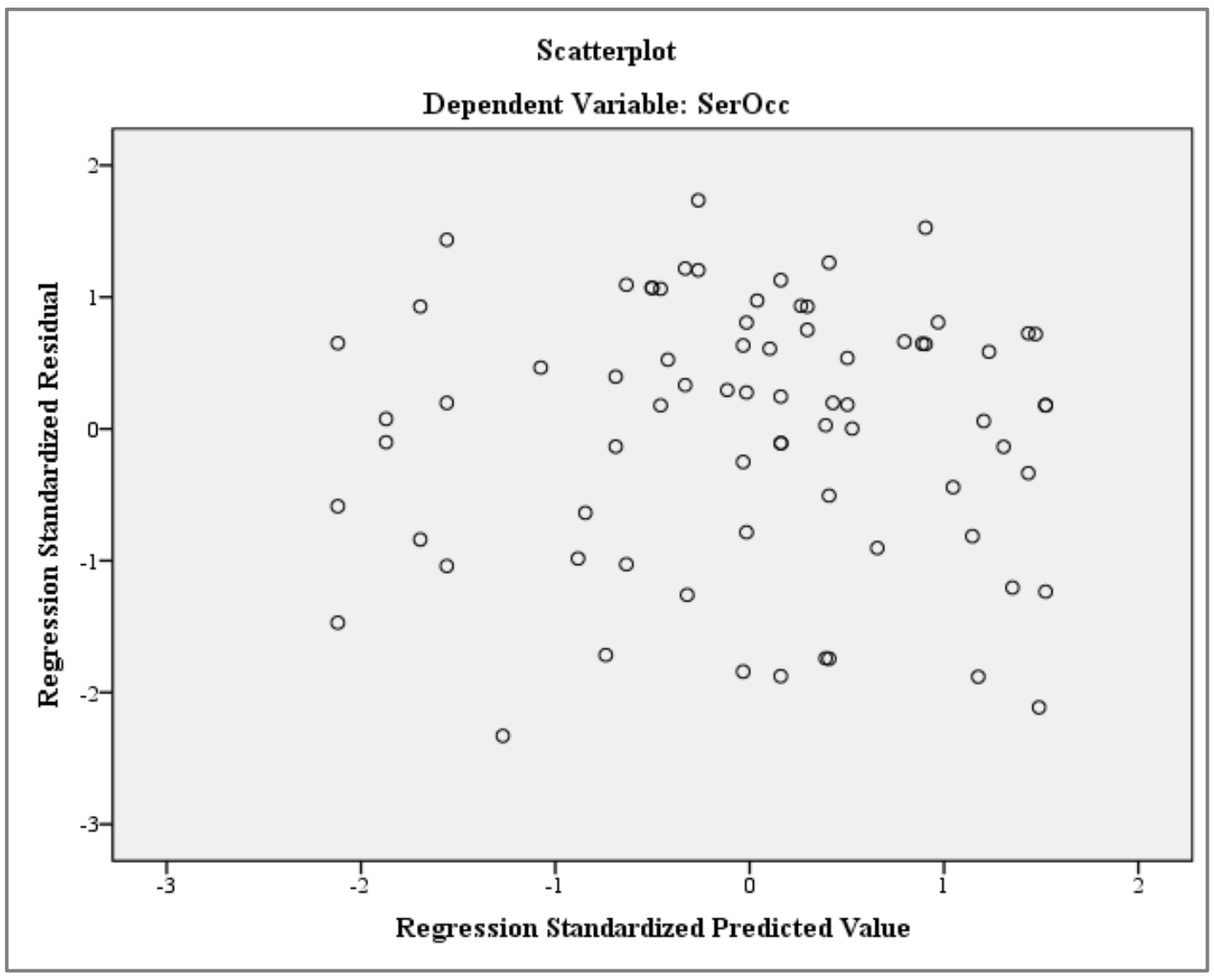

**Model 2b:**

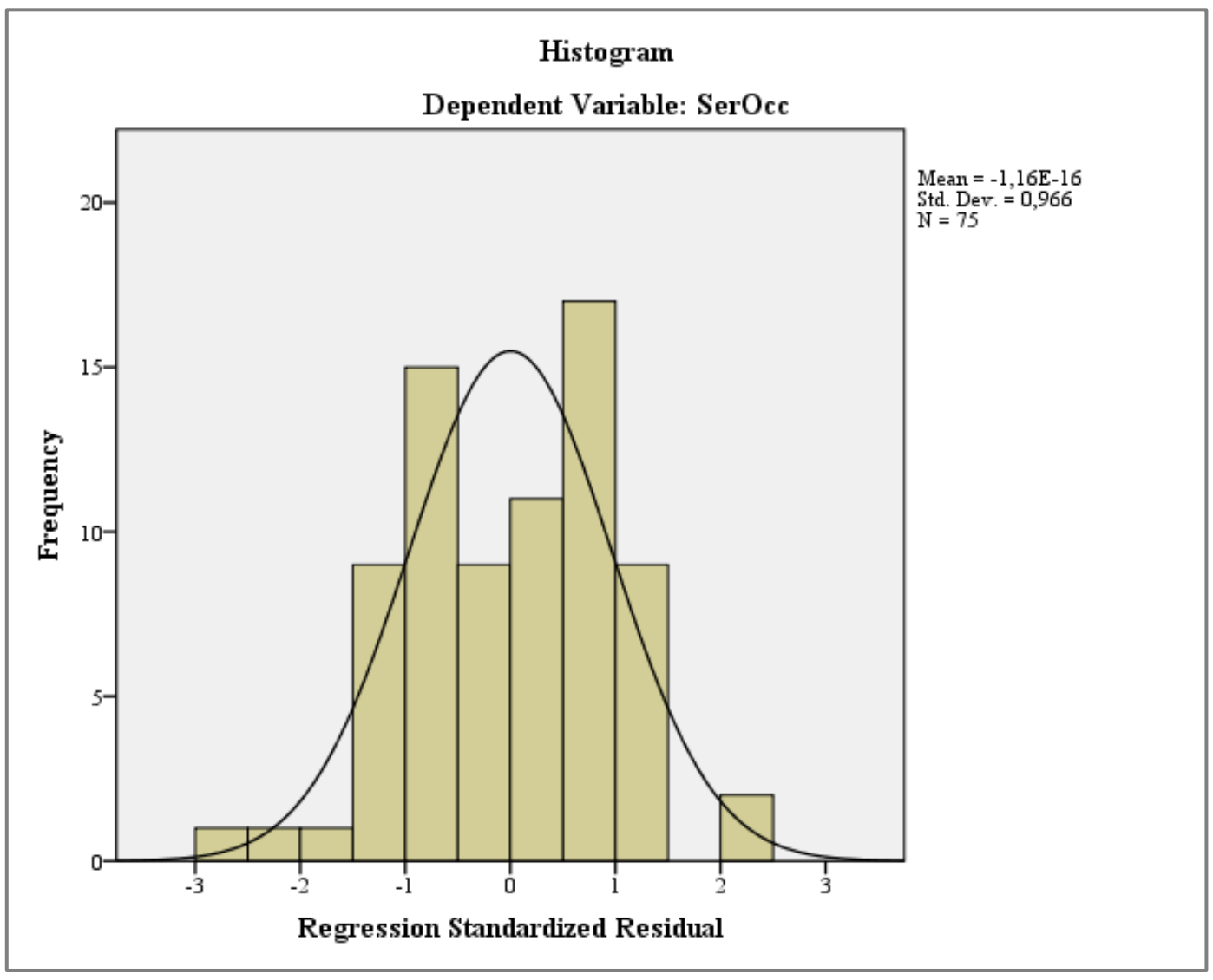

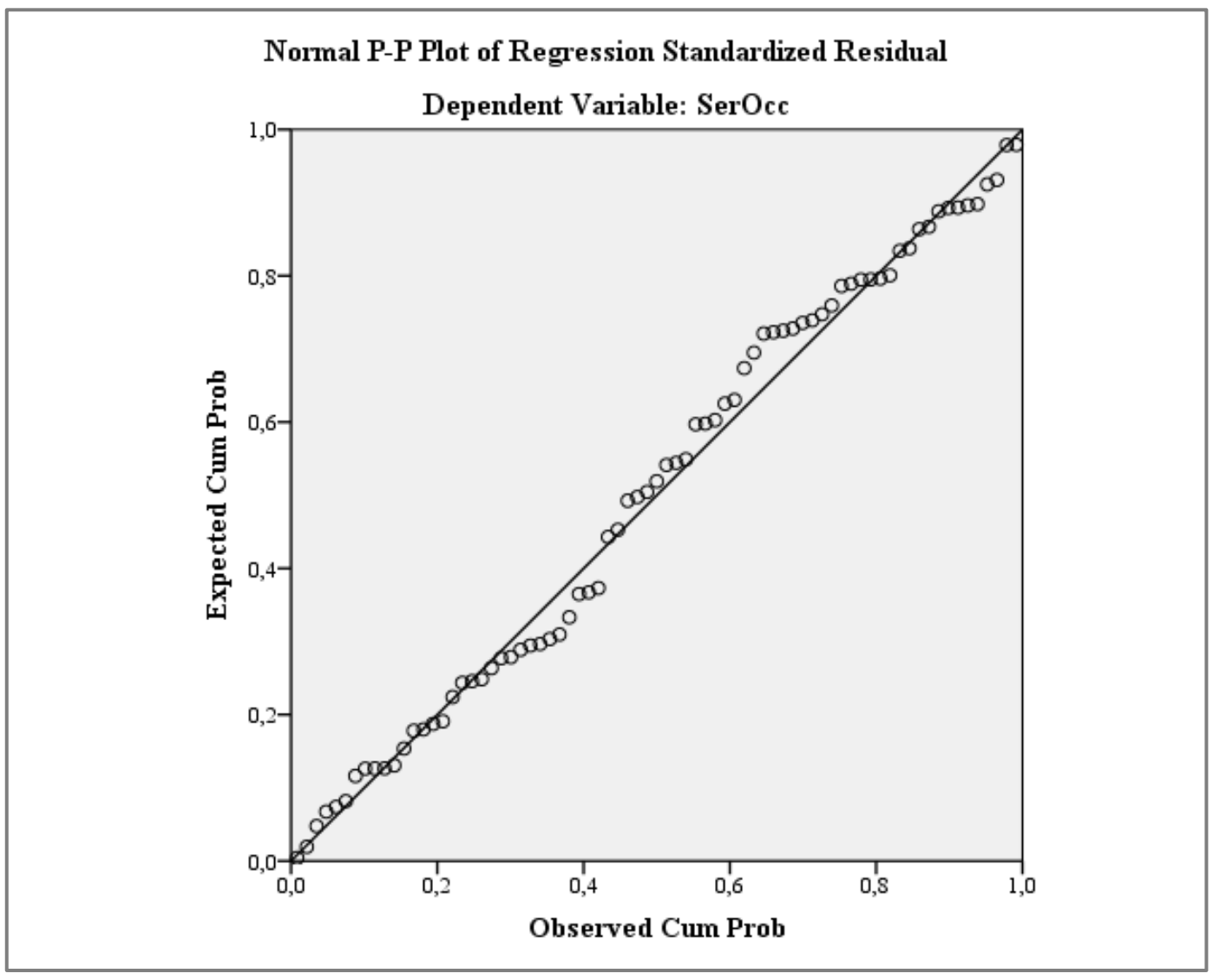
Normal P-P Plot of Regression Standardized Residual
Dependent Variable: SerOcc
Expected Cum Prob
Observed Cum Prob

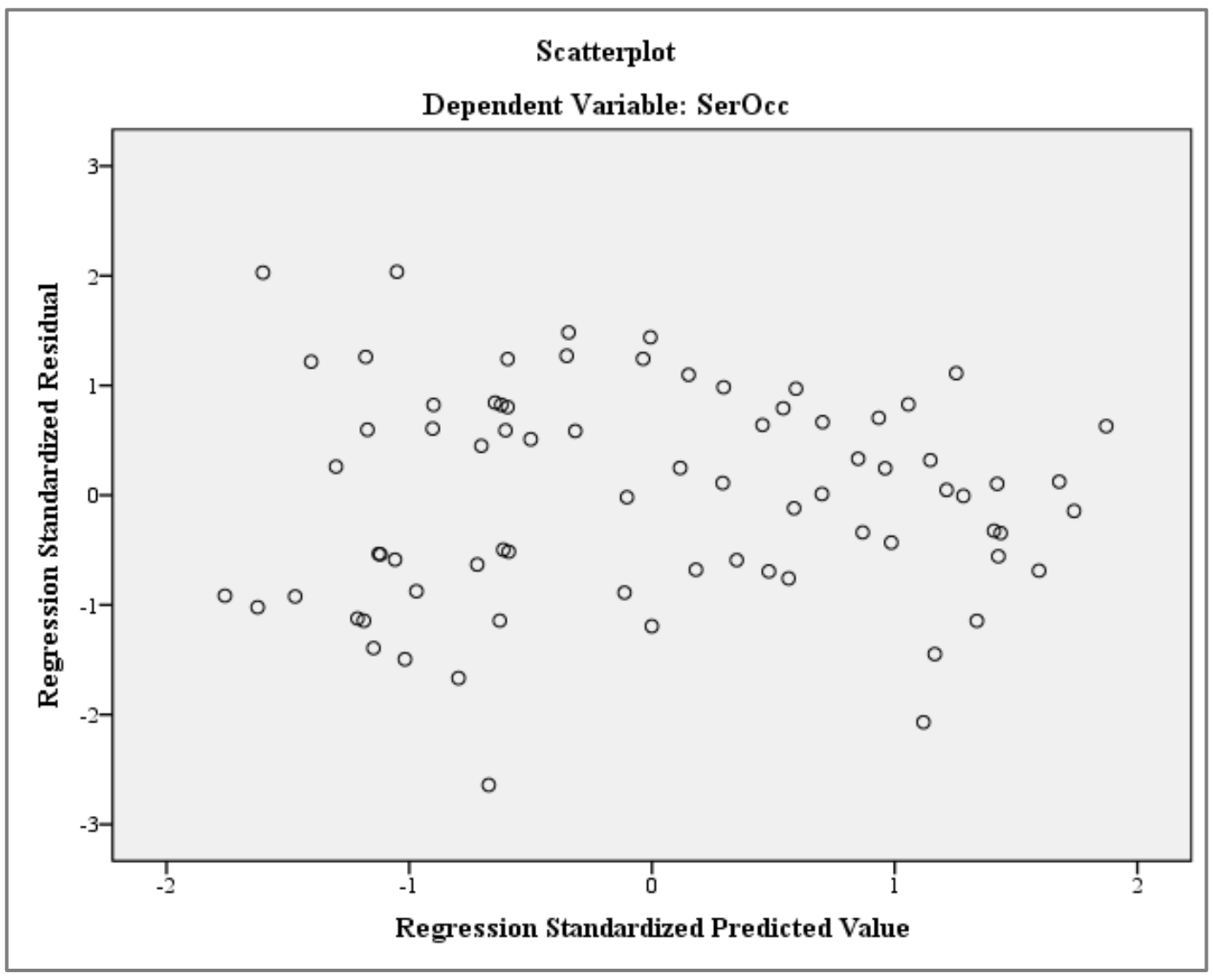
Scatterplot
Dependent Variable: SerOcc
Regression Standardized Residual
Regression Standardized Predicted Value

## Model 2c:

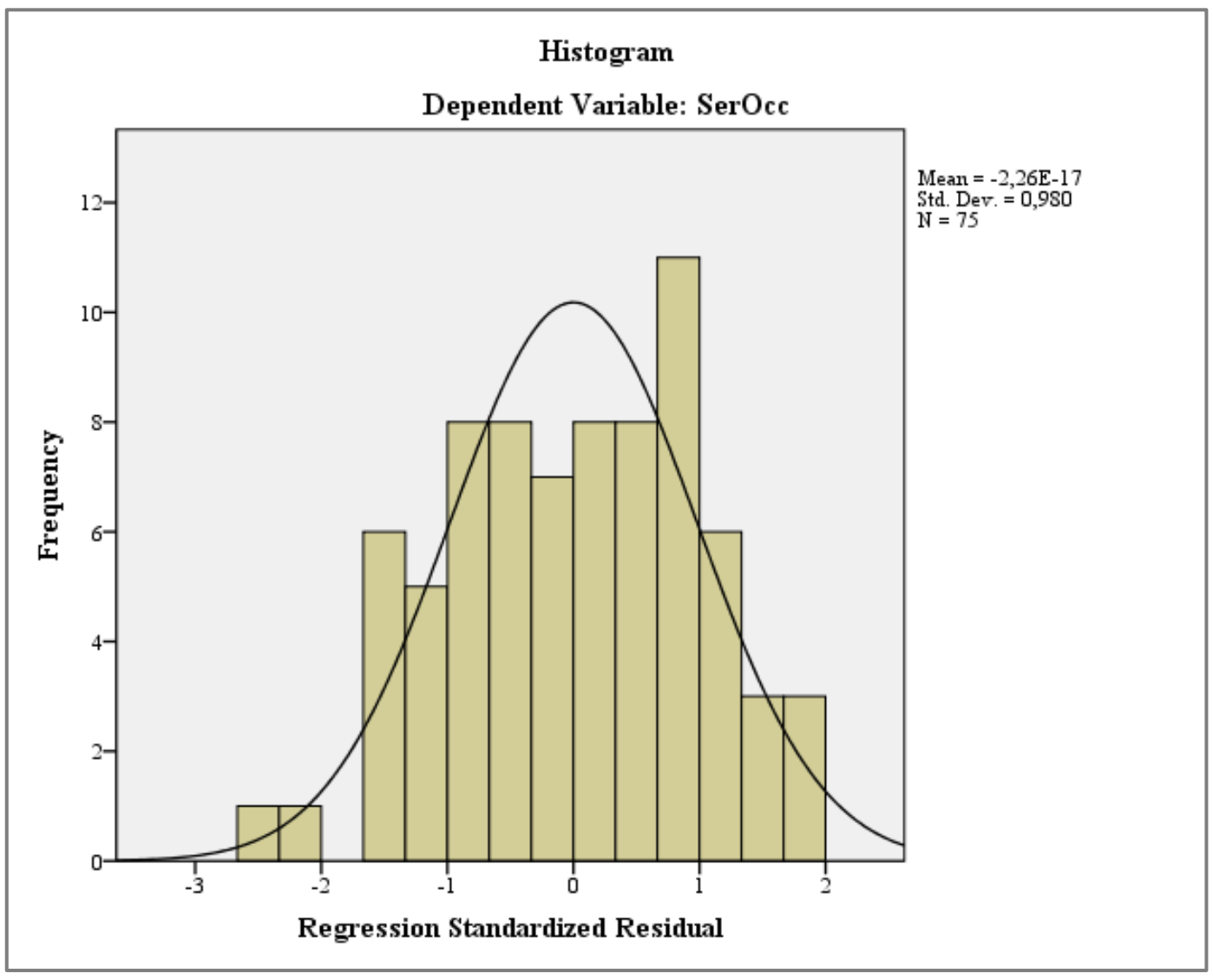

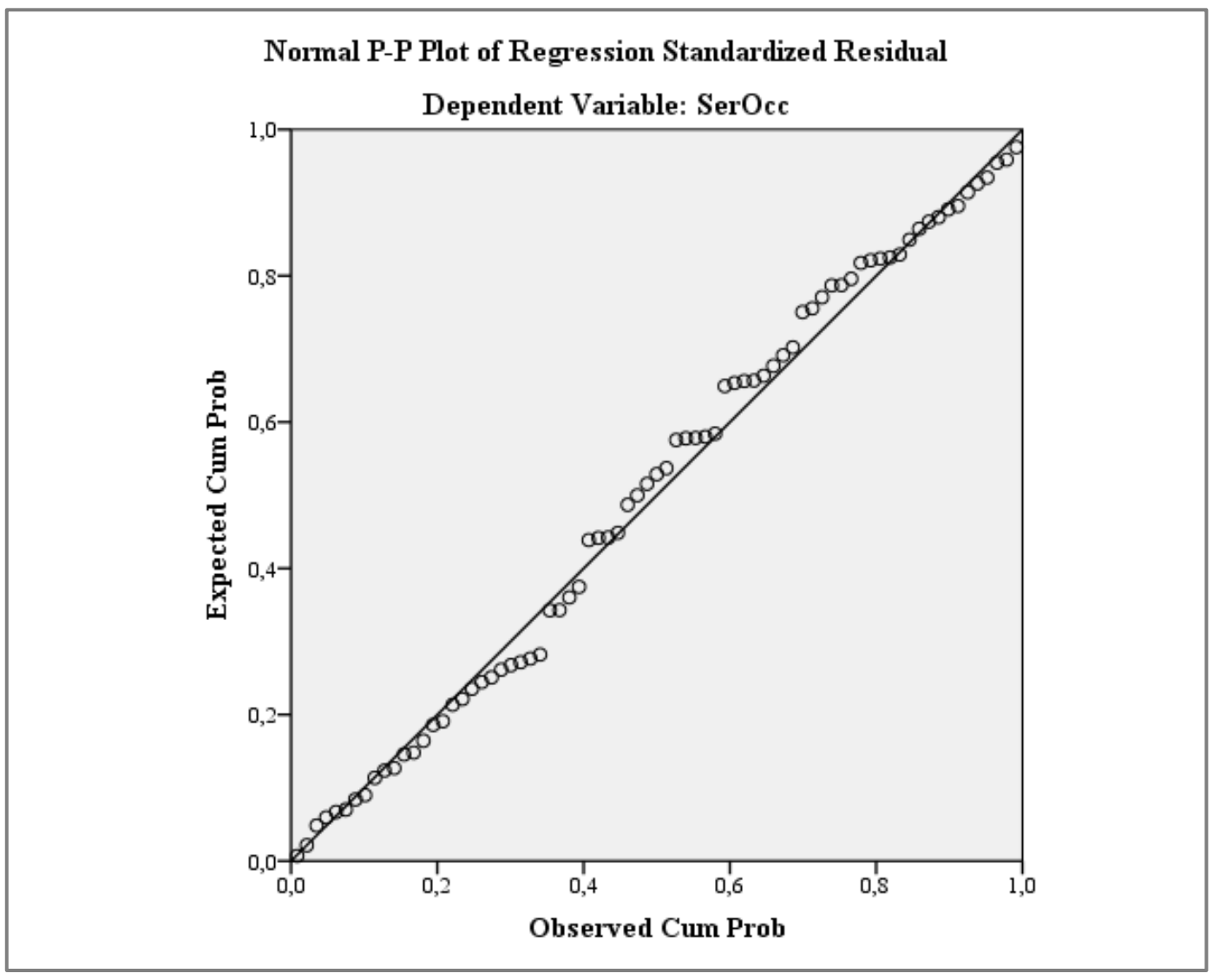
Normal P-P Plot of Regression Standardized Residual
Dependent Variable: SerOcc
Expected Cum Prob
Observed Cum Prob

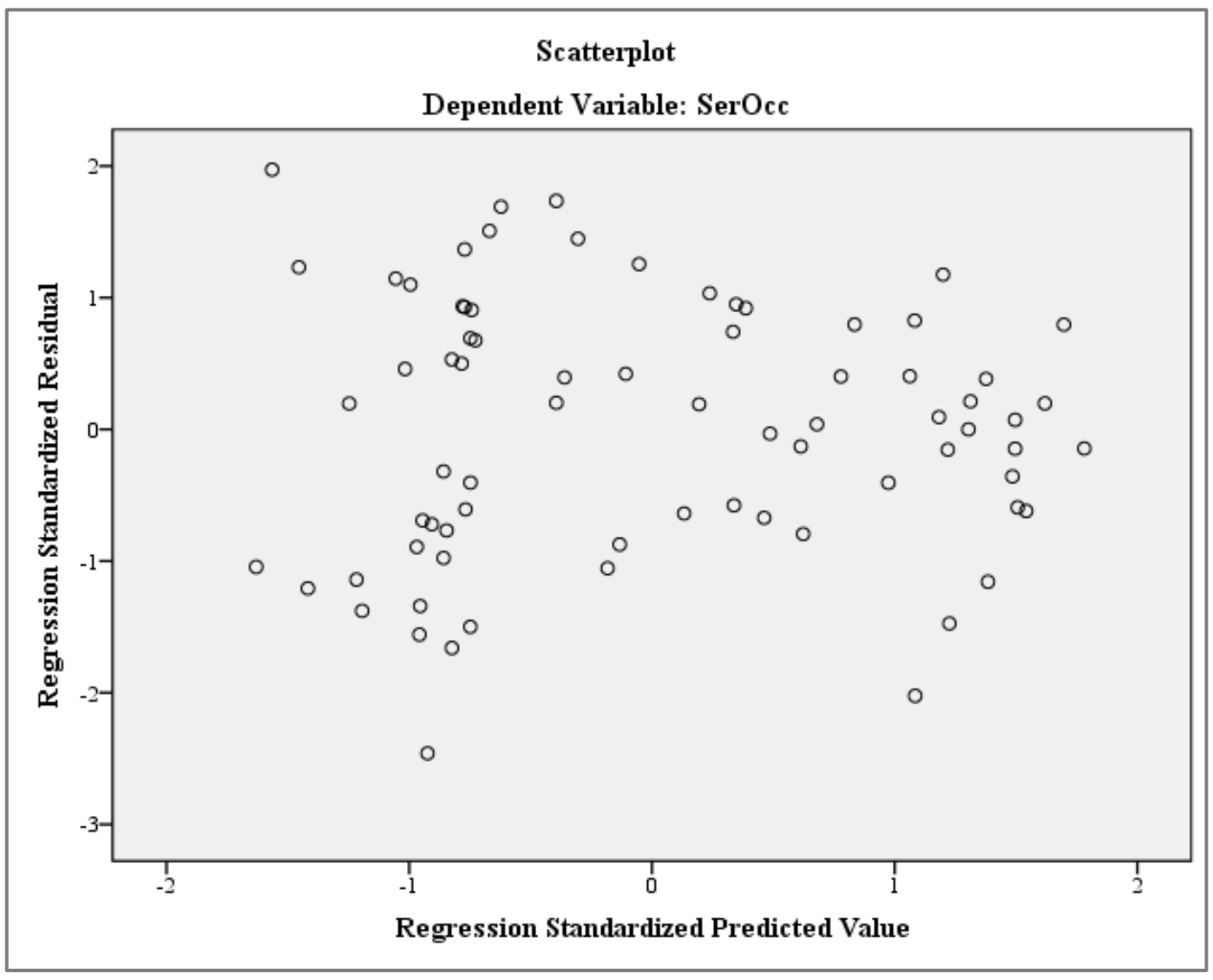
Scatterplot
Dependent Variable: SerOcc
Regression Standardized Residual
Regression Standardized Predicted Value

***Collaboration and the Impact on Business Models***

*Collaboration and the Impact on Value Creation Architecture Dimension (ValCreArc)*

**Model 4a:**

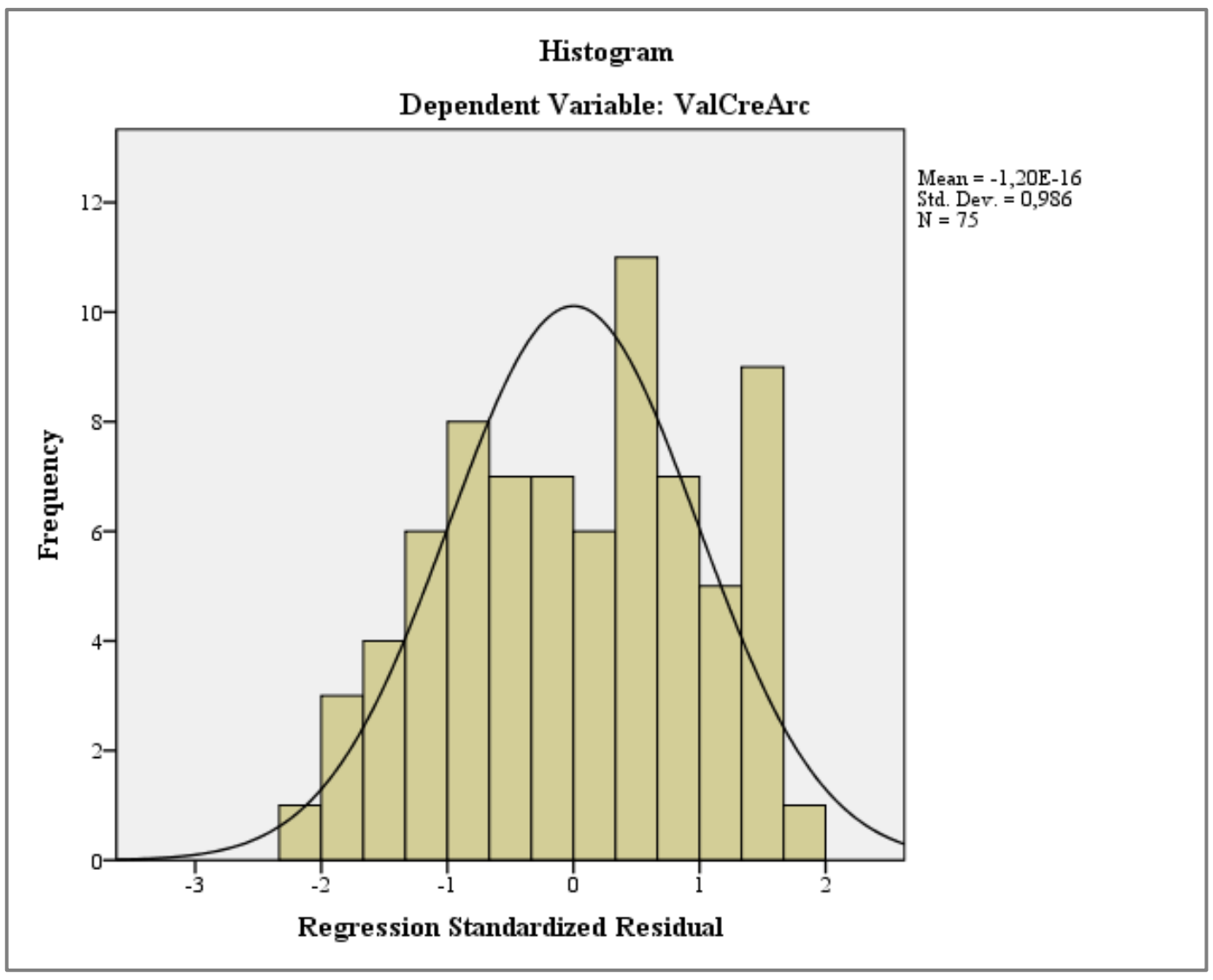

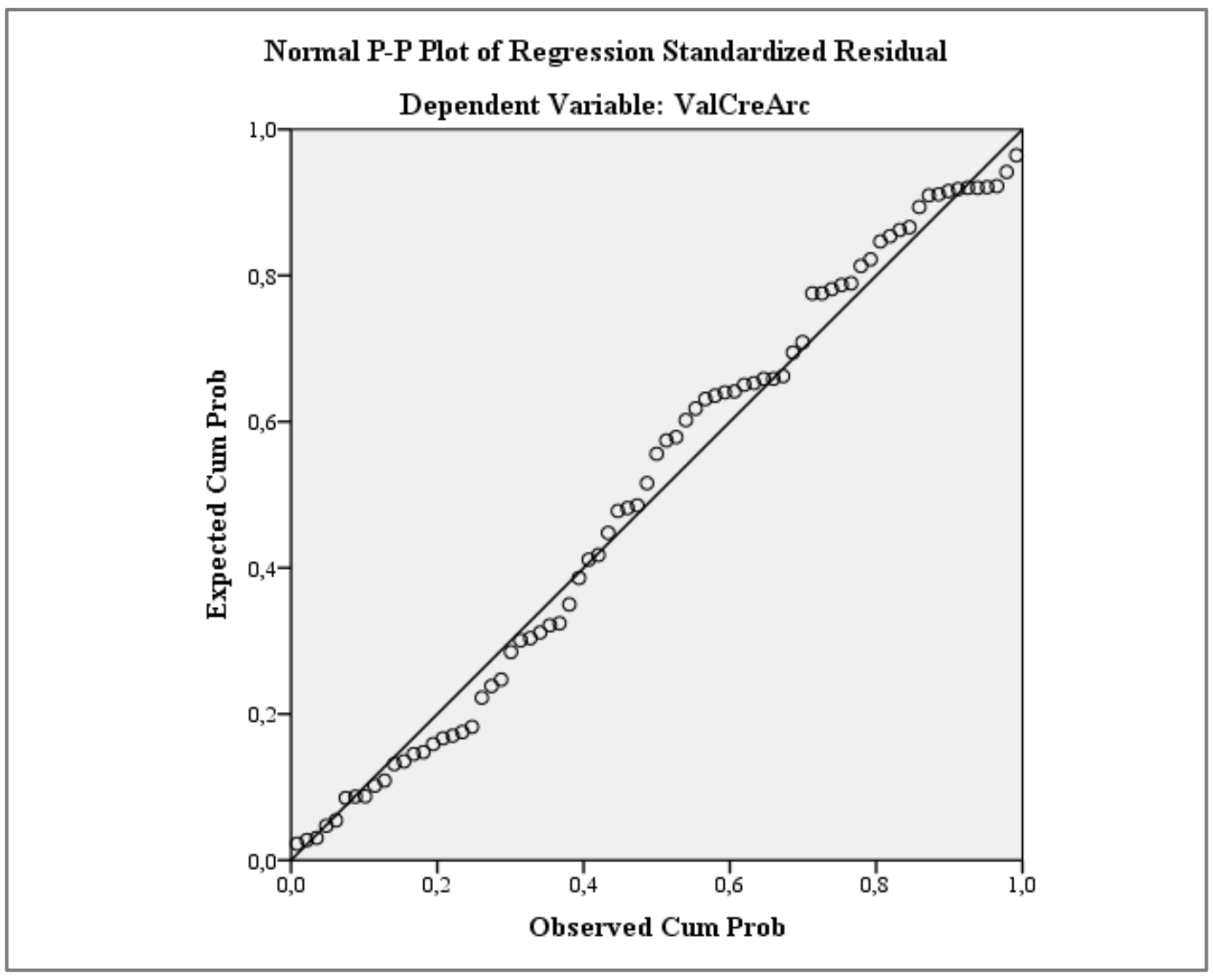

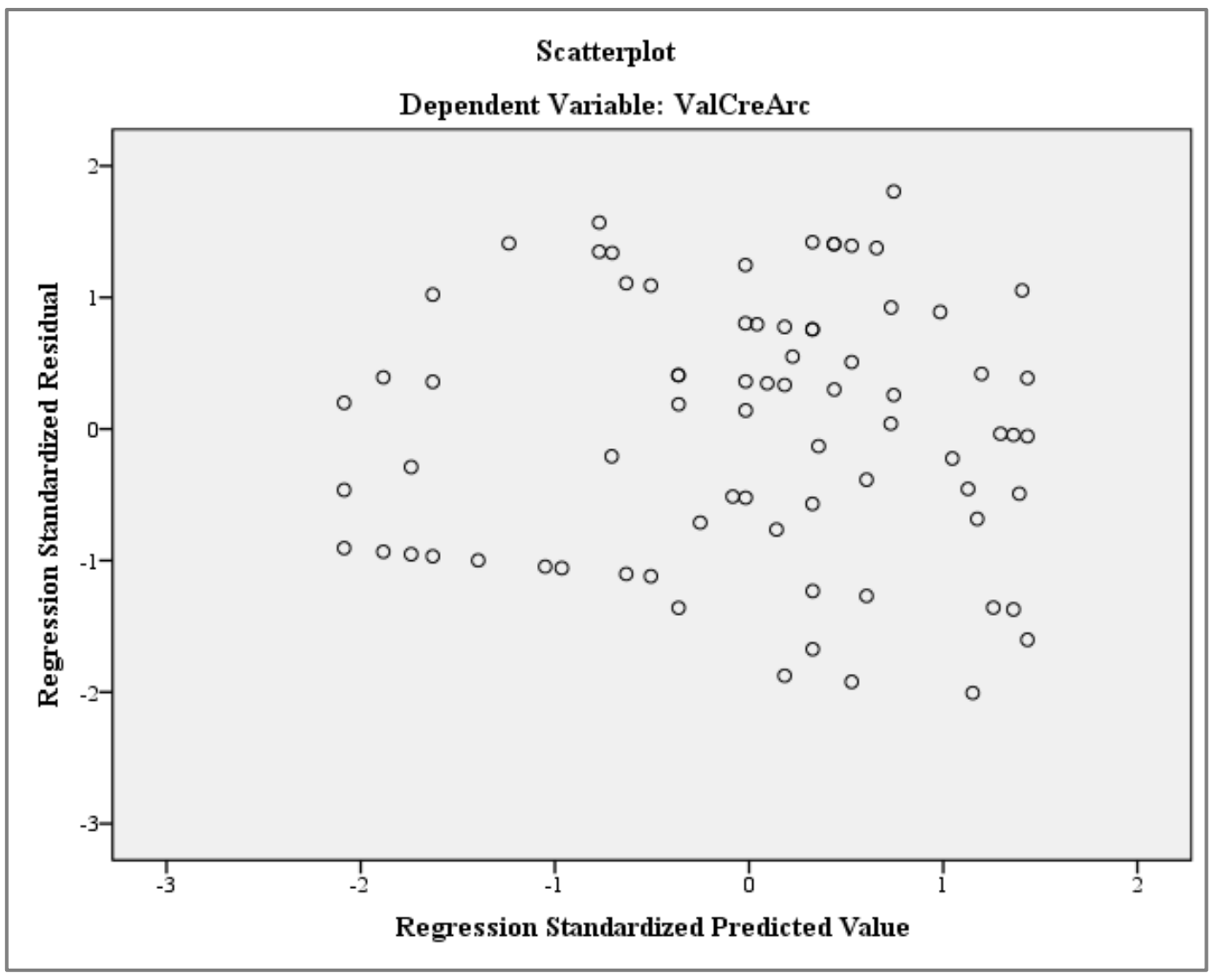

**Model 4b:**

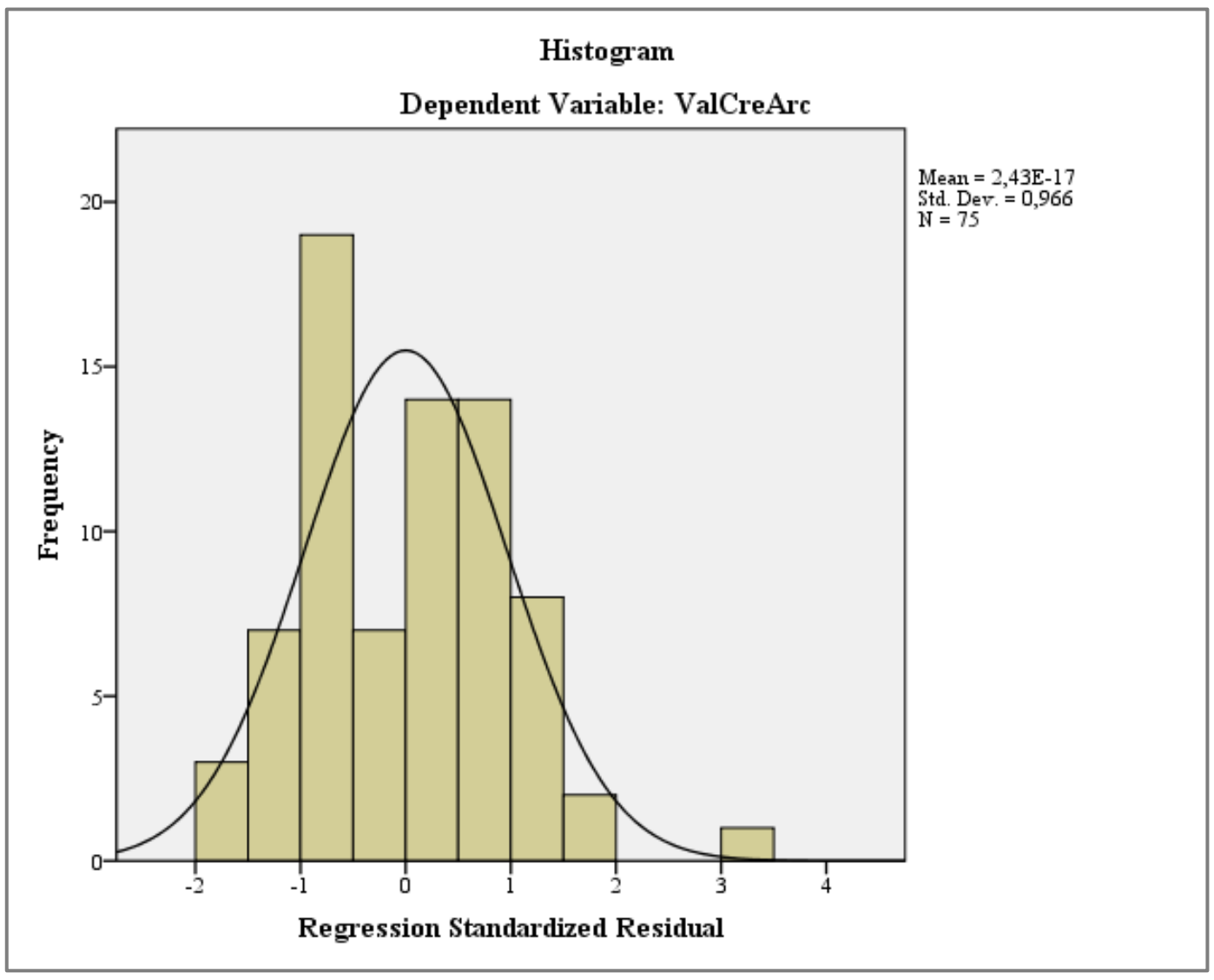

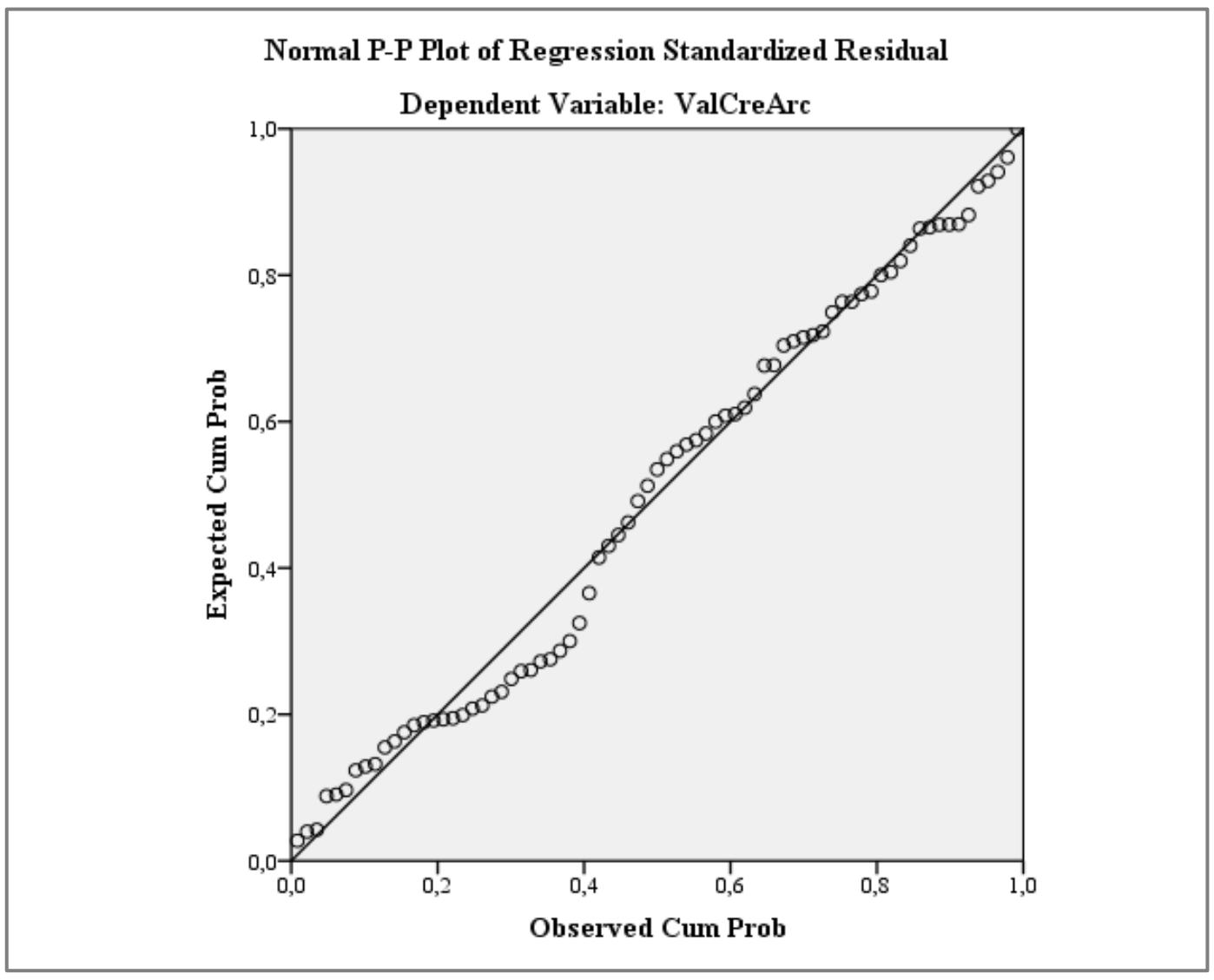
Normal P-P Plot of Regression Standardized Residual
Dependent Variable: ValCreArc
Expected Cum Prob
Observed Cum Prob

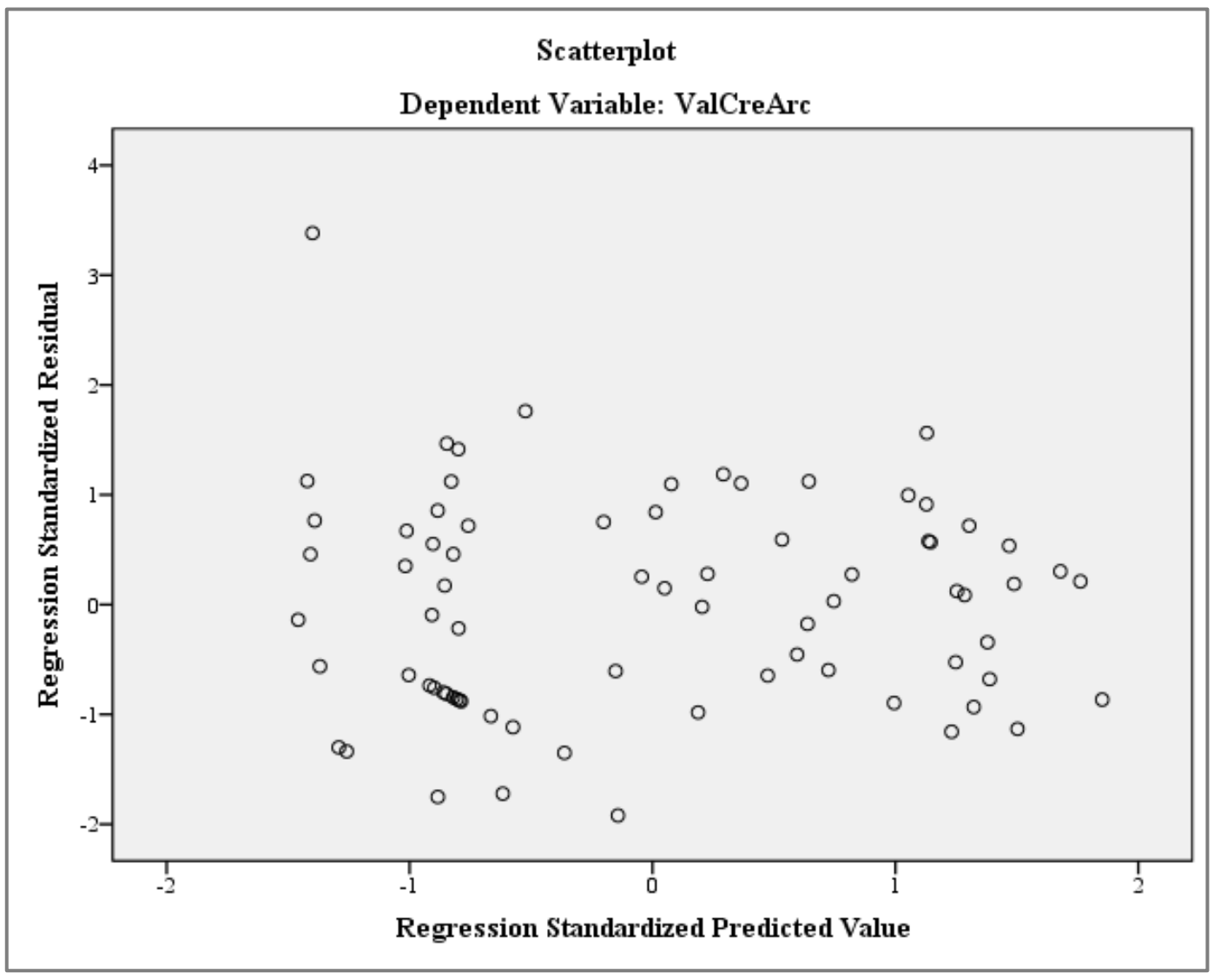
Scatterplot
Dependent Variable: ValCreArc
Regression Standardized Residual
Regression Standardized Predicted Value

## Model 4c:

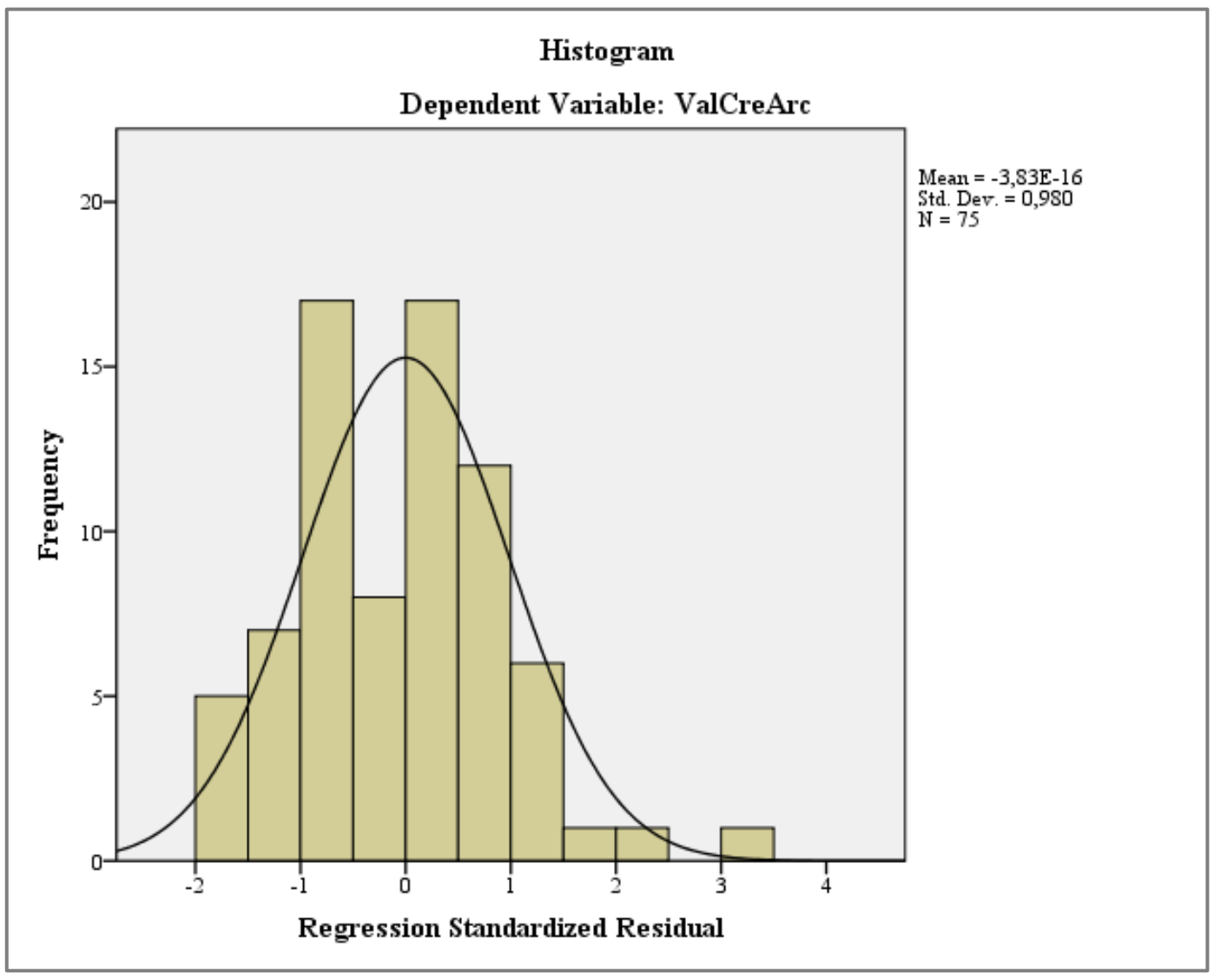

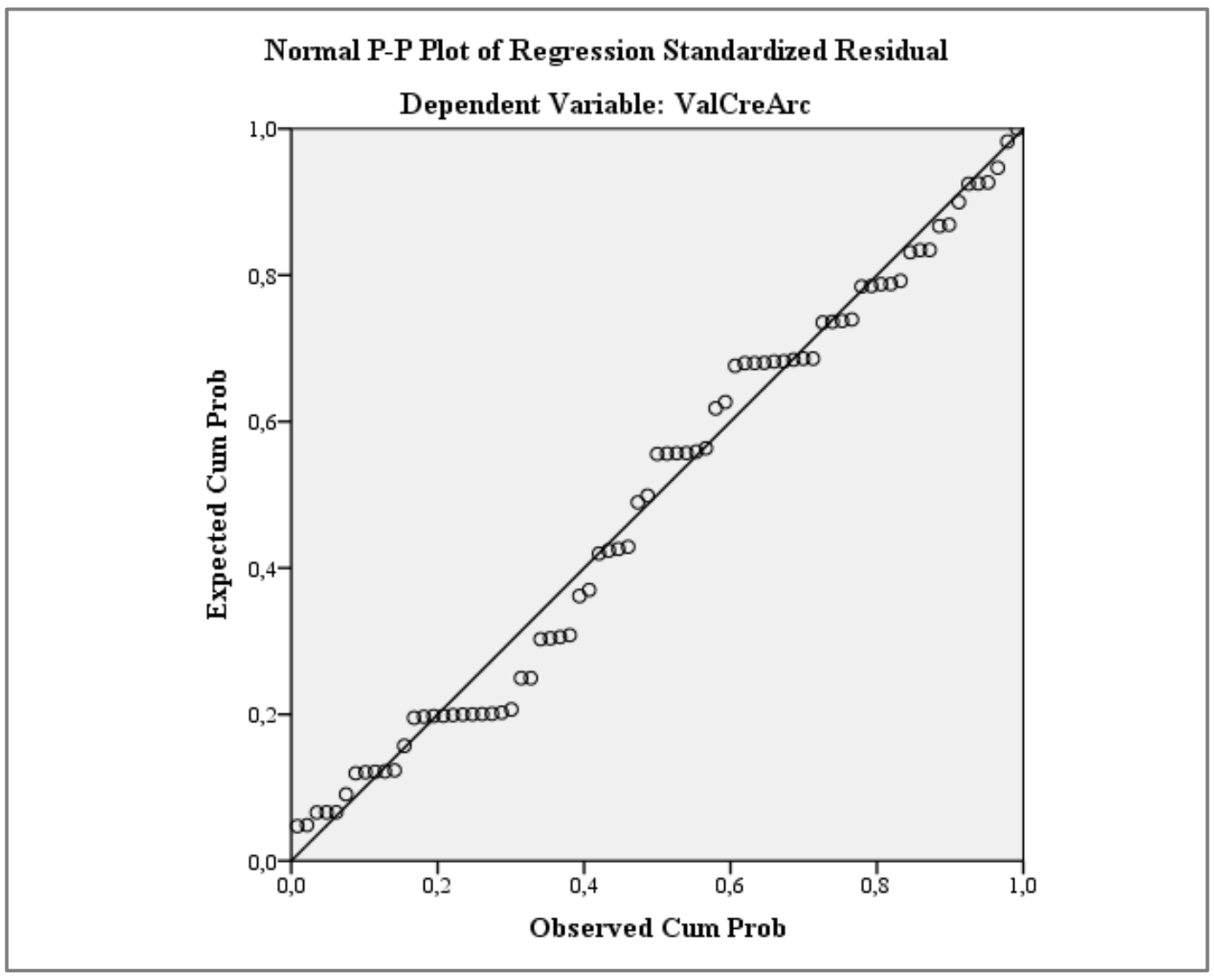
Normal P-P Plot of Regression Standardized Residual
Dependent Variable: ValCreArc
Expected Cum Prob
Observed Cum Prob

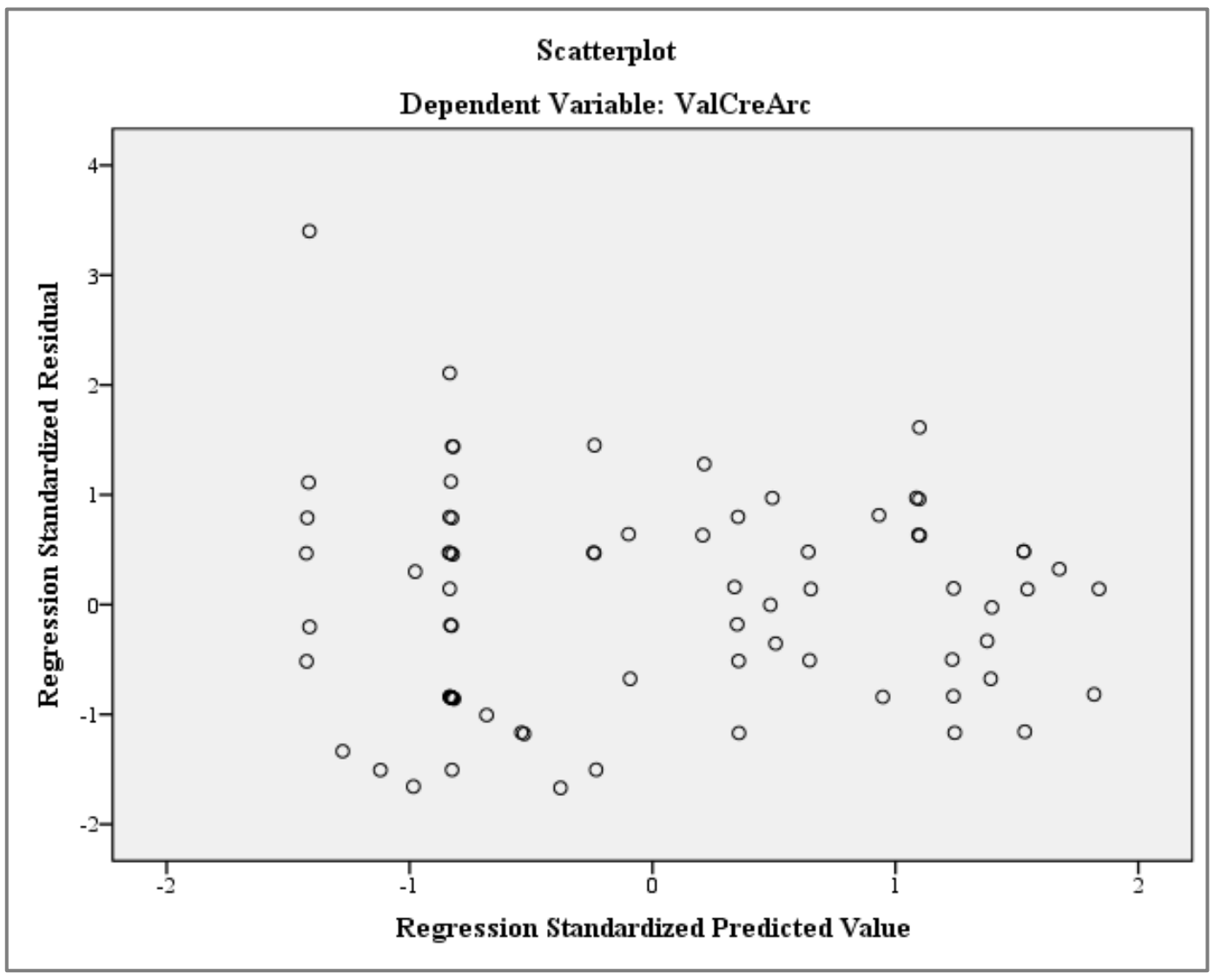
Scatterplot
Dependent Variable: ValCreArc
Regression Standardized Residual
Regression Standardized Predicted Value

*Collaboration and the Impact on the Value Offering Dimension (ValOff)*

**Model 4d:**

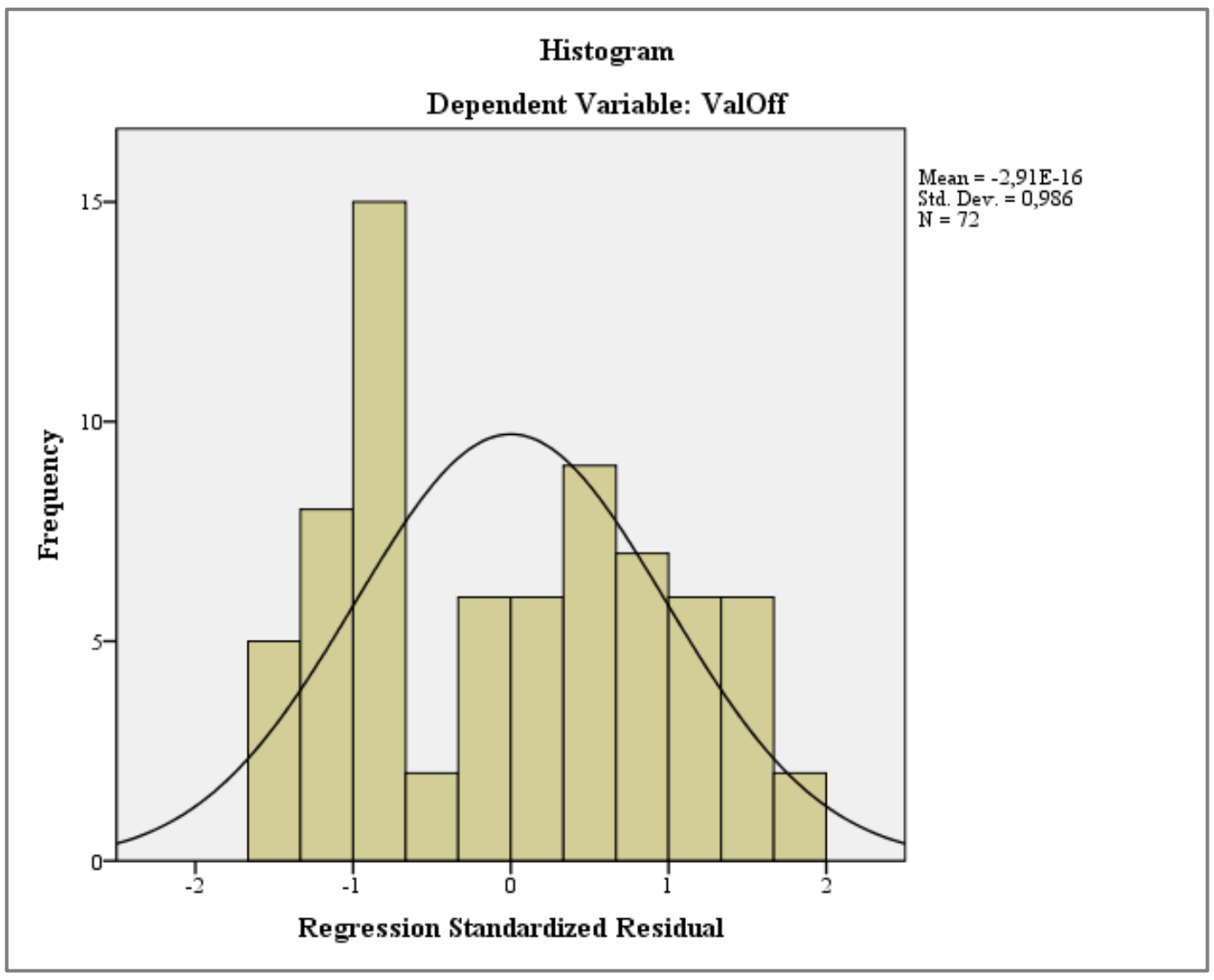

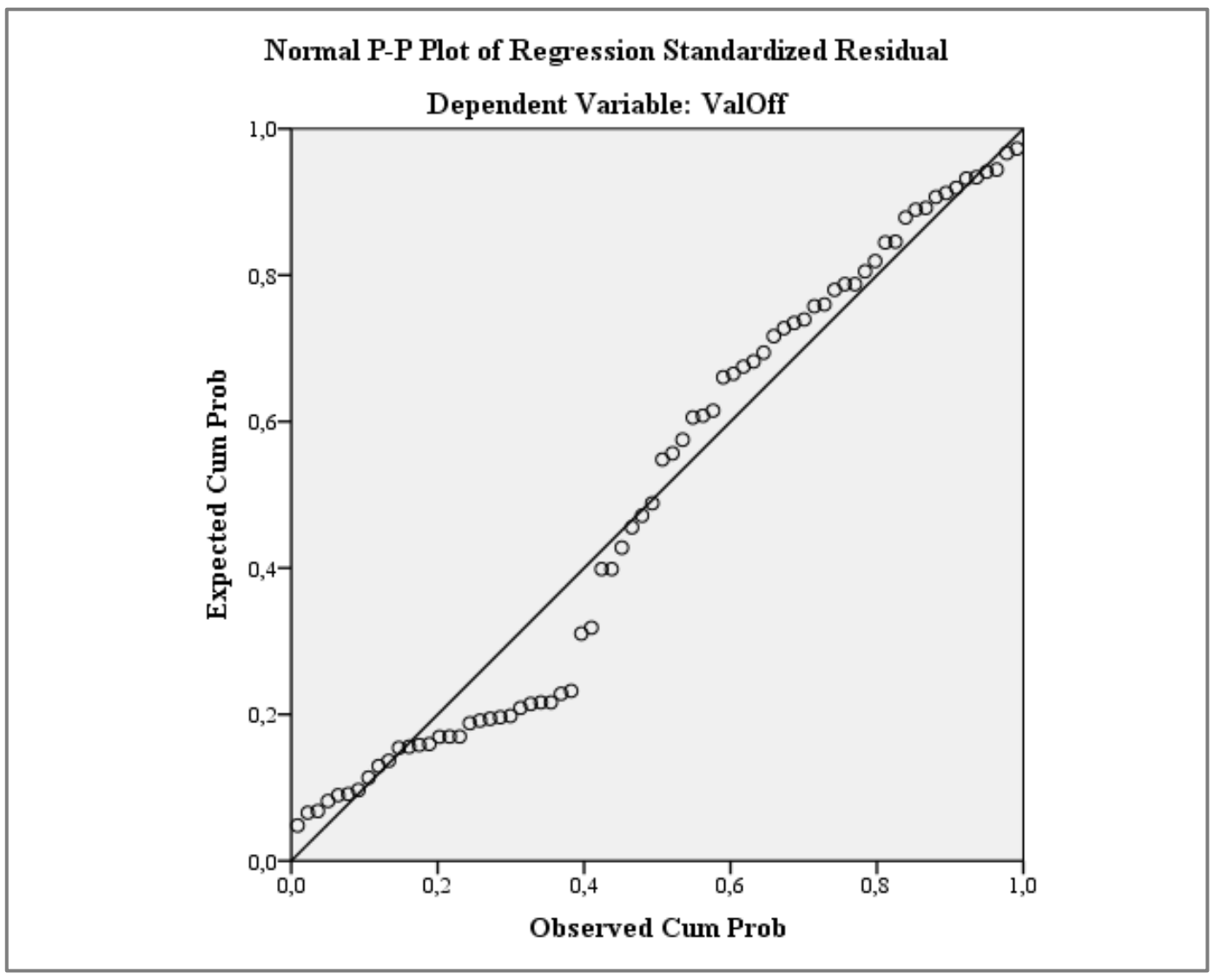

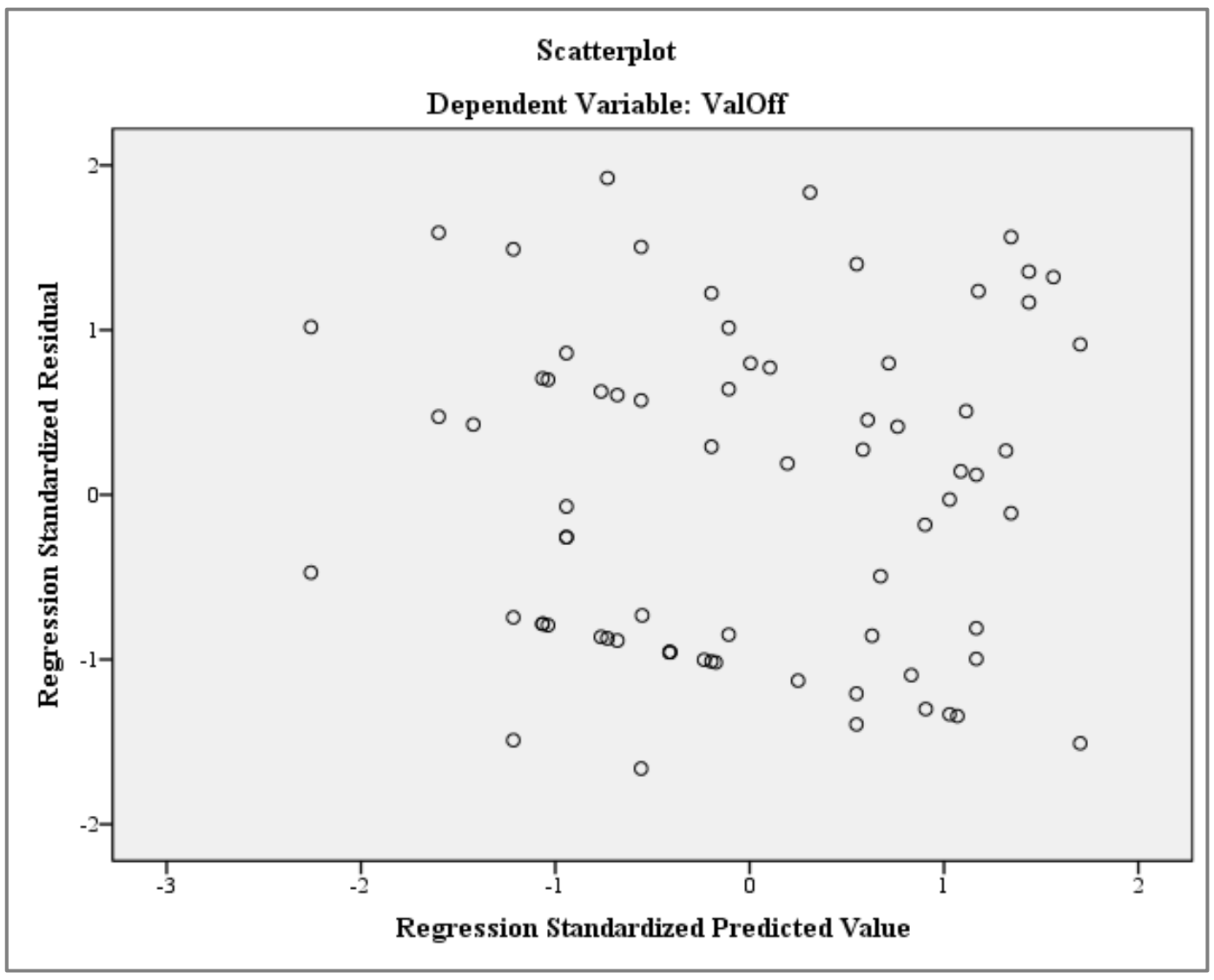

**Model 4e:**

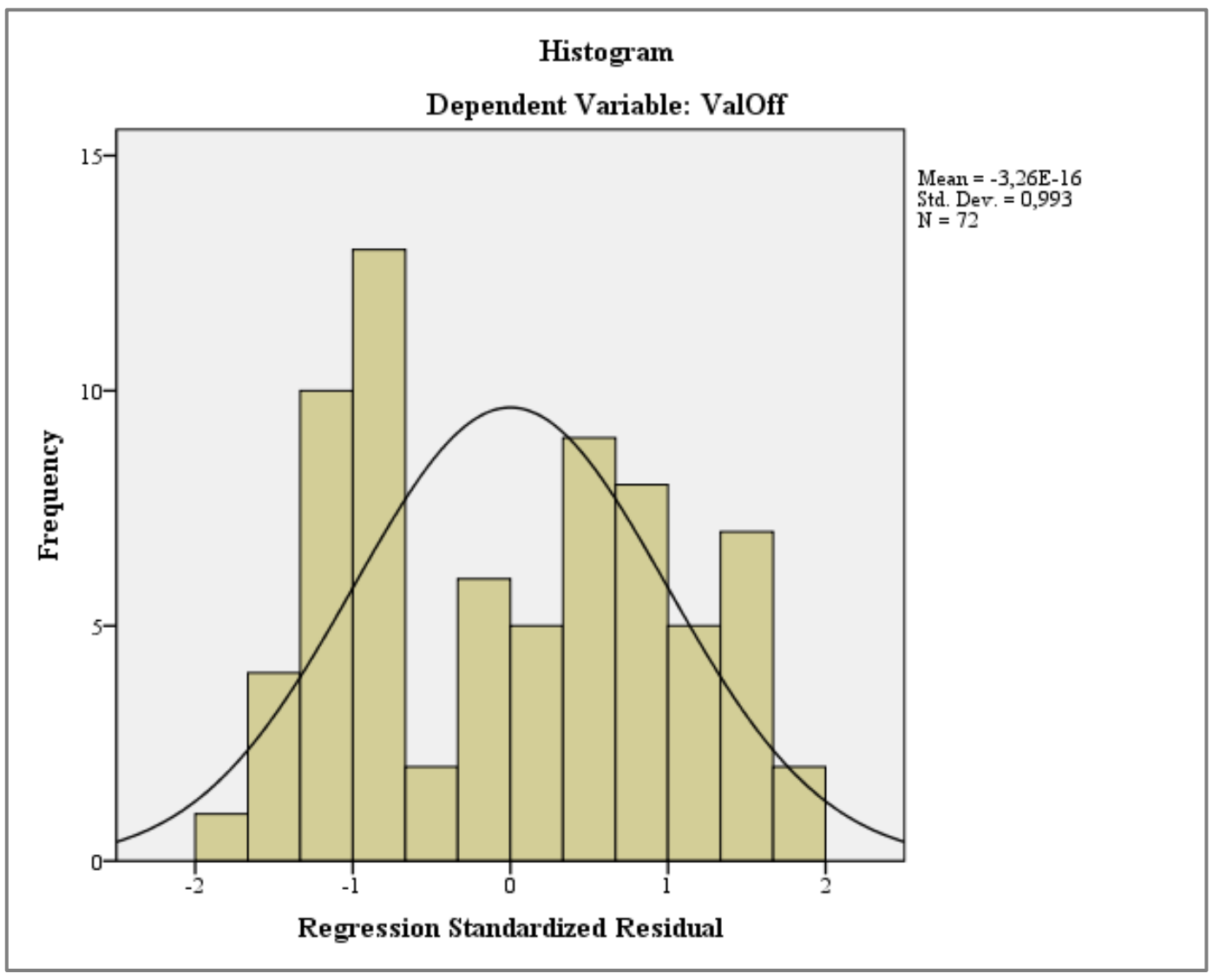

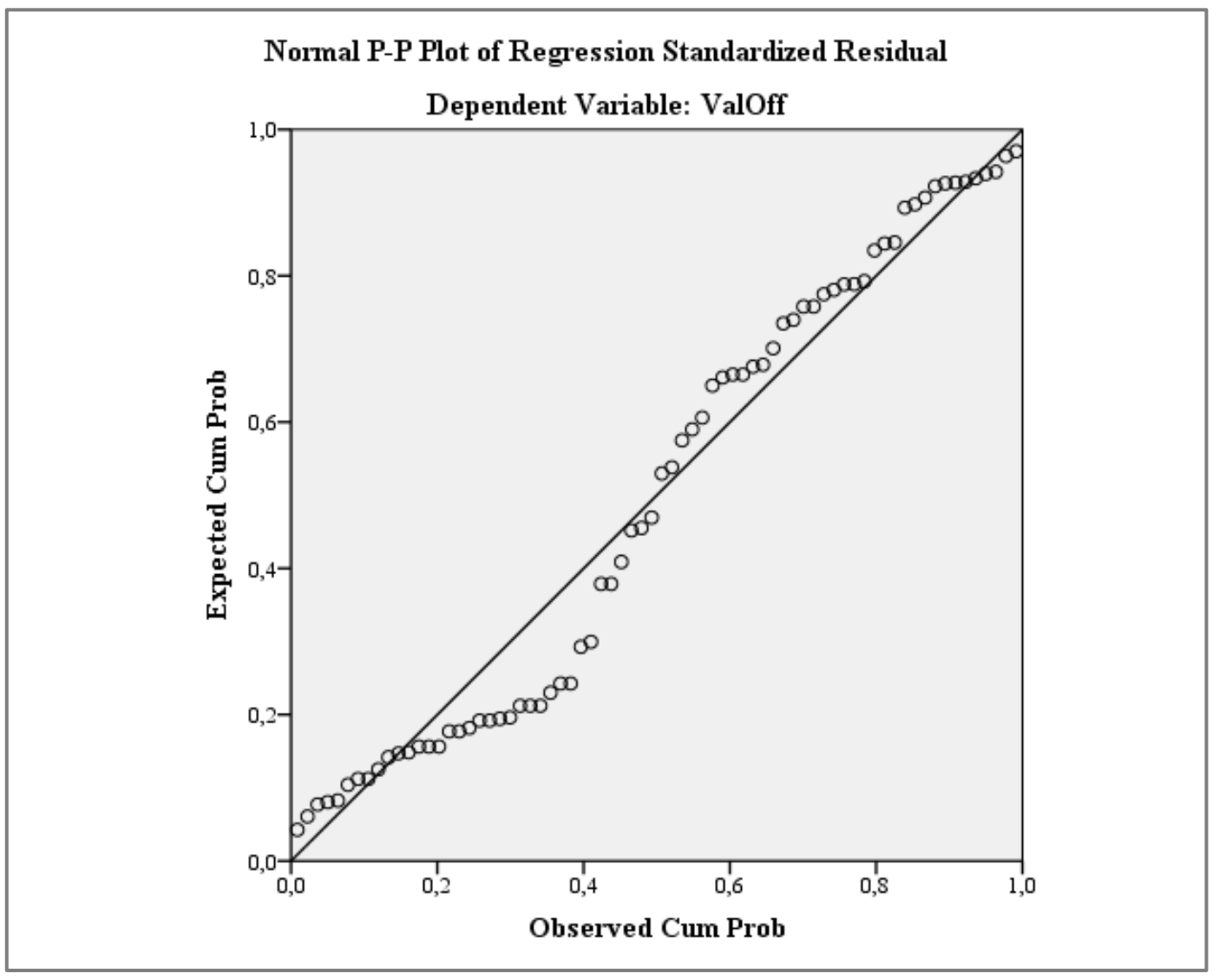

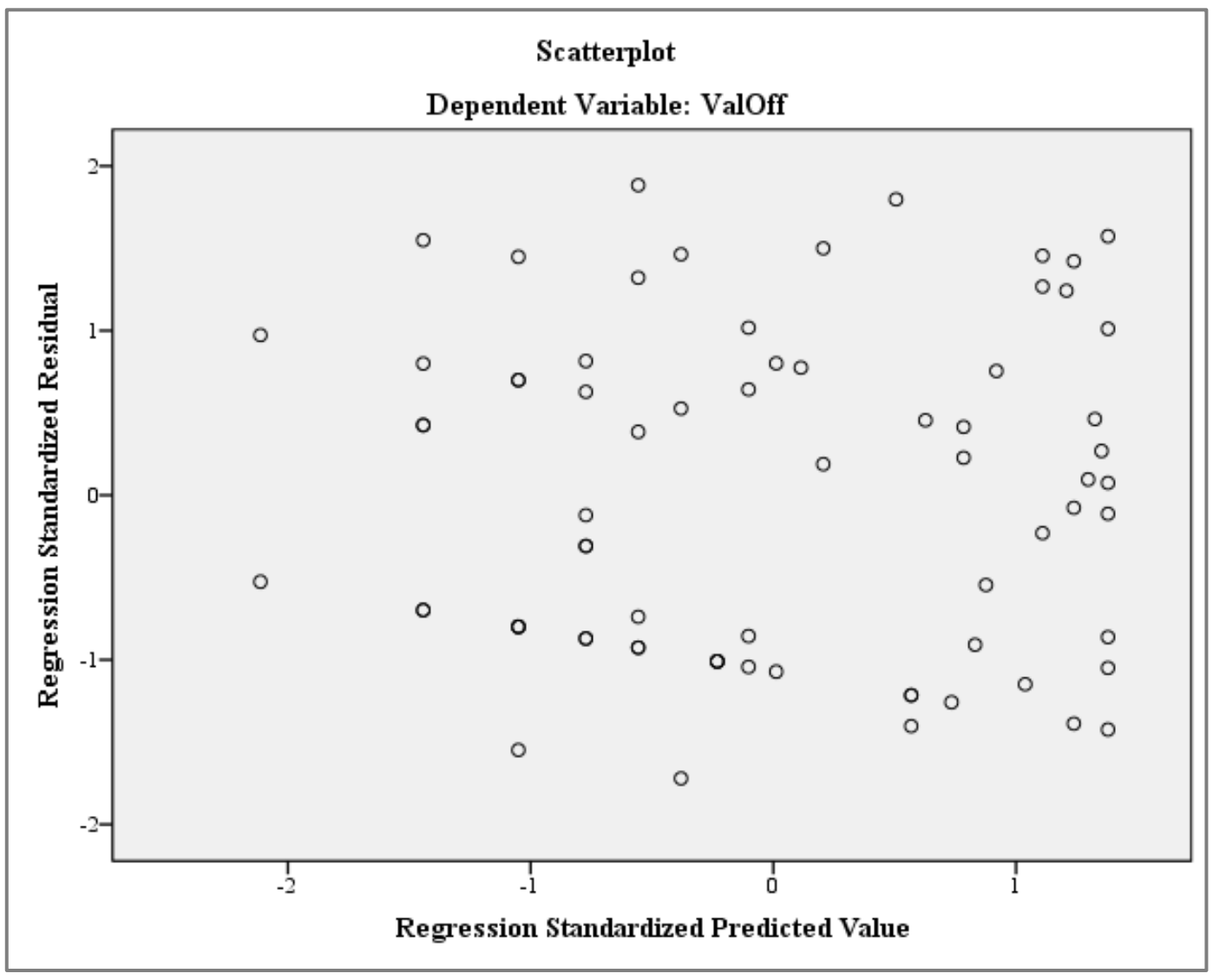

**Model 4f:**

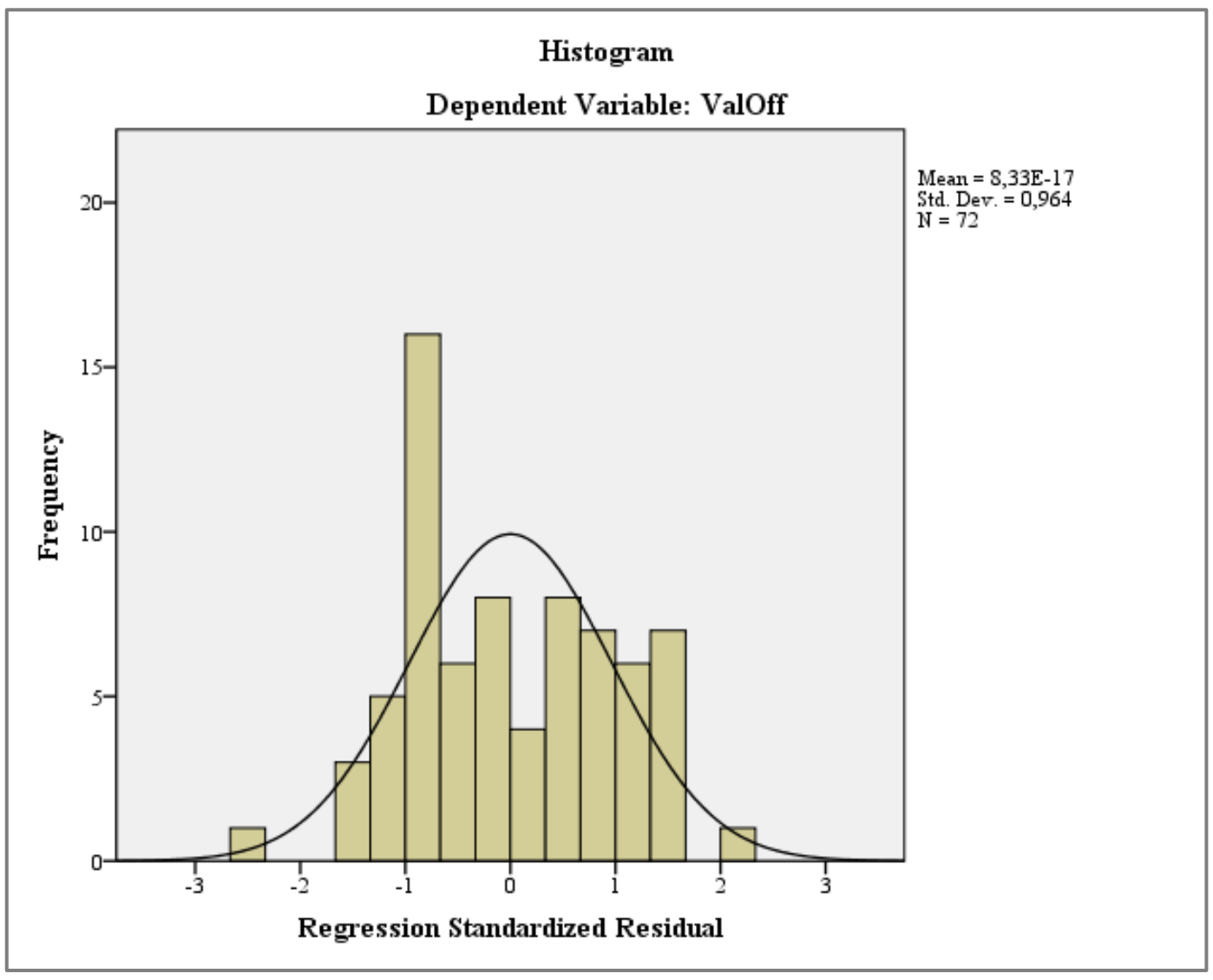

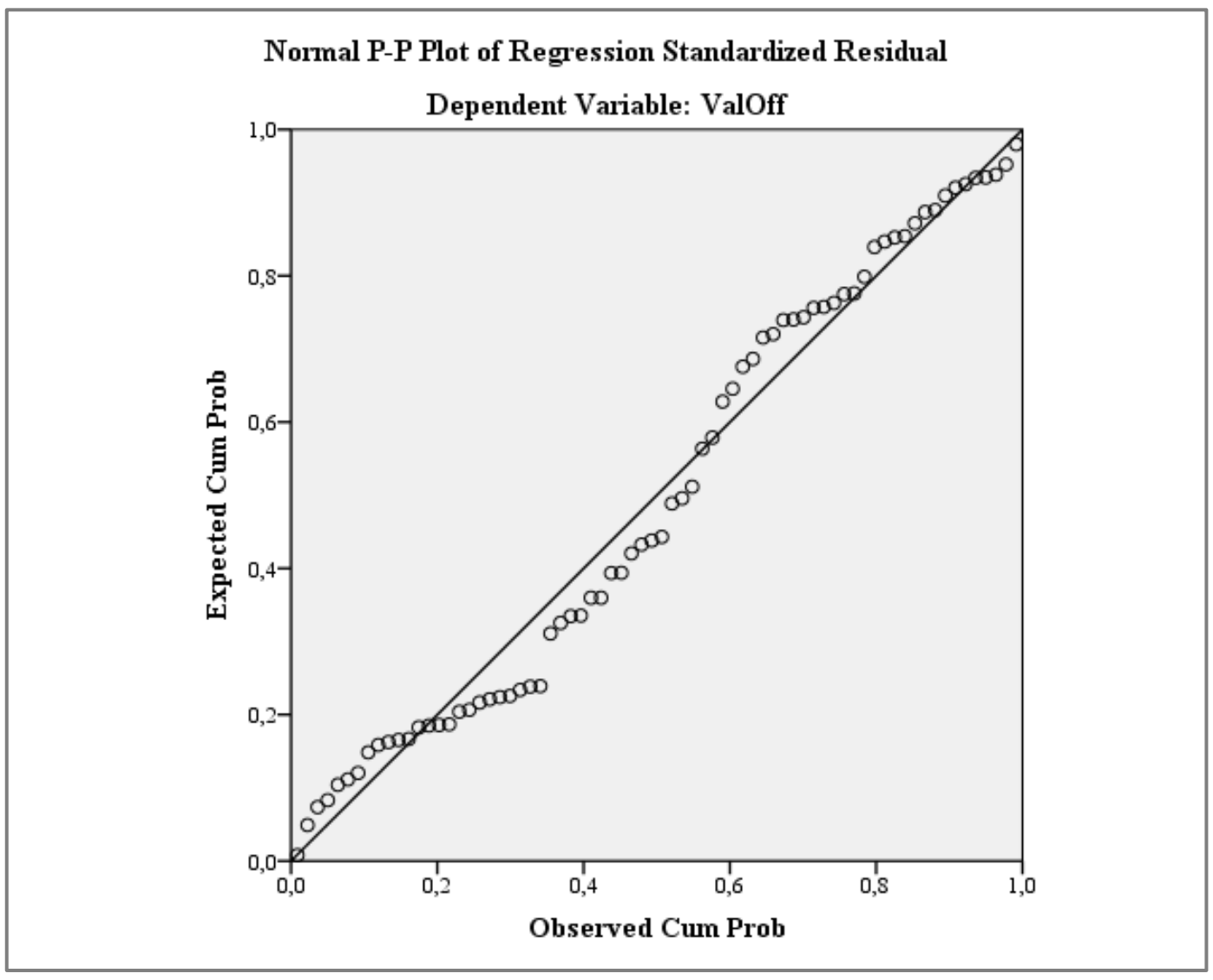
Normal P-P Plot of Regression Standardized Residual
Dependent Variable: ValOff
Expected Cum Prob
Observed Cum Prob

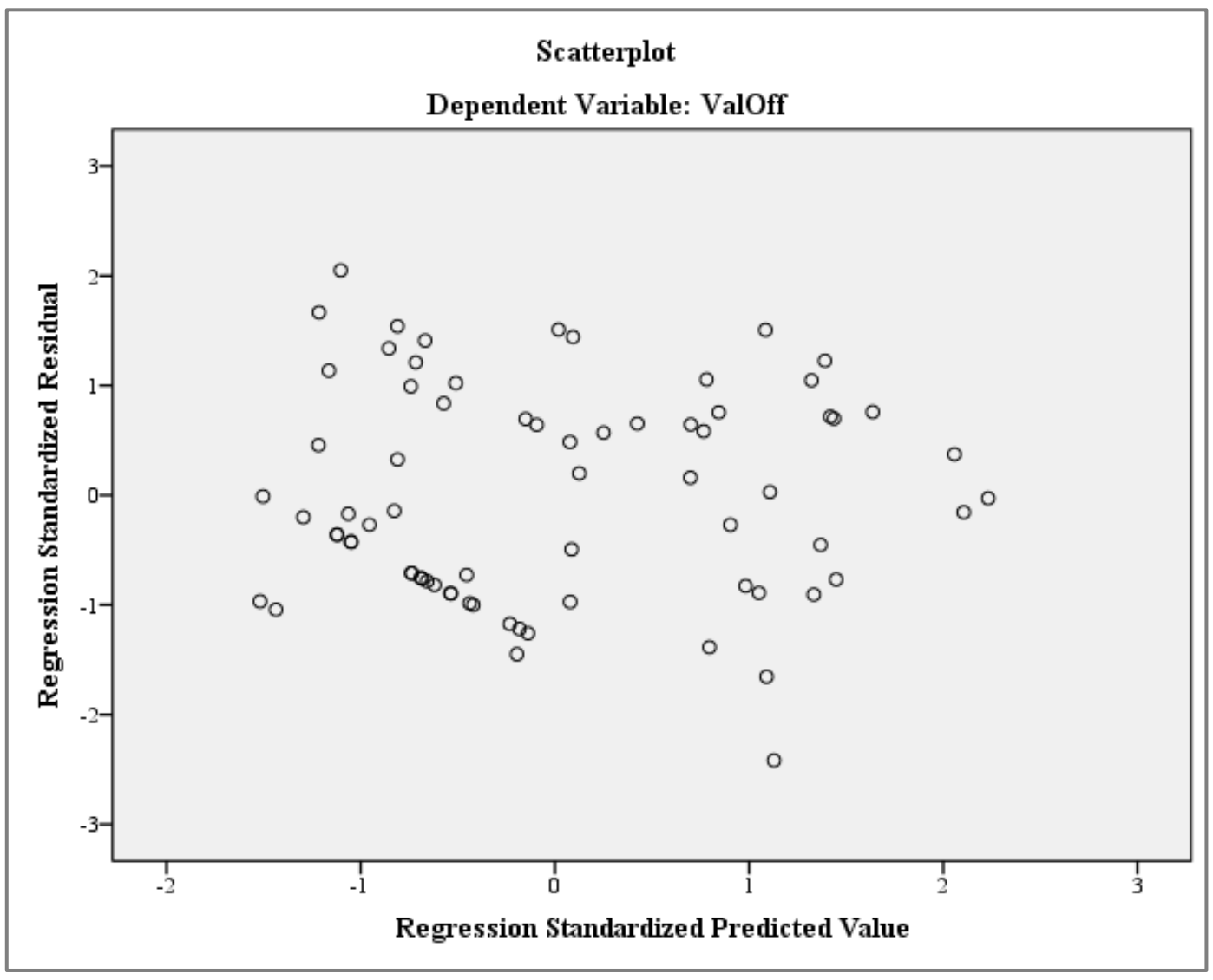
Scatterplot
Dependent Variable: ValOff
Regression Standardized Residual
Regression Standardized Predicted Value

**Model 4g:**

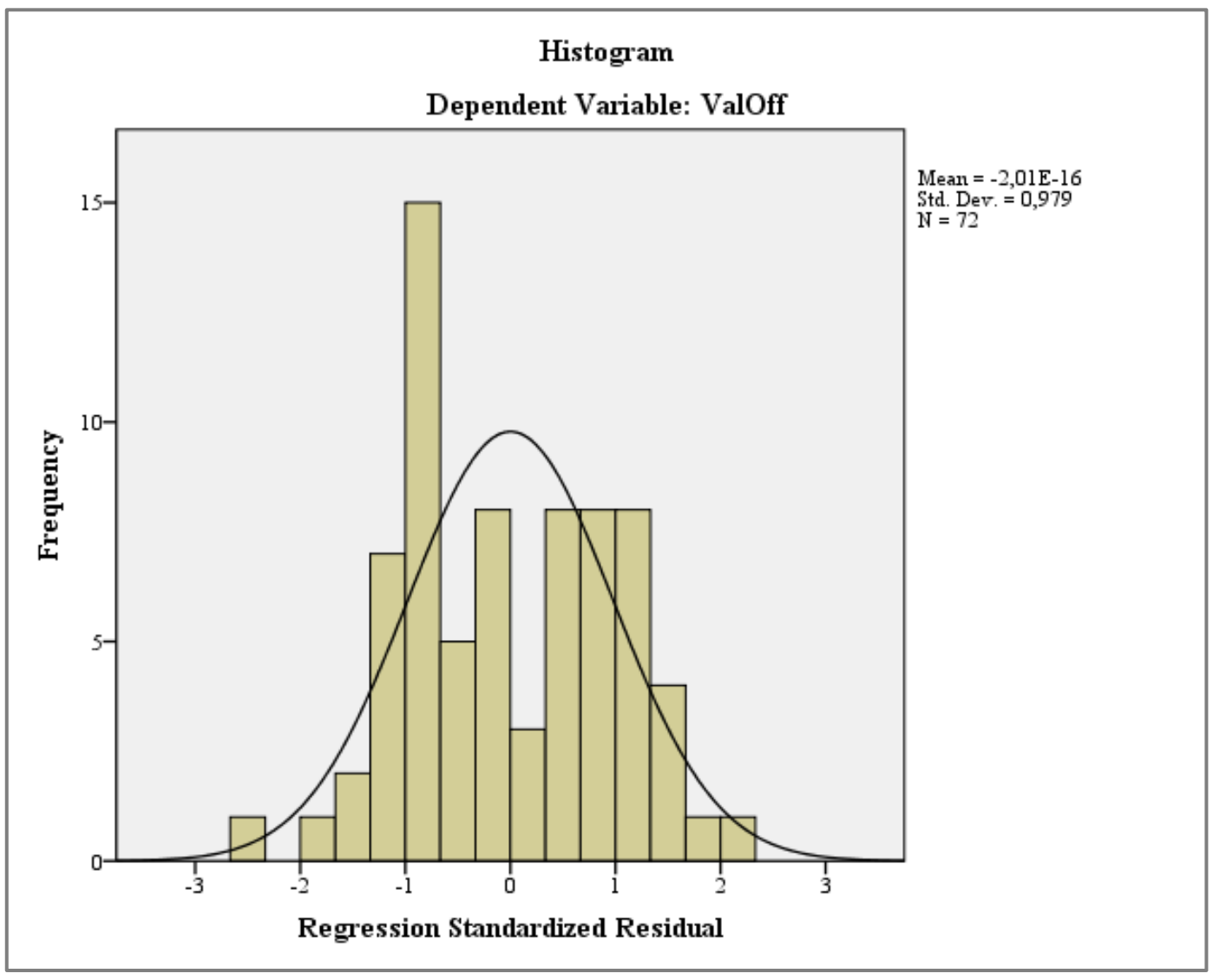

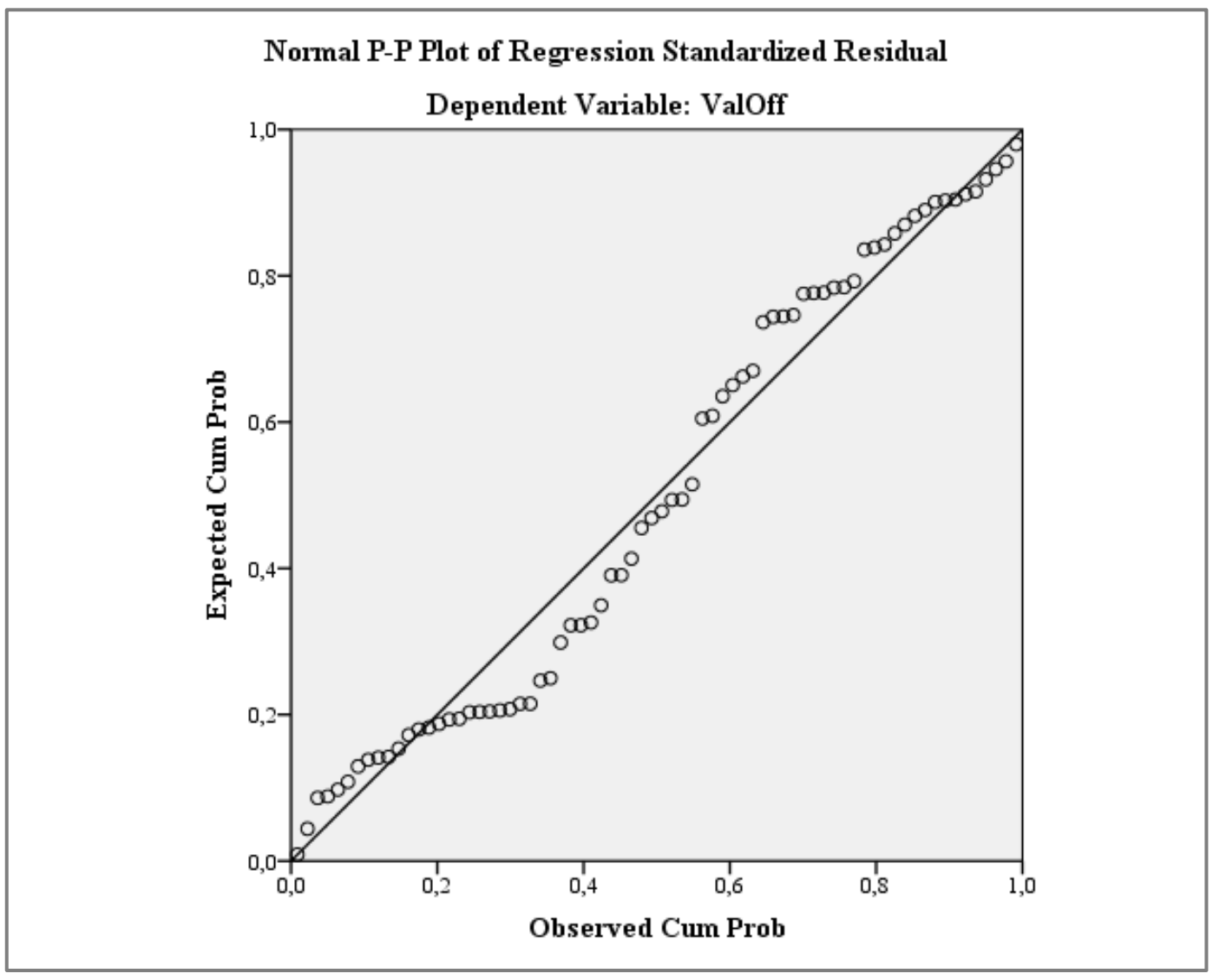

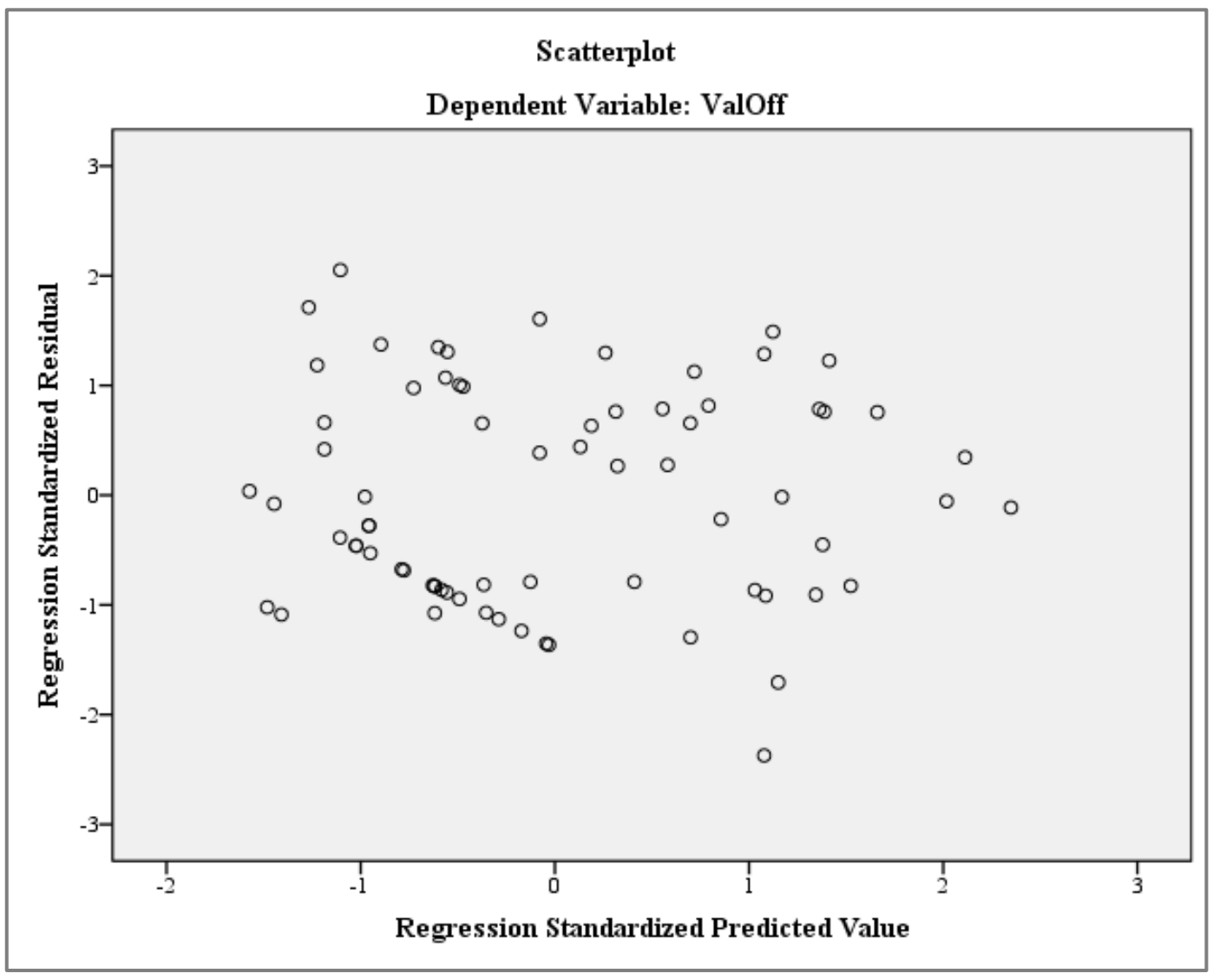

*Collaboration and the Impact on the Revenue Model Dimension (RevMod)*

**Model 4h:**

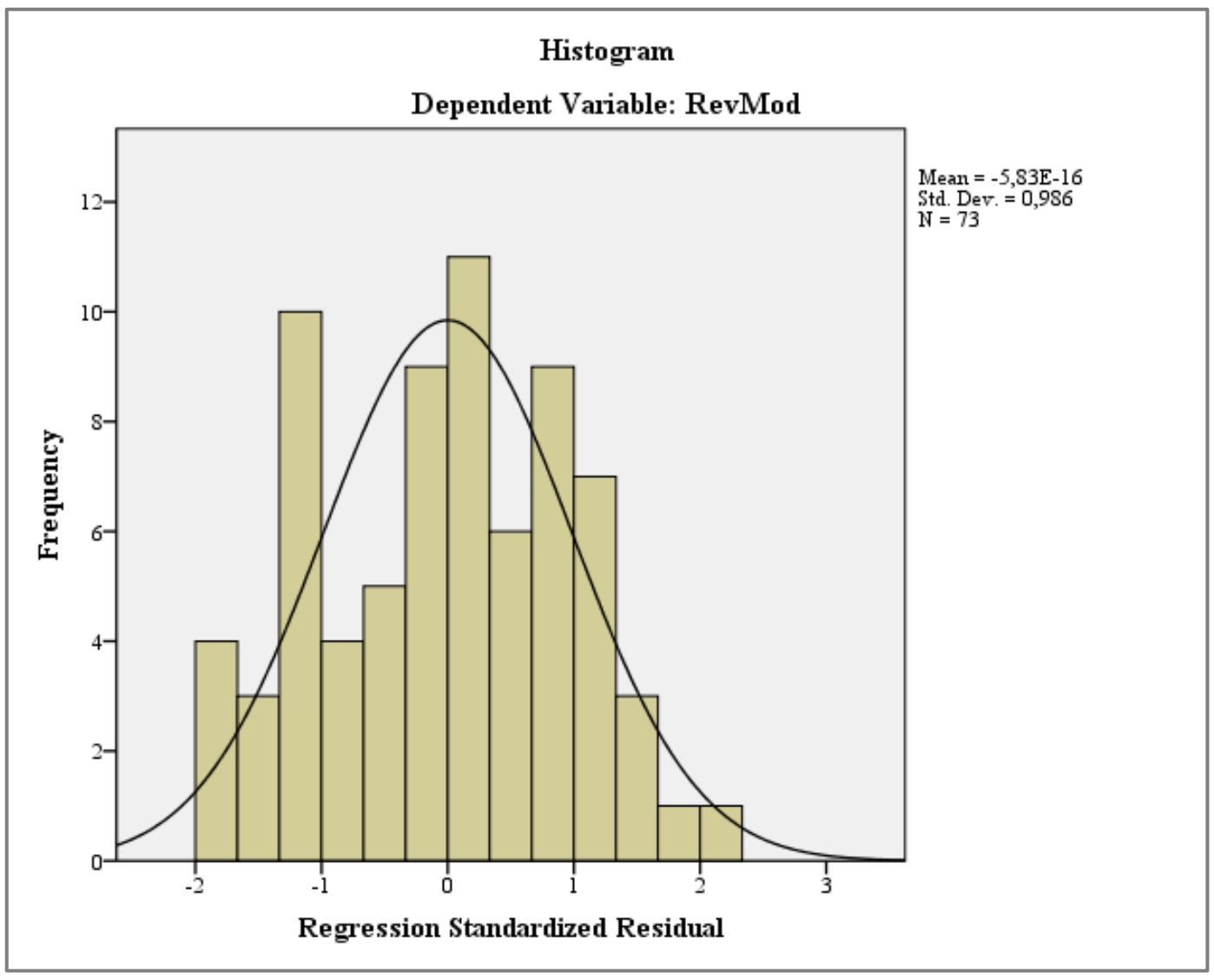

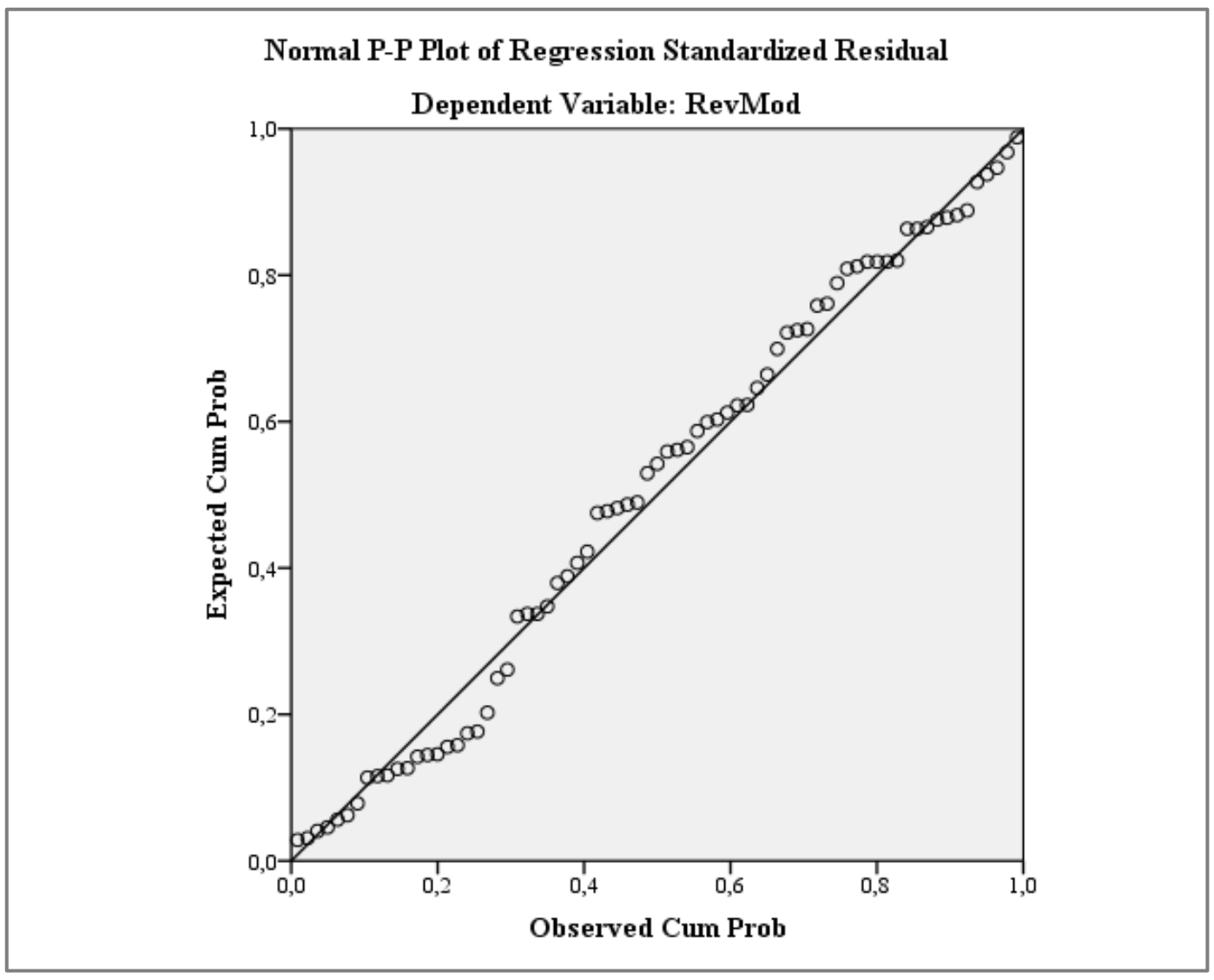

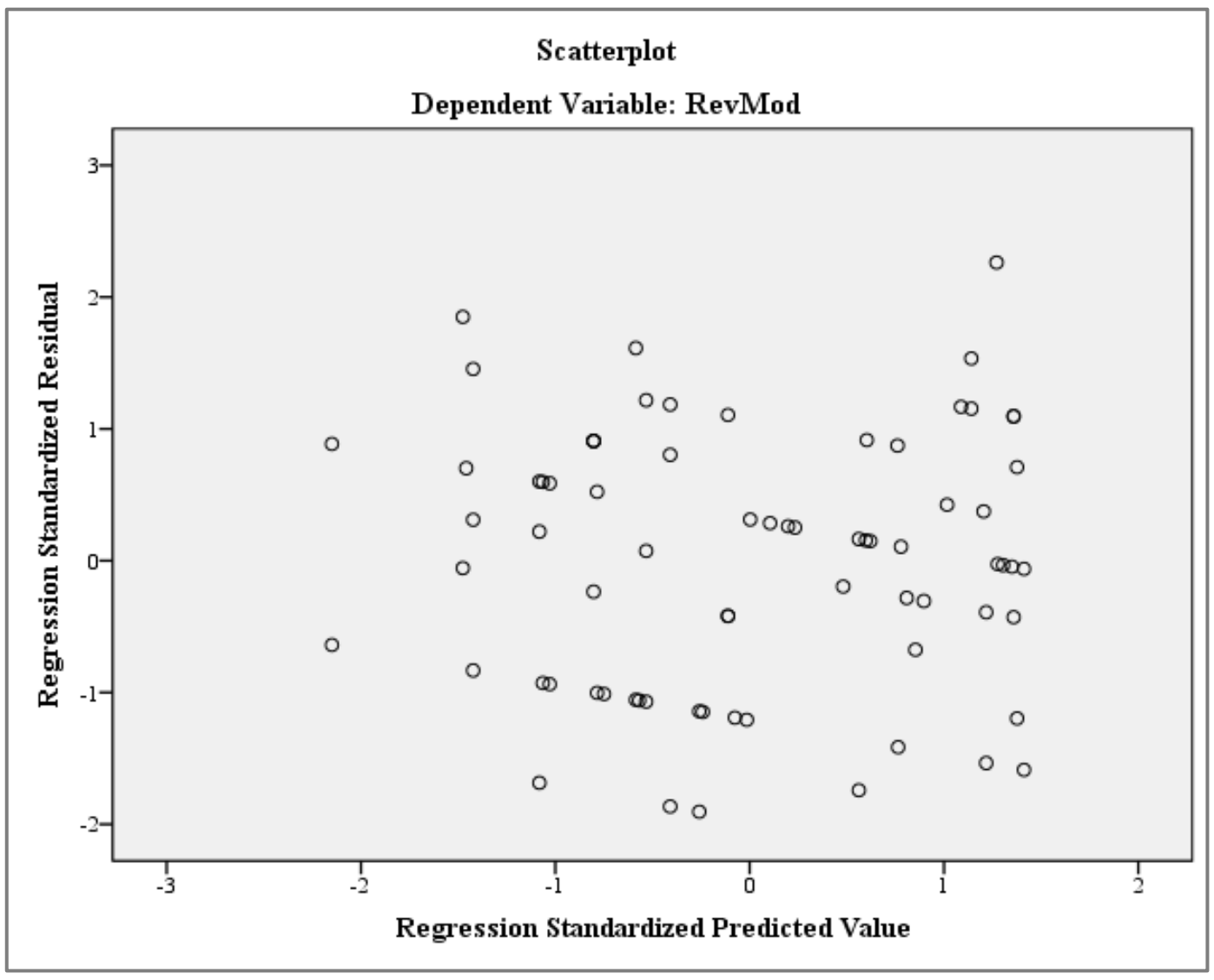

**Model 4i:**

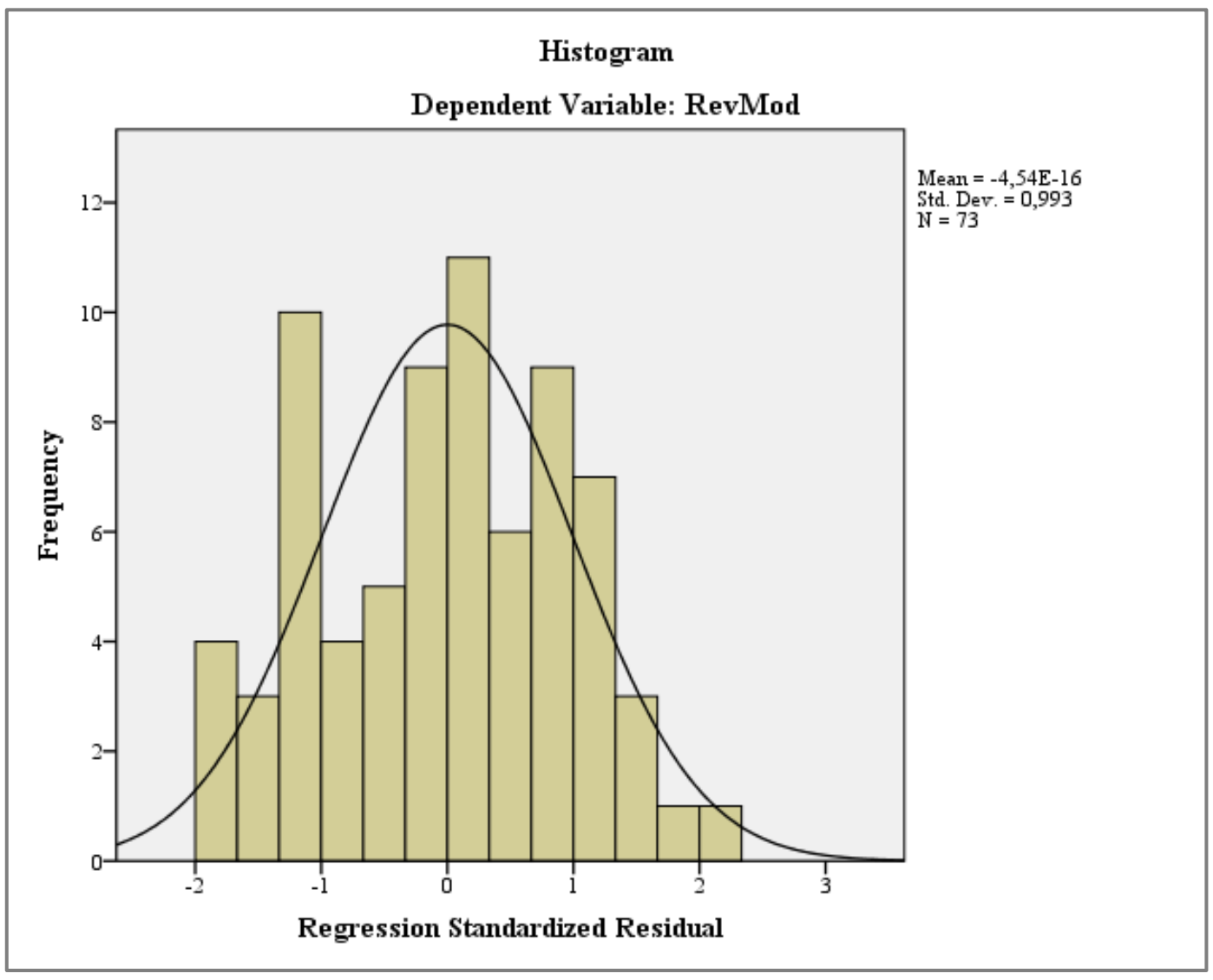

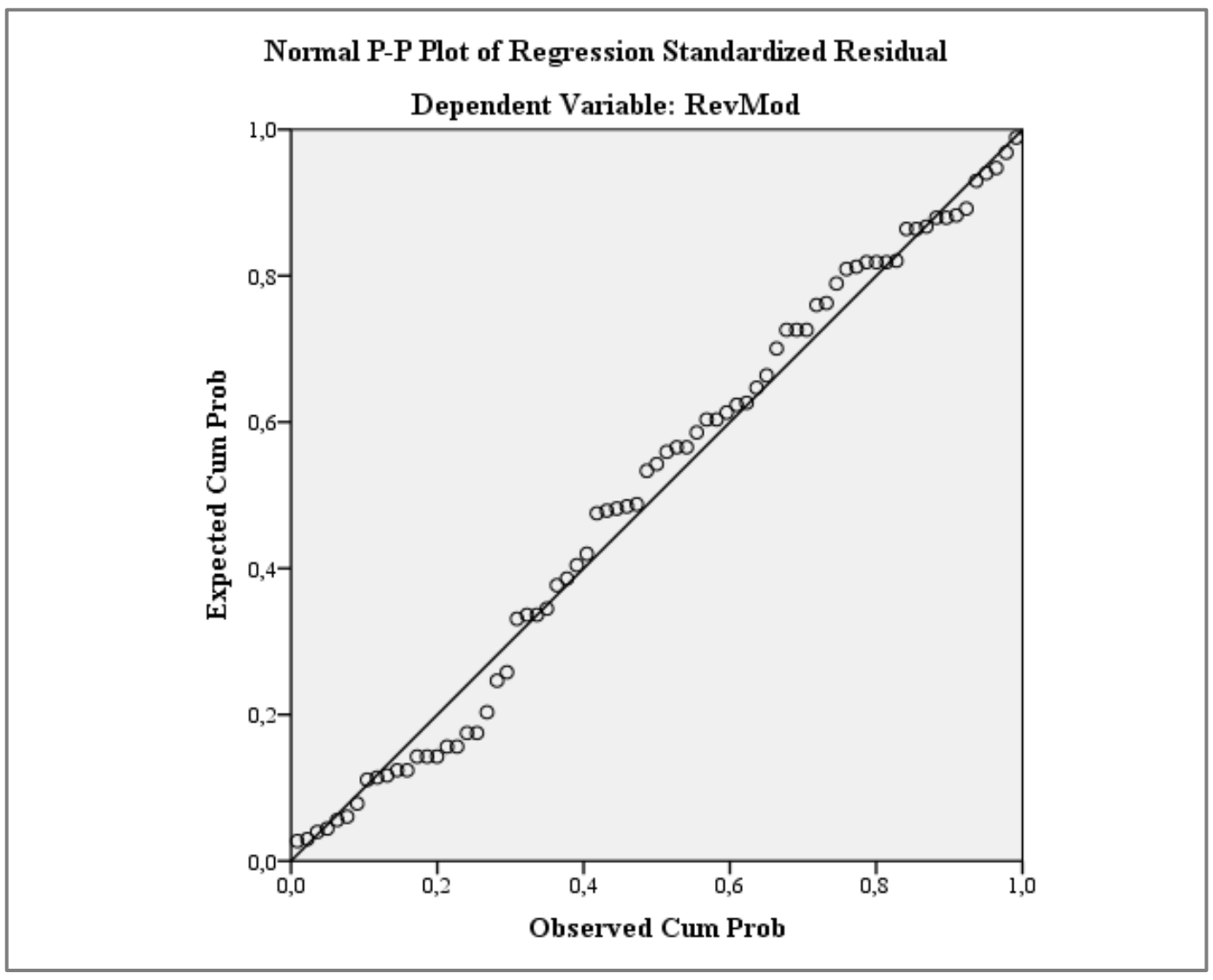
Normal P-P Plot of Regression Standardized Residual
Dependent Variable: RevMod
Expected Cum Prob
Observed Cum Prob

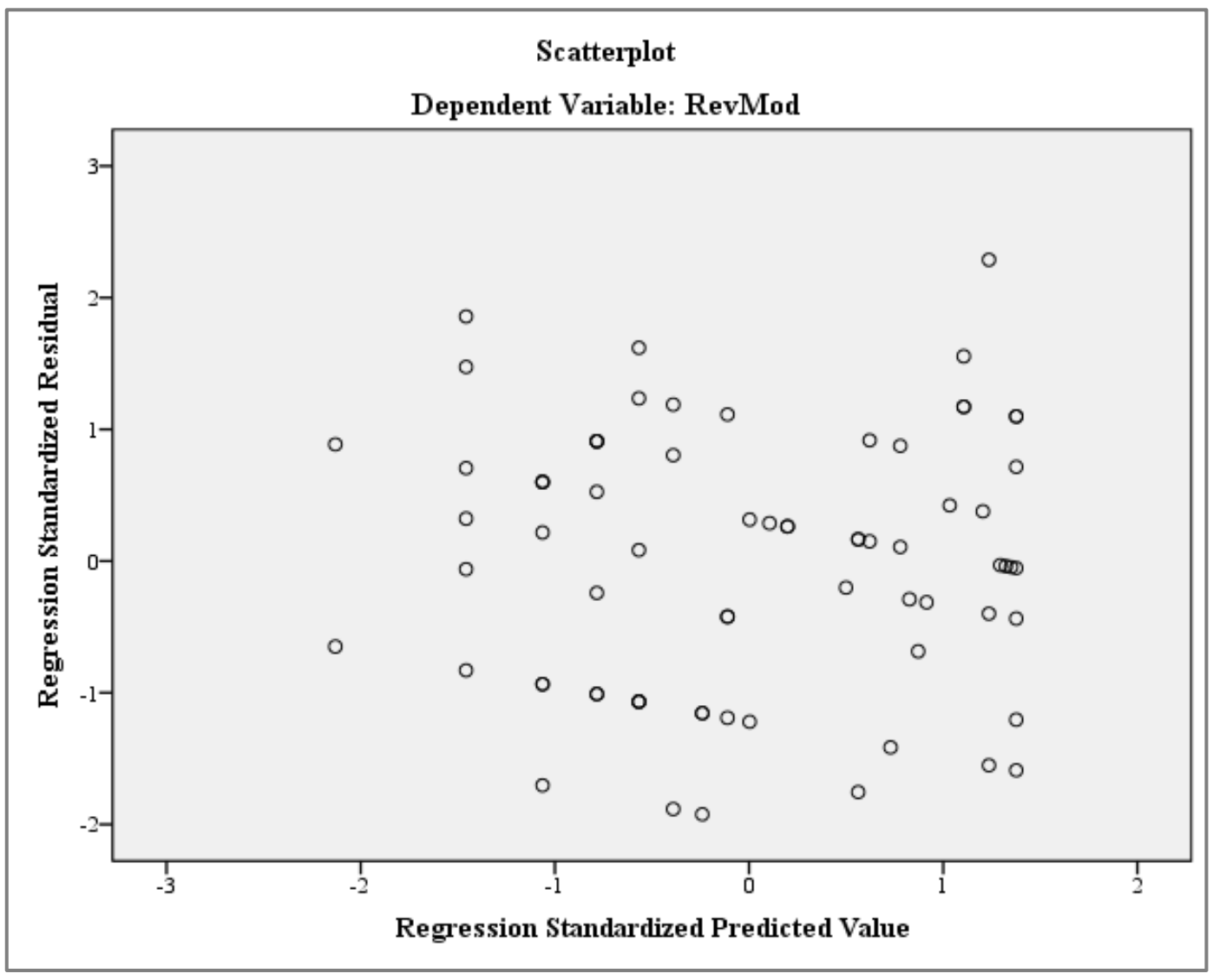
Scatterplot
Dependent Variable: RevMod
Regression Standardized Residual
Regression Standardized Predicted Value

**Model 4j:**

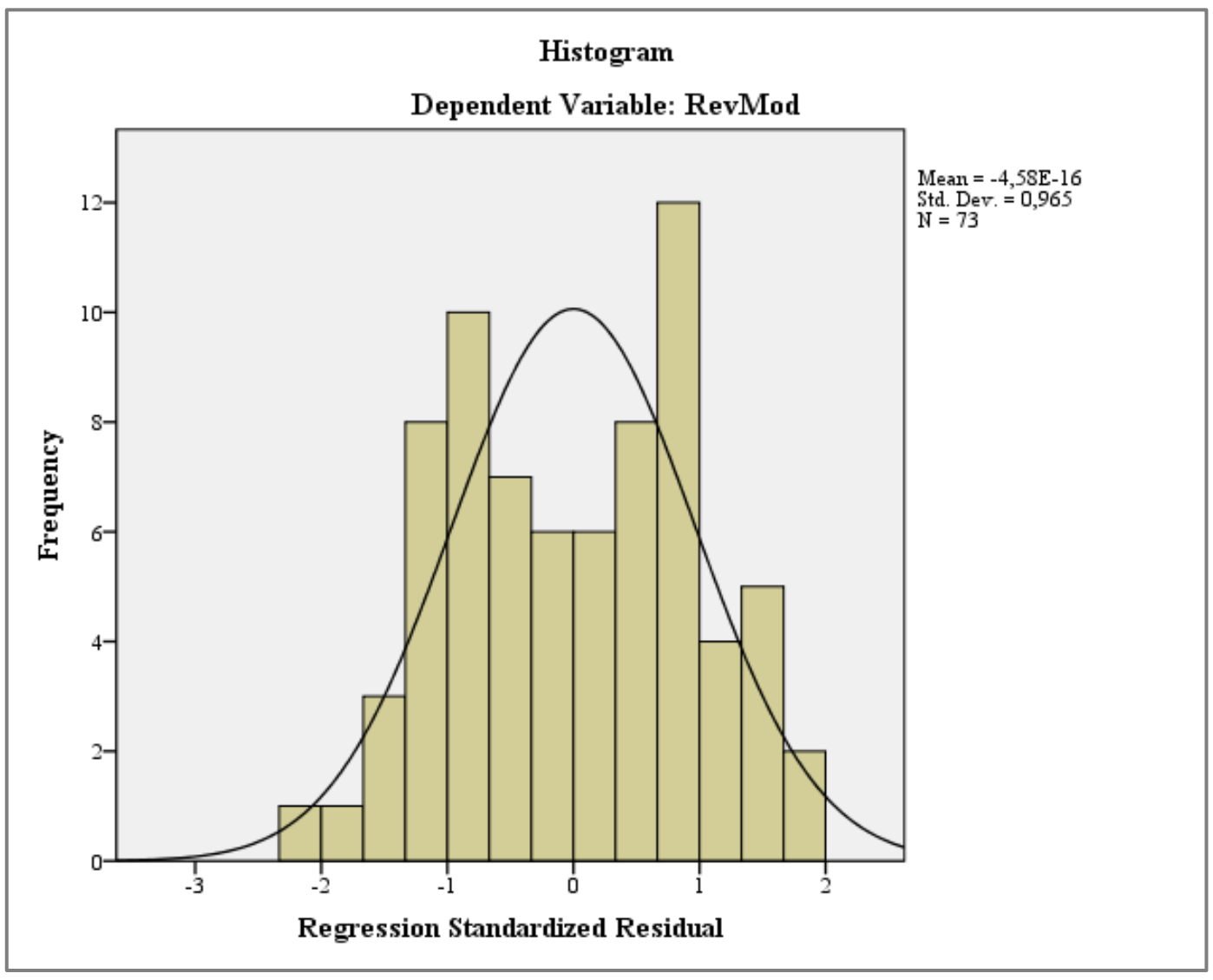

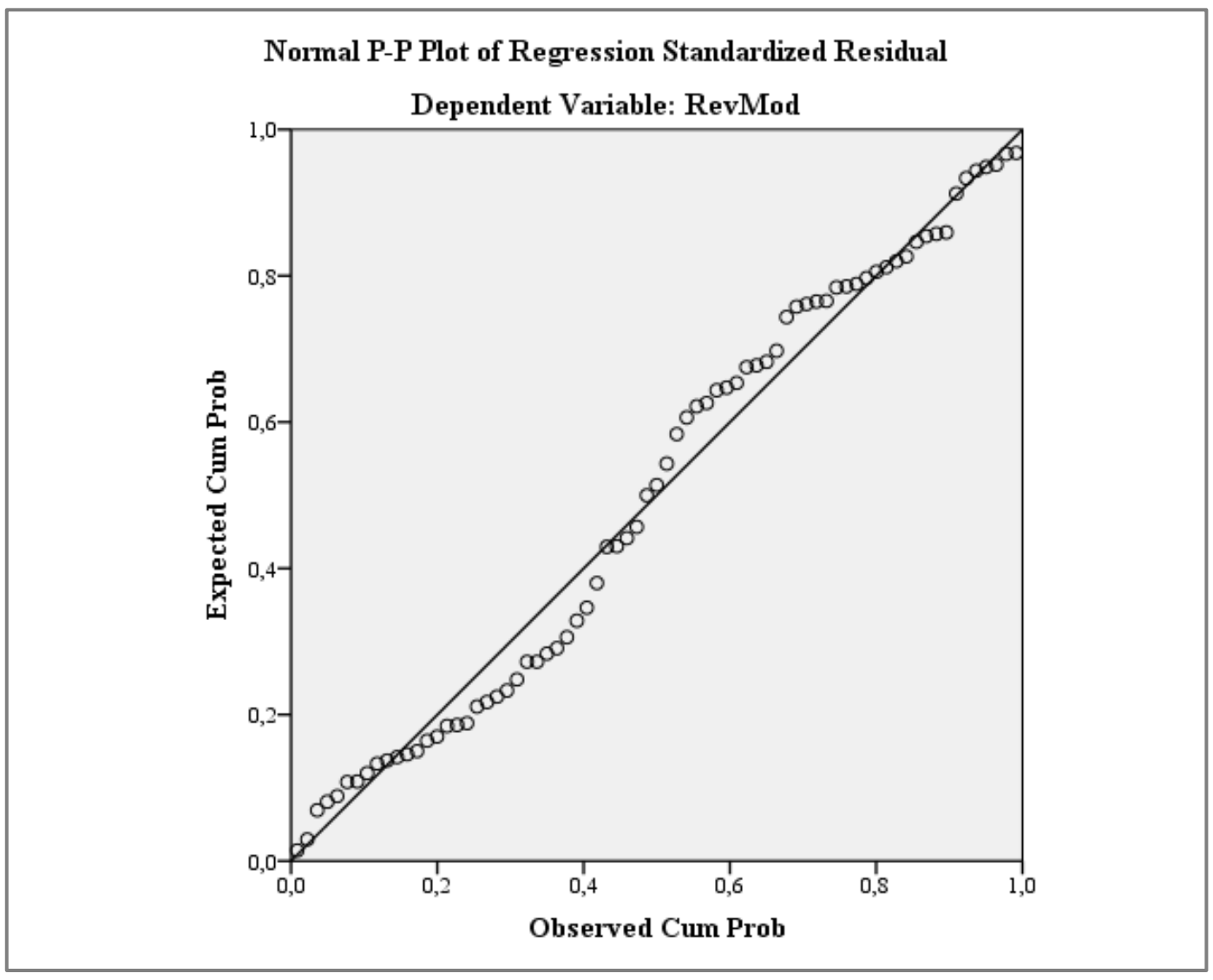
Normal P-P Plot of Regression Standardized Residual
Dependent Variable: RevMod
Expected Cum Prob
Observed Cum Prob

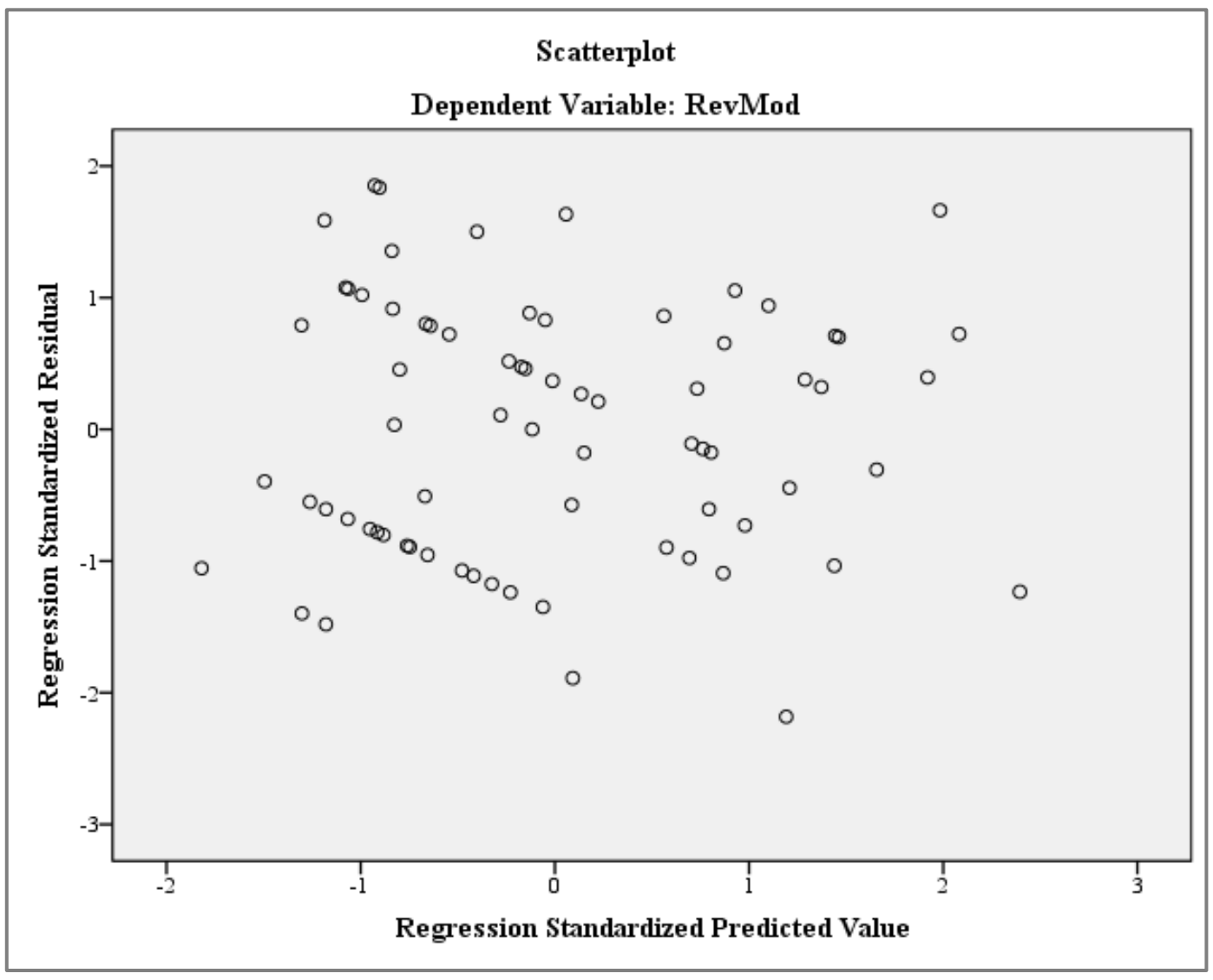
Scatterplot
Dependent Variable: RevMod
Regression Standardized Residual
Regression Standardized Predicted Value

**Model 4k:**

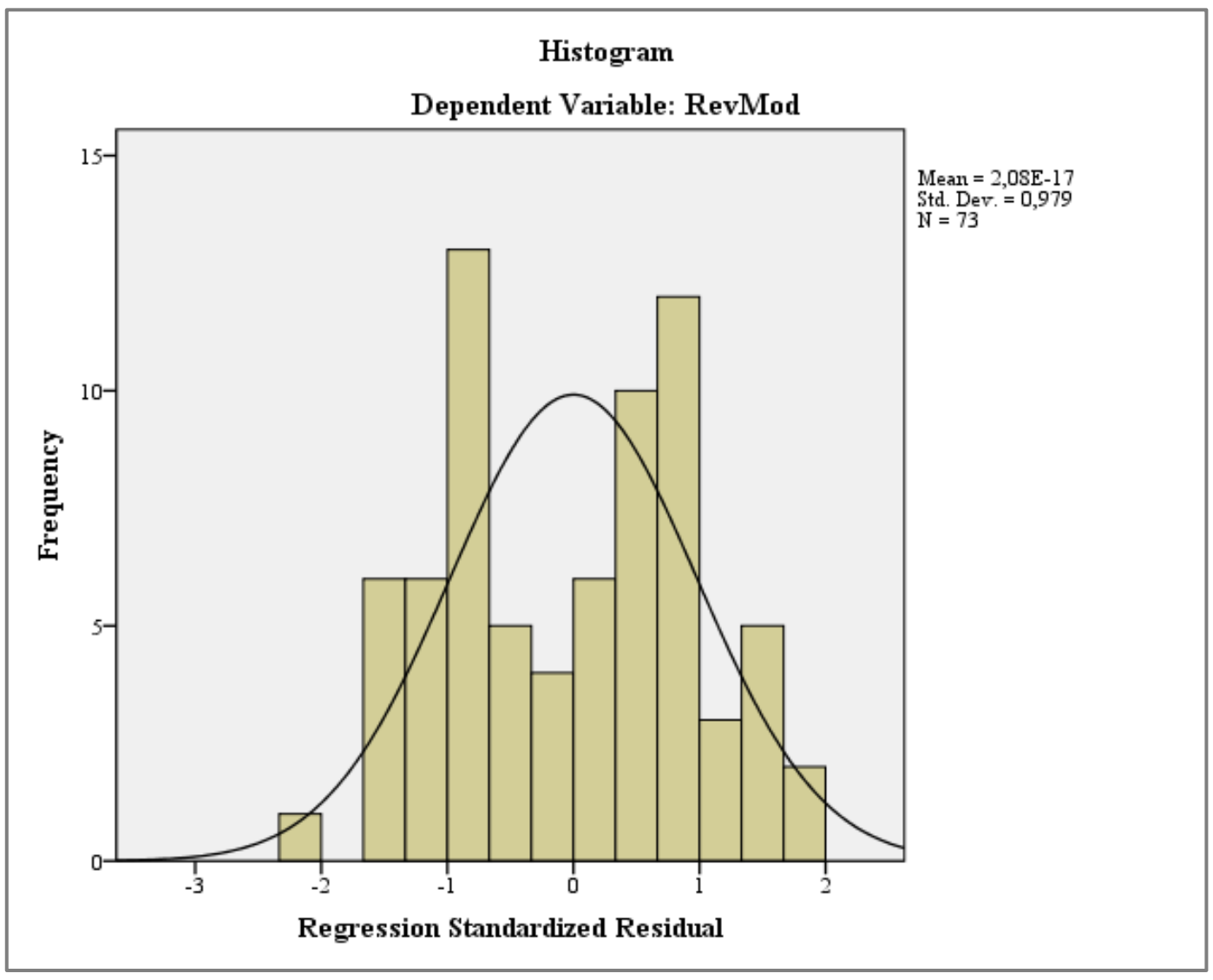

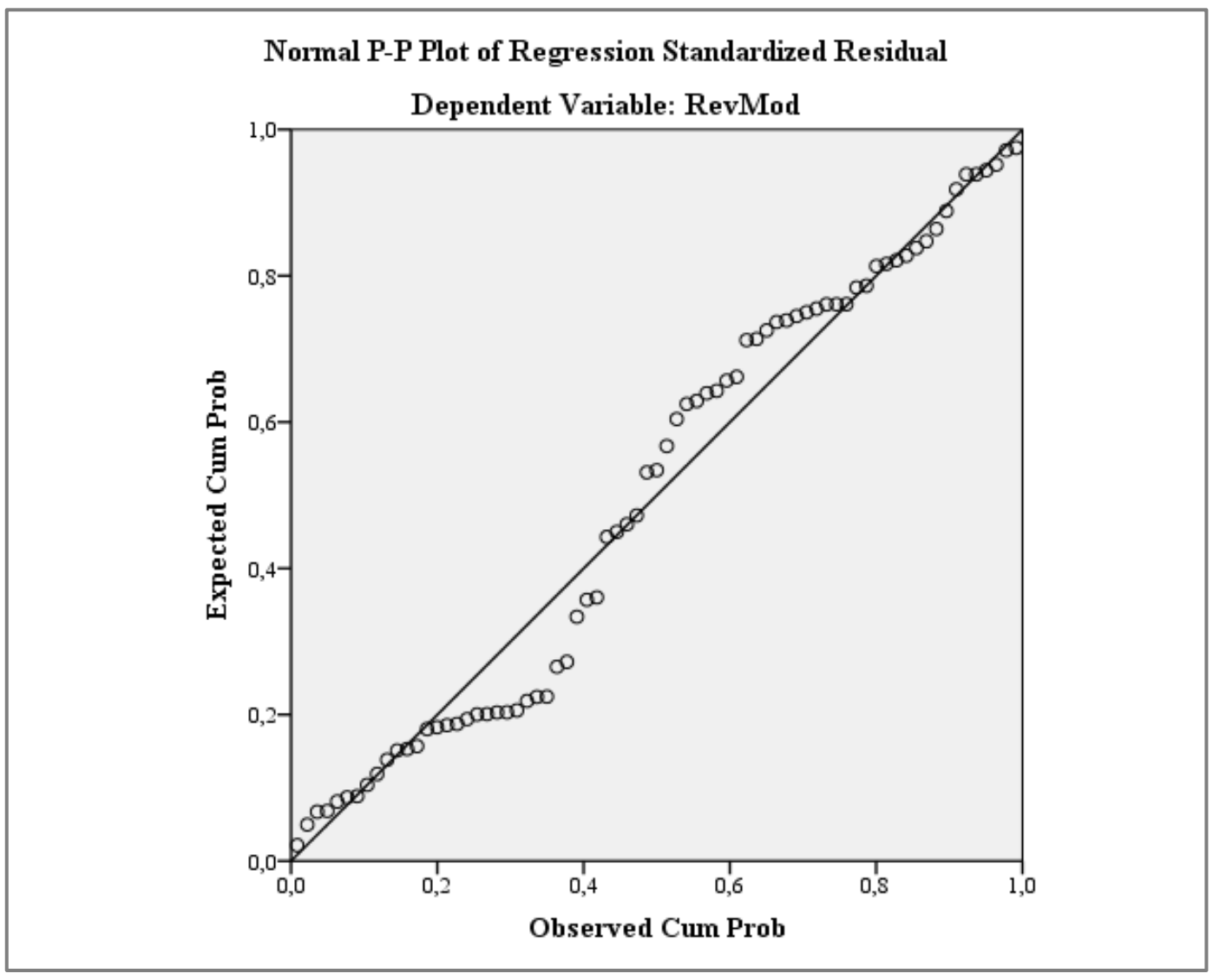
Normal P-P Plot of Regression Standardized Residual
Dependent Variable: RevMod
Expected Cum Prob
Observed Cum Prob

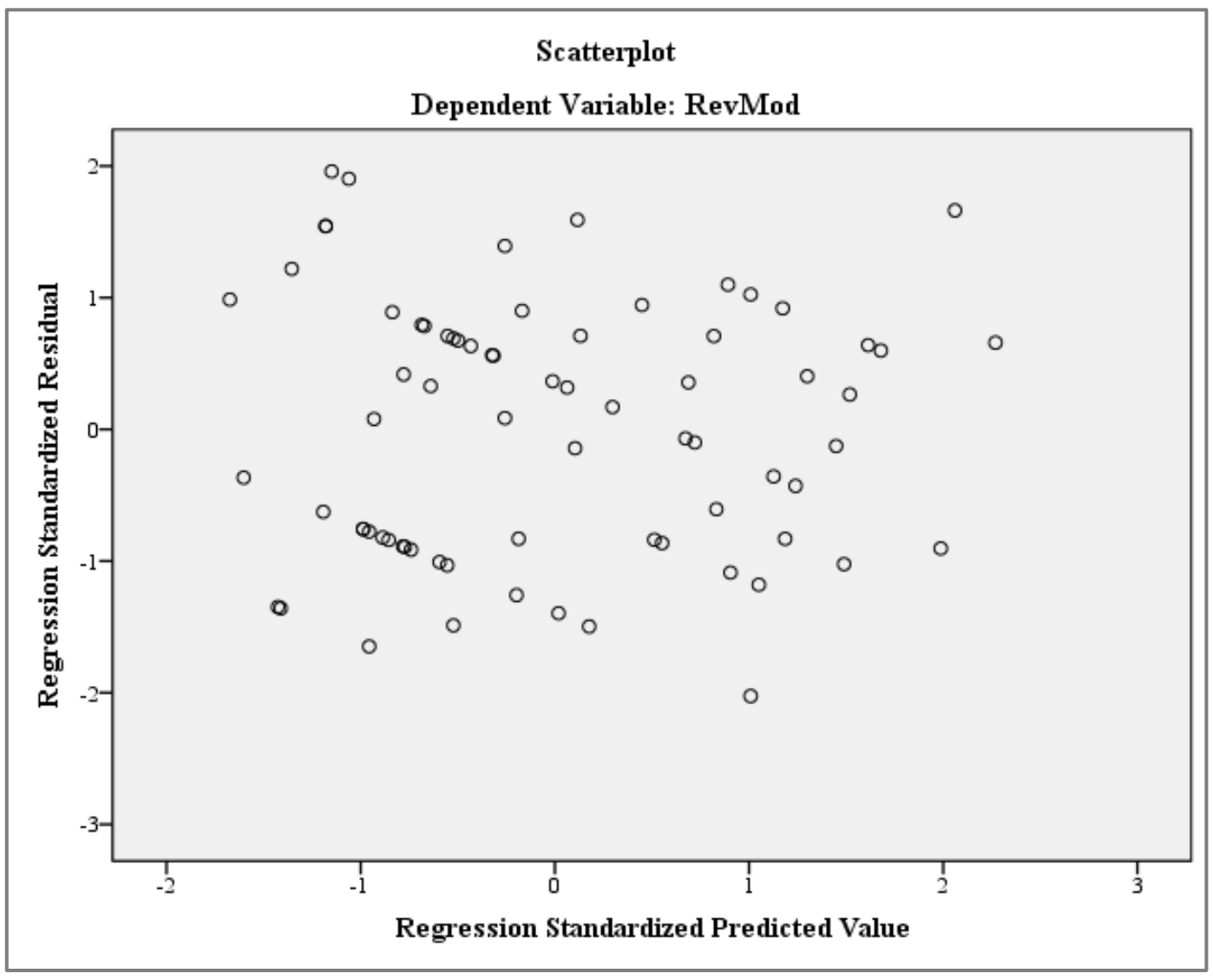
Scatterplot
Dependent Variable: RevMod
Regression Standardized Residual
Regression Standardized Predicted Value

*Collaboration and the Impact on the Business Model (BusMod)*

**Model 4l:**

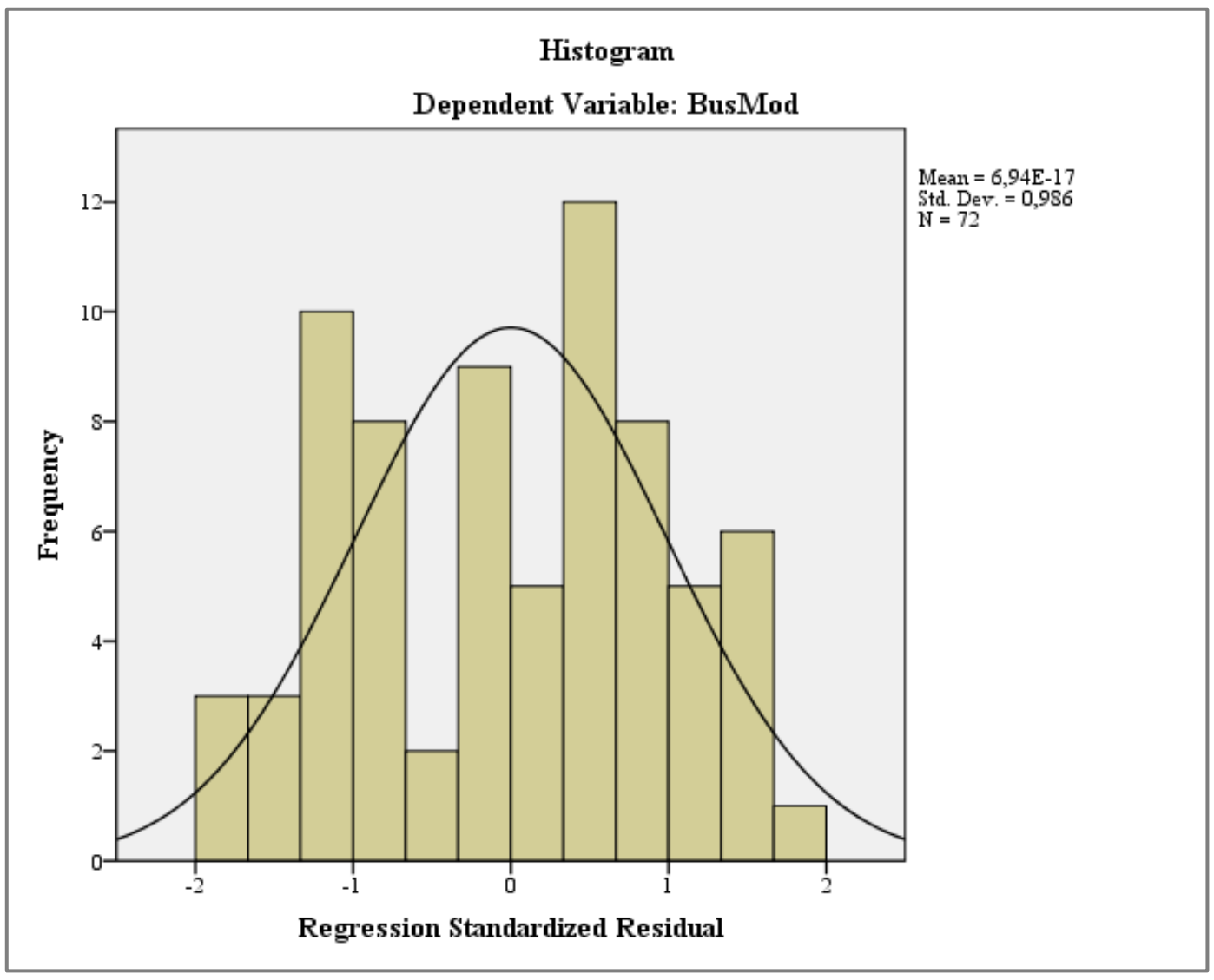

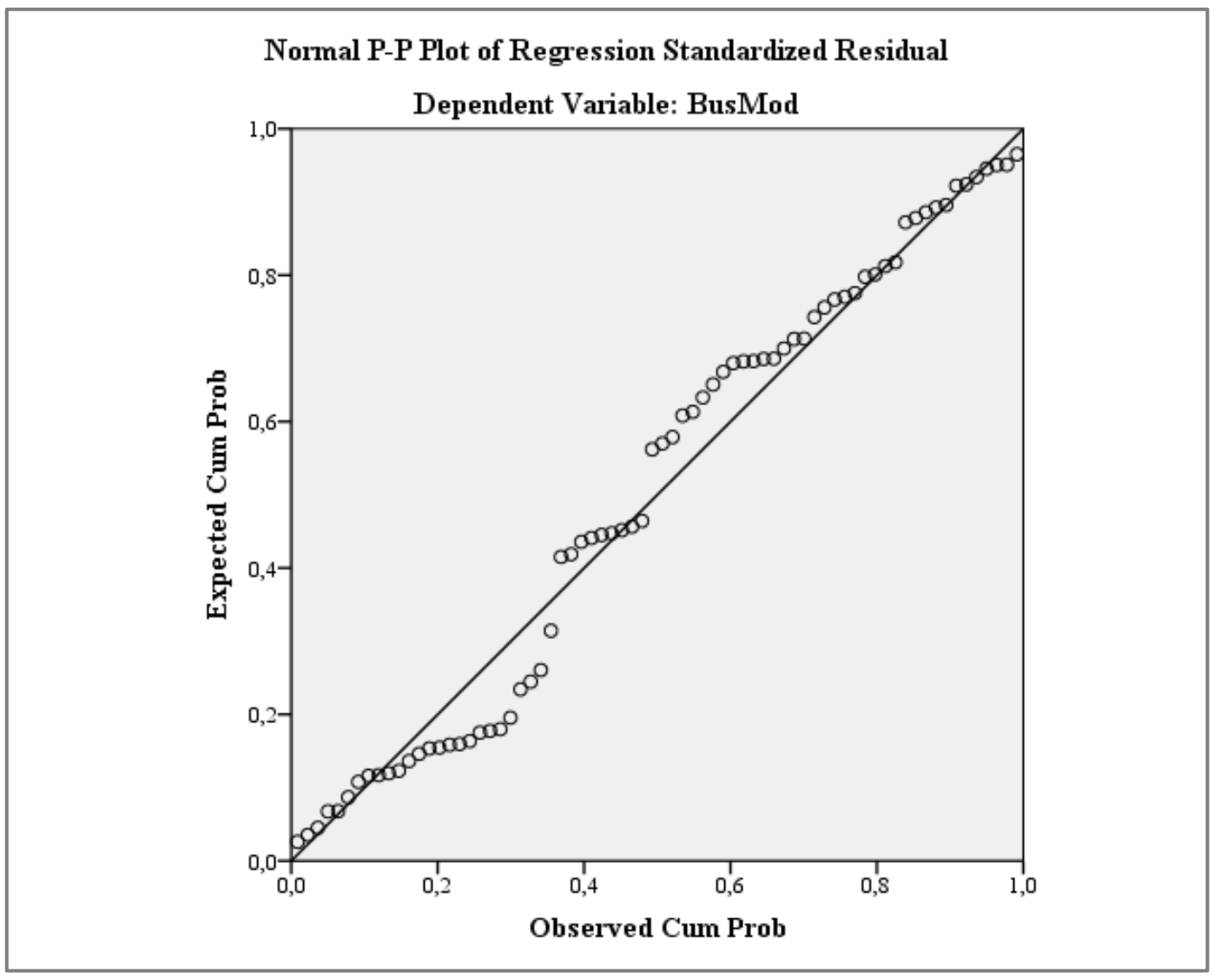

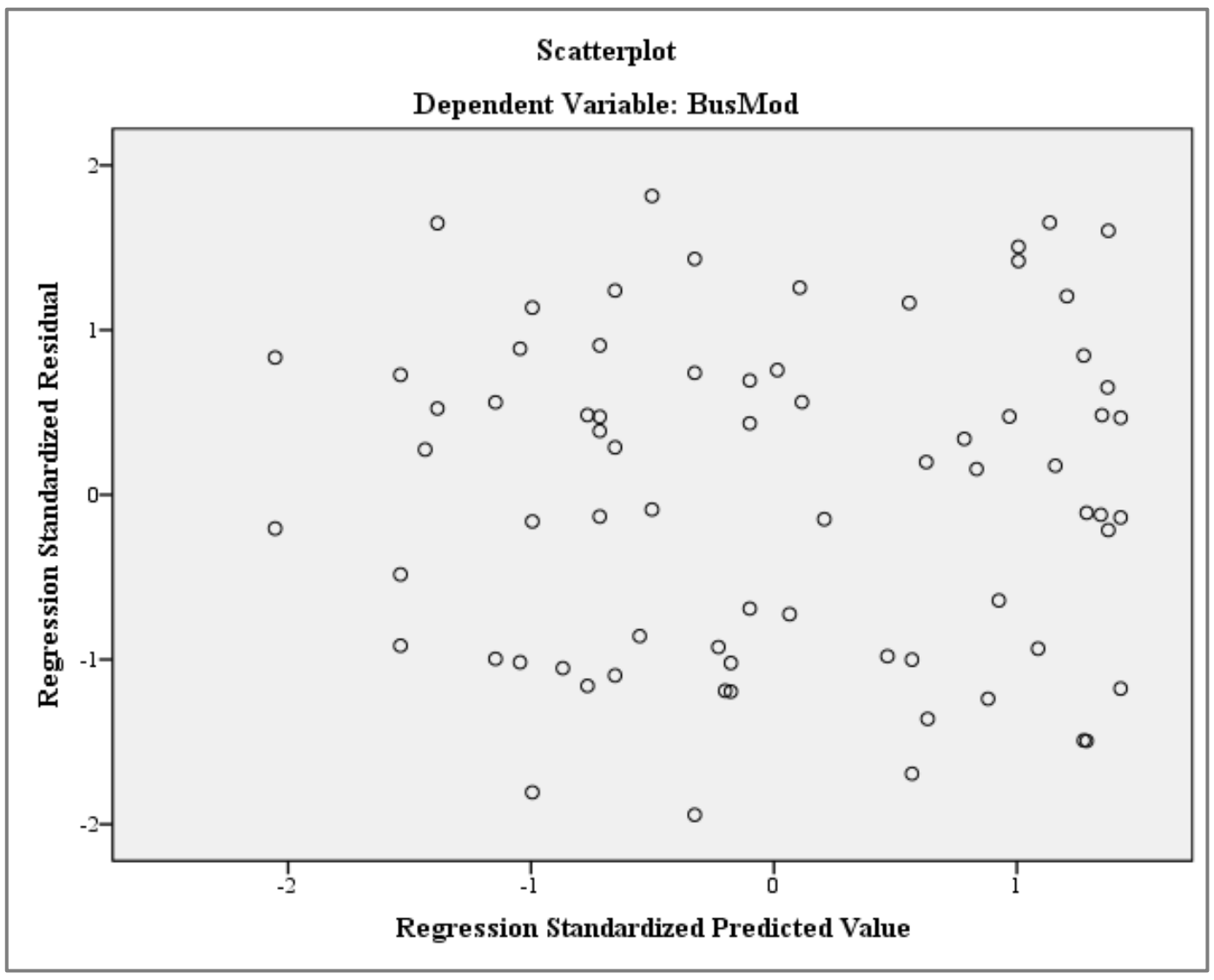

**Model 4m:**

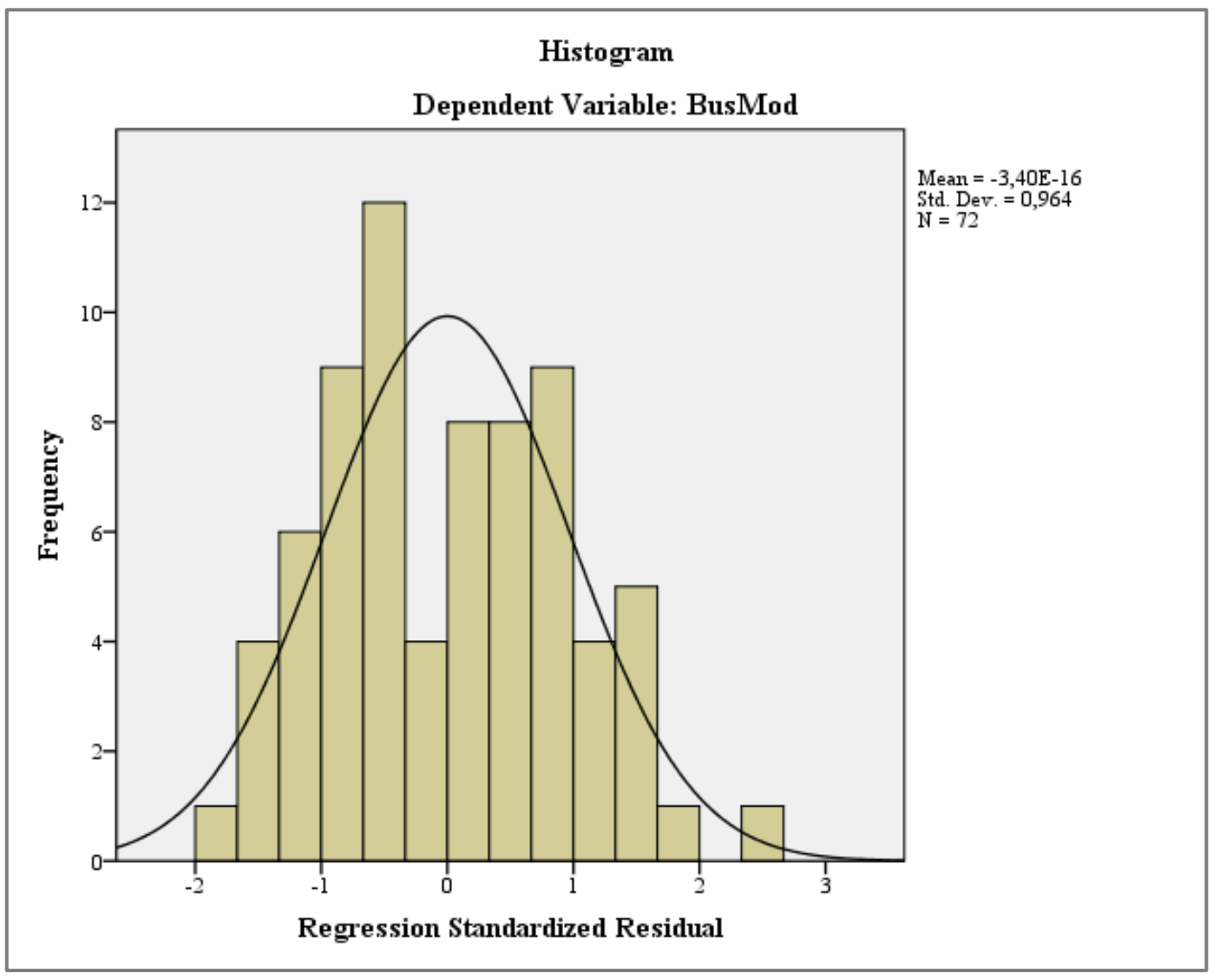

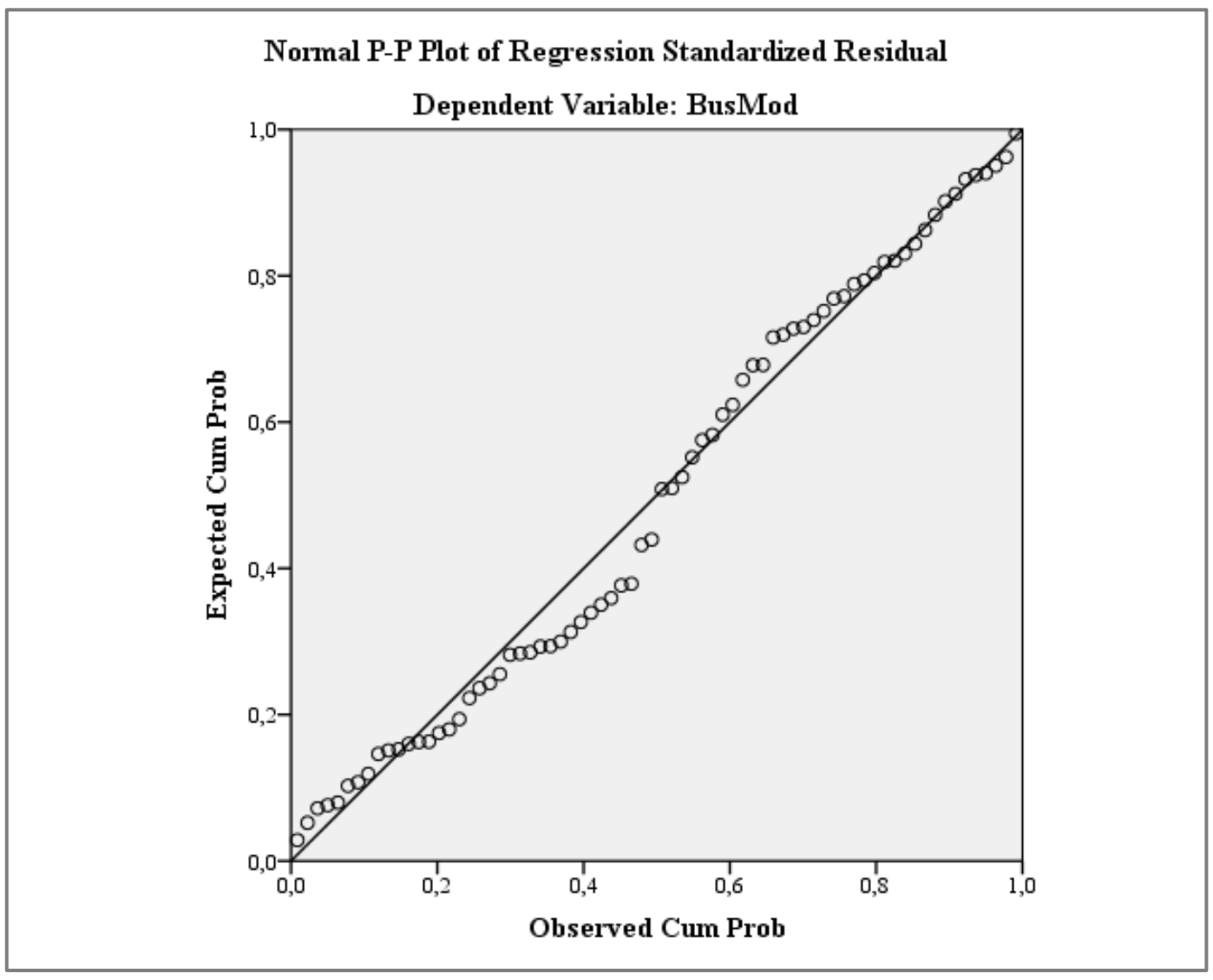
Normal P-P Plot of Regression Standardized Residual
Dependent Variable: BusMod
Expected Cum Prob
Observed Cum Prob

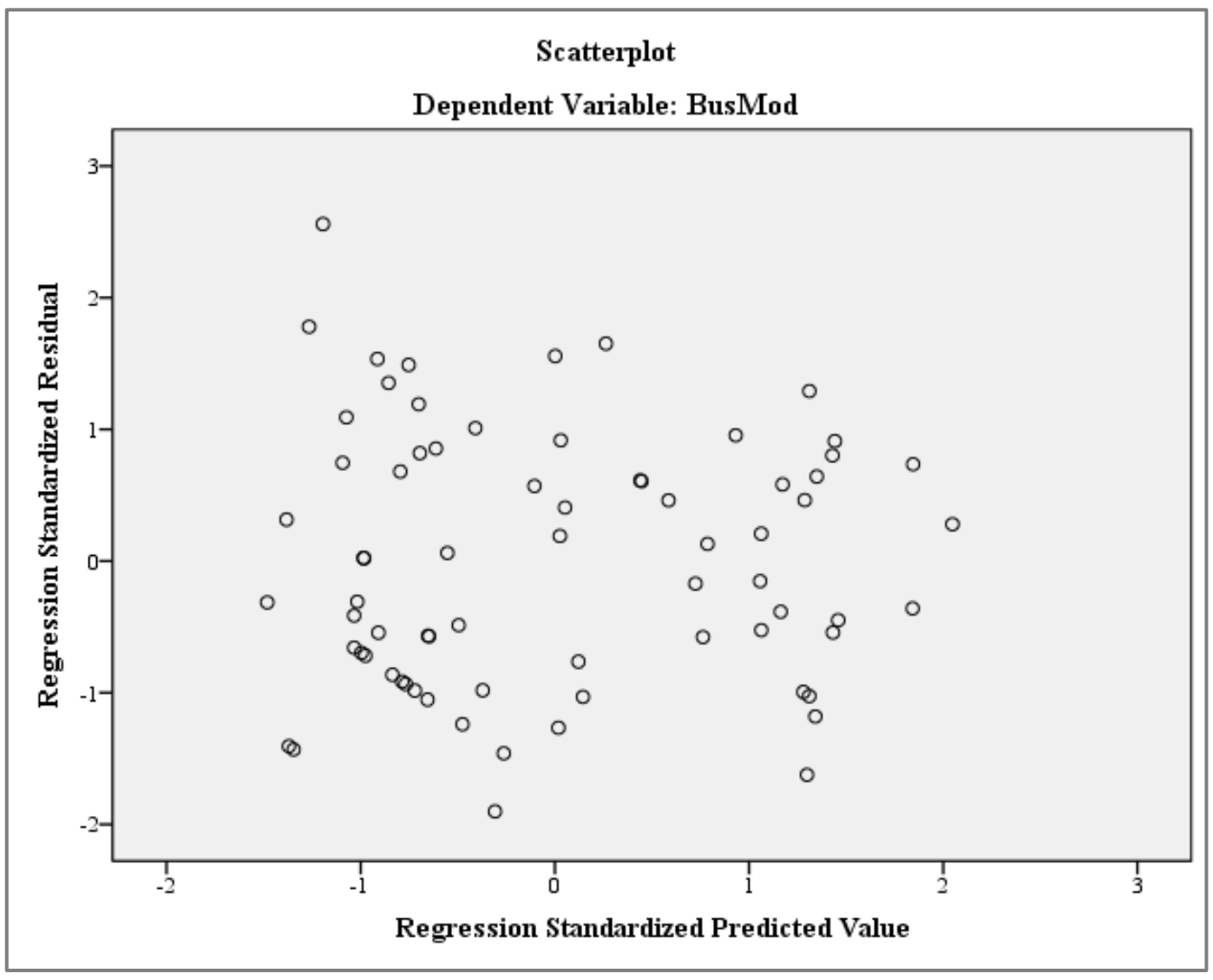
Scatterplot
Dependent Variable: BusMod
Regression Standardized Residual
Regression Standardized Predicted Value

## Model 4n:

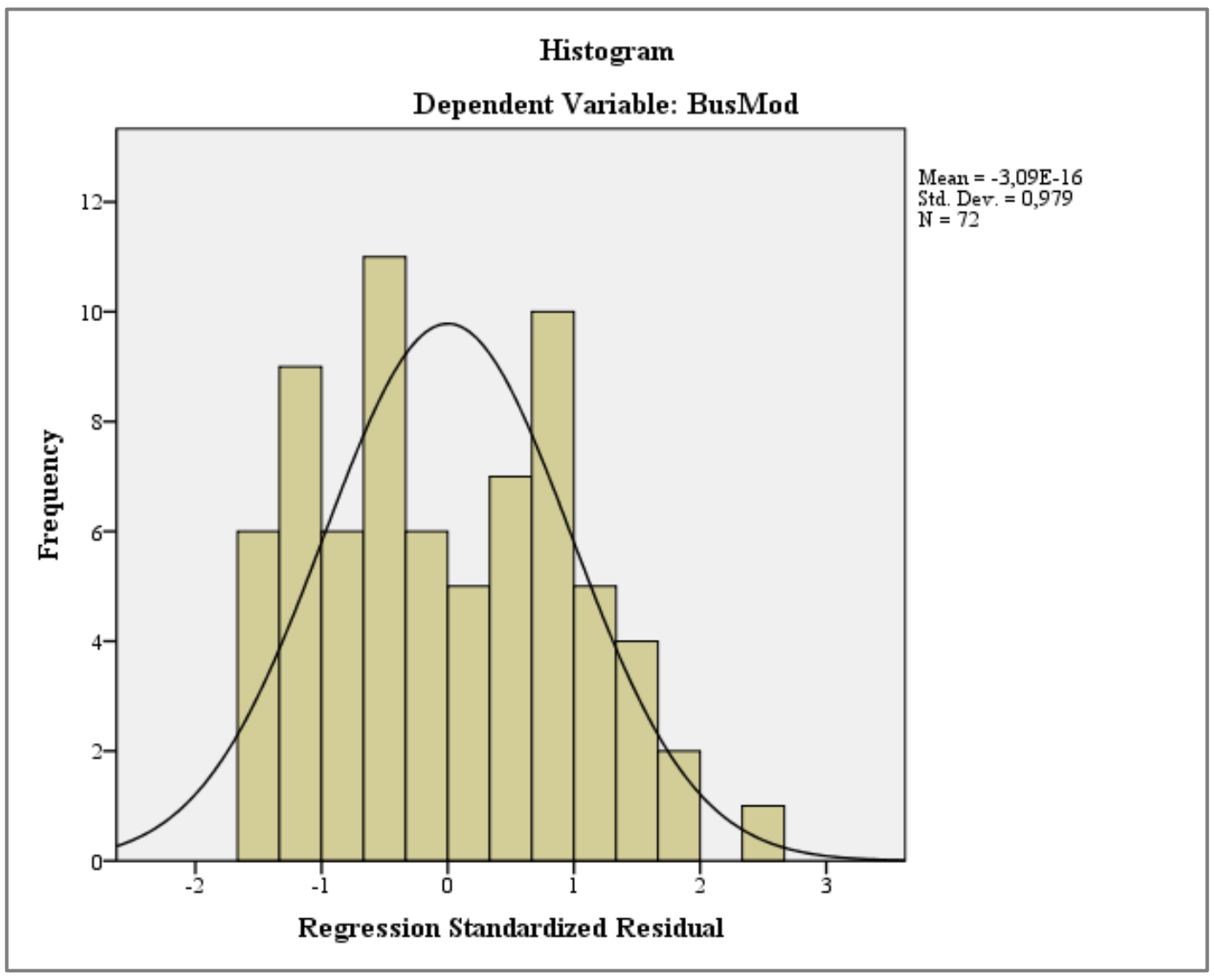

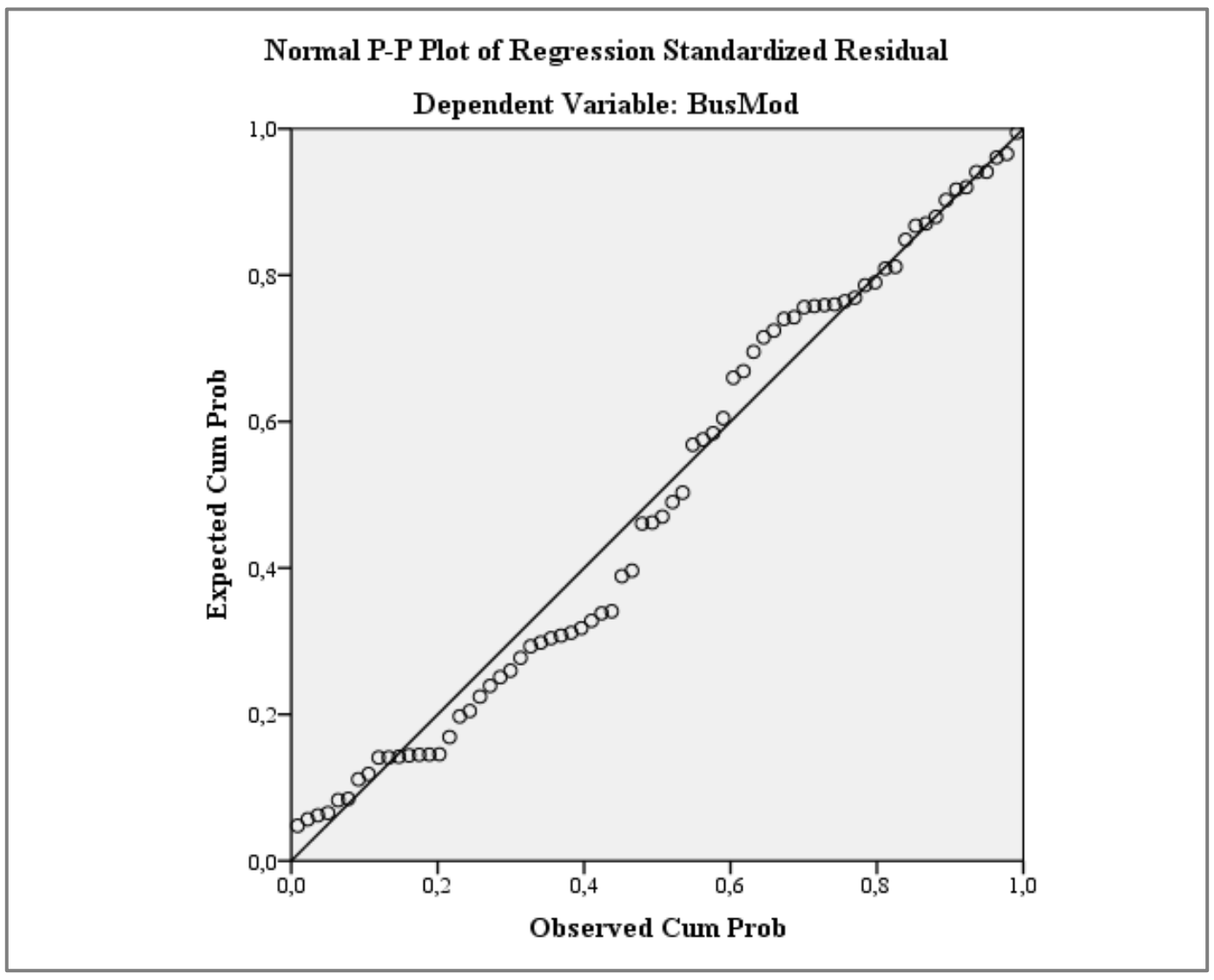
Normal P-P Plot of Regression Standardized Residual
Dependent Variable: BusMod
Expected Cum Prob
Observed Cum Prob

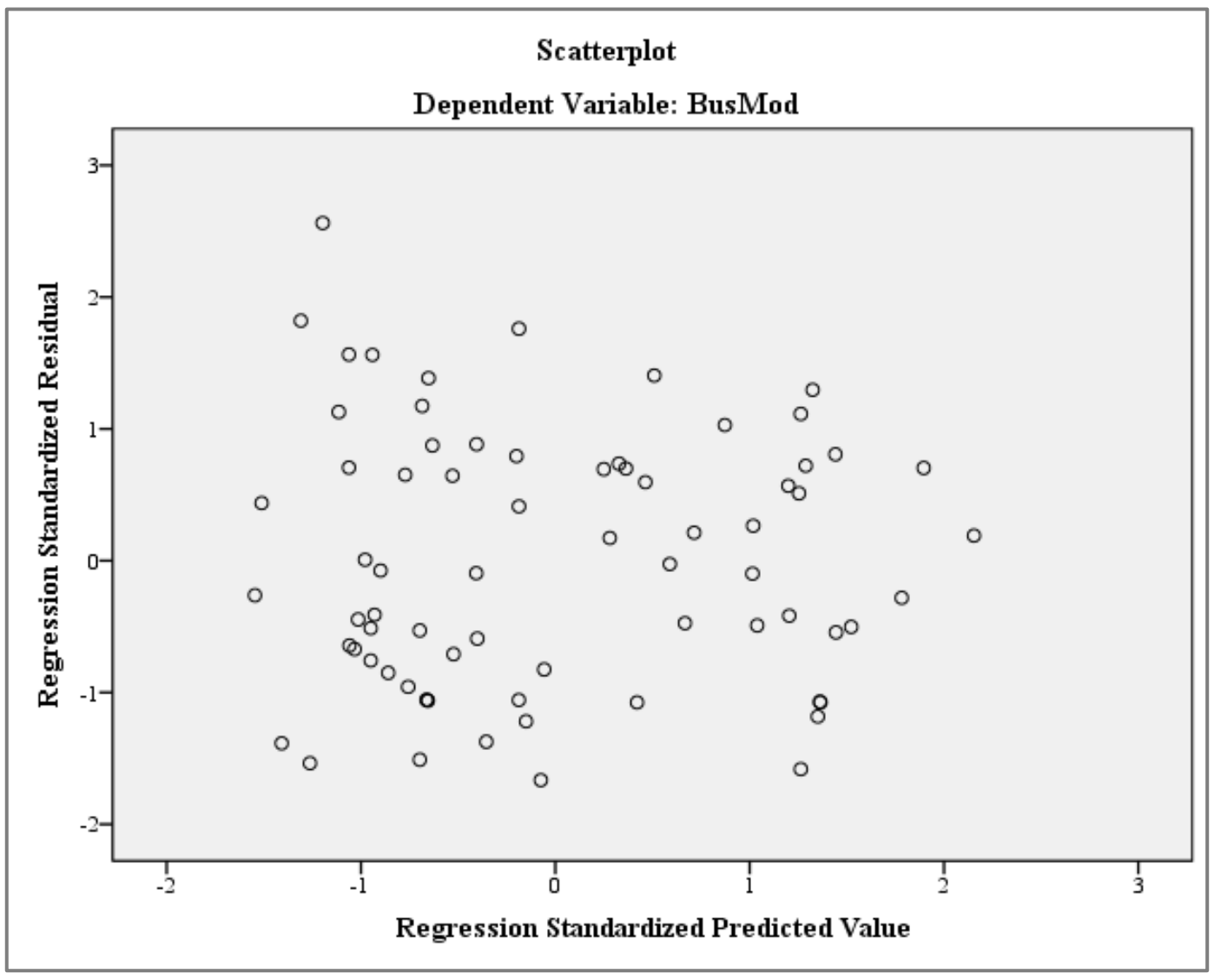
Scatterplot
Dependent Variable: BusMod
Regression Standardized Residual
Regression Standardized Predicted Value

## PLAGARSIM DECLARATION

I herewith formally declare that I have written the submitted thesis independently. I did not use any outside support except for the quoted literature and other sources mentioned in this paper.

I clearly marked and separately listed all of the literature and all of the sources, which I employed when producing this paper.

I am aware that the violation of this regulation will lead to failure of the thesis.

Steven Moore Berlin, 21/07/2017